\documentclass{article}

\usepackage{arxiv}
\usepackage{graphicx}%
\usepackage{multirow}%
\usepackage{amsmath,amssymb,amsfonts}%
\usepackage{amsthm}%
\usepackage{mathrsfs}%
\usepackage[title]{appendix}%
\usepackage{xcolor}%
\usepackage{textcomp}%
\usepackage{manyfoot}%
\usepackage{booktabs}%
\usepackage{algpseudocode}%
\usepackage{listings}%
\usepackage{wrapfig}
\usepackage{caption, subcaption}
\usepackage{verbatim}
\usepackage{graphicx}
\usepackage{cite}
\usepackage{stfloats}
\usepackage{amsmath}
\usepackage{algorithm}
\usepackage{color,soul}
\usepackage{placeins}
\usepackage{siunitx}
\usepackage{placeins}

\theoremstyle{thmstyleone}%
\theoremstyle{thmstyletwo}%

\theoremstyle{thmstylethree}%

\newcommand{\Qcarr}{Q_c}
\newcommand{\Qsh}{Q_s}
\newcommand{\Pcarr}{P_c}
\newcommand{\Psh}{P_s}
\newcommand{\PA}{I_A}
\newcommand{\lw}{L_w}
\newcommand{\ov}{L_o}
\newcommand{\Qink}{Q_m}
\newcommand{\Qdrop}{Q_d}
\newcommand{\DT}{ \Delta{r_T}}
\newcommand{\DN}{ \Delta{r_N}}
\newcommand{\DP}{\bar{d}_a}

\newcommand{\DA}{d_a}
\newcommand{\Vl}{V_l}
\newcommand{\AD}{\phi_A}
\newcommand{\RT}{\Delta r_T}
\newcommand{\RN}{\Delta r_N}
\newcommand{\LT}{L_T}
\newcommand{\Fdep}{P_{dep}}
\newcommand{\Fdiff}{P_{diff}}
\newcommand{\Fgrav}{P_{grav}}

\newcommand{\tubeR}{r_{T_0}}
\newcommand{\nozzleR}{r_{N_0}}
\newcommand{\tubeResistance}{R_T}

\newcommand{\inkdensity}{\phi_{m}}
\DeclareMathOperator*{\argmin}{arg\,min}

\begin{document}

\title{Digital Twin of Aerosol Jet Printing}
\newcommand*\samethanks[1][\value{footnote}]{\footnotemark[#1]}

\author{Aayushya Agarwal\thanks{Authors contributed equally.} \\ Department of Electrical and Computer Engineering \\ Carnegie Mellon University \\ Pittsburgh, USA \\ aayushya@andrew.cmu.edu \And Jace Rozsa \samethanks \\ Department of Electrical and Computer Engineering \\ Carnegie Mellon University \\ Pittsburgh, USA \\ jrozsa@andrew.cmu.edu \And Matteo Pozzi \\ Civil and Environmental Engineering Department \\ Carnegie Mellon University \\ Pittsburgh, USA \\ mpozzi@andrew.cmu.edu \And Rahul Panat \\ Department of Mechanical Engineering \\ Carnegie Mellon University \\ Pittsburgh, USA \\ rpanat@andrew.cmu.edu \And Gary K. Fedder \\ Department of Electrical and Computer Engineering \\ Carnegie Mellon University \\ Pittsburgh, USA \\ fedder@andrew.cmu.edu}

\date{}

\maketitle
\begin{abstract}

Aerosol Jet (AJ) printing is a versatile additive manufacturing technique capable of producing high-resolution interconnects on both 2D and 3D substrates. The AJ process is complex and dynamic with many hidden and unobservable states that influence the machine performance, including aerosol particle diameter, aerosol carrier density, vial level, and ink deposition in the tube and nozzle. Despite its promising potential, the widespread adoption of AJ printing is limited by inconsistencies in print quality that often stem from variability in these hidden states. To address these challenges, we develop a digital twin model of the AJ process that offers real-time insights into the machine's operations. The digital twin is built around a physics-based macro-model created through simulation and experimentation. The states and parameters of the digital model are continuously updated using probabilistic sequential estimation techniques to closely align with real-time measurements extracted from the AJ system's sensor and video data. The result is a digital model of the AJ process that continuously evolves over a physical machine's lifecycle. The digital twin enables accurate monitoring of unobservable physical characteristics, detects and predicts anomalous behavior, and forecasts the effect of control adjustments. This work presents a comprehensive end-to-end digital twin framework that integrates customized computer vision techniques, physics-based macro-modeling, and advanced probabilistic estimation methods to construct an evolving digital representation of the AJ equipment and process. While the methodologies are customized for aerosol jet printing, the process for constructing the digital twin can be applied for other advanced manufacturing techniques. 
\end{abstract}




\section{Introduction}
Aerosol jet (AJ) printing is a flexible direct-write manufacturing technique that offers digital control, contactless material deposition, and is compatible with a wide range of materials. This advanced additive manufacturing technique enables focused printing for rapid and precise fabrication at micro- and nanoscale resolutions, with feature sizes ranging from 20 to 200~\textmu{m}. Aerosol jet printing has shown its potential to fabricate intricate 2D and 3D structures for advanced applications, including complex antenna designs \cite{uya2023design,o2015aerosol}, micromechanical systems \cite{jahan2024aerosol,feng2019aerosol},  advanced brain-computer interfaces \cite{saleh2022cmu}, rapid disease detection devices \cite{ali2021sensing}, and 3D architectures for batteries \cite{deiner2017inkjet, paulsen2012printing, deiner2019aerosol, saleh20183d}.

\begin{figure}
    \centering
    \includegraphics[width=1\linewidth]{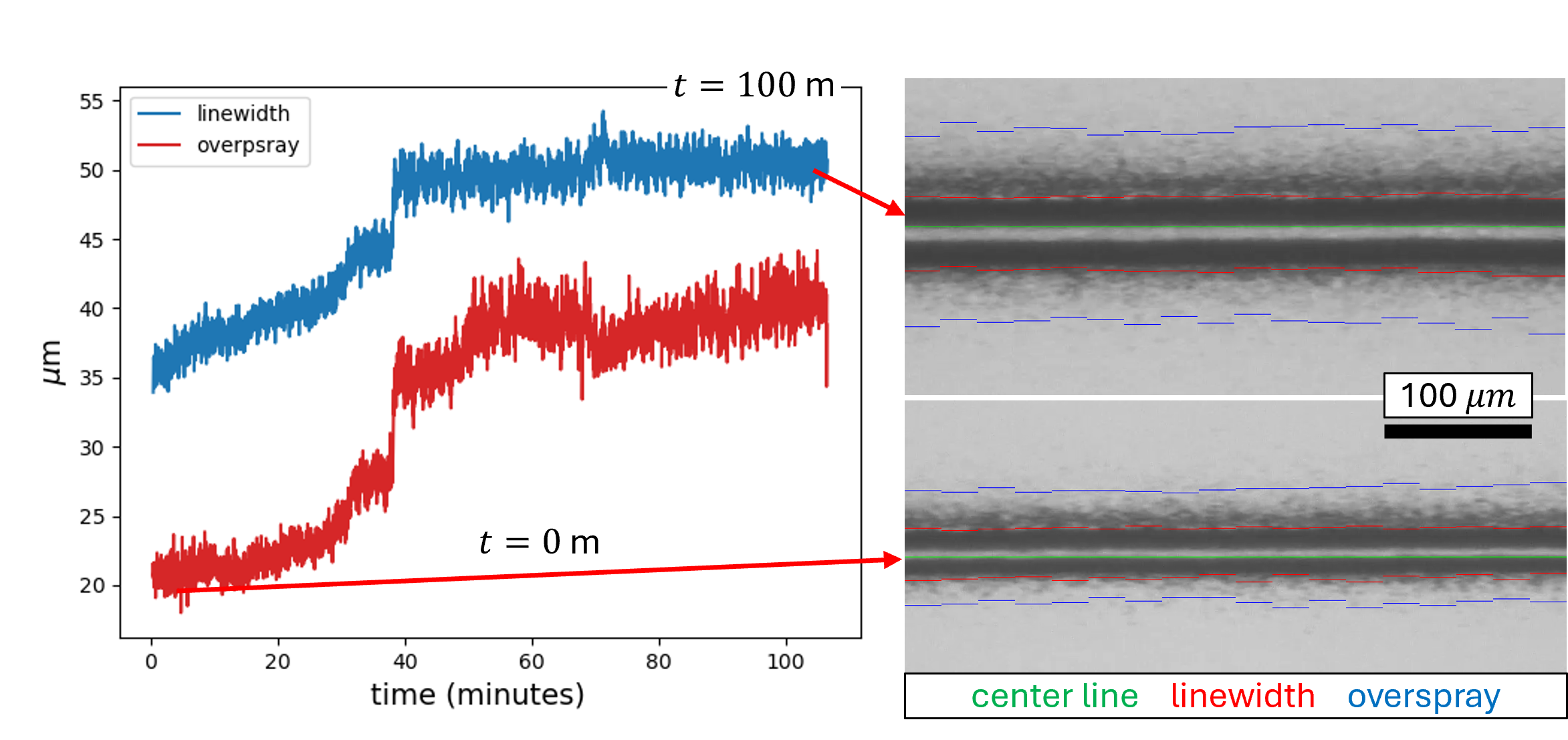}
    \caption{Example of drift that can occur in linewidth and overspray over the course of a single print.}
    \label{fig:bigdrift}
\end{figure}

Despite its potential, aerosol jet printing can exhibit large variations in print quality, hindering its widespread adoption in industry. A cross-sectional are increase of 100\% in just 20 minutes of printing has been reported \cite{smithdrift}. Figure \ref{fig:bigdrift} shows a 50\% drift over 45 minutes of printing. Similar issues of print inconsistency have been documented \cite{smithdrift, salary2021computational, tafoya2020understanding,chen2018effect}. The root cause of these variations has previously been attributed to changes in the internal operations of the ultrasonic atomizer and ink clogging within the machine \cite{smithdrift, salary2021computational}.

Various studies using computational fluid dynamics (CFD) have been performed to study and improve different aspects of the AJ process \cite{RAMESH2023312,full_process_ajp_modeling,ma17133179}. Open loop techniques for process improvement have included saturating the carrier gas with solvent vapor and heating the deposition head for greater droplet evaporation control \cite{secor_vapor,secor_heating}. There has also been some effort at implementing real-time processing monitoring and control \cite{zhang_aidriven}, including taking additional measurements, such as light scattering, to provide enhanced feedback for process control \cite{secor_light_scattering}. While these techniques offer significant improvements, advancing AJ printing to industrial scale requires identifying real-time changes in the internal physical characteristics that contribute to variations in the print output. 

Digital twins have garnered widespread interest across sectors including manufacturing, aerospace, energy, and healthcare for their ability to anticipate failures, optimize performance, and improve resiliency in operations \cite{grieves2017digital, tao2018digital}. Their strength lies in combining real-time sensor data, physics-based models, and machine learning algorithms to create a living model that reflects the actual system’s behavior under changing conditions. This capability makes digital twins especially powerful for complex, nonlinear, or partially observable systems, where static models fall short.

To improve real-time monitoring of the AJ process, we introduce a digital twin of the AJ equipment and process. A digital twin is a virtual replica of a physical system that mimics the structure, context and behavior of the system, continuously updates with data from the system, and has additional attributes of bidirectional interaction, predictive capability and utility in informing decisions \cite{national2023foundational}. Digital twins have gained significant attention across industries as a transformative tool for real-time monitoring, prediction, and optimization of complex systems. They have been successfully deployed in sectors such as aerospace for monitoring engine health monitoring \cite{grieves2016digital,li2021digital}, manufacturing for process optimization  and predictive maintenance \cite{kritzinger2018digital}, and monitoring the health of energy systems \cite{zolin2020digital}.
A key aspect of our digital twin is the bidirectional interaction between the virtual and physical AJ system, allowing data from the physical system to update the digital model and enabling the digital twin to inform decisions and control strategies in the physical world. In this work, the user is an integral component of the bidirectional loop, actively adjusting parameters, interpreting results, and implementing actions based on the digital twin’s feedback. Through this interaction, the user effectively closes the loop between the digital representation and real-world operation. Future developments point to automating this interaction in response to changing system conditions.

In the context of AJ, a digital twin allows for dynamic tracking of hidden variables, better predictive control, and fast root-cause analysis when output quality degrades. Our AJ digital twin employs probabilistic estimation techniques to infer hidden states and update model parameters. This digital replica provides physical insights into critical factors that significantly impact print quality, including ultrasonic atomizer operations and ink characteristics. With applications ranging from real-time performance assessment and anomaly detection to online control and accurate forecasting, the digital twin opens new capabilities for AJ manufacturing.

Our work presents a comprehensive framework for constructing a digital twin of the AJ equipment and process. This framework follows the ISO 23247 standard that establishes a systematic approach to developing and integrating the computational methodologies required to create a digital twin \cite{iso2020automation,song2023digital}.  The models and computational algorithms within the digital twin system are designed to address the operational constraints of the physical AJ machine.

At the core of the framework is a physics-based digital macro-model, informed through a combination of analytic analyses, simulation and experimental data. This model captures the intricate behaviors of the AJ machine that first-principles methods cannot address while providing critical physical insights often lacking in black-box approaches like machine learning-based methods. The model’s hidden states and parameters are continuously updated using Kalman filtering and expectation-maximization to ensure alignment with real-world observations from the AJ machine. These observations are obtained through alignment camera video data, analyzed by custom computer vision algorithms, including a deep neural network specifically trained to estimate the height profile of the printed trace.

The resulting digital twin continuously evolves with the lifecycle of each AJ machine. Building on this framework, the digital twin is used to forecast changes in line quality due to inputs and random processes, as well as detect anomalous behavior that may lead to issues such as process drift.  The digital twin bridges the gap between physical understanding and data-driven approaches in digital manufacturing systems and demonstrates the ability to improve the reliability of additive manufacturing processes.

\section{Background on Aerosol Jet Printing}

\begin{figure}
    \centering
    \includegraphics[width=\linewidth]{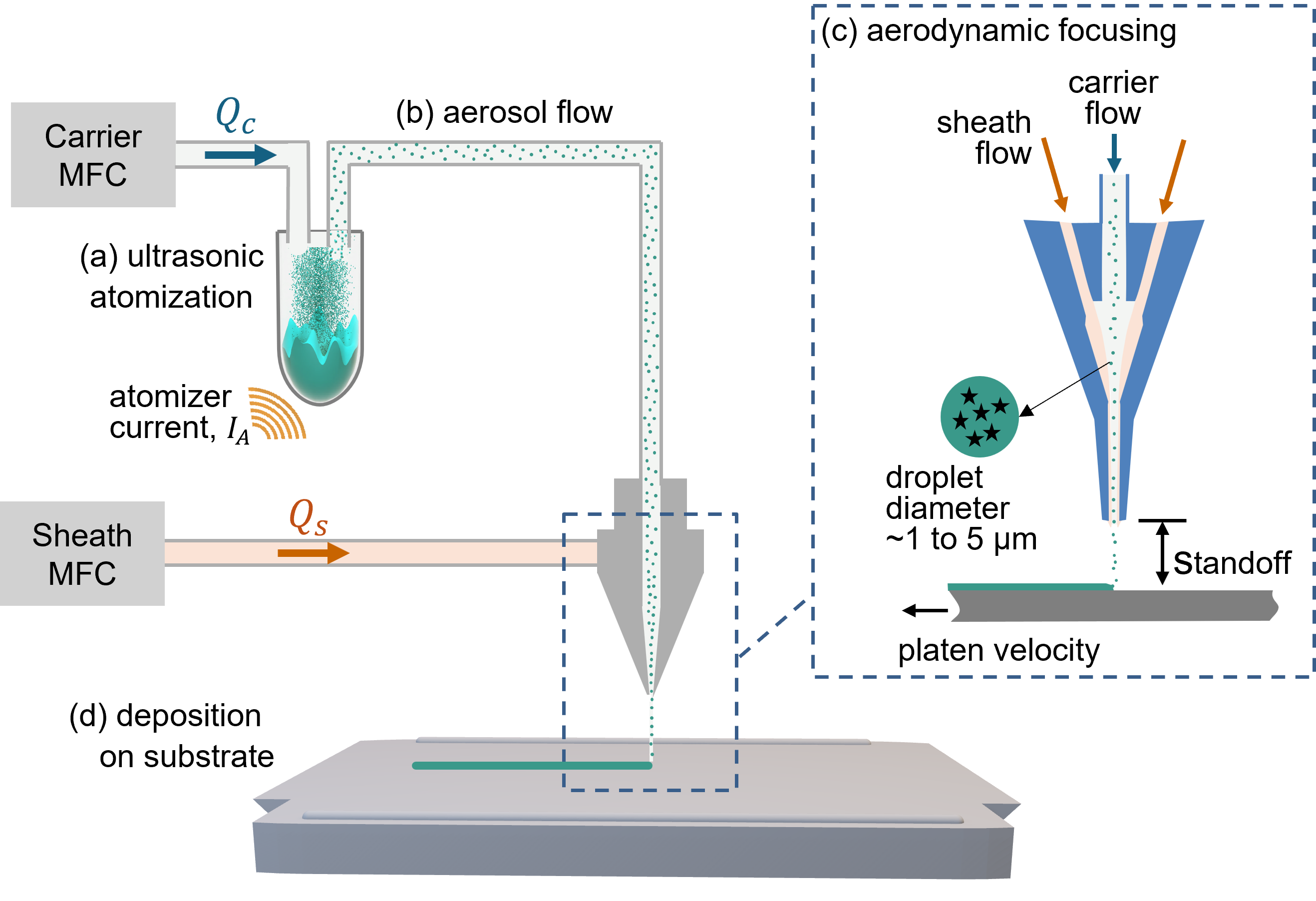}    
    \caption{The aerosol jet printer (a) initially aerosolizes an ink solution containing nanoparticles or polymer in the ultrasonic atomizer. The aerosol droplets are carried to the machine via the carrier gas flow and through the transport tube in (b). The flow of aerosol droplets are focused through a sheath gas flow in (c), and then deposited onto a moving platen with substrate in (d). The substrate is typically heated in order to evaporate the remaining solvent after droplet impact, leaving behind a profile of deposited material. 
    }
    \label{fig:AJP_operations}
\end{figure}

The AJ printing process is schematically shown in Figure \ref{fig:AJP_operations}. It begins with an ultrasonic atomizer, which generates aerosol droplets through the interaction between ultrasonic waves and the ink solution contained in the vial (Figure \ref{fig:AJP_operations}a). For viscous inks, a pneumatic atomizer (not shown) can be used to create aerosol droplets. In this work, inks typically consist of nanoparticles suspended in a liquid or polymers dissolved in solvent. The aerosol droplets are then carried through a transport tube by a stream of nitrogen (N$_2$) gas into the AJ nozzle (Figure \ref{fig:AJP_operations}b). Upon entering the nozzle, the droplets are focused by an angled flow of sheath gas (also N$_2$), illustrated in Figure \ref{fig:AJP_operations}c, that accelerates the flow of aerosol droplets onto the substrate. To create intricate patterns, the substrate is mounted on a programmable platen stage that moves in a controlled direction.
Upon impact with the (often heated) platen, the droplet solvent is evaporated, leaving behind a layer of deposited material as shown in Figure \ref{fig:AJP_operations}d.

Typically, AJ printing systems have two cameras for visual process monitoring. A process camera is positioned near the print head and provides an in-situ magnified video feed of the aerosol droplets being deposited on the print surface. An alignment camera, which is offset from the print head, provides an offline visual inspection of the printed structures after printing. Additional process monitoring comes from the pressure values recorded by the mass flow controllers (MFCs).  

The AJ printer provides the following control inputs and observed outputs:

\noindent \textbf{Control inputs}: The three primary control inputs that can be dynamically adjusted and monitored during the printing process are:
\begin{itemize}
    \item Ultrasonic atomizer current ($\PA$  [mA]): Controls the power supplied to the ultrasonic atomizer and is key in aerosol generation.
    \item Carrier gas flow rate ($\Qcarr$ [sccm]): The flow rate of carrier gas from the ultrasonic atomizer to the nozzle, set by a mass flow controller (MFC, Alicat Scientific MC-Series).
    \item Sheath gas flow rate ($\Qsh$ [sccm]): The gas flow used to focus the aerosol droplets into a narrow stream as they exit the nozzle, ensuring precise deposition onto the substrate, also set by an MFC (Alicat Scientific MC-Series).
\end{itemize}

\noindent \textbf{Monitored Outputs}: The output for the model comes from the following three data sources from the AJ printer:
\begin{itemize}
    \item 	Carrier gas flow pressure ($\Pcarr$ [Pa]): Pressure reading from the MFC controlling the carrier gas flow rate. 
    
	\item Sheath gas flow pressure ($\Psh$ [Pa]): Pressure reading from the MFC controlling the sheath gas flow rate.
	
    \item Alignment Camera video: The AJ setup provides a high-resolution video feed for the inspection of ink deposition after printing. Section 5 describes algorithms for processing this video feed to measure output features. (Processing video from a second in-situ process camera is difficult and less accurate because the camera is tilted, creating an image of non-uniform focus.)

\end{itemize}

Previous studies have attributed AJ process variability to changes in the ink level within the vial, ink deposition on the carrier tube walls, and ink clogging in the nozzle \cite{smithdrift, tafoya2020understanding}. In addition to these factors, successive prints also experience changes in the distribution of aerosol droplet sizes (Supplemental Material, Appendix \ref{app:variations_experiment}).
In this work, these sources of variability are treated as hidden states of the AJ system that are estimated using the digital twin framework to better understand and predict the behavior of the AJ machine during operation. 

\section{Experimental Methods}

Aerosol jet printing was performed using a commercial aerosol jet printer (Aerosol Jet 300, Optomec Inc., Albuquerque, NM, USA). This system enables high-resolution deposition of functional inks with a feature size down to 10 \textmu m,  suitable for microscale patterning applications. All structures were printed using a silver nanoparticle ink (Novacentrix JS-A426, Austin, TX), which was diluted with 1 part deionized water to 3 parts ink. The print surface was a standard 4" diameter silicon wafer, which was cleaned with acetone, IPA, and 3 minutes of oxygen plasma prior to printing. 

Post-print surface characterization was performed using an optical profilometer (model VK-X3000, Keyence Inc., Itasca, IL) to assess the topography and uniformity of printed features. The optimal profilometer scans were conducted with a lateral resolution of 80 nm on unsintered silver traces. The scans were used to validate profile measurements inferred from ML-based computer vision analysis of videos captured immediately after printing.  In typical applications, silver traces are sintered to enhance conductivity; however, in our case, the objective was to study the deposition of aerosol material on the substrate. For this reason, we focus on the unsintered traces, which more directly reveal the as-deposited topology. Additional profile scans were obtained using a surface profilometer (model P15, Tencor Inc., Milpitas, CA) .

Data analysis and image processing were carried out using OpenCV \cite{opencv_library} in Python, while digital twin processing was conducted using MATLAB and PyTorch \cite{paszke2019pytorch} for feature extraction, statistical analysis, and 3D reconstruction. CFD simulations were performed with ANSYS Fluent \cite{manual2009ansys}. 

\section{Digital Twin Framework}
\begin{figure*}[t]
    \centering
   \includegraphics[width=.7\linewidth]{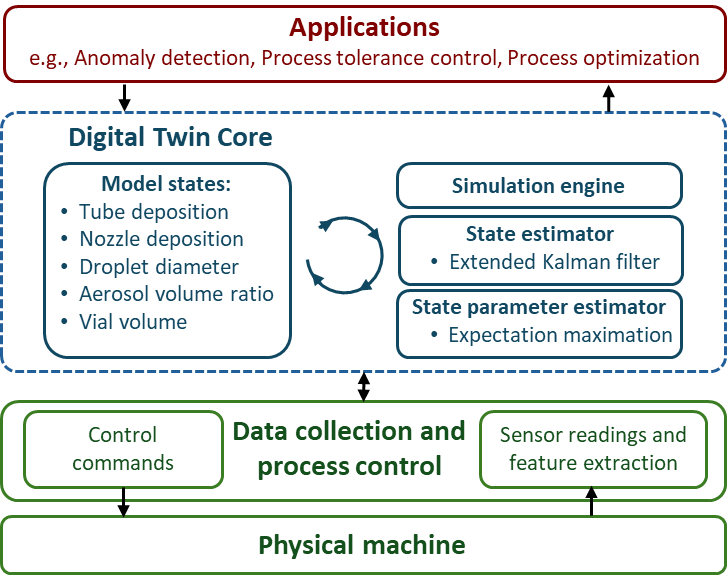}   
    \caption{The digital twin framework, inspired by the ISO 23247 standard \cite{iso2020automation}, provides a structured approach to developing algorithms that update and control the AJ printer. The present effort developed five modules: data collection and feature extraction module, the state-based model, state estimation module, state parameter estimation module, and the simulation engine. The digital twin supports applications such as anomaly detection, process tolerance control and process optimization. Control commands are left for future work.}
    \label{fig:digital_twin_framework}
\end{figure*}


Constructing the digital twin requires synchronization between many computational components that sense, estimate, and model the behaviors of the machine. The collection and interaction of these computational algorithms is represented through a digital twin framework. The proposed digital twin framework is shown in Figure \ref{fig:digital_twin_framework}.

The digital twin update process begins with collecting and synchronizing inputs and sensor measurements from the AJ, including carrier gas pressure, sheath gas pressure, and alignment camera video feeds. A sensor extraction module (Figure \ref{fig:digital_twin_framework}b), extracts quantitative features such as line quality metrics from the video. The processed data are then input to the estimation module in Figure \ref{fig:digital_twin_framework}c, which uses probabilistic estimation techniques to estimate the states and parameters of the digital model that closely align with observations. The updated digital model enables key processes such as simulation, anomaly detection, and control optimization to enhance production line quality (Figure \ref{fig:digital_twin_framework}d). The optimization framework accommodates human-in-the-loop systems by supporting adaptive control adjustments. This work establishes the foundational framework for physics-based AJ digital twin construction, with future research aimed at advancing automated controls. Section \ref{sec:measurement_data_extraction} presents algorithms for extracting and processing quantitative information from the machine and its output. Section \ref{sec:digital_model} introduces the underlying digital model of the AJ. Finally, Section \ref{sec:estimating_latent_states_parameters} details techniques for estimating and modeling changes in the digital model to ensure close alignment with the extracted measurements.

\section{Measurements and Data Extraction}
\label{sec:measurement_data_extraction}



The MFCs produce the carrier and sheath gas pressure necessary to maintain the flow rates at their defined set-points. Variation in these pressure values can reveal valuable information about the hidden states in the machine, such as material deposition in the tube and the nozzle. The Optomec AJ provides an interface to the pressure readings of the MFCs, which are exported directly for incorporation into the digital twin. 


\label{sec:extracting_line_quality_params}
Additionally, the AJ printer provides inspection videos from the alignment camera for a visual representation of aerosol droplet deposition on the substrate. A computer vision algorithm characterizes and quantifies the printed line from the alignment camera video feed. Specifically, the features of the printed line are linewidth ($\lw$), overspray ($\ov$), and cross-sectional area.  
 

\begin{figure}
    \centering
    \includegraphics[width=1\linewidth]{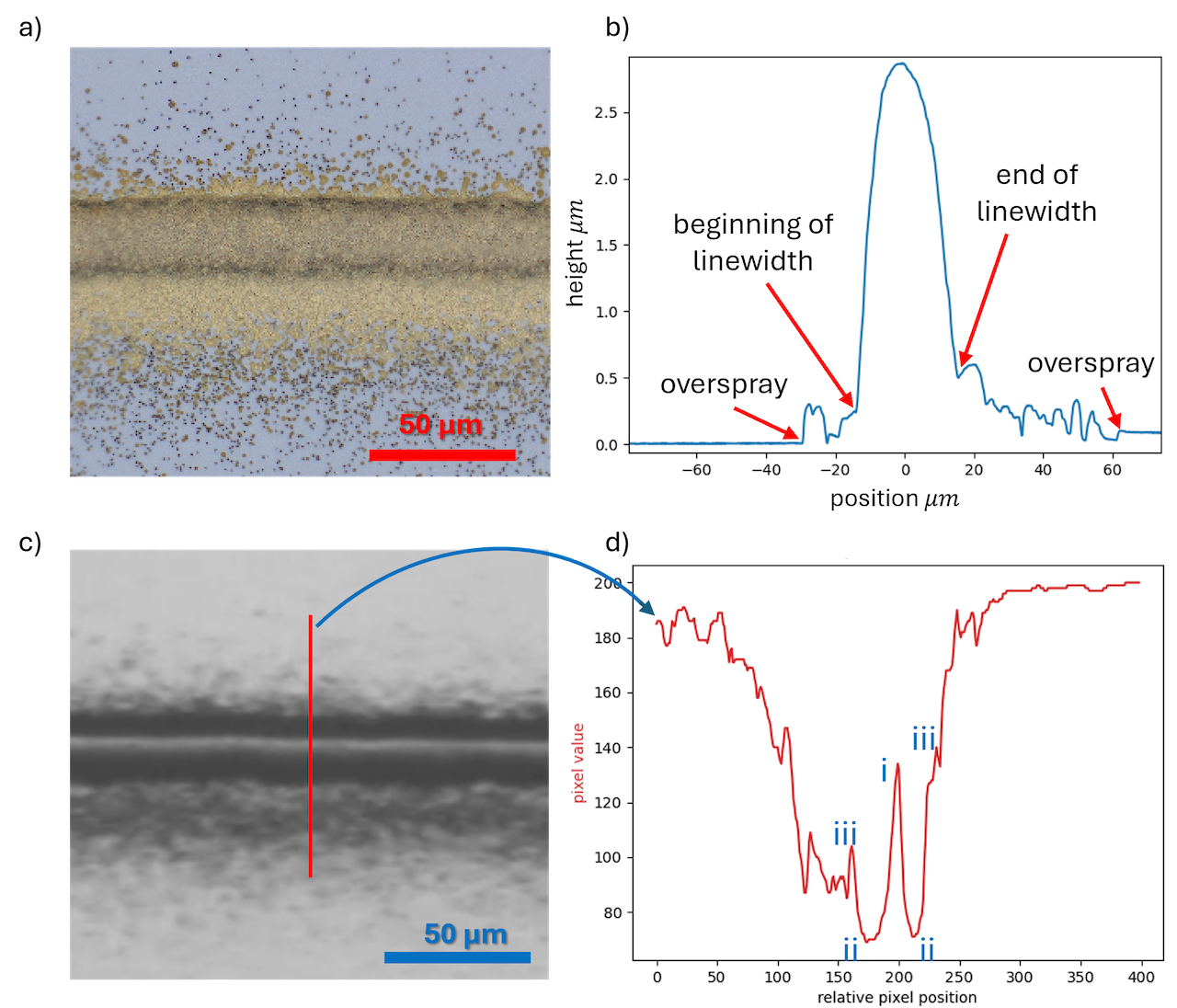}
    \caption{a) Optical image of printed silver line and b) surface profilometer scan, c) frame taken from alignment camera inspection video of print surface, d) cross section of the line shows the grayscale value as a function of position along the surface, i corresponds to center of line, ii are the points of maximum slope on either side, and iii is where the linewidth is defined.}
    \label{fig:lw_def}
\end{figure}


Linewidth, overspray, and cross-sectional area are key features that define the quality of the AJ print \cite{salary2021computational}. While a surface profilometer can provide precise, post-printing measurements of a printed silver line (for example in Figure  \ref{fig:lw_def}a and Figure \ref{fig:lw_def}b), these measurements are not accessible during printing. Instead, the printed line is monitored with a top-down grayscale video from an alignment camera. Consequently, these line characteristics are inferred from the visual features of the video feed.
Figure \ref{fig:lw_def}c shows a sample frame from the alignment camera inspection video. The red line marks the location of a cross-section, whose grayscale intensity is plotted in Figure \ref{fig:lw_def}d. The grayscale profile provides cues to the physical structure of the printed line. These cues are used to define linewidth and overspray.

Qualitatively, the grayscale value in Figure \ref{fig:lw_def}d is inversely proportional to the absolute value of the gradient of the printed feature. For the silver nanoparticle printed lines, flatter regions reflect more light and appear brighter in the image (higher grayscale value) while steeper regions will disperse the light away from the camera sensor, thus appearing darker. The relationship between the grayscale cross section and the physical profile, for example in Figure  \ref{fig:lw_def}d and Figure \ref{fig:lw_def}b, is as follows:
\begin{itemize}
    \item Point (i) marks the center of the line, where the profile is flat and brightness is highest relative to the rest of the line. This corresponds to the peak of the profile.
    \item Points (ii) are local minima in brightness, corresponding to maximum slope on the physical profile.
    \item 	Points (iii) are the first local maxima in the grayscale plot beyond (ii); these correspond to the end of the dark region in the frame from the alignment camera in Figure \ref{fig:lw_def}c.
\end{itemize}

\begin{figure}
    \centering
    \includegraphics[width=.9\linewidth]{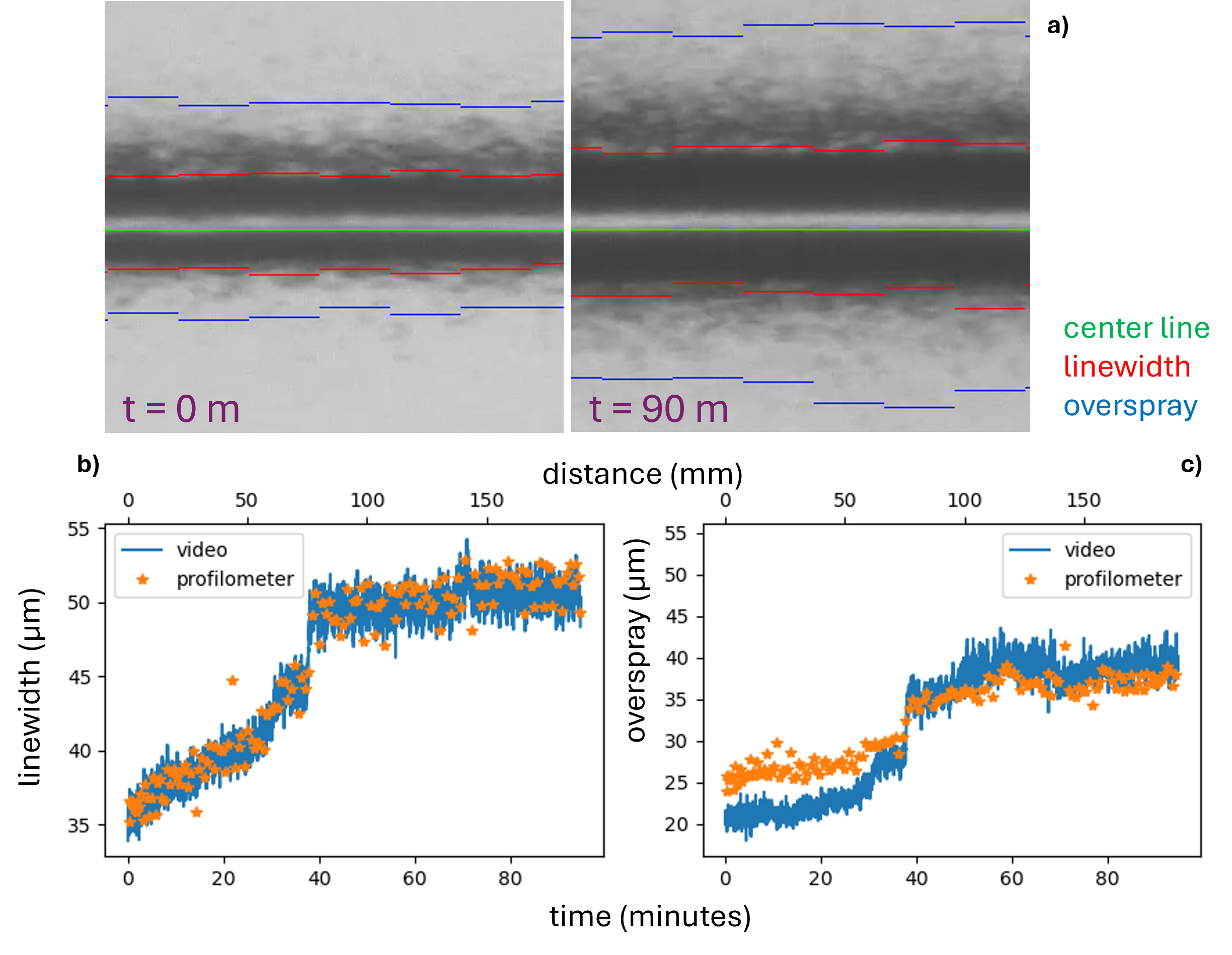}
    \caption{a) Annotated frame from alignment camera video, green is Hough transform fit, red is linewidth, blue is overspray. Averaging window size in this example is 50. Frames shown from t = 0 and t = 90 minutes. b) Plotting linewidth and c) overspray over time (and distance) for entire inspection video over wafer in comparison to optical profilometer results.}
    \label{fig:annotated_acam}
\end{figure}

Linewidth is defined as the distance between the points (iii) on either side of the center, which visually aligns with where the main lobe of the physical line ends in Figure  \ref{fig:lw_def}b. In the alignment camera video, this corresponds to the edge of the dark streak shown in Figure \ref{fig:annotated_acam}a. Overspray is defined as the region between each (iii) point and where the grayscale value returns to within 95\% of the substrate background intensity. In the alignment camera video, this correlates to the edges of the material deposited as highlighted by the blue line in Figure \ref{fig:annotated_acam}a.

The computer vision based definition of linewidth and overspray reflects real physical characteristics of the printed line. Profilometer data, shown in Figure \ref{fig:lw_def}b, confirm that the shoulders marked as (iii) in the grayscale profile correspond to inflection points in the height profile, located outside the central peak. Thus, the visually extracted linewidth is a proxy for the shoulders of the printed material’s main body, while overspray captures the extent of the peripheral droplet scattering.

The above analysis is implemented using a computer-vision algorithm (detailed in Appendix \ref{app:cv_algorithm}), which processes each frame of the alignment camera video to extract linewidth and overspray for every vertical column. To reduce noise and improve performance, columns are grouped into batches, and horizontal means and variances are computed for each batch. The number of columns per batch is an additional meta-parameter, typically set to $n = 50$ pixels, chosen to minimize measurement noise. Linewidth and overspray are then determined from the corresponding horizontal mean of each batch of columns. For the current work, the algorithm is only able to process data from prints of straight silver lines with single layers. We leave the development of more general algorithms to future work.

This method enables automated extraction of line quality metrics across entire prints. Figures \ref{fig:annotated_acam} b) and c) show the result of extracting linewidth and overspray from an inspection scan of a print. The drift in the linewidth and overspray is partially due to step changes in the gas flow inputs. Specifically, at around 40 minutes the carrier flow was increased. The plots also exhibit an apparent measurement oscillation of around 2~\textmu{m}. This is due to random fluctuations in material deposition at the line's edge and is validated with optical profilometry data (Appendix \ref{app:cv_algorithm}). 

Optical profilometer data of the sample print was further used to validate the linewidth and overspray values inferred by the computer vision algorithm (figures \ref{fig:annotated_acam} b) and c)) The two measurements for linewidth are in very close agreement and only differ on average by roughly a micron, which is less than the variation of the linewidth measurement. There is noticeable deviation between the overspray found by the computer vision algorithm and that of the profilometer. This is likely due to artifacts of the algorithm used for calculating overspray from the cross-sectional profile, as well as artifacts of the profile measurement itself. Despite this discrepancy, it is clear that the trend of the overspray through time is well captured by the proposed algorithm. Correctly capturing subtle changes in the print characteristics is crucial for making accurate state estimations. The proposed computer vision algorithm is therefore suitable for process monitoring.

In addition to linewidth and overspray, measuring the material volume flow rate, $\Qink$, can give greater insights into unobservable processes such as aerosol generation. The surface topology, which is directly related to material volume flow rate, is a complex function of the grayscale image from the alignment camera. To address this complexity, a convolutional neural network (CNN) is trained on grayscale images from the alignment camera to estimate the height profile. A ResNet-18 model was selected as the network architecture, with the input consisting of grayscale image columns of size (200, 1) \cite{resnet}. The output is the estimated height profile of the deposited material. As the standard ResNet-18 model is designed for classification, we modify the final layer with a dense linear layer to enable direct height regression, as described in \cite{lathuiliere2019comprehensive}.

\begin{figure}
    \centering
    \includegraphics[width=1\linewidth]{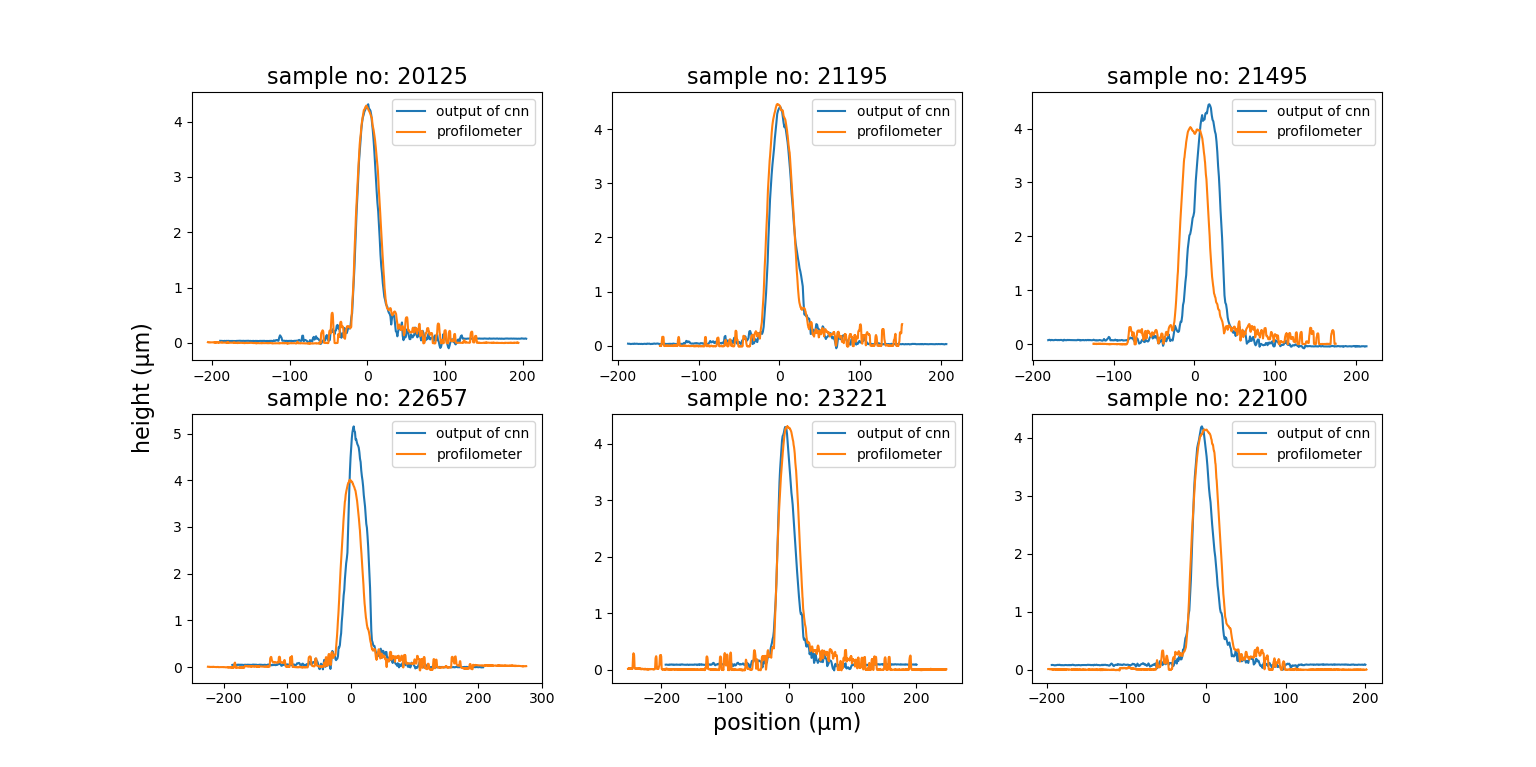}
    \caption{Comparison of CNN inferred profile with actual profile for a variety of random selected samples}
    \label{fig:nn_comp}
\end{figure}

\begin{figure}
    \centering
    \includegraphics[width=0.5\linewidth]{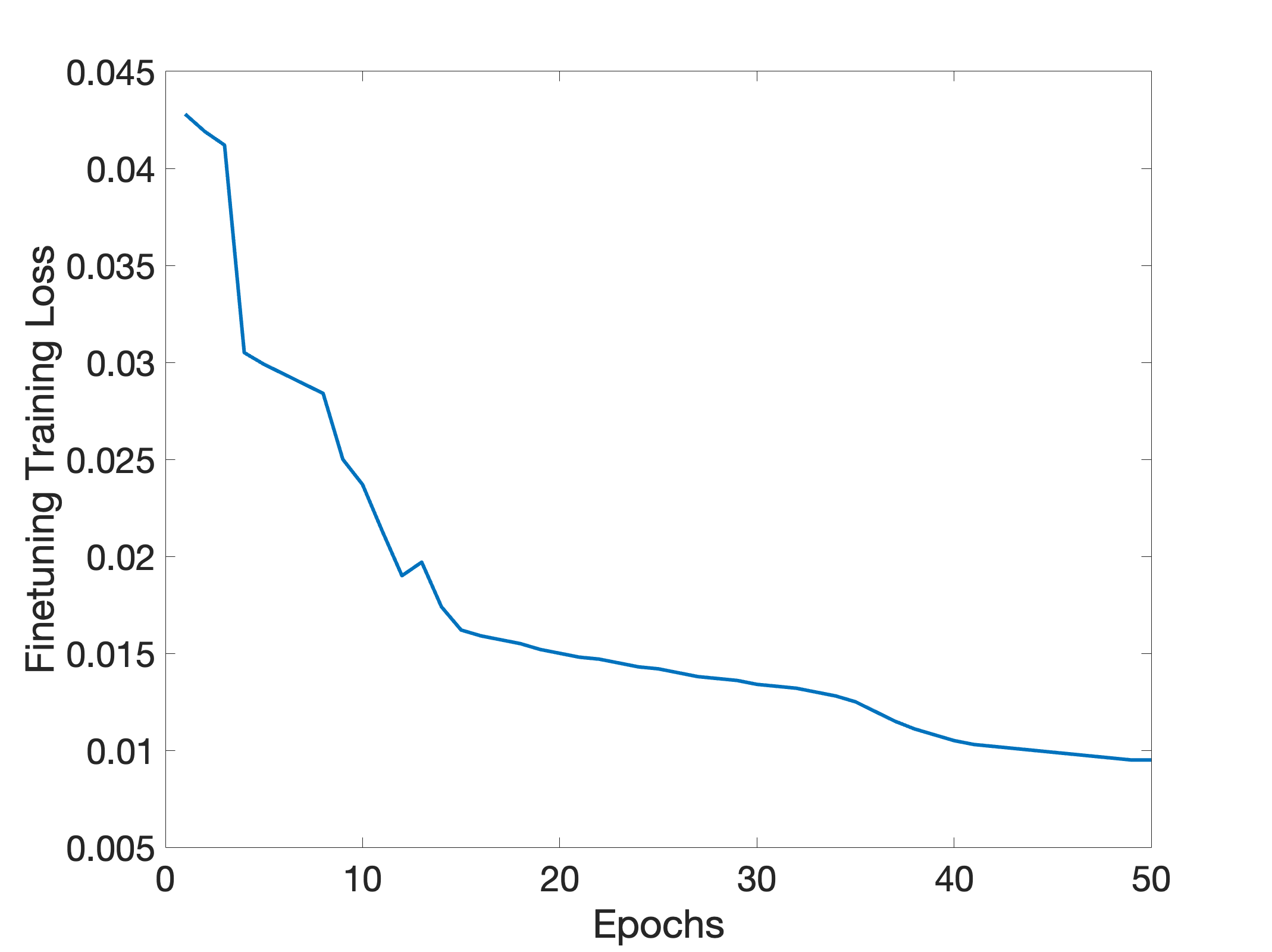}
    \caption{Training loss curve for fine-tuning the ResNet-18 model during secondary training, which minimizes the mean-squared error between linewidth and overspray derived from the CNN-predicted height profile and the ground-truth values from the computer-vision algorithm.}
    \label{fig:finetuning_training_loss}
\end{figure}

The CNN is trained using a dataset of grayscale images (e.g., Figure \ref{fig:lw_def}a) paired with corresponding ground truth measurements obtained from the optimal profilometer. A total of 125 two-dimensional profile scans were used in the training/validation process. However, the size of this dataset is small relative to the number of model parameters, raising concerns about overfitting and the ability to generalize to unseen conditions. 
To mitigate these challenges and improve accuracy, using a larger collection of data points from a secondary training dataset, fine-tunes the CNN model. The secondary dataset comprises of images from the alignment camera paired with ground-truth linewidth and overspray values extracted using the computer vision algorithm described later in this section.  After the initial training phase using profilometer scans, the weights of the CNN model are fine-tuned through a secondary objective that minimizes the mean-squared error loss between linewidth and overspray values extracted from the CNN-predicted height profiles and those obtained through the computer-vision algorithm described above.
By producing linewidth and overspray values consistent with real-world measurements, the fine-tuning process effectively grounds the CNN’s height predictions to larger data samples. This additional optimization step leads to a performance improvement of 20\%, leading to a more robust CNN model  for reconstructing height profiles from grayscale images. The result of the two-stage training process is shown in Figure \ref{fig:nn_comp}. During fine-tuning, the model was trained for 50 epochs using the secondary dataset, with the training loss function plot shown in Figure \ref{fig:finetuning_training_loss}. Evaluation on the full test dataset shows a significant improvement in performance as the mean absolute error for linewidth prediction decreased from 5.3 \textmu m to 1.3 \textmu m, and overspray prediction error dropped from 9.4 \textmu m to 6.1 \textmu m. 

Despite the improved performance (indicated by the low mean square error) from the two-stage training, the limited size of both training datasets leads to high variance in the reconstructed height profiles. The variance from a K-fold cross-validation (where $k=10$) after the two-stage training was 12 (\textmu m)\textsuperscript{2} and (15 \textmu m)\textsuperscript{2} for linewidth and overspray, respectively. This indicates that the CNN model parameters vary significantly when trained on new or unseen data, especially under different machine input conditions. The lack of generalization arises because key features of the predicted height profile—such as shoulder locations and tail endpoints—are highly sensitive to variations in input grayscale values. As a result, the high variance in these predictions limits the CNN’s effectiveness for consistent estimation of linewidth and overspray.

However, estimating the aerosolized material flow rate ($\Qink$) does not depend on specific local features of the predicted height profile. Instead, it is obtained by integrating over the entire profile, making it inherently less sensitive to localized variations. After performing K-fold cross-validation, a lower variance of $10^{-7}$ (sccm)\textsuperscript{2} was observed, confirming the stability of this metric. The flow rate $\Qink$ is computed as follows:

\begin{equation}
   \Qink(t) = \nu(t) \int h(r,t) dr
\label{eq:qink_measurement}
\end{equation}
where, at time $t$, $h(r,t)$ is the height profile predicted by the CNN at the coordinate $r$ perpendicular to the tool path direction at a cross-section of the trace, and $\nu(t)$ is the platen speed.

\section{Constructing the Digital Model of the AJ}
\label{sec:digital_model}

At the core of the digital twin is a representation of the AJ that captures its state dynamics and relates these to the measurable outputs. Previous studies, such as \cite{salary2021computational}\cite{ma17133179}, developed detailed computational fluid dynamics (CFD) models that simulate the flow of aerosol droplets through the machine and onto the substrate. The key distinction in our digital twin is the use of state and parameter estimation to continuously update the underlying digital model. This enables the system to incorporate measured data to align the digital model with the physical process conditions. 

Given the complexity of modeling the full physical process, a probabilistic dynamical macromodel of the AJ is developed. This model characterizes the relationships between inputs, latent states, and outputs. Unlike CFD-based methods, our probablistic model reduces the dimensionality of the state-space representation to a tractable number of state variables. The state variables of the proposed digital model (denoted $x$) are listed in Table \ref{tab:dt_state_variables} and correspond to the latent states of the AJ.

\begin{table*}
    \centering
    \begin{tabular}{|c|c|}
        \hline
         State Variable& Description  \\
         \hline
         $x_1 = \DP$ & Median aerosol droplet diameter [\textmu m]\\
         \hline
         $x_2 = \Vl$ & volume of the ink solution in the vial [mL] \\
         \hline
         $x_3 = \DT$ & Radial ink deposition in the tube assuming uniform deposition [\textmu m] \\
         \hline
         $x_4 = \DN$ & Radial ink deposition in the nozzle assuming uniform deposition [\textmu m]\\
         \hline
         $x_5 = \AD$ & Volume ratio of aerosol droplets to the carrier stream  [mL/mL] \\
         \hline
    \end{tabular}
    \caption{The digital model of the AJ is determined by a vector of state variables, $x=[\DP, \Vl, \DT, \DN, \AD]$. }
    \label{tab:dt_state_variables}
\end{table*}

The dynamics of the system are modeled using two relations: the transition function and the output function. The transition function, $f(\cdot)$, governs the evolution of latent states over time as:

\begin{subequations}
    \begin{equation}
         \dot{x}(t) = f(x(t), u(t)) + \xi(t)
        \label{eq:transition model}   
    \end{equation}
   
\end{subequations}

\noindent where $x(t)$ and $u(t)$ are the state variables and inputs to the AJ, respectively. The transition model is a probabilistic model with an added vector of white noise, $\xi(t)$.

The output function, $g(\cdot)$, establishes a relationship between the measured outputs, $y(t)$, defined in Table \ref{tab:dt_outputs}, the system’s inputs, $u(t)$, defined in Table \ref{tab:dt_inputs}, and latent states, $x(t)$ as:
\begin{subequations}
\begin{equation}
        y(t) = g(x(t), u(t)) + w(t) \label{eq:output_model} 
\end{equation}
\begin{equation}
        w(t) \sim N(0, \Sigma_{w})
\end{equation}
\end{subequations}
The output model includes a noise term, $w$, contributed by model and measurements uncertainty. The output noise has a zero-mean normal distribution, $N(0, \Sigma_{w})$, with a diagonal covariance matrix, $\Sigma_{w}$, implying that the output noises are independent. Although linewidth and overspray are extracted using the same computer-vision algorithm, treating them as independent simplifies the output model. In practice, shared sources of error can introduce correlated noise, and future work should incorporate this correlation to improve the fidelity of the transition model and overall estimation accuracy.

The equations \eqref{eq:transition model} and \eqref{eq:output_model} are discretized as follows:
\begin{align}
x_{k+1} &= f_d(x_k, u_k) + \xi_k, \\
y_k &= g_d(x_k, u_k) + w_k,
\end{align}
where $f_d(x_k,u_k)$ and $g_d(x_k, u_k)$ are discretized versions of the continuous-time models in \eqref{eq:transition model}, \eqref{eq:output_model}, and the subscript $k$ indicates the discrete-time index. For example, using a first-order Euler discretization yields $x_k =x_k + \Delta t f(x_k, u_k)$ and $g_d(x_k, u_k) = g(x_k, u_k)$ with a discretization time step $\Delta t$ (experimentally set to 1 s to match measurement sampling interval). When time is discretized at time-step $k$, the noise vector, $\xi_k$, is a Gaussian random variable with  distribution $N(0,\Sigma_{\xi})$, with zero mean and a covariance matrix $\Sigma_{\xi}$. It is assumed that state-variable transitions exhibit independent model uncertainty and as a result reduce $\Sigma_{\xi}$ to a diagonal matrix, at least approximately. 

The dynamical model offers a macro-level representation of the AJ system that captures the interactions between the observable behavior, controllable inputs and the latent states. The model effectively abstracts complex physical interactions within the physical system for a more computationally efficient and interpretable representation of the AJ printer. 

\begin{table*}[]
    \centering
    \begin{tabular}{|c|p{10cm}|}
        \hline
         Input name& Description  \\
         \hline
         $u_1 = \PA$ &  Current for ultrasonic atomizer that controls the ultrasonic frequency to generate aerosol droplets [mA]\\
         \hline
         $u_2 = \Qcarr$ & Carrier gas flow rate [mL/s] \\
         \hline
         $u_3 = \Qsh$ & Sheath gas flow rate [mL/s] \\
         \hline
    \end{tabular}
    \caption{The digital model of the AJ is controlled by a vector of inputs, $u=[\PA, \Qcarr, \Qsh]$. }
    \label{tab:dt_inputs}
\end{table*}

\begin{table*}[]
    \centering
    \begin{tabular}{|c|c|}
        \hline
        Output name& Description  \\
         \hline
         $y_1 = \lw$ & Linewidth [\textmu m]\\
         \hline
         $y_2 = \ov$ & Overspray [\textmu m]\\
         \hline
         $y_3 = \Pcarr$ & Carrier gas pressure [Pa] \\
         \hline
         $y_4 = \Psh$ & Sheath gas pressure [Pa]\\
         \hline
         $y_5 = \Qink$ & Flow rate of aerosolized material deposited on the substrate [mL/s] \\
         \hline
    \end{tabular}
    \caption{The measurable outputs of the digital model of the AJ is determined by a vector $y=[\lw, \ov, \Pcarr, \Psh, \Qink]$. }
    \label{tab:dt_outputs}
\end{table*}

\subsection{Transition Model Derivation}
The transition model characterizes the time-varying dynamics of the latent states presented in Table \ref{tab:dt_state_variables} and is constructed through a combination of analytical models derived from prior studies and empirical data obtained from experimental results.

\subsubsection{Median Droplet Diameter}
The ultrasonic atomizer generates ink droplets whose diameter, $d_a$, follows the log-normal probability distribution in \eqref{eq:aerosol_droplet_distribution} with a fixed coefficient of variance of 25\% \cite{secor2018principles}. The model assumes that the droplet size distribution is governed by the frequency of the ultrasonic atomizer, while remaining independent of the atomization current and the ink level within the vial. Experimental observations (Appendix \ref{app:variations_experiment}) reveal that the aerosol droplet size distribution fluctuates throughout the printing process. Due to a lack of physical equations that can explain these changes, it is assumed that changes in the median droplet diameter, denoted by $\DP$, are stochastic and can be modeled as a Gaussian random process:

\begin{equation}
\frac{\partial}{\partial t} \DP = \theta_{da} \DP +  \xi_{da}(t).
\end{equation}

 where $\xi_{da}(t)$ is a scalar of white noise that is modeled, when time is discretized, as $\xi_{da_k} \sim N(0,\sigma^2_{da})$, where $\sigma_{da}$ represents the standard deviation of the fluctuation. To ensure that the transition model is not purely stochastic, we introduce a parameter $\theta_{da}$ into the median aerosol droplet size dynamics. This provides a deterministic drift term that accounts for systematic effects such as nozzle temperature changes, solvent evaporation, or material buildup that bias droplet formation beyond random fluctuations. In the nominal case, parameter $\theta_{da}$ is zero, and the transition model reduces to a purely stochastic fluctuations in droplet size. In practice, however, the digital twin estimates $\theta_{da}$ from measurements to account for complex effects not captured by physical equations but evident in observed behavior.

\subsubsection{Ink Volume}

The volume of ink solution in the vial has been experimentally shown to influence key line quality parameters \cite{smithdrift}. According to mass conservation, any decrease in the ink volume in the vial is due to aerosol droplets being transported out of the vial by the carrier gas stream. This process can be analytically modeled by:

\begin{equation}
\frac{\partial}{\partial t} \Vl = -\AD \Qcarr +\theta_{\Vl} \Vl + \xi_{\Vl}(t).
\label{eq:aerosol_density_transition}
\end{equation}
where $\Qcarr$ is the input carrier gas flow rate, and $\AD$ is the latent state modeling the volume ratio of ink to carrier gas flowing through the tube and out the nozzle. $\xi_{\Vl}(t)$ is a scalar of white noise that when discretized at iteration $k$ results in a zero-mean Gaussian, $\xi_{V_{l_k}}\sim N(0,\sigma^2_{\Vl})$ where $\sigma_{\Vl}$ is the standard deviation relating to the vial level. To account for any deviations from the physical model, $\theta_{\Vl}$ is introduced as a tunable parameter set through the digital twin parameter estimation. In the nominal case, $\theta_{\Vl}$ is zero and the transition function of the ink volume is determined by $\AD$ and $\Qcarr$.

\subsubsection{Deposition of Ink in the Tube and Nozzle}

Aerosolized ink droplets have been observed to deposit onto the inner walls of the carrier tube and nozzle (Appendix \ref{app:variations_experiment}).
Although the build-up caused by this deposition is slow, it can lead to gradual changes in the print process that impact linewidth and overspray. This rate of ink deposition in the carrier tube and nozzle is modeled using physics derived in \cite{secor2018principles}. The deposited ink is assumed to form a uniform layer along the inner walls of the carrier tube or nozzle, thereby altering its effective radius. The digital twin model tracks the change in effective radius of the carrier tube, $\RT$ and nozzle, $\RN$,  which evolves over time according to the following analytical functions \cite{secor2018principles}: 

\begin{equation}
\begin{aligned}
 \frac{d \RT}{dt} =&  \theta_{\RT}\RT + \frac{\Qdrop(t)}{2 \pi(\tubeR - \RT (t)) \LT}\times\\&\int_{d_a} \Fdep^T(d_a, \RT, \Qcarr) p(d_a)\; d (d_a) + \xi_{\DT}(t),
 \label{eq:dr_dt_carrier}
\end{aligned}
\end{equation}

\begin{equation}
\begin{aligned}
 \frac{d\RN}{dt} =& \theta_{\RN}\RN +  \frac{\Qdrop(t)}{2\pi(\nozzleR - \RN (t))\LT}\times \\ &\int_{d_a}{\Fdep^N\left(d_a,\RN, \Qsh+\Qcarr\right) p(d_a) \; d(d_a)} + \xi_{\DN}(t)
\end{aligned}
\end{equation}

\noindent where $d_a$ is the diameter of a single aerosol droplet, $L_T$ and $L_N$ are the lengths of the tube and nozzle, $\Fdep^{T,N}$ is the probability that an aerosol droplet contacts with the wall of the tube or nozzle, respectively, $p(d_a)$ is the log-normal probability distribution of the aerosol droplet diameter in \eqref{eq:aerosol_droplet_distribution}, $\nozzleR$ and $\tubeR$ are the beginning radii for the nozzle and tube, respectively, and $\Qdrop = \AD \Qcarr$ is the volume flow rate of aerosolized droplets. $\xi_{\DT}(t)\sim N(0,\sigma^2_{\DT})$ and $\xi_{\DN}(t)\sim N(0,\sigma^2_{\DN})$ are white noise scalars for tube and nozzle deposition that is sampled from a zero-mean Guassian distribution with a standard deviation of $\sigma_{\DT}$ and $\sigma_{\DN}$, respectively. The full derivation of the physics-based transition equations is provided in Appendix~\ref{app:aerosol_generation}-\ref{app:nozzle_deposition} . 

Additional terms, ($\theta_{\RT}$ and $\theta_{\RN}$), are incorporated into the transition model to account for nonlinear effects such as fluid-fluid interactions during aerosol accumulation in the tube and nozzle, respectively, which are not captured in the simplified physics-based formulation.
 In the nominal physics-based case, the parameters $\theta_{\RT}$ and $\theta_{\RN}$ are zero, but the digital twin estimates its value from data to reflect trends observed in practice.

\subsubsection{Volume fraction of aerosolized ink}

\begin{figure}
\centering
    \includegraphics[width=0.8\linewidth]{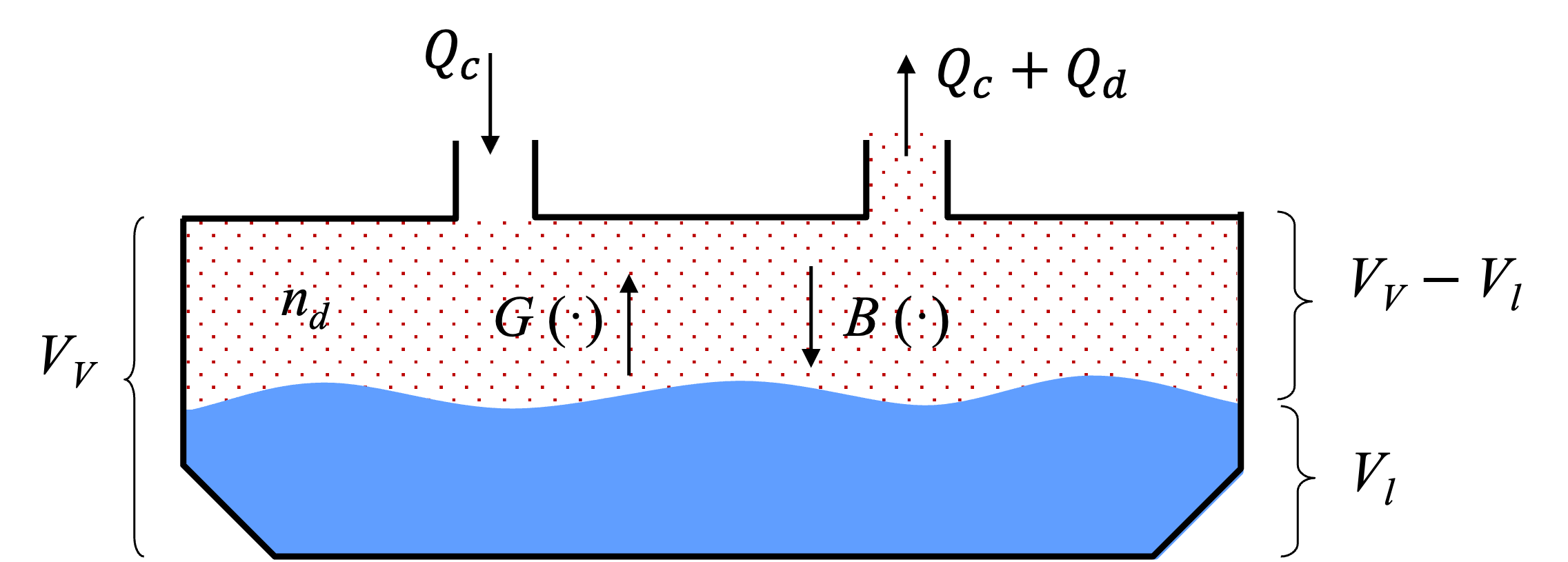}
\caption{ The aerosol generation of the ultrasonic atomizer is modeled by a generation, $G(\cdot)$, and recombination, $B(\cdot)$, functions that are determined experimentally in Appendix \ref{app:effective_droplet_generation}.
}    
    \label{fig:atomizer_model}
\end{figure}

The transition model of the aerosol volume fraction, $\AD$, depends on the physics and geometry of the ultrasonic atomizer. Accurate modeling of $\AD$ requires an analytical, multi-physics model that has not been fully developed. As an alternative, the aerosol generation process is behaviorally modeled by fitting experimental data from observations of the ultrasonic atomizer to an analytical function. The aerosol generation macromodel is shown in Figure \ref{fig:atomizer_model}. 

The vial contains a liquid ink solution of volume $\Vl$, while the remaining volume, $V_v - \Vl$, is occupied by aerosolized droplets. This derivation assumes that aerosol droplet generation by the ultrasonic atomizer is in equilibrium with the rest of the system such that the aerosol volume fraction in the vial equals that in the carrier tube. The resulting aerosol droplet flow rate out of the vial is given by $\AD \Qcarr$.

The number of aerosol droplets in the empty region of the vial, $n_d$, is governed by two competing processes: a generation function  $G(\cdot) $, which describes the rate of droplet formation from the liquid ink solution, and a recombination function $B(\cdot) $, which represents the rate at which aerosol droplets recombine with the liquid ink solution and the vial sidewalls. The net droplet generation rate is then expressed as:

\begin{equation}
    H(\cdot) = G(\cdot) - B(\cdot).
\end{equation}
An analytical form for $H(\cdot)$ is derived by fitting parameters of a quadratic model using a three-level Box Behnken design of experiments \cite{ferreira2007box} shown in Appendix \ref{app:effective_droplet_generation}. This experiment reveals that the recombination function, $H(\cdot)$, is a function of ultrasonic atomizer current, vial volume and carrier gas flow rate.

Given this empirical estimate for $H(\cdot)$, the derivation for the transition model is as follows. By conservation of mass, the change in the volume of aerosolized droplets, $V_a$, is given by:
\begin{equation}
    \frac{d V_a}{dt} = H(\cdot) - \AD \Qcarr.
        \label{eq: dnd_dt_def1}
\end{equation}

\noindent where $\AD \Qcarr$ is the volume rate of aerosolized droplets being carried out of the atomizer. 
The aerosol droplet  volume fraction in the vial,  $\AD$, is:
\begin{equation}
    \AD = \frac{V_a}{V_v-V_l}.
    \label{eq:Ad_definition}
\end{equation}

\noindent Where $V_v$ and $V_l$ are the volume of the vial and liquid in the vial. Rearranging equation  \eqref{eq:Ad_definition} and taking the time-derivative:
\begin{align}
    \frac{dV_a}{dt} &= \frac{d}{dt} (\AD(V_v - V_l)) \\
     &= \frac{d \AD}{dt} (V_v - V_l) -\AD \frac{d V_l}{dt}
     \label{eq:dVa_dt_def2}
\end{align}

\noindent Equating \eqref{eq: dnd_dt_def1} and \eqref{eq:dVa_dt_def2}, the resulting expression for the time derivative of the aerosol volume fraction is:

\begin{equation}
    \frac{d \AD}{dt} = \frac{1}{V_v - V_l} \left(H(\cdot) - \AD \Qcarr + \AD \frac{d V_l}{dt}\right).
    \label{eq:dAd_dt_def1}
\end{equation}

This formulation macromodels the aerosol generation within the ultrasonic atomizer whose parameters are experimentally determined.

To account for systematic trends not explicitly modeled by the generation function $H$, an additional term $\theta_{\AD}\AD$ is introduced into the transition model according to:
\begin{equation}
    \frac{d \AD}{dt} = \frac{1}{V_v - V_l} \left(H(\cdot) - \AD \Qcarr + \AD \frac{d V_l}{dt}\right) + \theta_{\AD}\AD + \xi_{\AD}(t).
    \label{eq:dAd_dt_def}
\end{equation}
  In practice, aerosol generation can be influenced by complex phenomena such as nonlinear secondary droplet breakup, coalescence, or variations in transducer efficiency, which may not be accurately captured by effective generation function alone. The inclusion of $\theta_{\AD}$ provides a deterministic drift component that compensates for these unmodeled effects. Nominally, $\theta_{\AD}=0$ recovers the physics-based model, but in the digital twin framework this parameter is estimated from measurement data to capture systematic trends observed during operation. The term $\xi_{\AD}(t)\sim N(0,\sigma^2_{\AD})$ is white noise scalar added for modeling the randomness in the transition and is sampled from a zero-mean Gaussian distribution with a standard deviation of $\sigma_{\AD}$.

\subsection{Deriving the Output Model}
The output model relates the measurable outputs in Table \ref{tab:dt_outputs} to the latent variables and inputs. Although outputs can be experimentally measured, their dependence on latent states and inputs is not directly observable. To address this, analytical models of these outputs are developed using a combination of experimental results and macromodels derived from CFD simulations.

\subsubsection{Linewidth and Overspray Definitions}

\begin{figure}
    \centering
    \includegraphics[width=\linewidth]{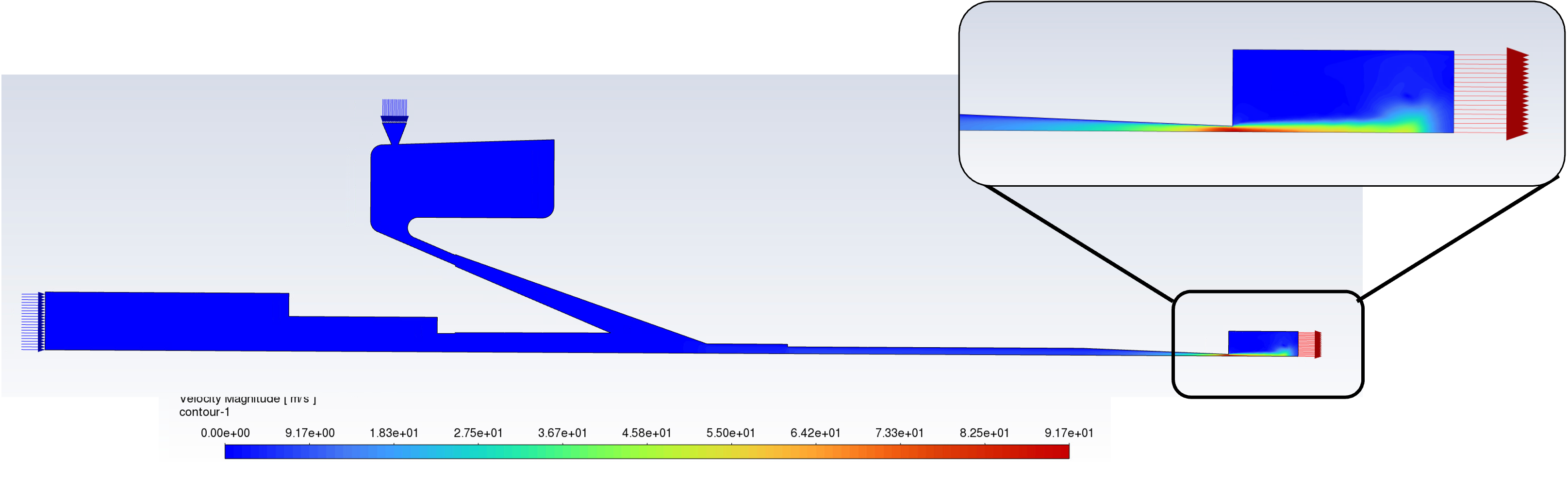}
    \caption{Velocity contour plot of $N_2$ gas through the AJ aerodynamic focusing chamber and nozzle, simulated using ANSYS Fluent with carrier flow of 25 sccm and sheath flow of 50 sccm. Boundary conditions of the CFD simulation include the AJ walls with an inlet for the carrier gas with aerosolized droplets, an inlet for the angled sheath flow, and an outlet representing the substrate located 2 mm from the nozzle orifice. Color map minimum (blue) is 0 m/s and maximum (red) is 91.7 m/s.}
    \label{fig:ansys_velocity}
\end{figure}

CFD simulations (Figure \ref{fig:ansys_velocity}) are used to analyze the effects of state variables (Table \ref{tab:dt_state_variables}) and inputs (Table \ref{tab:dt_inputs}) on linewidth and overspray. The CFD model, detailed in Appendix \ref{app:cfd_simulation}, is an axisymmetric representation of the AJ that simulates deposition of discrete aerosol droplets onto the substrate from the carrier inlet. Note, this model does not account for interactions between droplets once they reach the substrate. The simulation output provides the density distribution of deposited aerosol droplets as a function of radial distance from the center, as illustrated in Figure \ref{fig:atomizer_diameter_distribution}.

Using this distribution, linewidth and overspray are defined in a manner consistent with the parameters extracted from the alignment camera in Section \ref{sec:extracting_line_quality_params} (further details provided in Appendix \ref{app:cfd_simulation}). A parameter sweep is performed in the CFD simulation, systematically varying states and inputs to measure the effect on the simulated linewidth and overspray.

To establish an analytical relationship between these parameters and outputs, we construct a relational model using a polynomial regression assuming independence between state variables. This assumption is justified because the latent state variables influence separate stages of aerosol generation and transport, allowing their effects on linewidth and overspray to be treated as approximately independent. The resulting macromodel relates the linewidth and overspray to state variables and inputs without requiring direct CFD simulations when executing the digital twin.

\begin{figure}
\centering
\begin{subfigure}{.5\textwidth}
  \centering
  \includegraphics[width=1.15\linewidth]{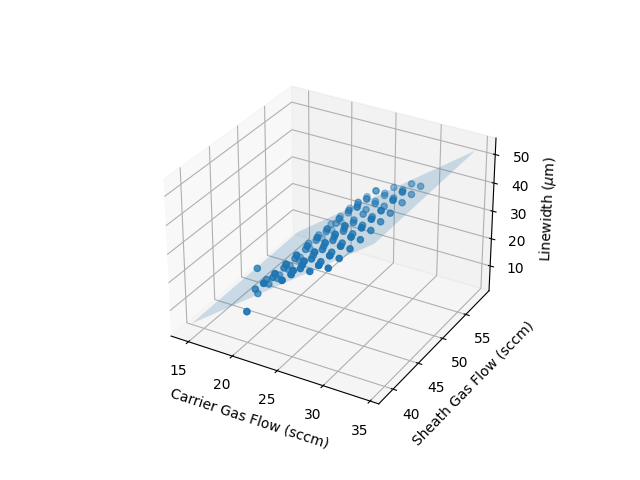}
  \caption{}
  \label{fig:linewidth_vs_flow_rate}
\end{subfigure}%
\begin{subfigure}{.5\textwidth}
  \centering
  \includegraphics[width=1.15\linewidth]{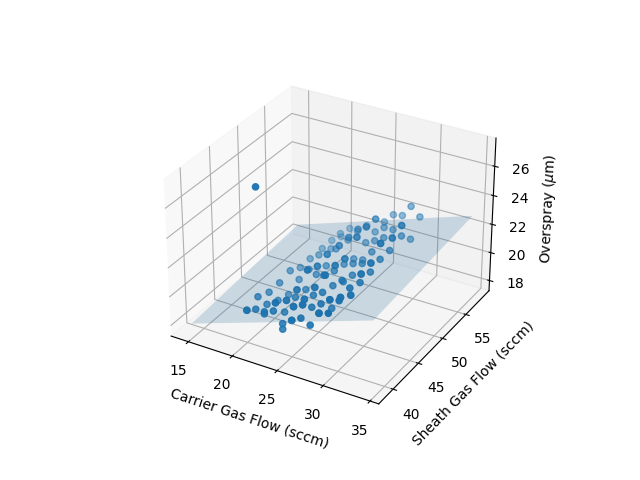}
    \caption{}
  \label{fig:overspray_vs_flow_rate}
  \end{subfigure}
    \caption{CFD simulation (shown in blue plane) of the effect of gas flow rates (median aerosol droplet diameter of 3~\textmu{m}) on a) linewidth $\lw$ and b) overspray extent $\ov$. The CFD simulation results match the sensitivities from experimental results shown in blue dots. }
    \label{fig:line_params_vs_flow_rate}
\end{figure}

Linewidth and overspray exhibit a nearly linear relationship with the carrier and sheath gas flow rates, as shown in Figure \ref{fig:line_params_vs_flow_rate}. This trend is experimentally validated by sweeping the gas flow rates within operational bounds. The CFD-derived macromodel accurately predicts the response within 7\% of experimental results.

\begin{figure}
\centering
\begin{minipage}{.45\textwidth}
  \centering
  \includegraphics[width=\linewidth]{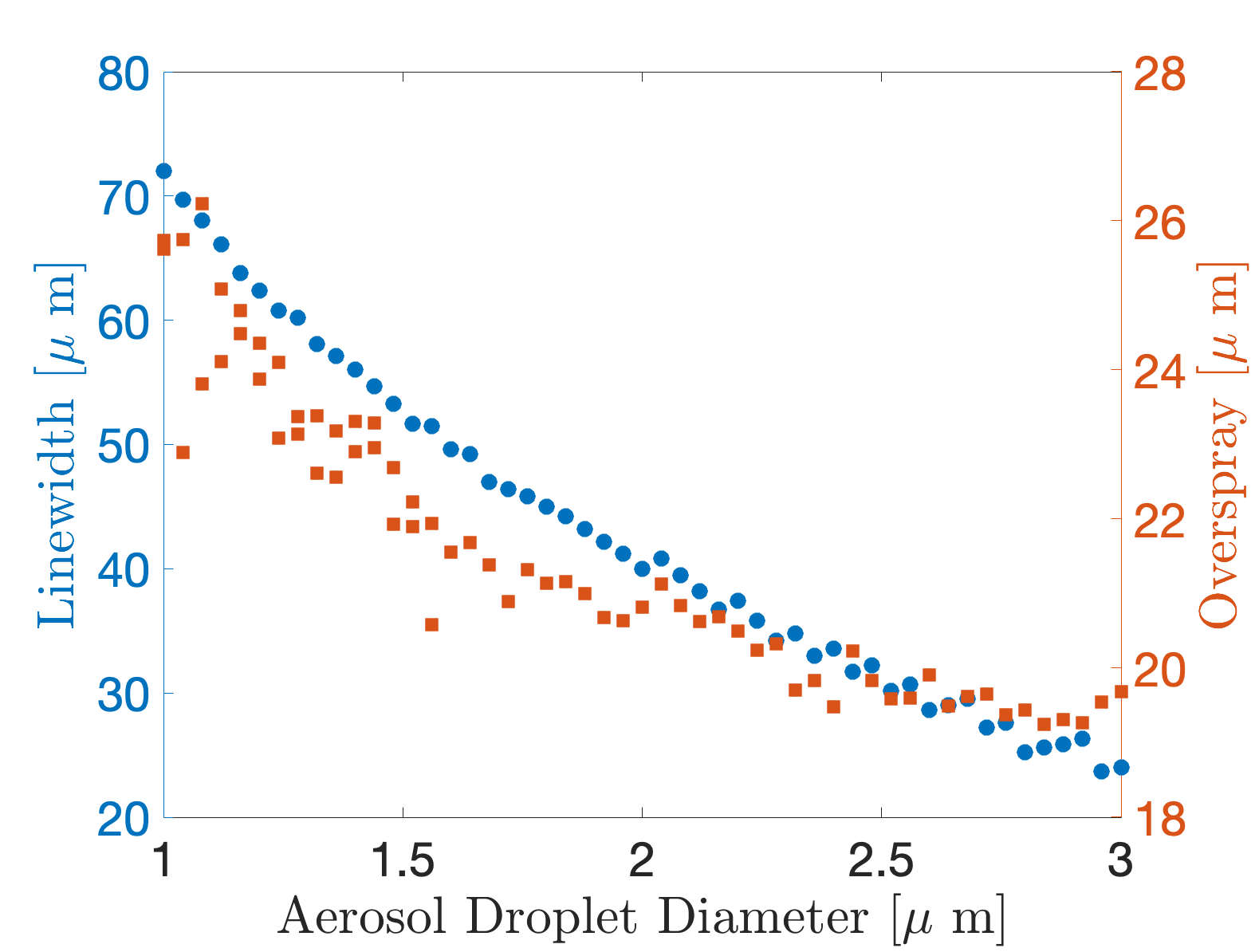}
  \captionof{figure}{CFD simulation of the effect of ink droplet diameter (constant carrier gas of 22 sccm and sheath gas flow of 58 sccm) on linewidth $\lw$ and overspray extent $\ov$.}
  \label{fig:line_params_vs_droplet_size}
\end{minipage}%
\hfill
\begin{minipage}{.45\textwidth}
  \centering
  \includegraphics[width=\linewidth]{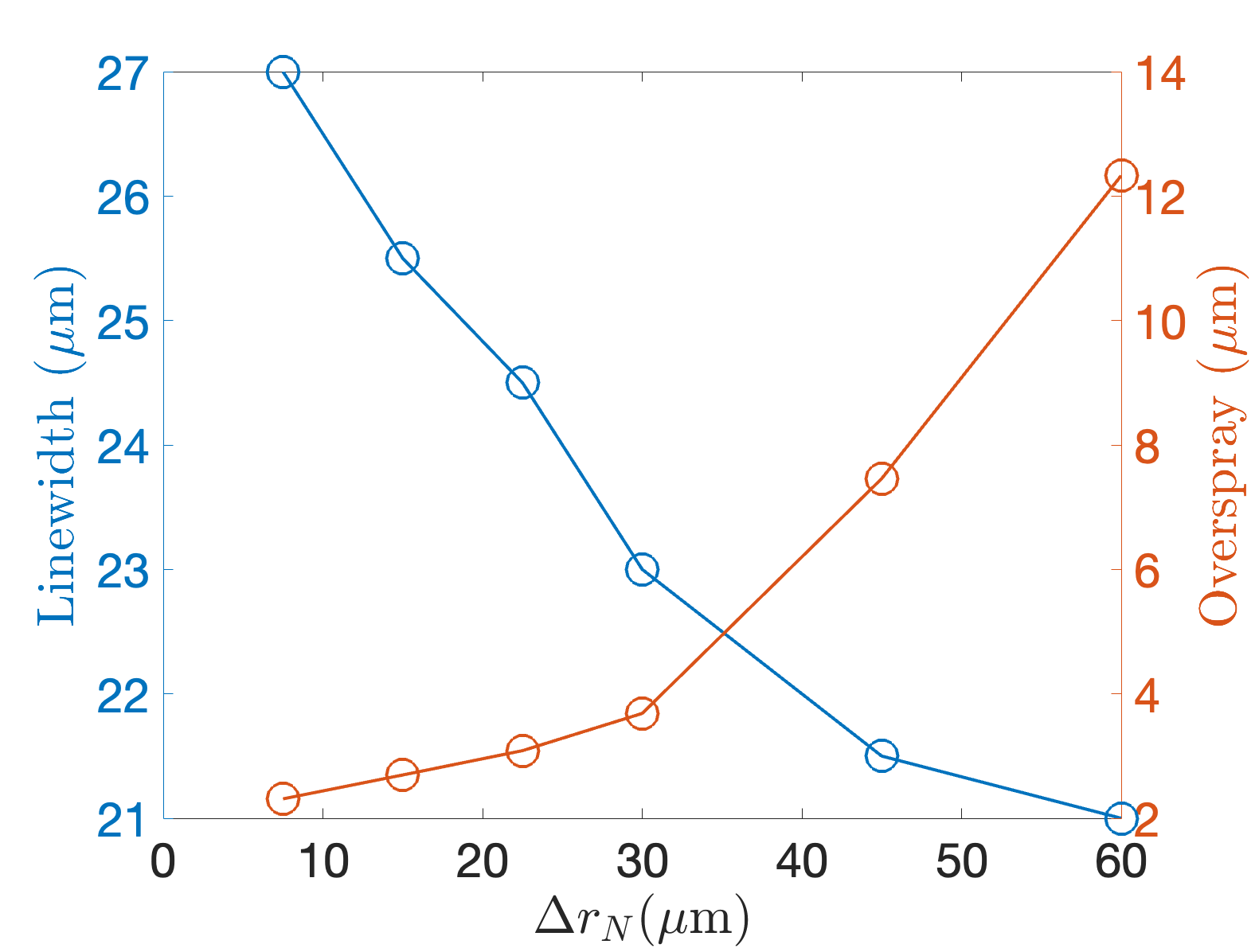}
  \captionof{figure}{CFD simulation (with a nominal nozzle radius of 35~\textmu{m} corresponding to $\DN=0$) of the effect of ink deposition in the nozzle, $\RN$, on linewidth $\lw$ and overspray extent $\ov$.}
\label{fig:line_quality_v_nozzle_deposition}
\end{minipage}
\end{figure}

Additionally, linewidth and overspray are negatively correlated with median aerosol droplet size, $\DP$, as shown in Figure \ref{fig:line_params_vs_droplet_size}. This aligns with the findings in \cite{secor2018principles}, which suggest that larger, heavier droplets tend to remain near the center of the aerosol stream, while smaller droplets diffuse outward toward the edges. As a result, larger median droplet sizes promote more concentrated line formations and reduce overspray.

The CFD-derived macromodel also provides insights into the sensitivity of linewidth and overspray to ink deposition within the nozzle and carrier tube. Since ink accumulation in the carrier tube has minimal impact on particle distribution, the model shows no significant sensitivity of linewidth and overspray to carrier tube deposition. However, ink deposition within the nozzle alters the internal geometry, disrupting the flow of the output gas streams. These changes lead to deviations in the droplet trajectories, ultimately affecting the linewidth and overspray. This effect becomes more pronounced as deposition accumulates, further constricting the flow path of the aerosolized droplets. CFD simulations (Figure \ref{fig:line_quality_v_nozzle_deposition}) reveal that the linewidth and overspray
follow a third-order polynomial relationship with respect to nozzle deposition, $\DN$, as
increasing nozzle deposition distorts the gas flow out of the nozzle and degrades line
quality.

The complete models for linewidth and overspray are:
\begin{equation}
\begin{aligned}
    \lw = \alpha_{lw}^{\DP}\DP &+ \alpha_{lw}^{\AD}\AD + \alpha_{lw}^{\DN,1}\DN + \alpha_{lw}^{\DN,2}\DN^2 + \alpha_{lw}^{\DN,3}\DN^3 \\ &+  \beta_{lw}^{c}\Qcarr + \beta_{lw}^{s}\Qsh +\gamma_{lw} + w_{\lw}(t)
\end{aligned}
\label{eq:linewidth_output_model}
\end{equation}
\begin{equation}
\begin{aligned}
    \ov = \alpha_{ov}^{\DP}\DP &+ \alpha_{ov}^{\AD}\AD + \alpha_{ov}^{\DN,1}\DN + \alpha_{ov}^{\DN,2}\DN^2 + \alpha_{ov}^{\DN,3}\DN^3 \\ &+  \beta_{ov}^{c}\Qcarr + \beta_{ov}^{s}\Qsh + \gamma_{ov} + w_{\ov}(t)
\label{eq:overspray_output_model}
\end{aligned}
\end{equation}

Here, the scalar model parameters represent how the output variables (linewidth and overspray) are influenced by latent states and inputs. Specifically, the $\alpha$ parameters ($\alpha_{lw}^{\DP}, \alpha_{lw}^{\AD}, \alpha_{lw}^{\DN,1},\alpha_{lw}^{\DN,2},\alpha_{lw}^{\DN,3}$ for linewidth and $\alpha_{ov}^{\DP}, \alpha_{ov}^{\AD}, \alpha_{ov}^{\DN,1},\alpha_{ov}^{\DN,2},\alpha_{ov}^{\DN,3}$ for overspray) quantify the sensitivity between the outputs and the latent state variables. The $\beta$ parameters ($\beta_{lw}^{c}, \beta_{lw}^{s}$ and $\beta_{ov}^{c}, \beta_{ov}^{s}$) capture the effect of observable input variables on the outputs. Finally, the $\gamma$ terms ($\gamma_{\lw}$ and $\gamma_{\ov}$) represent scalar offsets in the output measurements.
These parameter values are obtained by fitting to the CFD simulation results described in Appendix \ref{app:estimating_line_params_from_ansys}. Additionally, $w_{\lw}(t)\sim N(0, \sigma^2_{\lw})$ and $w_{\ov}(t)\sim N(0, \sigma^2_{\ov})$ represent the noise terms contributing to the probabilistic models of linewidth and overspray, respectively (which are modeled using zero-mean normal distributions with standard deviations of $\sigma_{\lw}$ and $\sigma_{\ov}$).

\subsubsection{Carrier and Sheath Gas Pressure Models}


To relate measured carrier and sheath pressures to the inputs and latent states in the AJ system, a fluidic resistance model is developed. This model is built by deriving the geometric resistance of the individual components and connections of the AJ system. A simplified diagram is shown in Figure \ref{fig:plumbing_simple} (with more detail given in Appendix \ref{app:fluidic_resistance}). 

\begin{figure}
    \centering
    \includegraphics[width=0.8\linewidth]{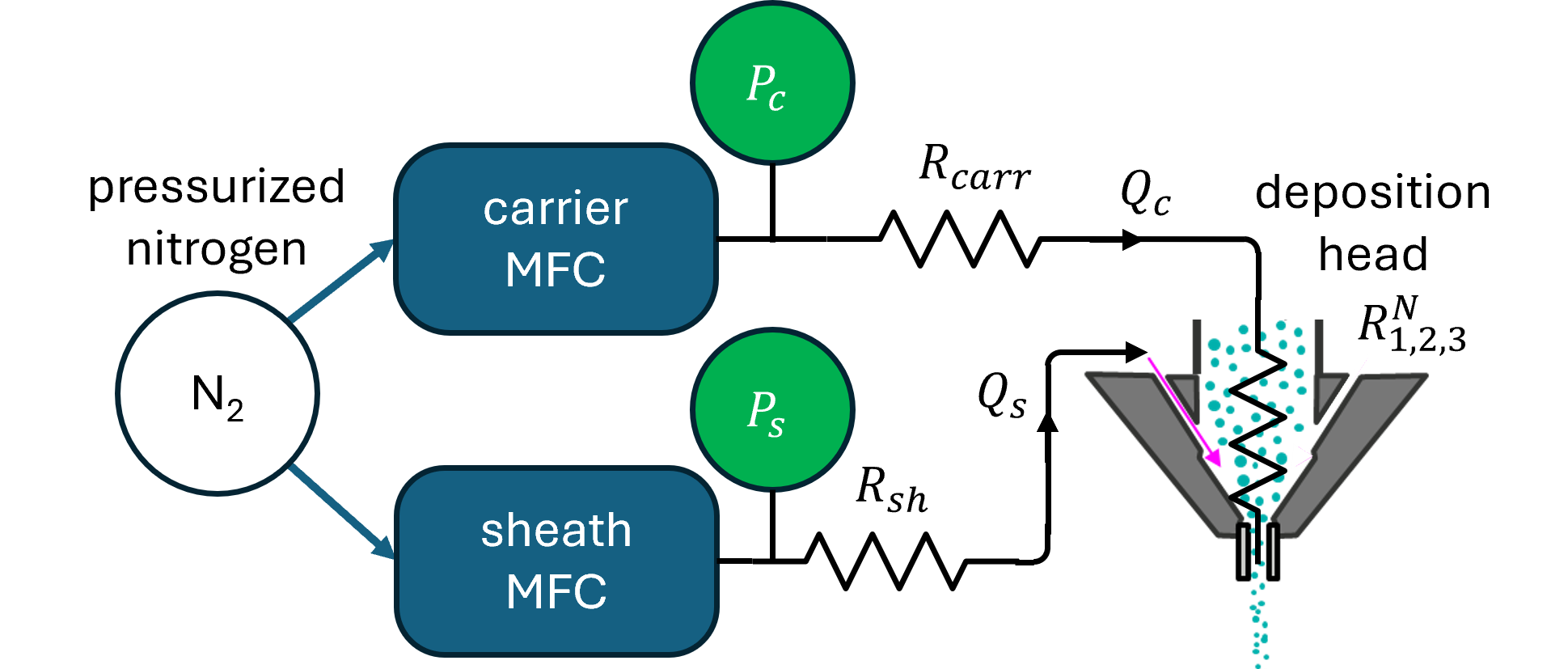}
    \caption{Simplified diagram of gas flow system of AJ printer, with fluidic resistances placed according to where they appear in the flowpath (more detailed diagram of resistances in nozzle is shown in figure \ref{fig:resistive_model_AJ}).}
    \label{fig:plumbing_simple}
\end{figure}

Ink deposition within the carrier tube and nozzle alters the geometry of the system and thus changes fluidic resistances during printing. The fluidic resistance model explicitly accounts for variations in these hidden state variables ($\DT$ and $\DN$), which represent buildup of ink in the carrier tube and nozzle, respectively. As ink deposition increases, the corresponding rise in fluidic resistance requires higher pressure to maintain the same flow rate. Consequently, the gas pressures measured by the MFCs are highly sensitive to changes in ink flow behavior.

This results in a mapping between the measured gas pressures (carrier gas  $\Pcarr$ and sheath gas  $\Psh$), the input flow rate ($\Qcarr$ and $\Qsh$), and the ink deposition in the carrier tube and nozzle ($\DT$  and $\DN$), expressed as:
\begin{equation}
\begin{aligned}
    \Pcarr = &\Qcarr \cdot  R_{c}(\DN, \DT)\\& + (\Qsh+\Qcarr)(R_1^N + R_2^N + R_3^N(\DN)) + w_{\Pcarr}(t)
\end{aligned}
\end{equation}
\begin{equation}
\begin{aligned}
\Psh = &\Qsh \cdot R_{s}(\DN) \\ &+ (\Qcarr+\Qsh)(R_1^N + R_2^N + R_3^N(\DN)) + w_{\Psh}(t)
\end{aligned}
\end{equation}
where $R_{c},R_{s}, R_1^N, R_2^N, R_3^N$ are the individual fluidic resistances of the chambers in the AJ. \(w_{\Pcarr}(t)\sim N(0, \sigma^2_{\Pcarr})\) and \(w_{\Psh}(t)\sim N(0, \sigma^2_{\Psh})\) represent noise terms arising from measurement noise in the MFC sampled from a zero-mean Gaussian with standard deviations of $\sigma_{\Pcarr}$ and $\sigma_{\Psh}$.

This formulation provides an explainable model that relates the pressure measurements extracted by the AJ with the hidden states and inputs.

\subsubsection{Material Flow Rate Output Model}

By mass conservation, the measured aerosol flow rate is:

\begin{equation}
\Qink(t) = \inkdensity \AD(t) \Qcarr(t) + w_{\Qink}(t)
\label{eq:qink_output_model}
\end{equation}
where $w_{\Qink}(t)\sim N(0, \sigma^2_{\Qink})$ represents noise arising from the CNN-based estimation and sampled from a zero-mean Gaussian with standard deviation of $\sigma_{\Qdrop}$ and $\inkdensity$ is the volume percentage of silver in the ink and is defined as $\inkdensity = V_{silver} \big/ \left(V_{silver}+V_{solvent}\right)$. 

\section{Estimating Latent States and Parameters}
\label{sec:estimating_latent_states_parameters}

Digital twins continuously evolve with the physical system to enable accurate predictions under varying conditions and detect anomalous behaviors. For the AJ digital twin, this is achieved by updating the latent states and parameters of its transition and output models through a two-stage estimation process. First, the values of latent states are inferred from output measurements using a state estimation method. Then, the parameters of the transition and output models are periodically updated to reflect changes in the AJ’s operating dynamics.



\subsection{Inferring Latent States of the AJ}

 An Extended Kalman Filter (EKF) is employed to estimate the latent states, denoted by $\hat{x}$, using observable measurements, $\hat{y}$, and inputs, $\hat{u}$. The EKF is a recursive estimator designed for nonlinear systems with Gaussian noise that reconstructs the state vector, $\hat{x}(t)$, by minimizing the expected mean squared estimation error:

\begin{equation}
    \hat{x}(t) = \argmin_{\hat{x}(t)} \mathbb{E}\left[ \|x(t) - \hat{x}
    (t)\|^2 \; \Big| \; y(\tau) \forall \tau \leq t\right],
\end{equation}
where $x(t)$ represents the true and unobservable AJ states and $y(\tau) \forall \tau \leq t$ are the previous measurements.

Before the EKF can begin tracking state evolution during the printing process, it requires an initial state estimate at $t = 0$. For new systems without a known initial condition, the latent state vector must be estimated. For subsequent prints, the initial state is initialized using the latent states from the end of the previous print.

\subsubsection{Estimating Initial Conditions}
For a new system, the initial states are estimated using a constrained optimization method, with the assumption that the hidden states remain approximately constant during the first few time steps given the slow dynamics of the machine. A key motivation for estimating the initial states is that the EKF algorithm is sensitive to initial conditions as poor initialization may place the linearization point in regions where the nonlinear transition and output models are poorly approximated, leading to improper or divergent estimates.  To mitigate this, the initial latent states, $\hat{x}_0$, are estimated by treating them as static over a window of the first $n$ time points (each time step corresponds to one second), where $n=10$ offers a good tradeoff between filtering measurement noise and maintaining the constant-state assumption. This constrained optimization provides a physically consistent initialization, ensuring that the EKF begins with a reliable linearization point for accurate state estimation.

The initial state, $\hat{x}_0$, is estimated by minimizing the mean squared error between the predicted outputs from the digital twin’s output model and the actual observed measurements:

\begin{gather}
\hat{x}_0 = \argmin_{\hat{x}_0} \sum_{k=1}^{n}\left\| \hat{y}_k - g_d(\hat{x}_0, u_k) \right\|^2_{\Sigma_w} \\
A\hat{x}_0=0
\end{gather}
where $g_d(x, u)$ is the discretized nonlinear output model, $u_k$ are the known inputs, $\hat{y}_k$ are the corresponding observed outputs, and $\|v\|^2_{\Sigma}=v^\top\Sigma^{-1}v$. Additionally, the initial values of certain latent variables are assumed to be known (namely deposition in the tube and deposition in the nozzle are zero). This is captured by the added constraint $A\hat{x}_0 = 0$, where $A \in \mathbb{R}^{2\times 5}$ is a matrix of zeros except for ones at the positions corresponding to the tube and nozzle deposition variables, thereby enforcing $\DT = 0$ and $\DN = 0$. 


\subsubsection{Dynamically Estimating Hidden States}

Once the initial state is determined, the Extended Kalman Filter (EKF) is used to estimate the hidden states at all subsequent time steps $k > 1$, based on the system’s nonlinear dynamics with additive Gaussian noise. At each time step, the EKF linearizes the transition and output models around the current estimate to compute the updated state, $\hat{x}(t)$, and the covariance, $\hat{P}(t)$. This recursive process enables real-time tracking of the system’s internal behavior as it evolves. The full EKF estimation workflow, outlined in Algorithm \ref{alg:ekf_alg} \cite{ribeiro2004kalman}, begins by taking the initial state estimate, $\hat{x}_0$, and the initial state covariance, $P_0$, along with the definitions of the discretized state-transition and measurement functions, and their corresponding process and measurement noise covariance matrices. 
At each time step, the EKF first performs a prediction step, which predicts the state (line 4) and the associated error covariance (line 6). It then performs the update step, where the incoming measurements are used to compute the Kalman gain (line 9), which is subsequently used to update both the state estimate (line 10) and the error covariance (line 11).

\begin{algorithm}
\caption{Extended Kalman Filter (EKF)}
\label{alg:ekf_alg}
\begin{algorithmic}[1]
\State{Input: Initial state estimate $\hat{x}_0$, error covariance $P_0$, transition function $f_d(x, u)$, output function $g_d(x, u)$, transition noise covariance $\Sigma_\xi$, output noise covariance $\Sigma_w$, inputs $\{u_k\}$, measurements $\{y_k\}$}

\State{For each time step $k = 1, 2, \dots$}
    \State{ \quad \textbf{Prediction Step:}}
    \State \quad \quad $\hat{x}_{k|k-1} = f_d(\hat{x}_{k-1}, u_{k-1})$ 
    \State{ \quad \quad $F_{k-1} = \left.\frac{\partial f_d}{\partial x}\right|_{\hat{x}_{k-1}, u_{k-1}}$}
    \State{ \quad \quad $P_{k|k-1} = F_{k-1} P_{k-1} F_{k-1}^\top + \Sigma_\xi$}

    \State \quad \textbf{Update Step:}
    \State \quad \quad $H_k = \left.\frac{\partial g_d}{\partial x}\right|_{\hat{x}_{k|k-1}, u_k}$
    \State \quad \quad $K_k = P_{k|k-1} H_k^\top (H_k P_{k|k-1} H_k^\top + \Sigma_w)^{-1}$ 
    \State \quad \quad $\hat{x}_k = \hat{x}_{k|k-1} + K_k \left( y_k - g_d(\hat{x}_{k|k-1}, u_k) \right)$ 
    \State \quad \quad $P_k = (I - K_k H_k) P_{k|k-1}$ 
\end{algorithmic}
\end{algorithm}

\subsection{Updating the AJ Model Parameters}

The EKF approach described above can estimate the hidden states when the transition and output models are probabilistically defined.
But the digital twin framework can also be used for updating the parameters of these models, to ensure accurate tracking of the changes in the real system.
To calibrate the parameters and estimate the hidden states, an Expectation-Maximization (EM) algorithm is adopted in this work.

In EM, jointly updating all parameters of the transition and output functions is challenging because the high-dimensional parameter space can lead to convergence toward poor local optima. This issue is more pronounced when the number of model parameters greatly exceeds the number of outputs, as in the AJ model.
To make the model updates tractable, this work focuses on refining the transition model parameters, specifically the key components where the physically derived formulation diverges from observed behavior:

\begin{itemize}
    \item Median aerosol droplet diameter
    \item Vial level
    \item Tube Deposition
    \item Nozzle deposition
    \item Aerosol volume fraction
\end{itemize}

Although the derivation for these transition models were grounded in first-principles physics, we observe in Appendix \ref{app:variations_experiment} that their values fluctuate significantly over the course of a single print. The variability observed in these transition functions deviates from the physical model, which  suggests that the initial physics-based parameters are either incomplete or influenced by unmodeled system dynamics, such as complex fluidic interactions or coupling effects. 
To address these discrepancies, the digital twin dynamically updates the corresponding parameters in the transition model, represented by the vector $\theta = \left[\theta_{da}, \theta_{\Vl},\theta_{\RT}, \theta_{\RN}, \theta_{\AD}\right]$, where $\theta_{da}, \theta_{\Vl},\theta_{\RT}, \theta_{\RN}, \theta_{\AD}$ are the partial terms of the transition model for median aerosol droplet diameter, vial level, tube deposition, nozzle deposition, and aerosol volume fraction, respectively, as defined above.

The EM algorithm seeks to maximize the likelihood estimate of the model parameters, $\theta$, by iteratively applying two steps. First, the expectation step uses the EKF and its smoothing counterpart (Rauch–Tung–Striebel smoother \cite{rauch1965maximum}) to compute the expected value of the hidden variables given the model parameter estimates $\theta^{(i)}$. Then, in the maximization step, the model parameters, $\theta^{(i+1)}$, are updated by maximizing the complete log-likelihood, fixing the hidden variables at the value computed in the expectation step. The resulting EM step is performed for a set time-window and is shown in Algorithm \ref{alg:em_algorithm}.

\begin{algorithm}[h!]
\caption{Expectation-Maximization for EKF Parameter Estimation}
\label{alg:em_algorithm}
\begin{algorithmic}[1]
\State{Input: Initial parameter estimates $\theta^{(0)}$, observations $\{y_j\}_{j=k-T}^k$, inputs $\{u_j\}_{j=k-T}^k$, time-step $k$, time-window $T$}
\State{Set iteration index $i = 0$}
\State{Repeat:}
    \State{\textbf{Expectation-step:}}
    \State{\quad Compute smoothed estimates $\{\hat{x}_k\}$ and covariances $\{P_k\}$ via EKF (Algorithm \ref{alg:ekf_alg}) and RTS smoother (Algorithm \ref{alg:rts_smoother}) with parameters $\theta^{(i)}$}
    \State{\quad }

    \State{\textbf{Maximization-step:}}
    \State{\quad Update transition model parameters $\theta^{(i+1)}$ by minimizing:
    \[
        \sum_{j=k-T}^{k} \left\|  \hat{x}_{j+1} - f_d(\hat{x}_j, u_j; \theta) \right\|^2_{\Sigma_\xi}
    \]}
    \State{ Increment iteration index: $i \gets i + 1$}
\State{ Until convergence or maximum iterations}
\State{Return: Final parameter estimates: $\theta^{(i+1)}$}
\end{algorithmic}
\end{algorithm}

\section{Experimental Validation}

The performance of the AJ digital twin is evaluated through three experiments that demonstrate its ability to identify the root causes of unobservable output variations and evolve the twin with the physical system by updating its parameters to improve its prediction accuracy. The values of the covariance matrix used in the transition and output functions are provided in Appendix \ref{app:covariance}.

\subsection{Experiment 1: Inferring AJ States Through Digital Twin Estimation}
\label{sec:exp1_estimation}

In the first experiment, we compare the predictive performance of a static, physics-based digital model with that of the closed-loop state estimation in the digital twin (images from print shown in Figure \ref{fig:sup_dat_8_1}). 

\begin{figure*}[t!]
    \centering
    \begin{subfigure}[t]{0.3\textwidth}
        \centering
        \includegraphics[height=1.2in]{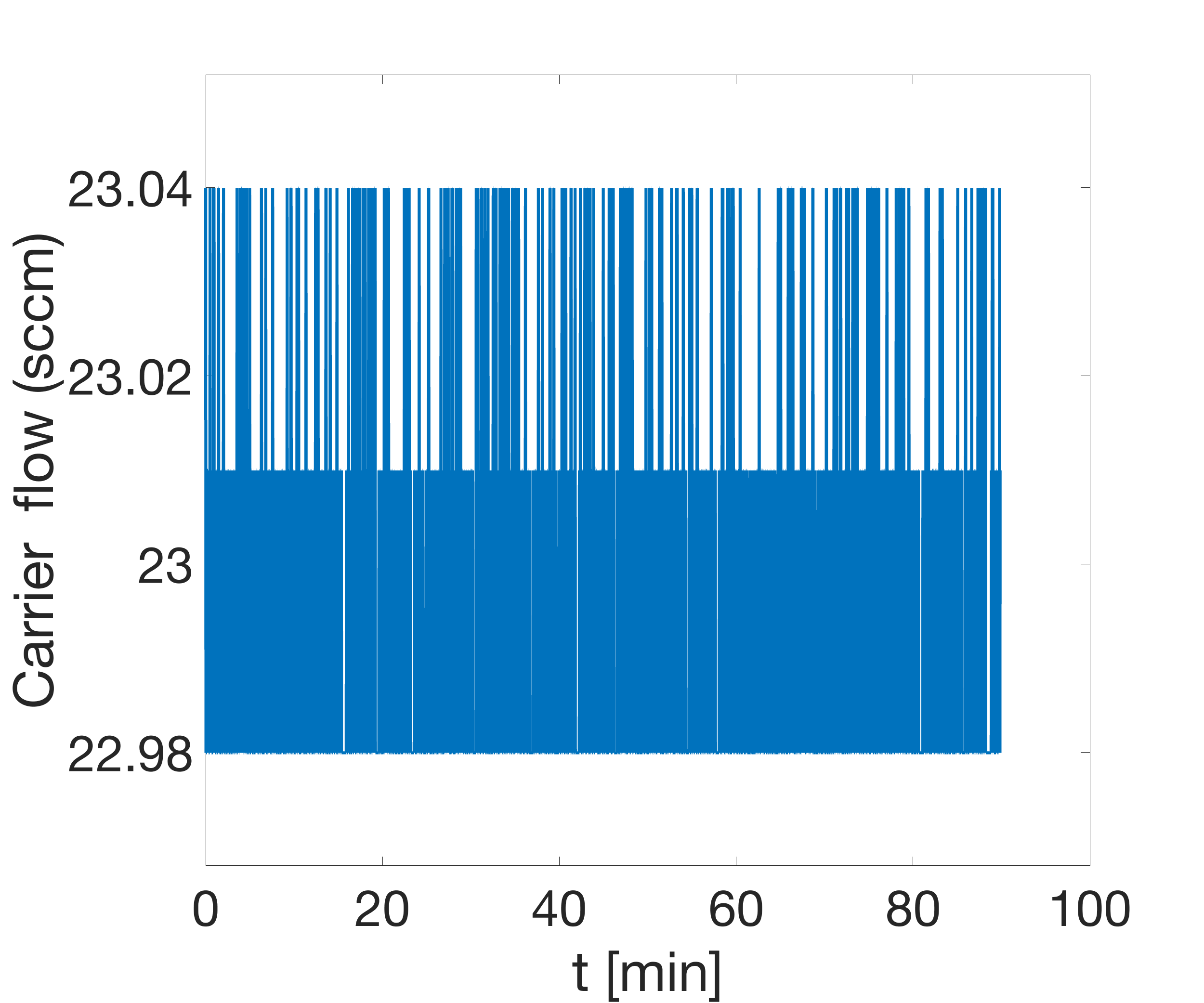}
        \caption{}
    \end{subfigure}%
    ~ 
    \begin{subfigure}[t]{0.3\textwidth}
        \centering
        \includegraphics[height=1.2in]{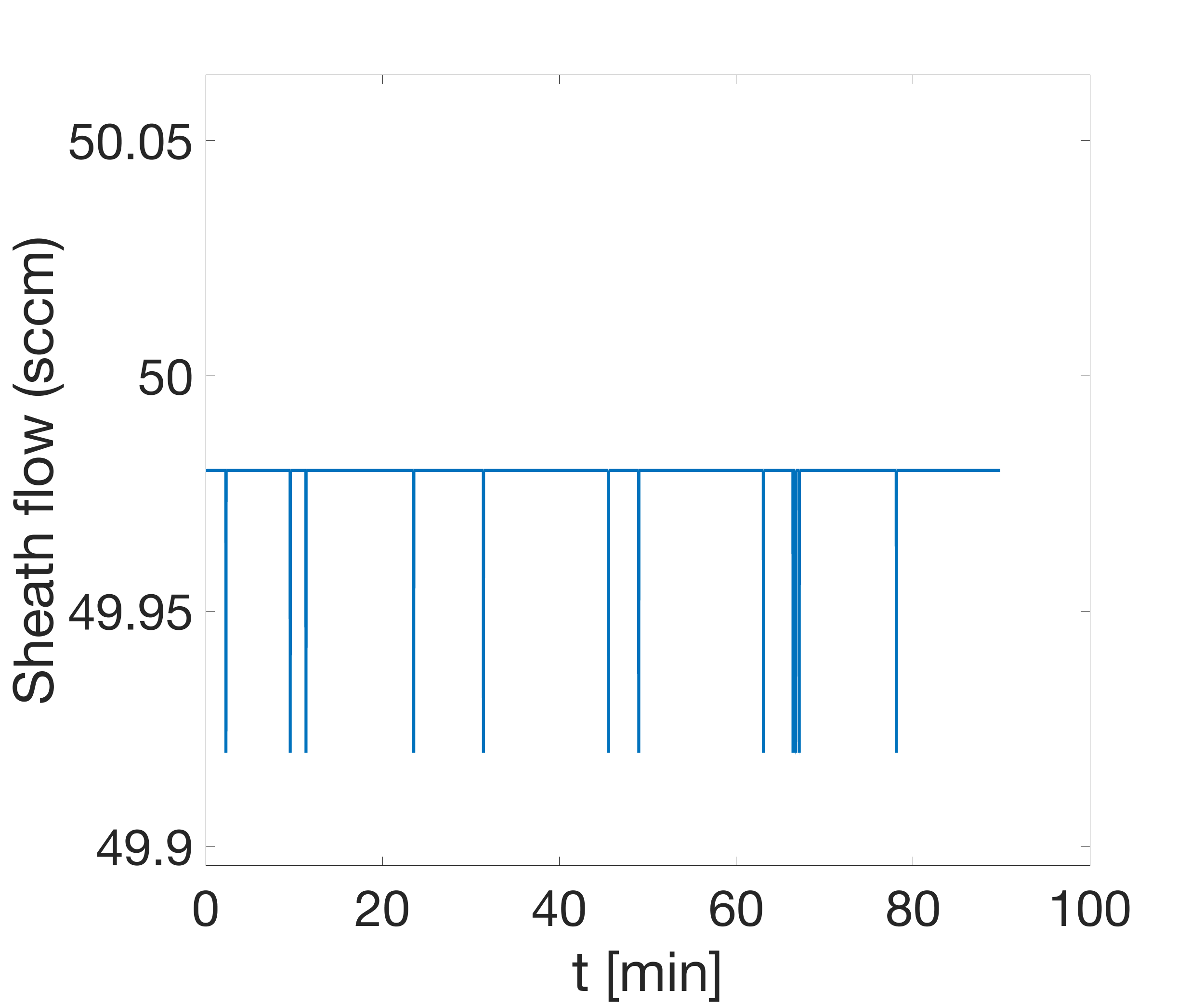}
        \caption{}
    \end{subfigure}
        ~ 
    \begin{subfigure}[t]{0.3\textwidth}
        \centering
        \includegraphics[height=1.2in]{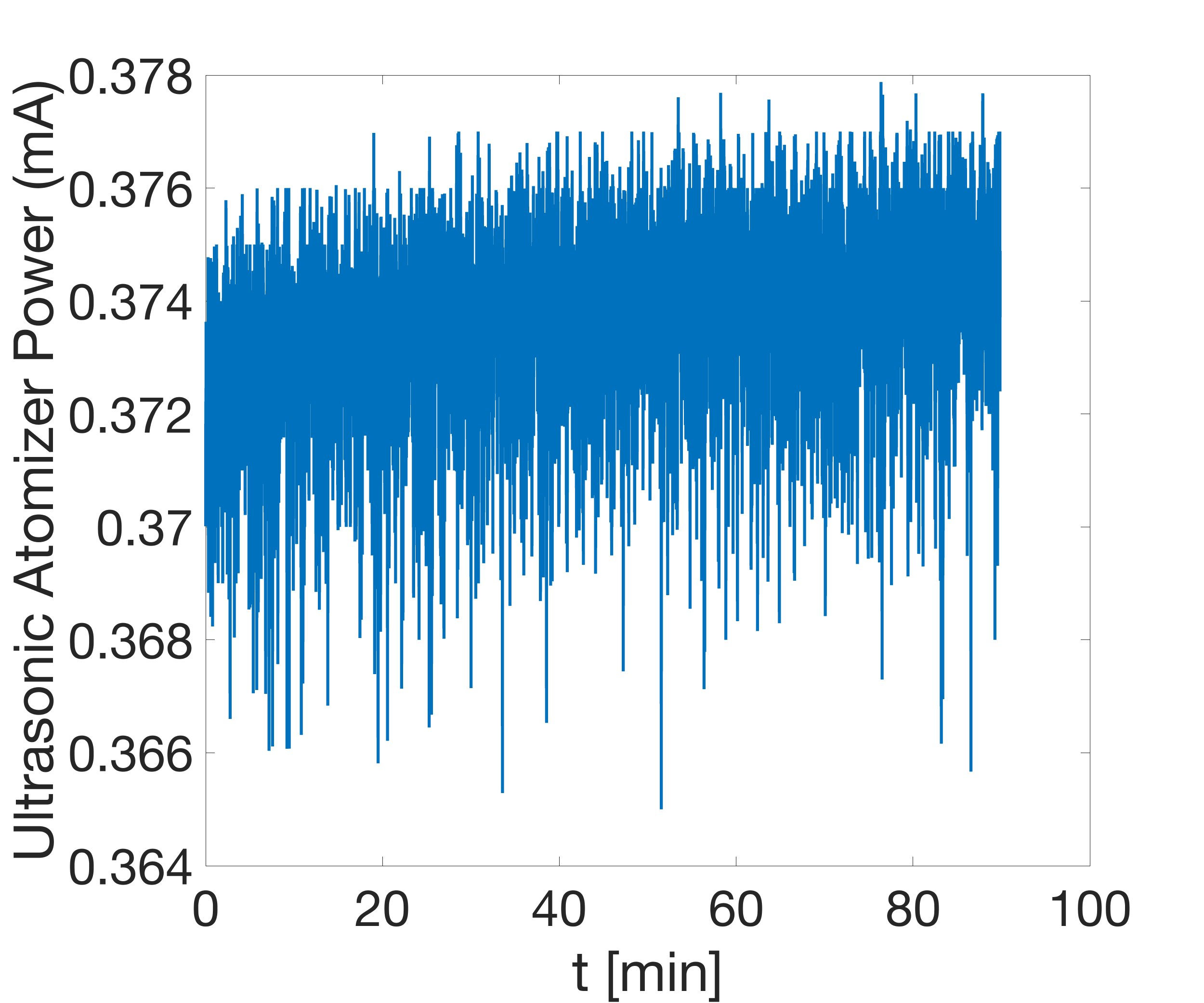}
        \caption{}
    \end{subfigure}
    \caption{Inputs used for the experiment in Section \ref{sec:exp1_estimation} include: (a) measured carrier flow rate, $\Qcarr$, set to a nominal value of 23 sccm; (b) measured sheath flow rate, $\Qsh$, set to a nominal value of 50 sccm; and (c) measured ultrasonic atomizer current, $\PA$, set to a nominal value of 0.375 mA. Variations in the flow rates and atomizer current arise from the limited precision of the mass flow controllers (MFCs) and the inherent limitations of the atomizer in maintaining a constant current.}
    \label{fig:exp1_inputs}
\end{figure*}

\begin{figure*}[t!]
    \centering
    \begin{subfigure}[t]{0.3\textwidth}
        \centering
        \includegraphics[height=1.2in]{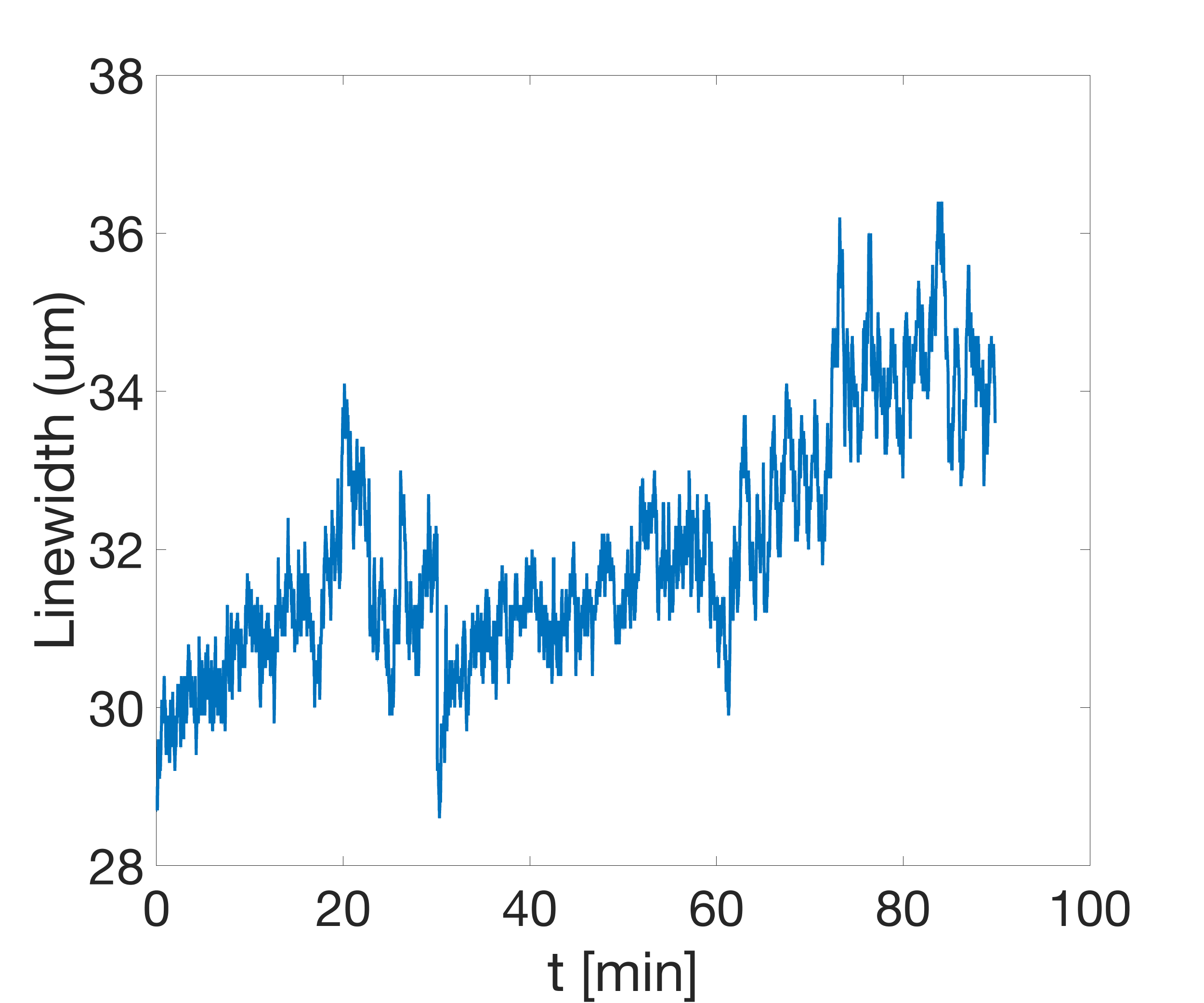}
        \caption{}
    \end{subfigure}%
    ~ 
    \begin{subfigure}[t]{0.3\textwidth}
        \centering
        \includegraphics[height=1.2in]{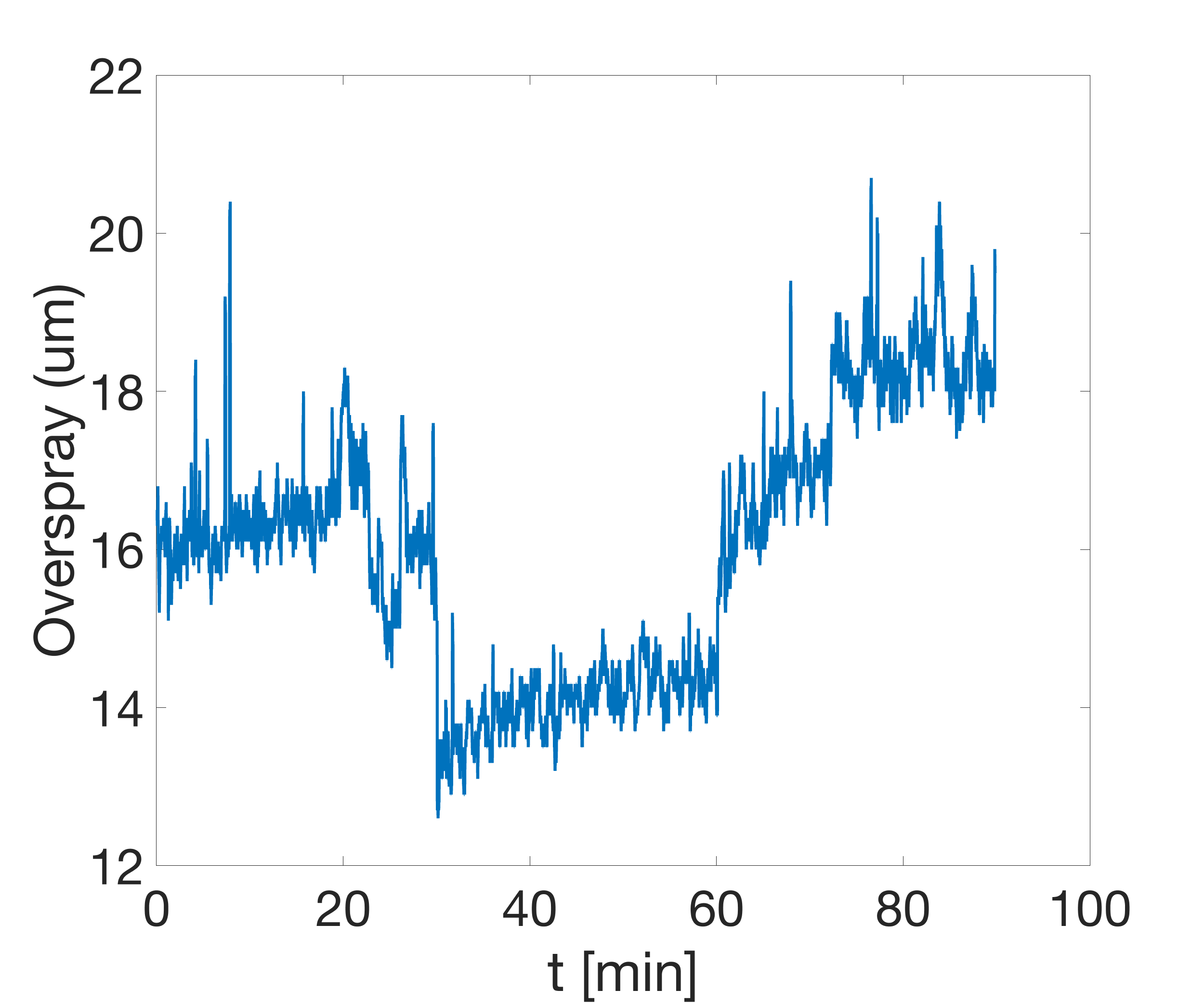}
        \caption{}
    \end{subfigure}
        ~ 
    \begin{subfigure}[t]{0.3\textwidth}
        \centering
        \includegraphics[height=1.25in]{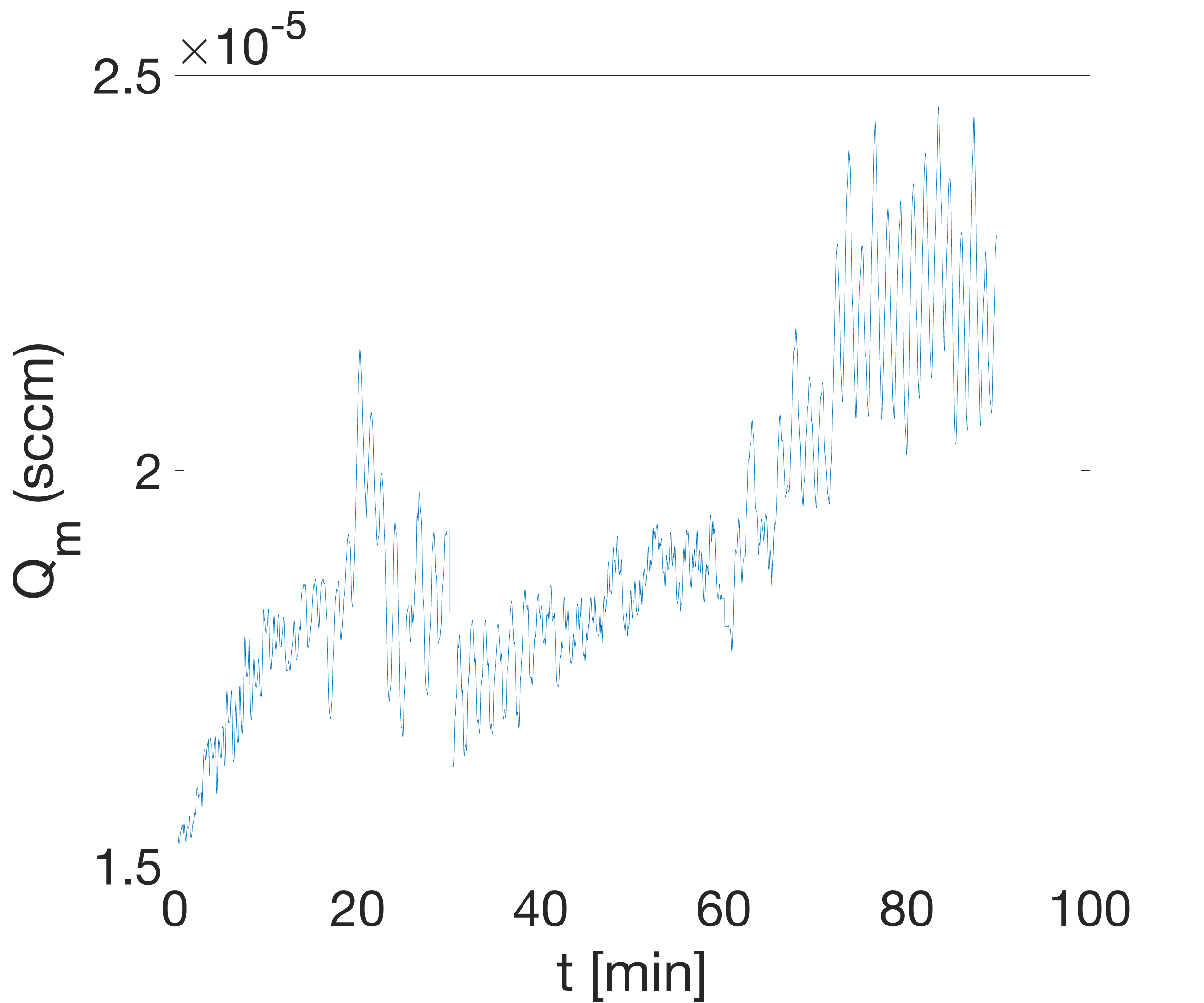}
        \caption{}
    \end{subfigure}
    \newline
    \begin{subfigure}[t]{0.3\textwidth}
        \centering
        \includegraphics[height=1.2in]{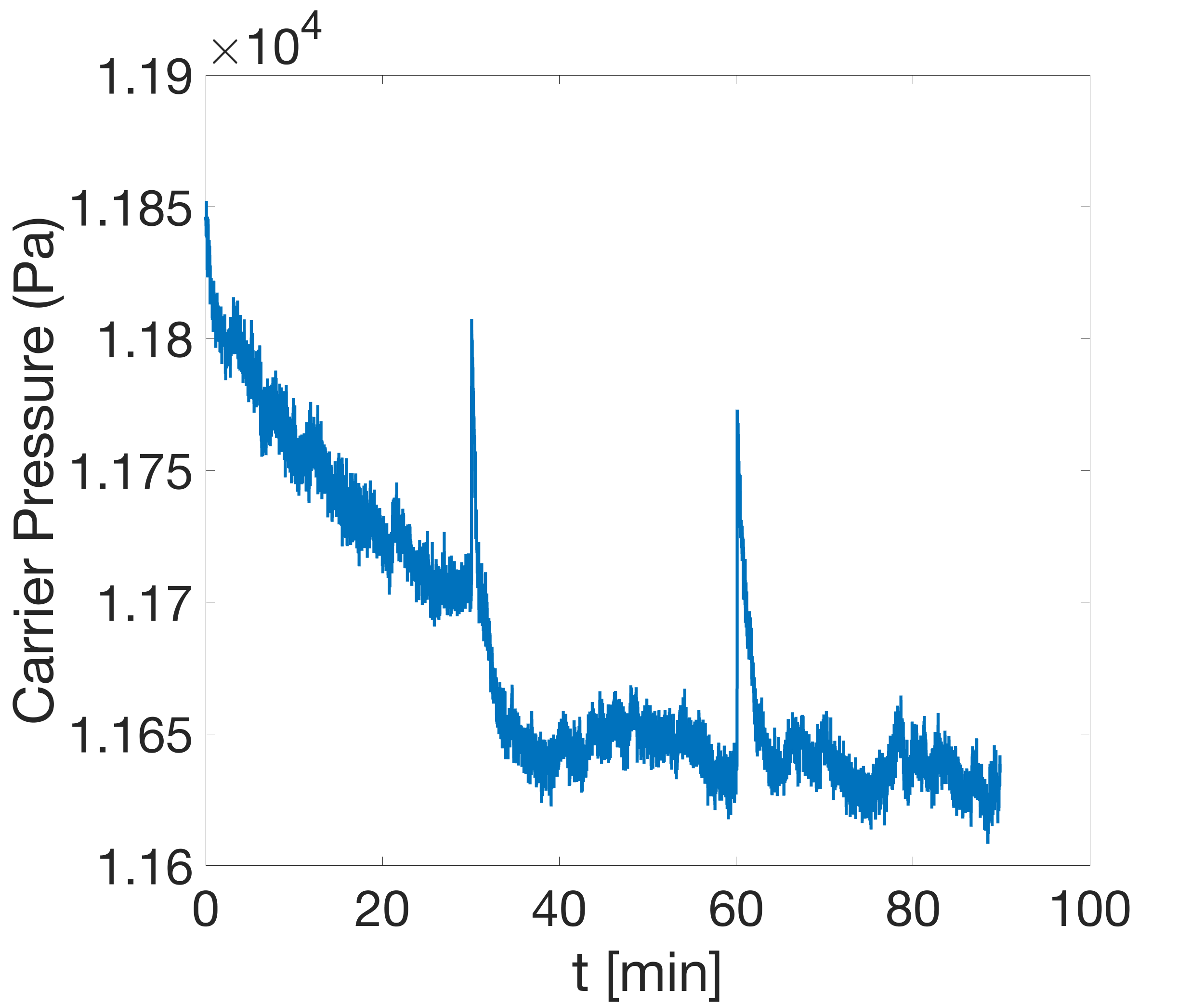}
        \caption{}
        \label{fig:exp1_carrier_pressure}
    \end{subfigure}
        \begin{subfigure}[t]{0.3\textwidth}
        \centering
        \includegraphics[height=1.2in]{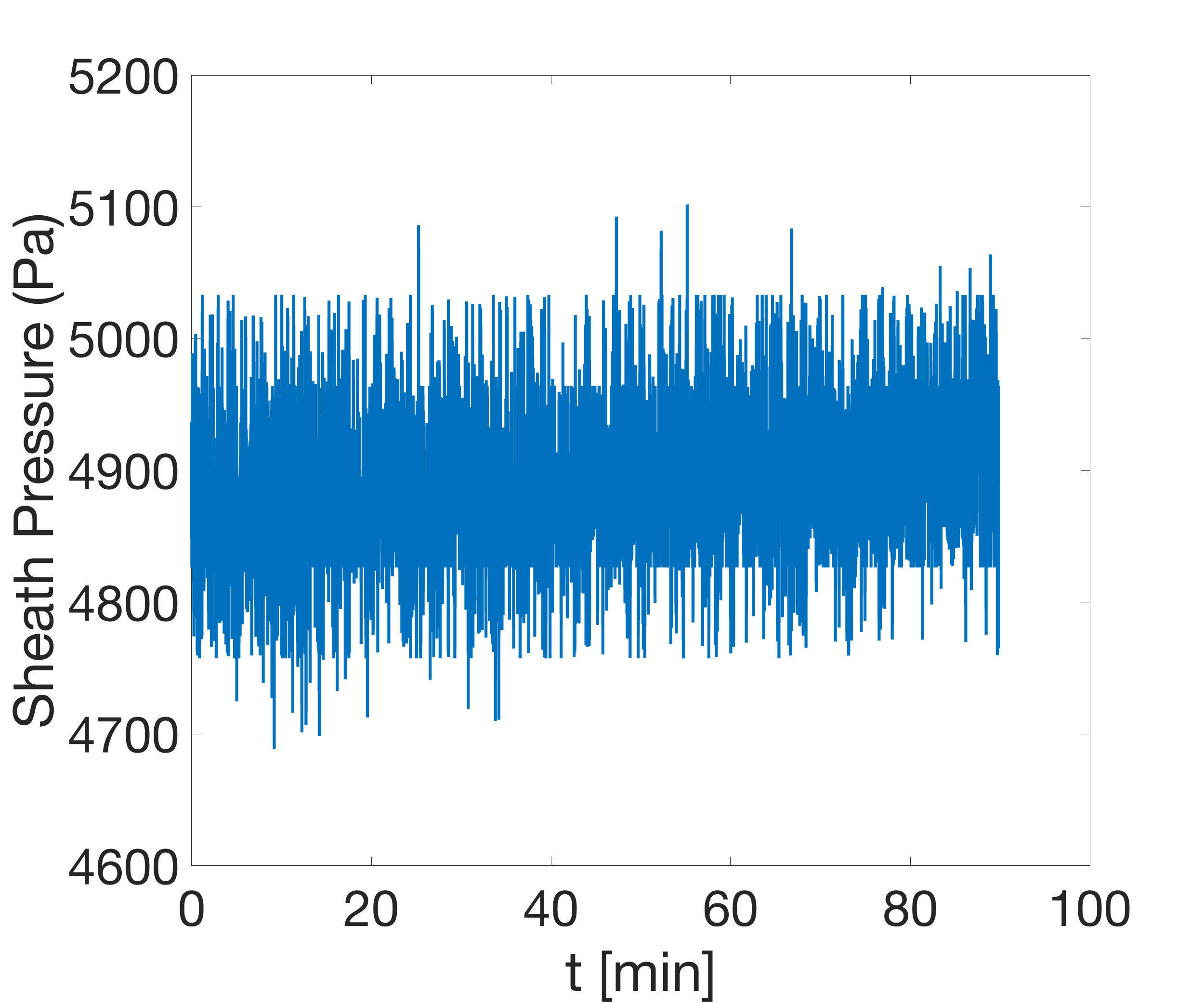}
        \caption{}
    \end{subfigure}
    \caption{Recorded outputs for the experiment in Section \ref{sec:exp1_estimation} include: (a) linewidth, $\lw$, (b) overspray, $\ov$, (c) deposited material flow rate, $\Qink$ (filtered with a 40-second moving average), (d) output carrier pressure, $\Pcarr$ and (e) output sheath pressure, $\Psh$.}
    \label{fig:exp1_outputs}
\end{figure*}

The experiment is driven by a set of inputs shown in Figure \ref{fig:exp1_inputs}, with the corresponding measured outputs presented in Figure \ref{fig:exp1_outputs}. Although the inputs are set to be constant throughout the print, variations in the outputs are still observed. This highlights the need to infer changes in latent system states that contribute to the observed process variations. Specifically, two approaches are compared for predicting the evolution of these internal states:
\begin{enumerate}
    \item Open-loop prediction using the physics-derived transition and output models without incorporating observational feedback.
    \item Closed-loop state estimation using the digital twin which continuously updates latent states and parameters (using EM and EKF) based on observed outputs.
\end{enumerate}

 To validate the accuracy of the inferred latent states using both methods, direct measurements of the latent states are taken at discrete time points. These ground truth values (listed in Appendix~\ref{app:experiment_1_details}, Tables \ref{tab:median_diameter_ex1}-\ref{tab:aerosol_density_ex1}) measure median aerosol droplet diameter, vial level, ink deposition in the tube and nozzle, and aerosol volume fraction and are taken at discrete time points as described in Appendix \ref{app:experimental_setup_validation}. However, the invasive procedures used to measure latent states alter the machine’s configuration, introducing step changes in its output. These changes appear as shifts in linewidth and overspray in Figure \ref{fig:exp1_outputs}, aligning with the time points at which the measurements for the latent states occur. Additionally, high variation in the measured aerosolized material flow rate ($Q_m$) required that the data be smoothed using a simple averaging filter of length 40.

\begin{figure*}[h!]
    \centering
    \begin{subfigure}[t]{0.3\textwidth}
        \centering
        \includegraphics[height=1.2in]{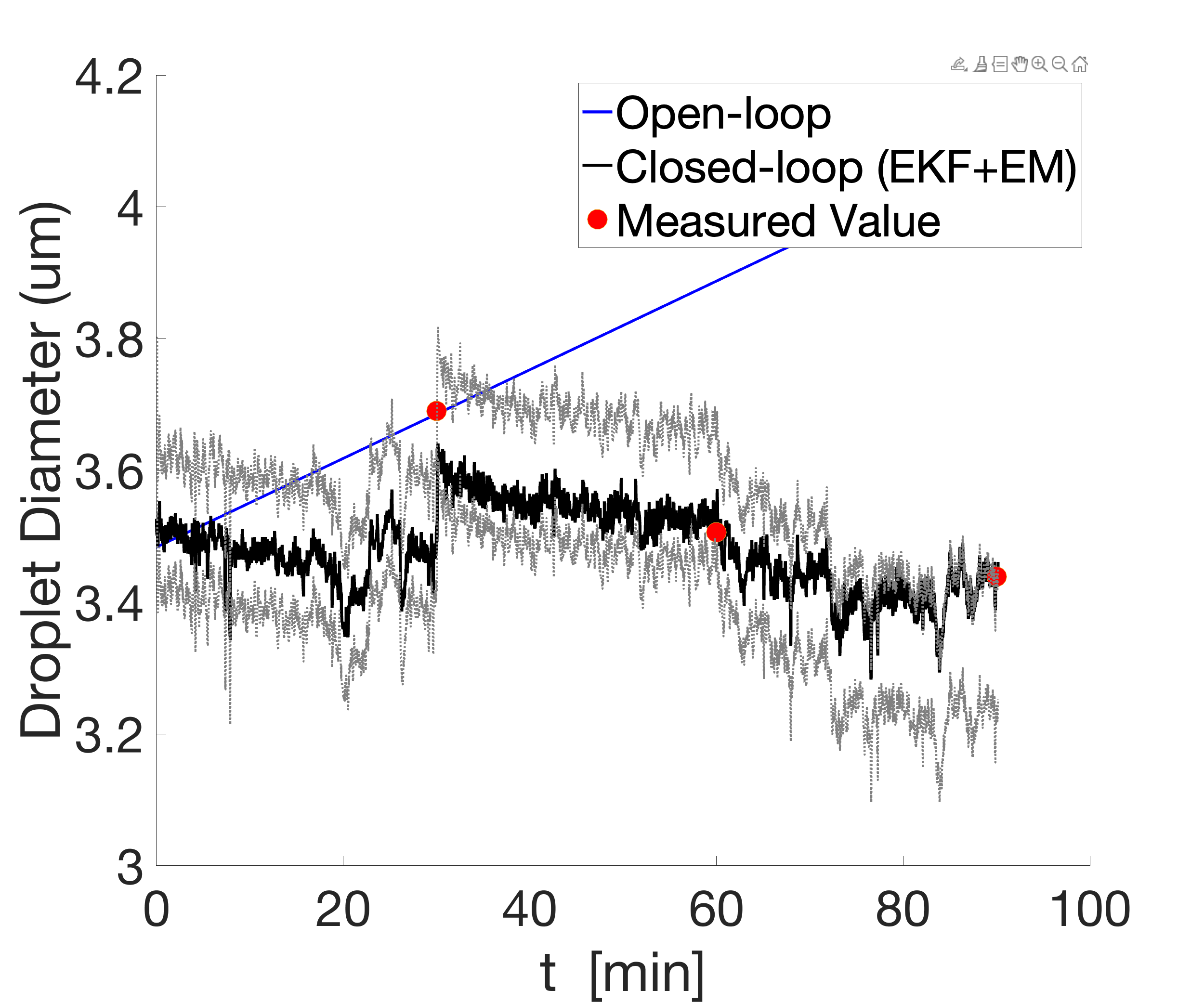}
        \caption{}
    \end{subfigure}%
    ~ 
    \begin{subfigure}[t]{0.3\textwidth}
        \centering
        \includegraphics[height=1.2in]{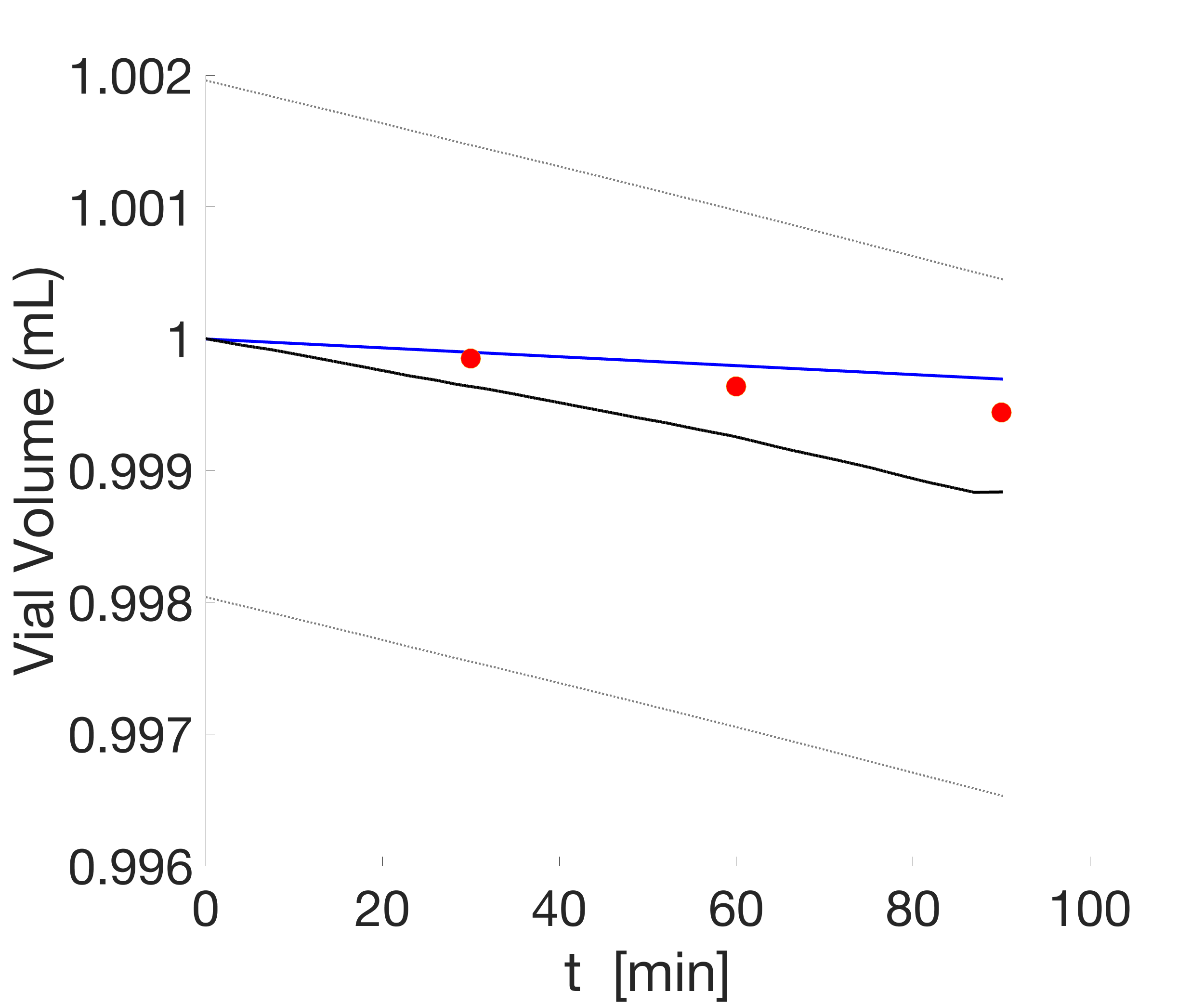}
        \caption{}
    \end{subfigure}
        ~
    \begin{subfigure}[t]{0.3\textwidth}
        \centering
        \includegraphics[height=1.2in]{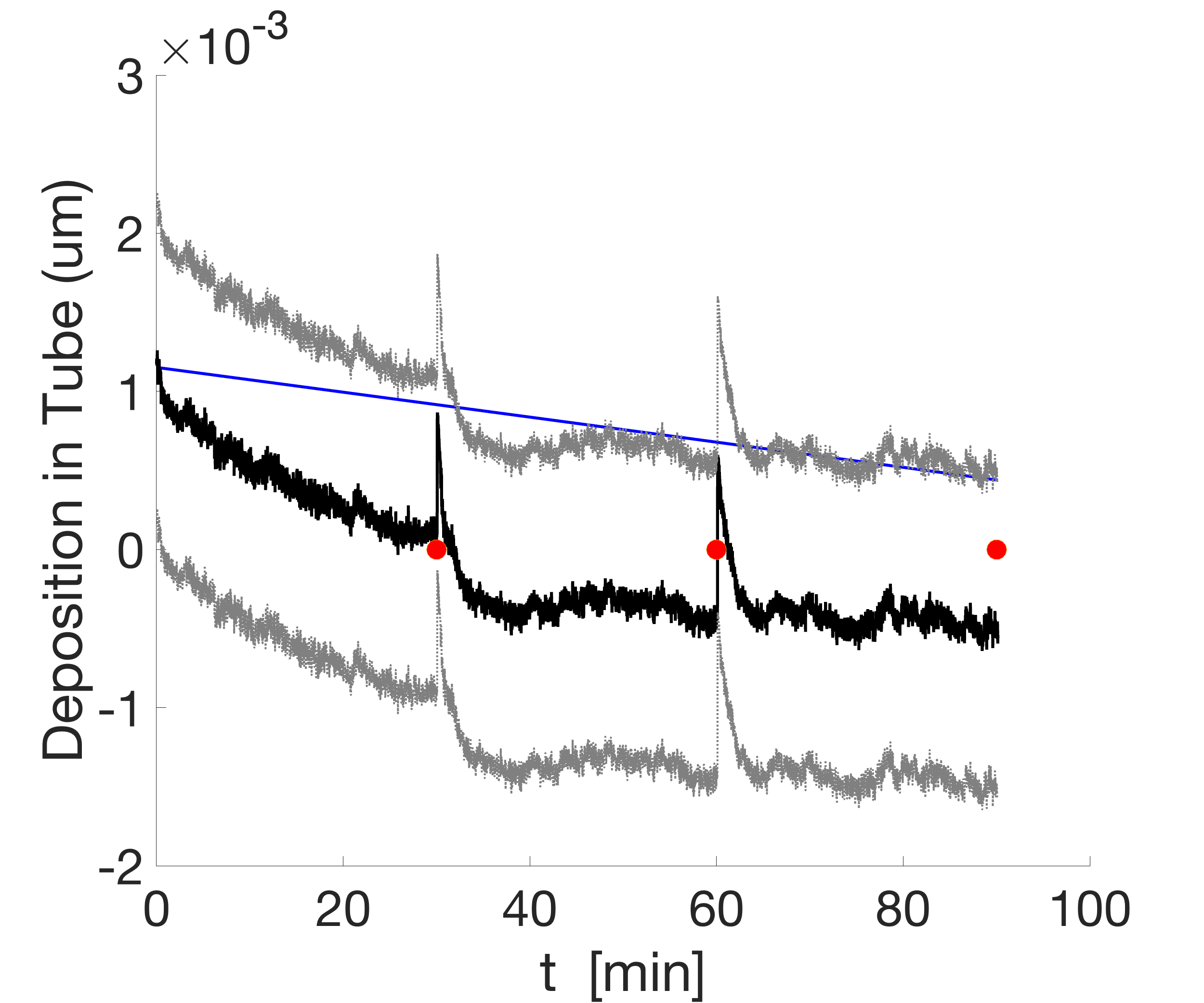}
        \caption{}
        \label{fig:open_loop_prediction_exp1_tube_dep}
    \end{subfigure}
    
        \begin{subfigure}[t]{0.3\textwidth}
        \centering
        \includegraphics[height=1.2in]{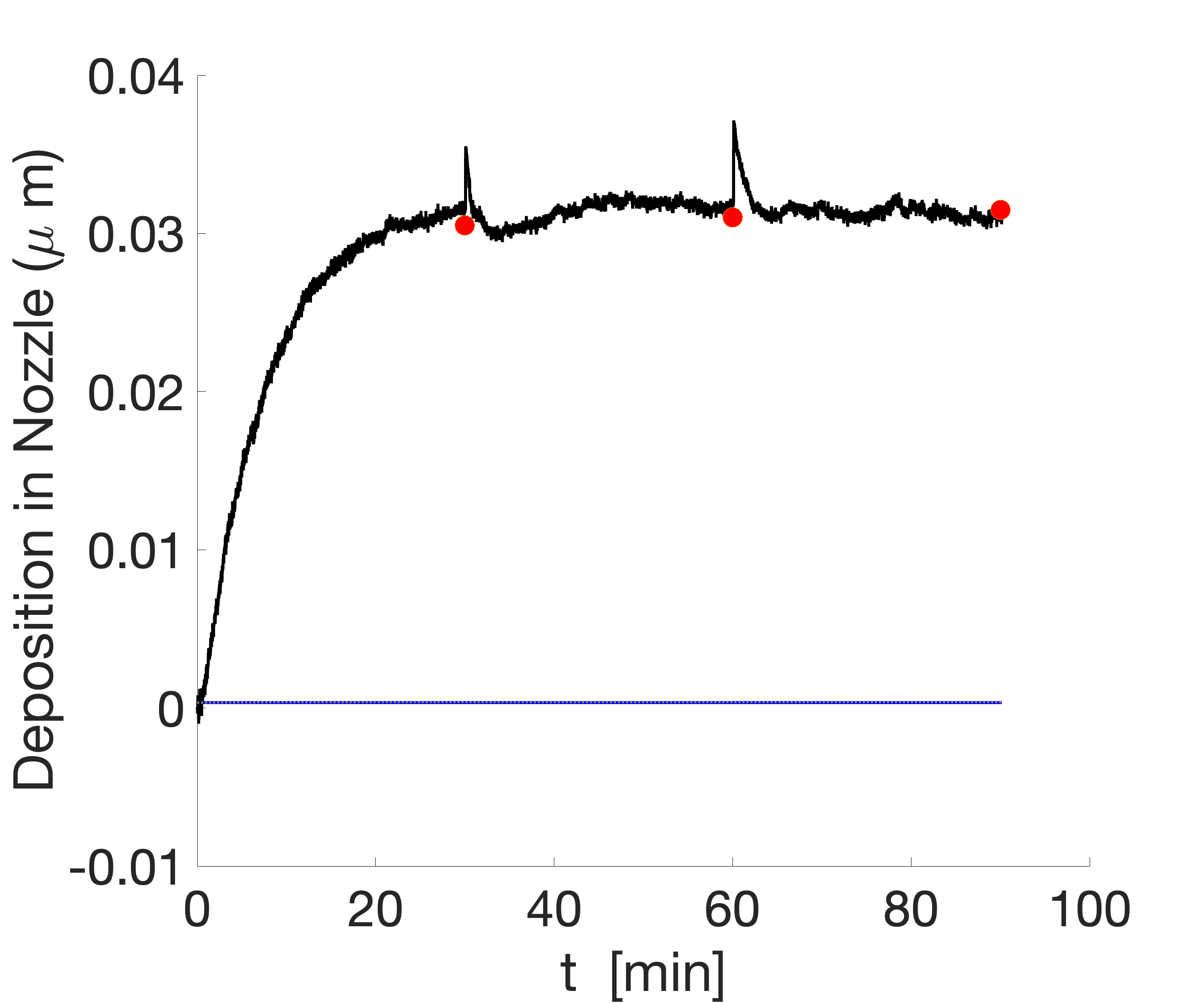}
        \caption{}
    \end{subfigure}
        ~ 
    \begin{subfigure}[t]{0.3\textwidth}
        \centering
        \includegraphics[height=1.2in]{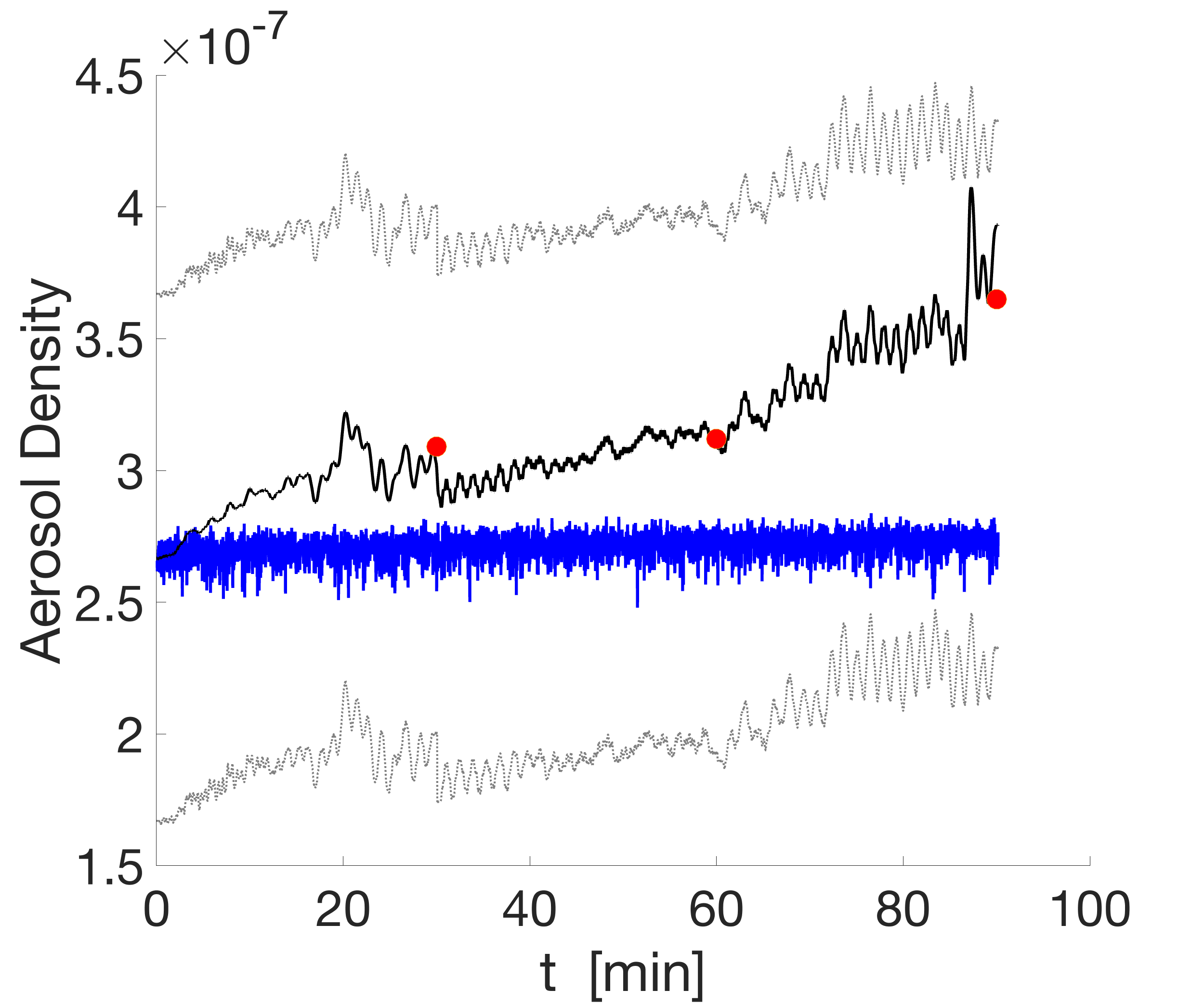}
        \caption{}
    \end{subfigure}
    \caption{Open-loop prediction for the experiment in Section \ref{sec:exp1_estimation} using static physics-derived model of the latent states compared to the closed-loop prediction of the digital twin updated using Expectation-Maximization and the measured values of latent states at discrete time-points. The measured latent states include (a) median aerosol droplet diameter out of the atomizer, $\DA$, (b) ink solution volume in vial, $\Vl$, (c) ink deposition in the tube, $\DT$, (d) ink deposition in the nozzle, $\DN$, (e) fraction of aerosol droplets in the carrier stream, $\AD$. The measurement of the latent states are within the confidence bounds ($\pm 2\sigma$) of the closed-loop estimation, shown by the dotted lines.}
    \label{fig:open_loop_prediction_exp_1}
\end{figure*}

The results of the open-loop and closed-loop latent state inference are shown in blue and black, respectively, in Figure \ref{fig:open_loop_prediction_exp_1}, with direct validation measurements marked by red dots.

\begin{figure*}[h!]
    \centering
    \begin{subfigure}[t]{0.3\textwidth}
        \centering
        \includegraphics[height=1.2in]{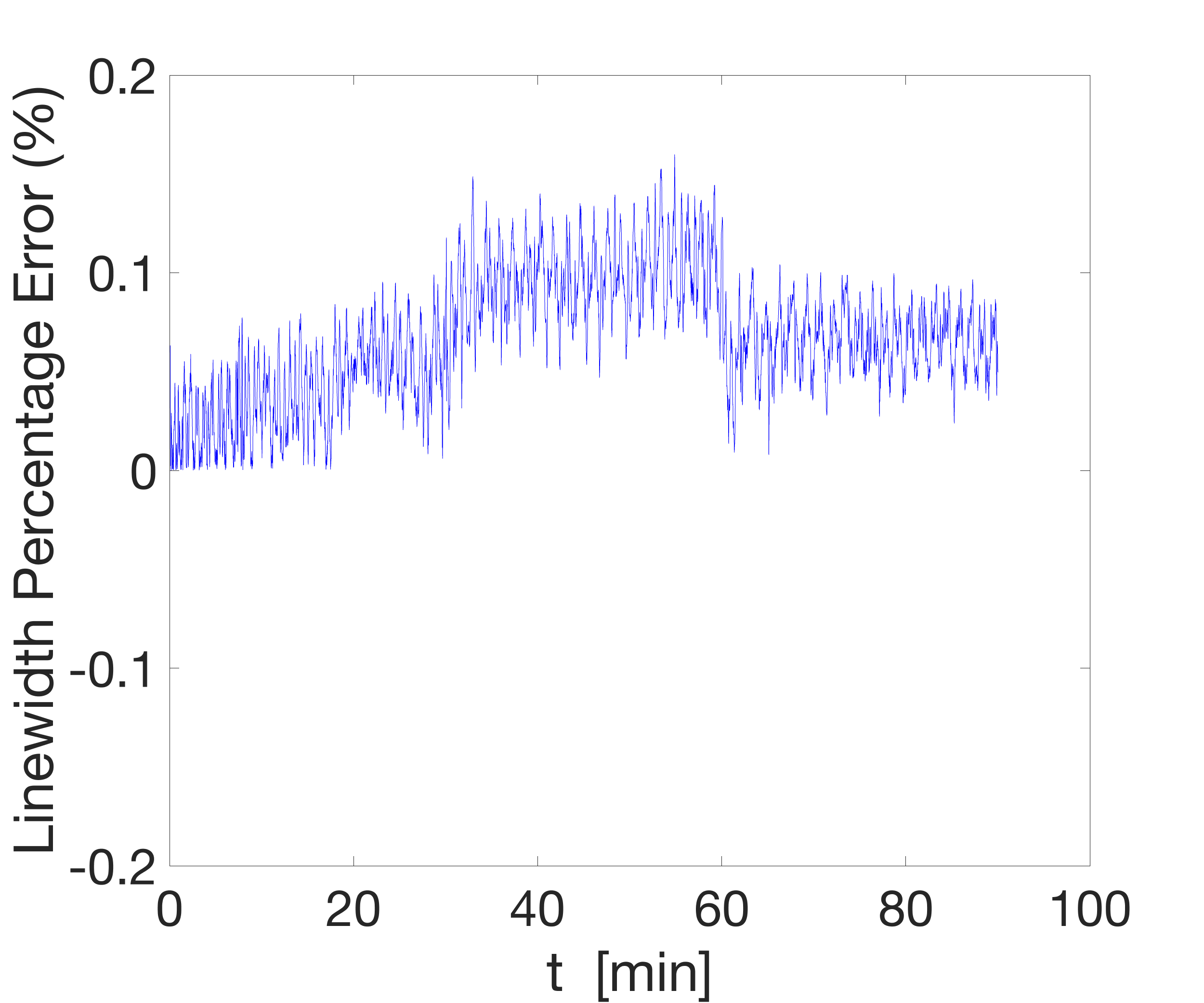}
        \caption{}
    \end{subfigure}%
    ~ 
    \begin{subfigure}[t]{0.3\textwidth}
        \centering
        \includegraphics[height=1.2in]{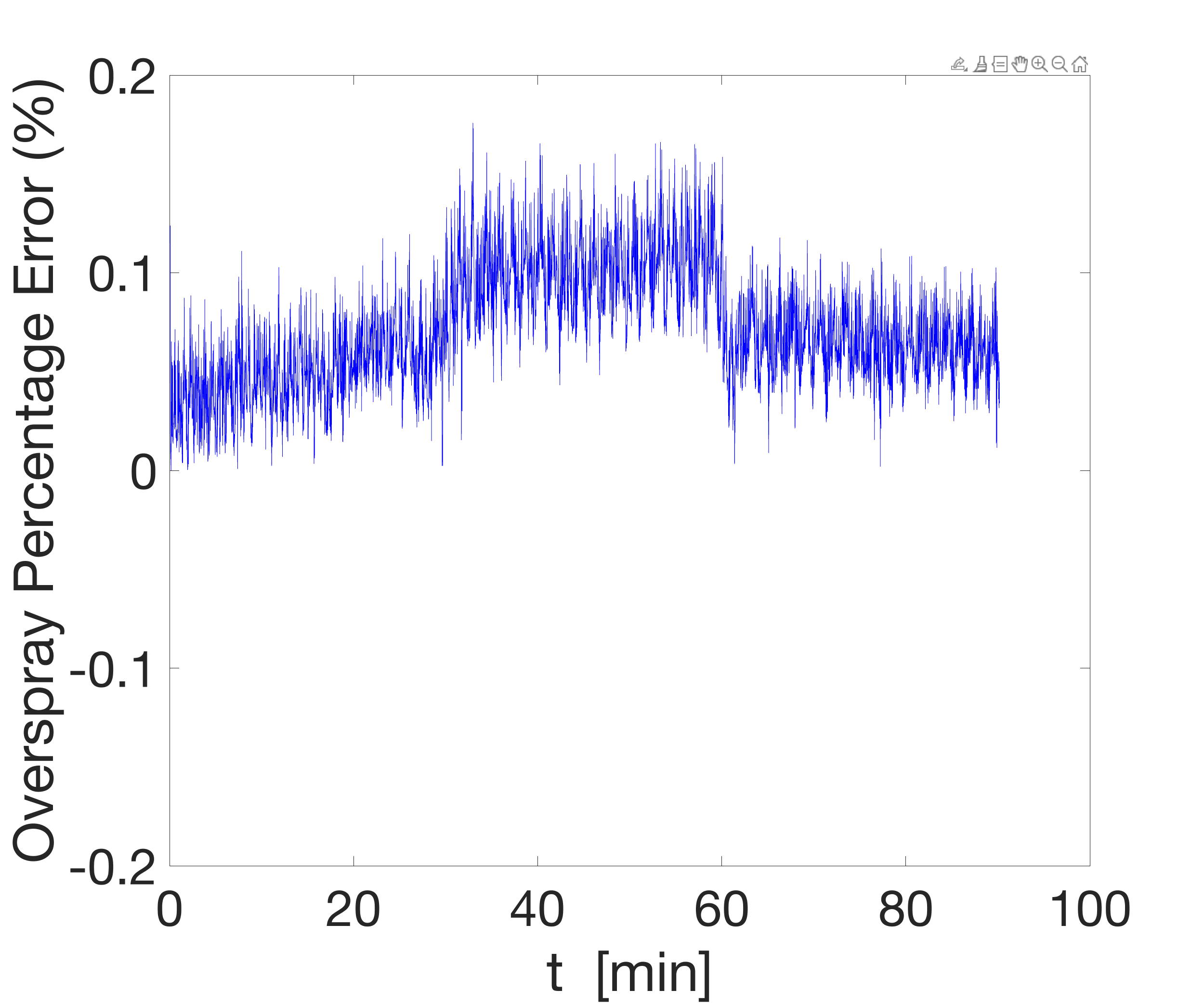}
        \caption{}
    \end{subfigure}
        ~ 
    \begin{subfigure}[t]{0.3\textwidth}
        \centering
        \includegraphics[height=1.2in]{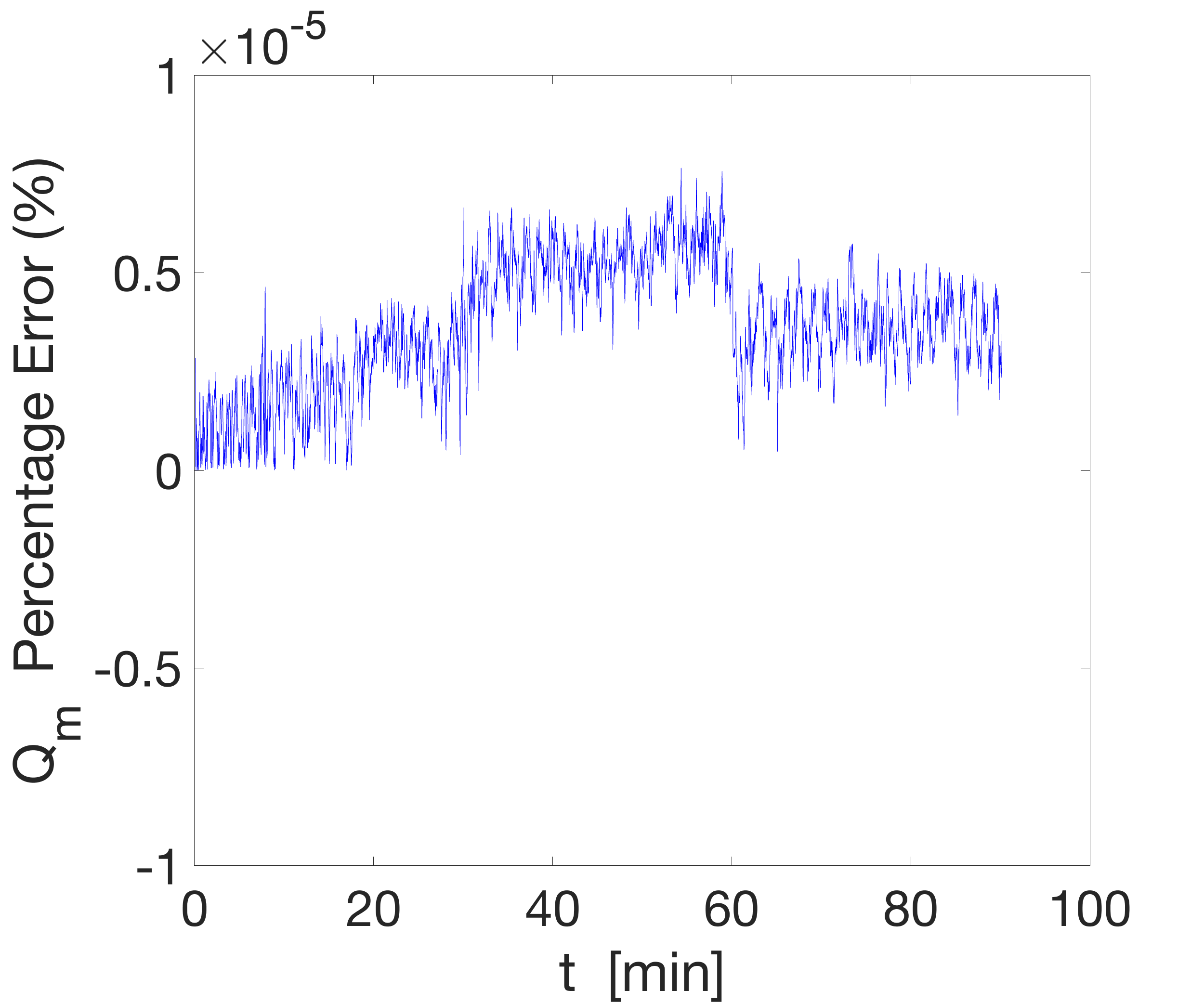}
        \caption{}
    \end{subfigure}
    \newline
    \begin{subfigure}[t]{0.3\textwidth}
        \centering
        \includegraphics[height=1.2in]{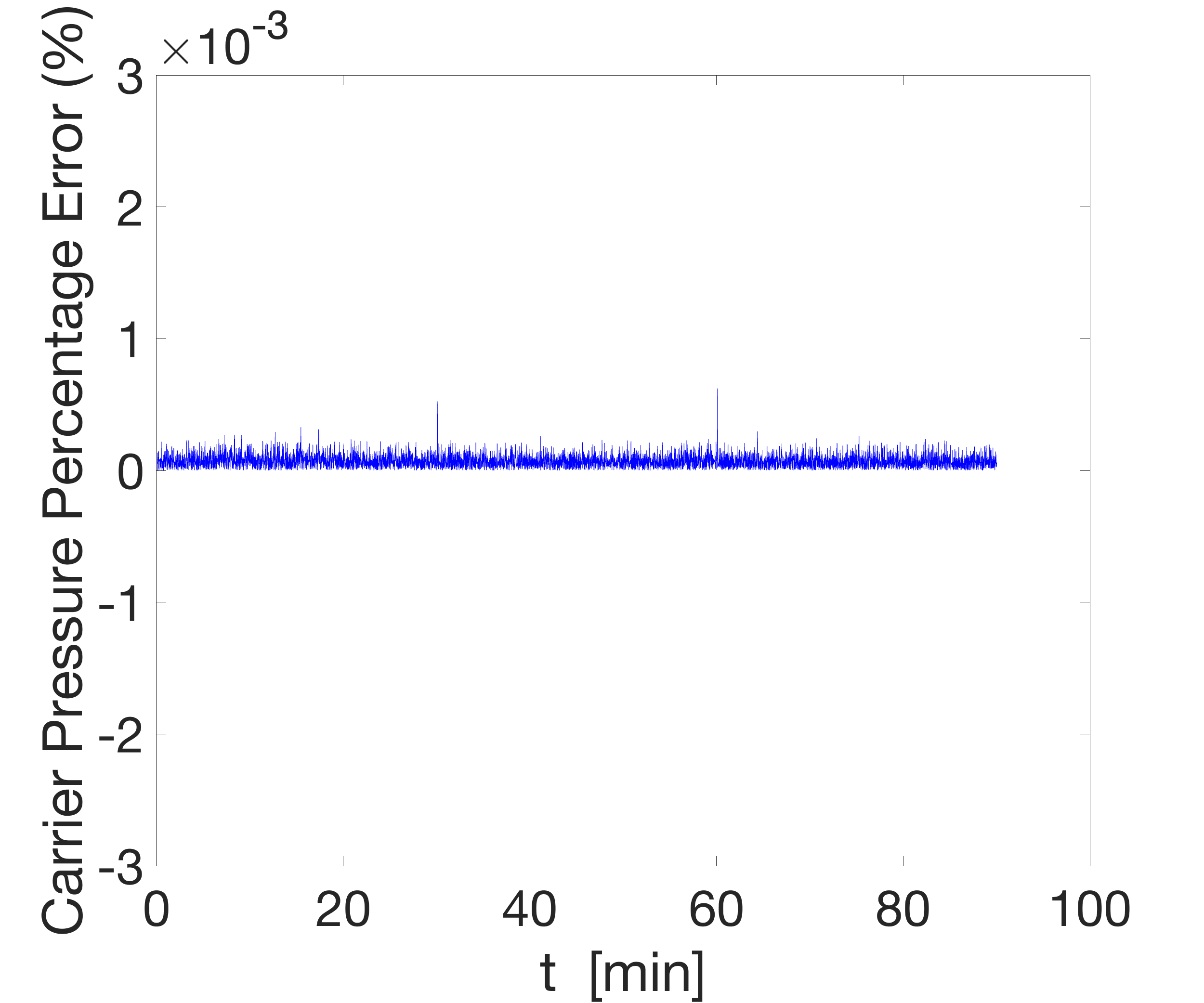}
        \caption{}
    \end{subfigure}
        \begin{subfigure}[t]{0.3\textwidth}
        \centering
        \includegraphics[height=1.2in]{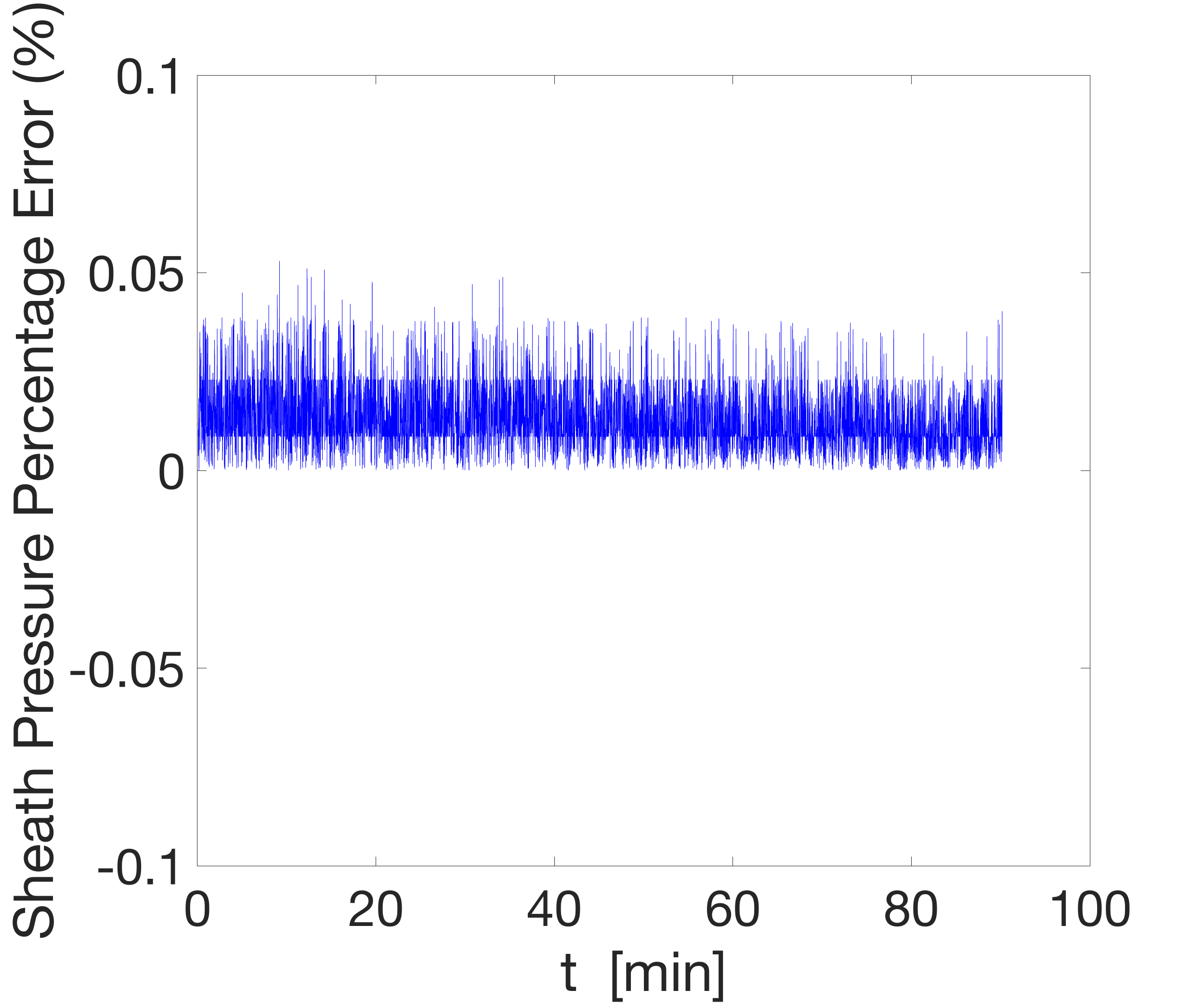}
        \caption{}
    \end{subfigure}
    \caption{The percentage error for the experiment in Section \ref{sec:exp1_estimation} between the measured system output and the output predicted from states determined by the closed-loop Extended Kalman Filter for (a) linewidth, $\lw$, (b) overspray, $\ov$, (c) deposited material flow rate, $\Qink$, (d) output carrier pressure, $\Pcarr$, (e) sheath output pressure, $\Psh$.}
    \label{fig:digital_twin_outputs_exp1}
\end{figure*}

To validate that the inferred latent states produce consistent results with the measurements, the predicted outputs are computed using the nonlinear output model. Figure \ref{fig:digital_twin_outputs_exp1} presents the percentage error between the output predictions from the closed-loop estimation and the corresponding measurements. The strong agreement between predictions and measurements indicates that the inferred latent states accurately capture the system’s underlying behavior, as they reproduce the observed outputs despite potential discrepancies introduced by modeling errors and process noise.

In the open-loop prediction, the model uses only the initial state and known inputs to simulate the latent state evolution via the transition model.
Due to the incomplete transition model for droplet size, aerosol volume fraction, and nozzle deposition parameters, the open-loop approach fails to accurately capture process drifts observed in the physical system. These models cannot capture complex physical processes; for example, variations in the median aerosol droplet diameter can arise from fluctuations in changes in ink viscosity, temperature, or stochastic aerosol-liquid interactions in the vial. Likewise, aerosol density may fluctuate due to inconsistencies in carrier gas flow rates or aerosol agglomeration. Additionally, ink deposition inside the nozzle can be caused by solvent evaporation and particulate buildup. Consequently, the predicted outputs diverge significantly from the measured data, particularly for linewidth, overspray, and aerosolized ink flow rate.

In contrast, the digital twin framework incorporates closed-loop estimation that infers the most likely latent states at each time step by integrating observed measurements with model dynamics. The estimated state variables reveal clear trends (most notably the median aerosol droplet diameter and aerosol volume fraction), which show drift over time. In each case, the inferred states from the digital twin fall within 10\% of the measured values, confirming the reliability of the EKF-based estimation. Using these updated latent states, the output model produces predictions that closely match the measured outputs, as illustrated in Figure \ref{fig:digital_twin_outputs_exp1}.


One notable result is the deposition thickness in the nozzle, which is inferred to be approximately 30 nm different from the nominal initial setting. However, small ink depositions result in noticeable changes in linewidth and overspray. This highlights the importance of accurately inferring the nozzle dimensions through the closed-loop digital twin. This offset is detected by leveraging observations of the carrier and sheath pressures through the EKF. In contrast, the open-loop model cannot capture such discrepancies and therefore fails to account for their impact on the process.

By tracking the latent states, the digital twin enables deeper insight into variations of the AJ outputs. For example, deviations in linewidth and overspray correlate strongly with gradual changes in the estimated droplet diameter, which is unobservable by the open-loop physics model. Additionally, the digital twin captures variations in the material flow rate, $\Qink$, which is highly sensitive to complex fluid dynamics and material interactions within the system. These unpredictable fluctuations stem from changes in ink properties such as viscosity, ink deposition that modifies flow paths, and variations in environmental conditions like temperature and humidity. By integrating measurement data, the digital twin continuously estimates these latent state changes, enabling accurate modeling of aerosol density fluctuations that the open-loop model cannot represent.

The states inferred by the digital twin are subject to epistemic uncertainty, with regards to both the output and transition models. This can result in states which do not have a direct physical interpretation. An example of this is shown in Figure \ref{fig:open_loop_prediction_exp1_tube_dep}, wherein the tube appears to be expanding. The reason for this state behavior is a drift in carrier pressure (shown in Figure \ref{fig:exp1_carrier_pressure}, and verified in Figure \ref{fig:carrier_drift}). This carrier drift occurs independent of input flow changes, and is therefore a result of either MFC drift or a leak in the AJP system. The closed-loop estimation compensates for this behavior through a reduction in deposition within the tube (Figure \ref{fig:open_loop_prediction_exp_1}c), which can be interpreted as a decrease in fluidic resistance in the carrier line. Future work will incorporate MFC related errors into the updated model parameters, $\theta$, to better capture such effects; however, this closed-loop estimation already helps operators identify potential faults in the machine.

To demonstrate the advantage of updating digital model parameters using EM compared to the standard EKF, an ablation study is conducted on this experiment as shown in Appendix~\ref{app:exp1_ablation_experiment}. 

\subsection{Experiment 2: Using Digital Twin to Explain Changes in Latent States}
\label{sec:exp2_estimation}

\begin{figure*}[t!]
    \centering
    \begin{subfigure}[t]{0.3\textwidth}
        \centering
        \includegraphics[height=1.2in]{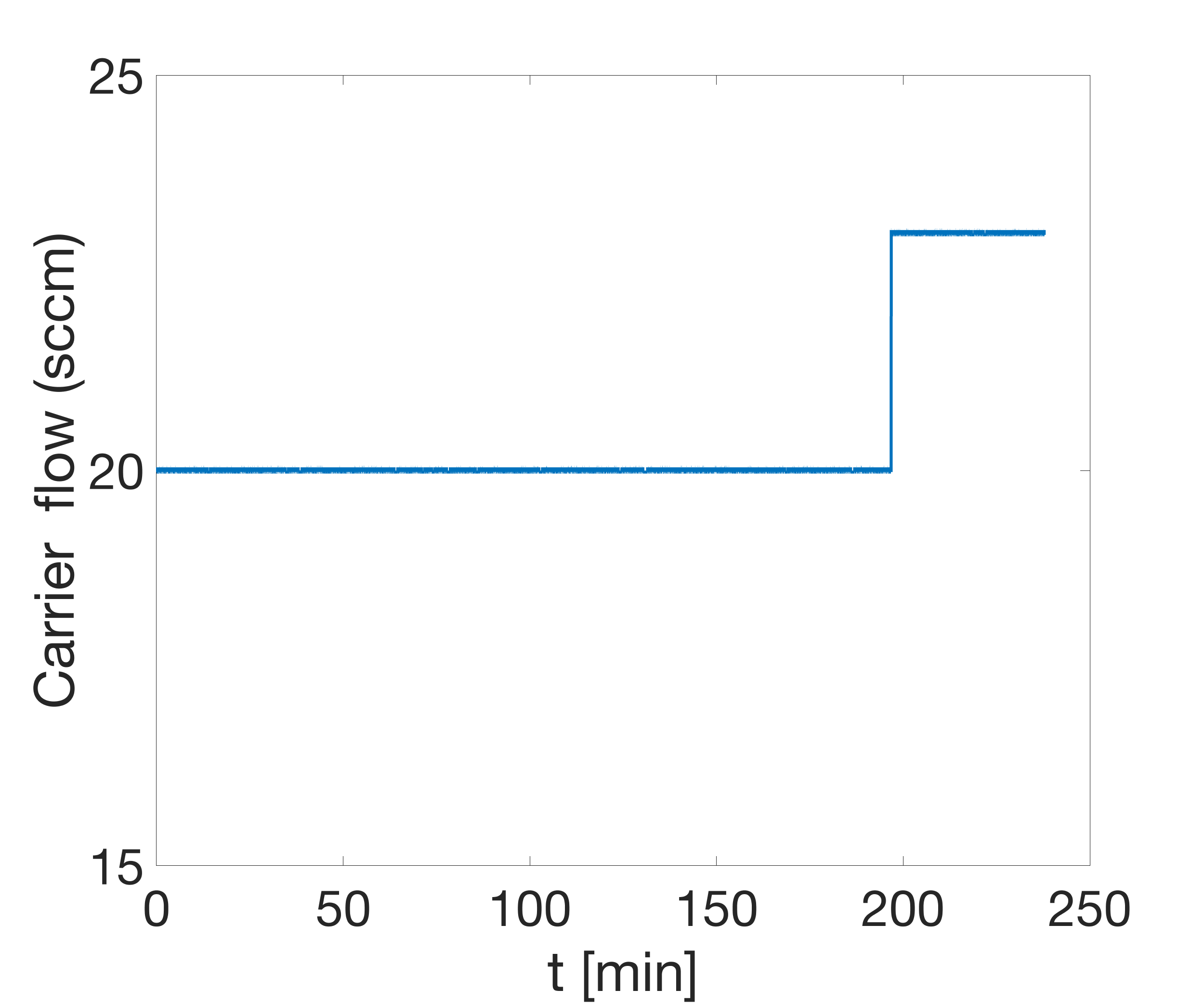}
        \caption{}
    \end{subfigure}%
    ~ 
    \begin{subfigure}[t]{0.3\textwidth}
        \centering
        \includegraphics[height=1.2in]{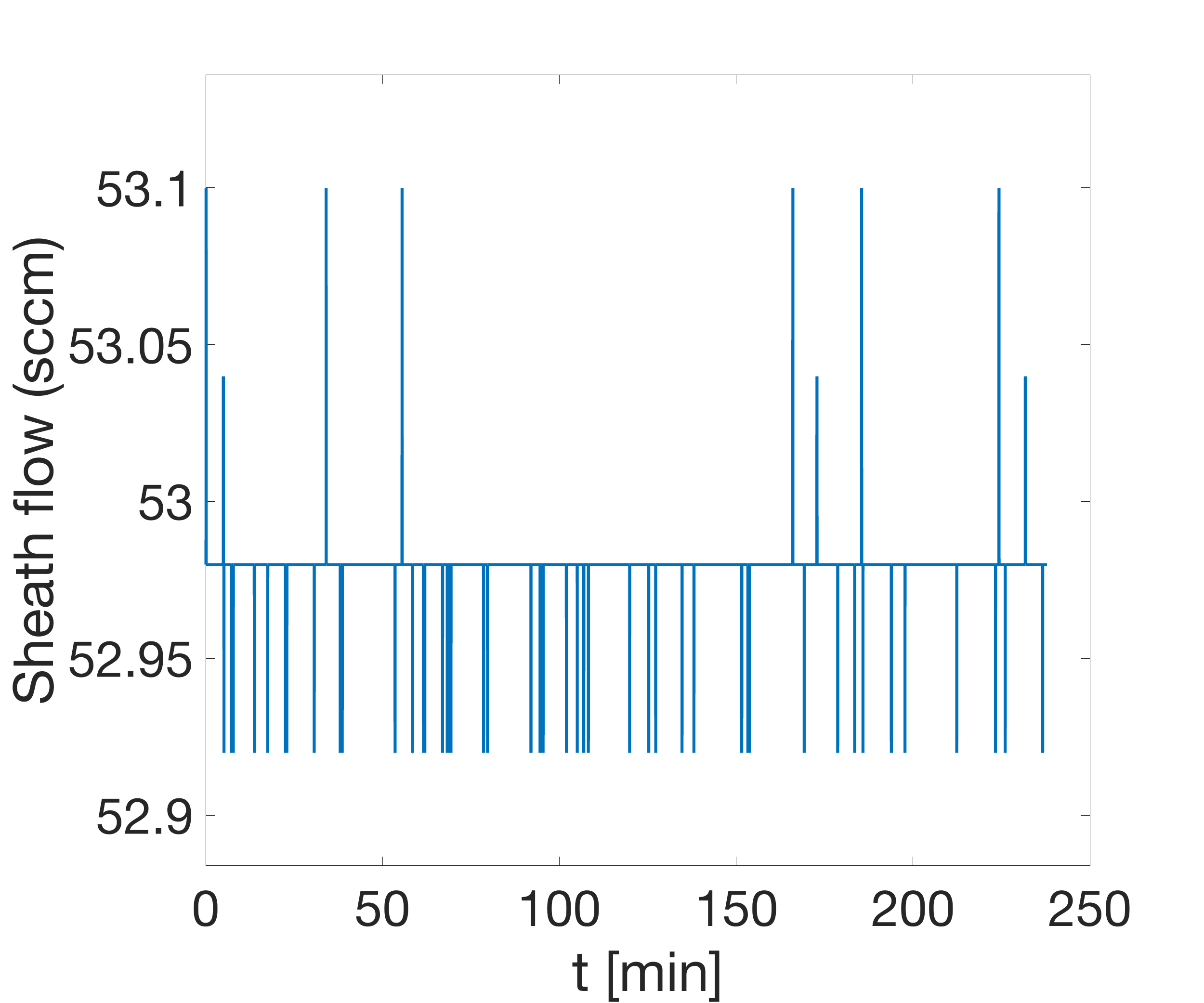}
        \caption{}
    \end{subfigure}
        ~ 
    \begin{subfigure}[t]{0.3\textwidth}
        \centering
        \includegraphics[height=1.2in]{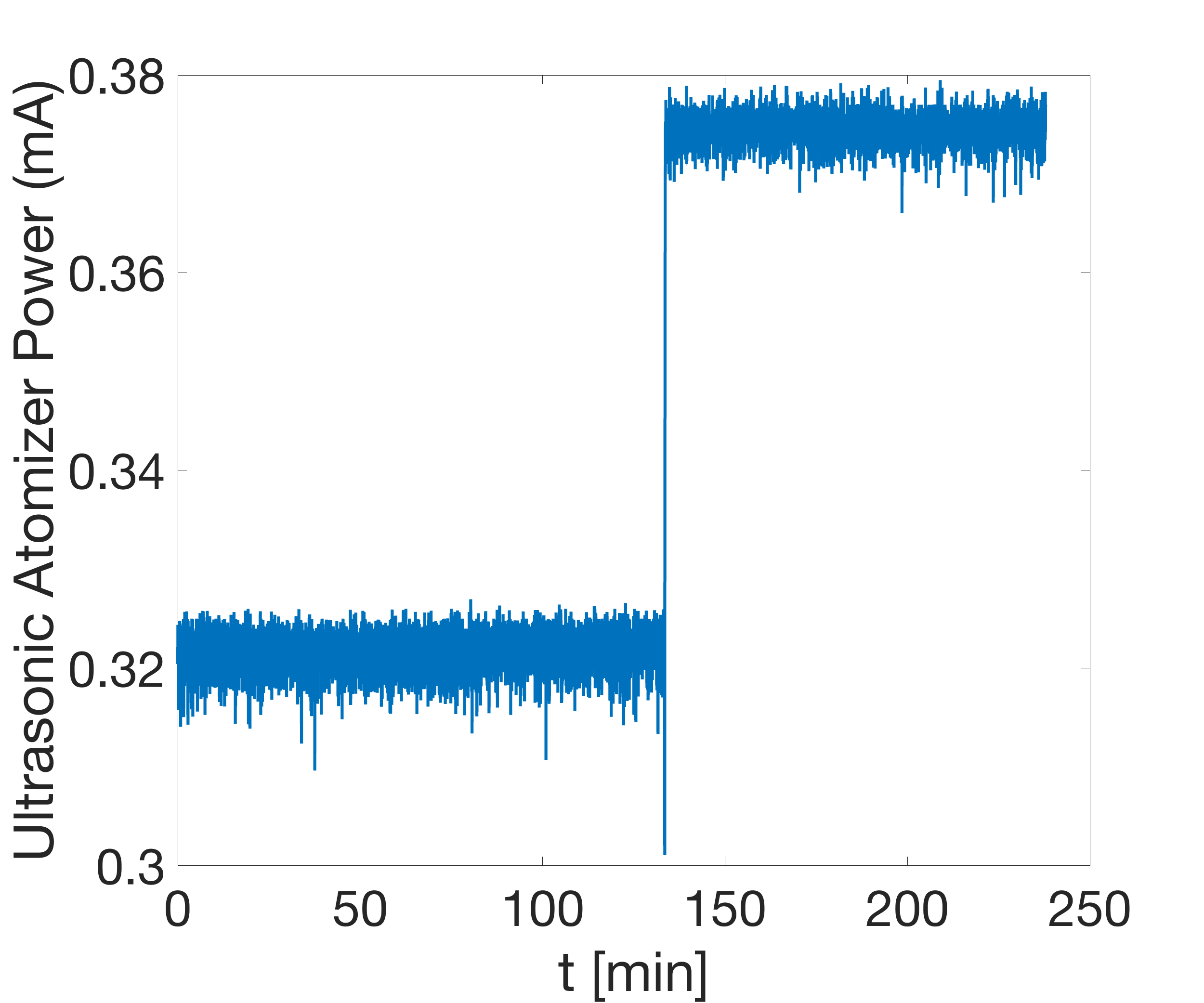}
        \caption{}
    \end{subfigure}
    \caption{Inputs of the estimation for the experiment in Section \ref{sec:exp2_estimation}: (a) measured carrier flow rate, initially set to 20~sccm and is adjusted to 23~sccm at $t=196$~min, (b) measured sheath gas flow rate. The MFC aims to maintain a constant set-point of 53~sccm. The observed variability is due to the limited precision of the MFC, (c) measured ultrasonic atomizer current, initially set at 0.33~mA and is adjusted to 0.375~mA at 133~min.}
    \label{fig:experiment2_inputs}
\end{figure*}

\begin{figure*}[t!]
    \centering
    \begin{subfigure}[t]{0.3\textwidth}
        \centering
        \includegraphics[height=1.2in]{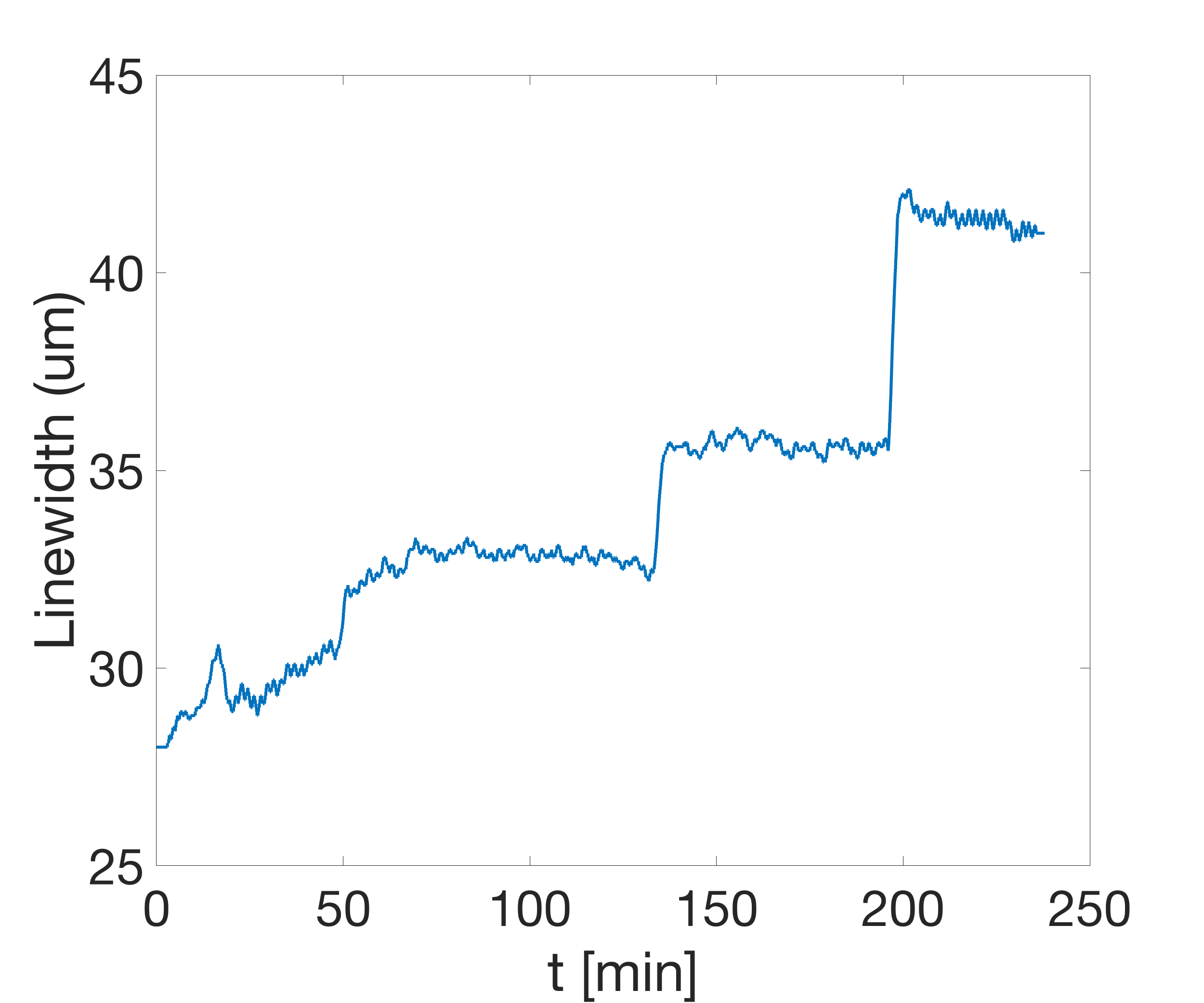}
        \caption{}
    \end{subfigure}%
    ~ 
    \begin{subfigure}[t]{0.3\textwidth}
        \centering
        \includegraphics[height=1.2in]{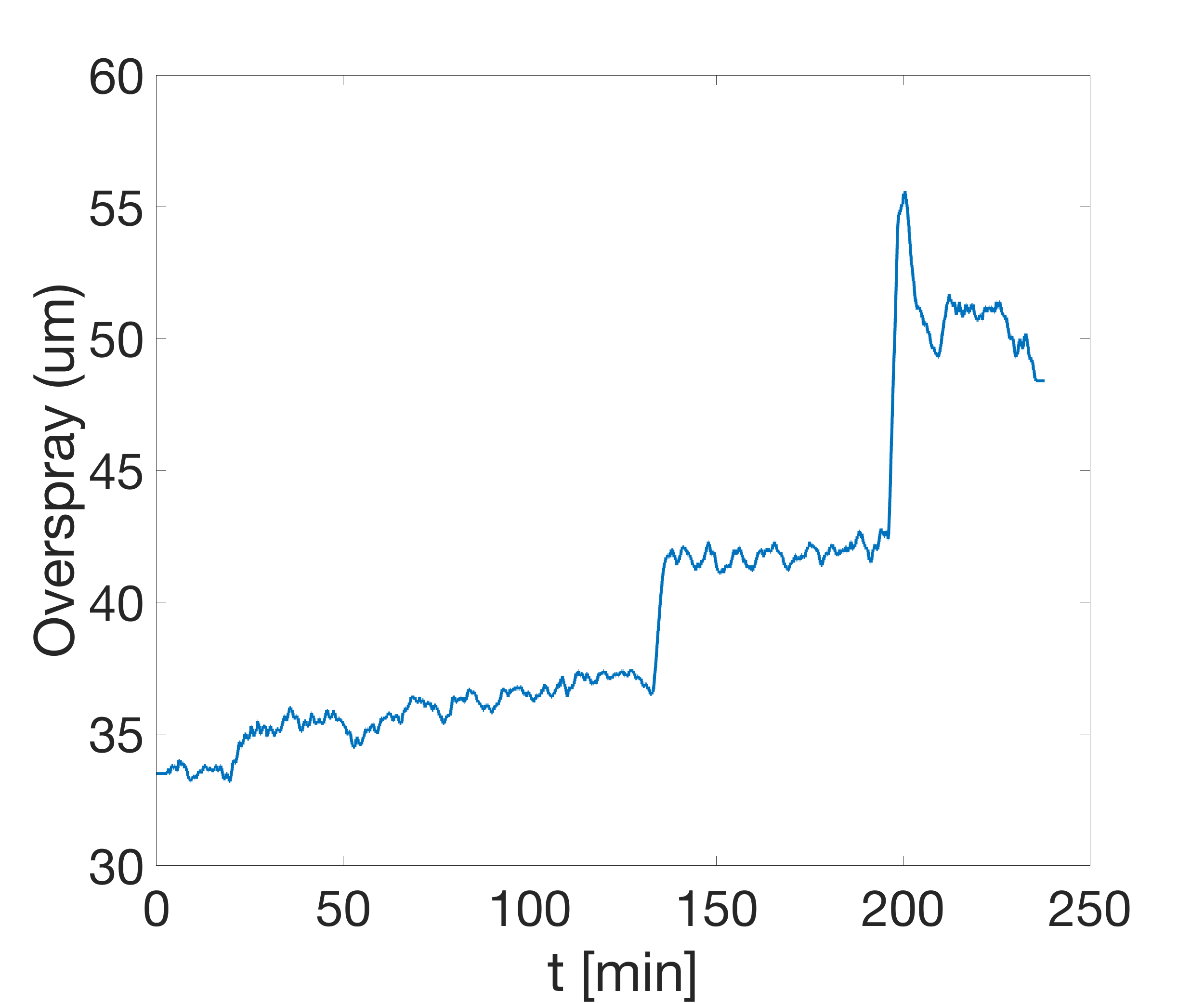}
        \caption{}
    \end{subfigure}
        ~ 
    \begin{subfigure}[t]{0.3\textwidth}
        \centering
        \includegraphics[height=1.2in]{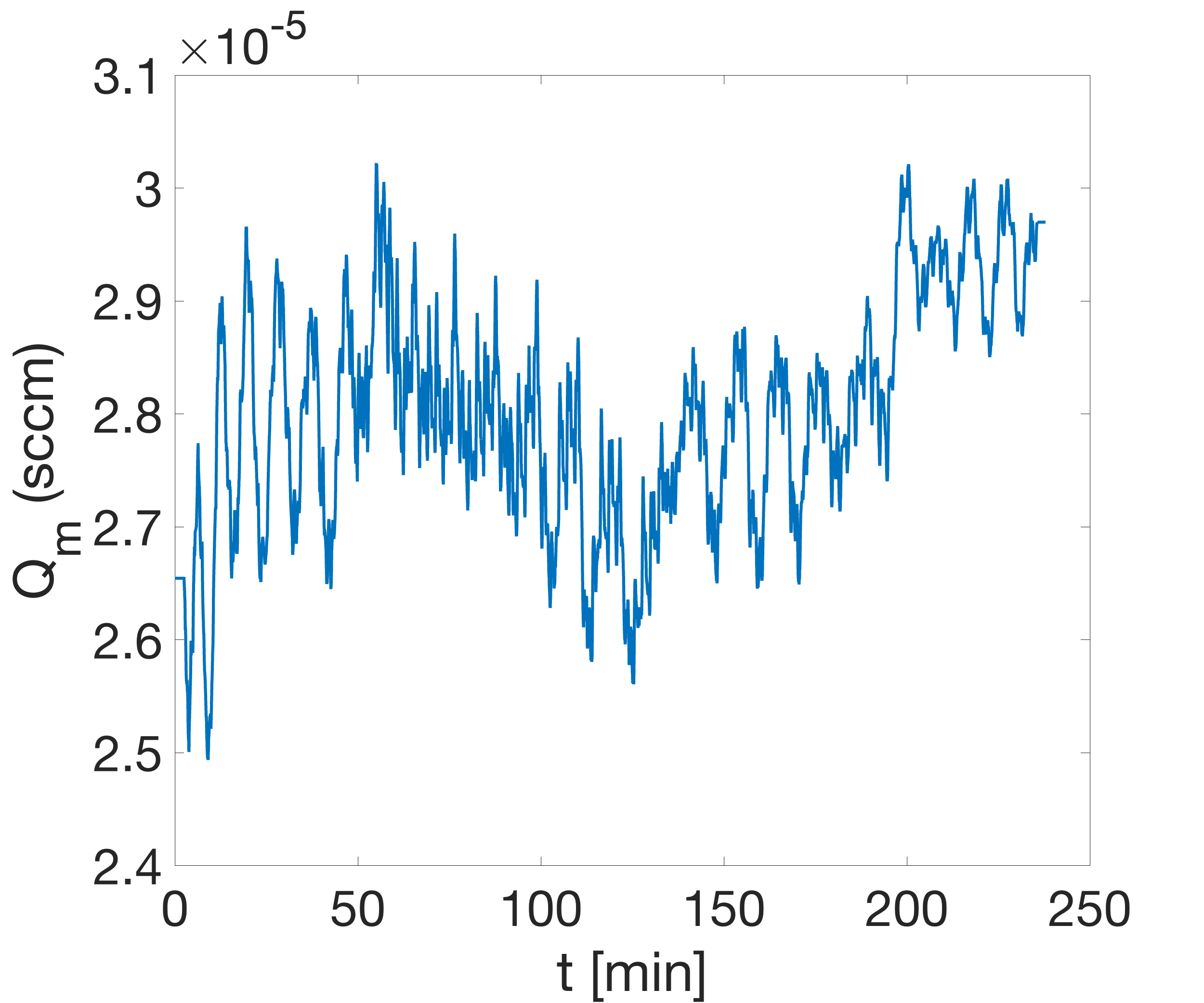}
        \caption{}
    \end{subfigure}
    \newline
    \begin{subfigure}[t]{0.3\textwidth}
        \centering
        \includegraphics[height=1.2in]{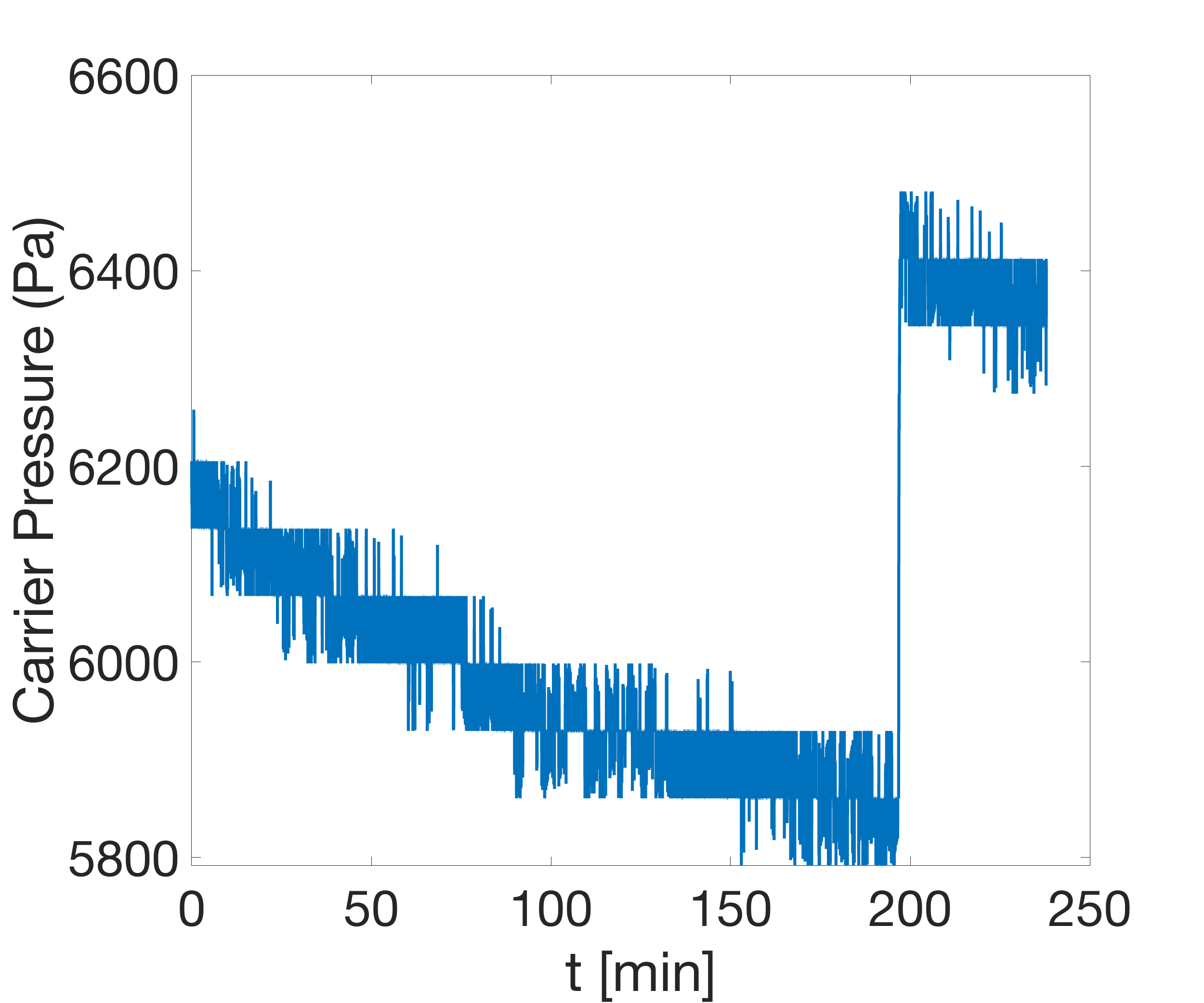}
        \caption{}
    \end{subfigure}
        \begin{subfigure}[t]{0.3\textwidth}
        \centering
        \includegraphics[height=1.2in]{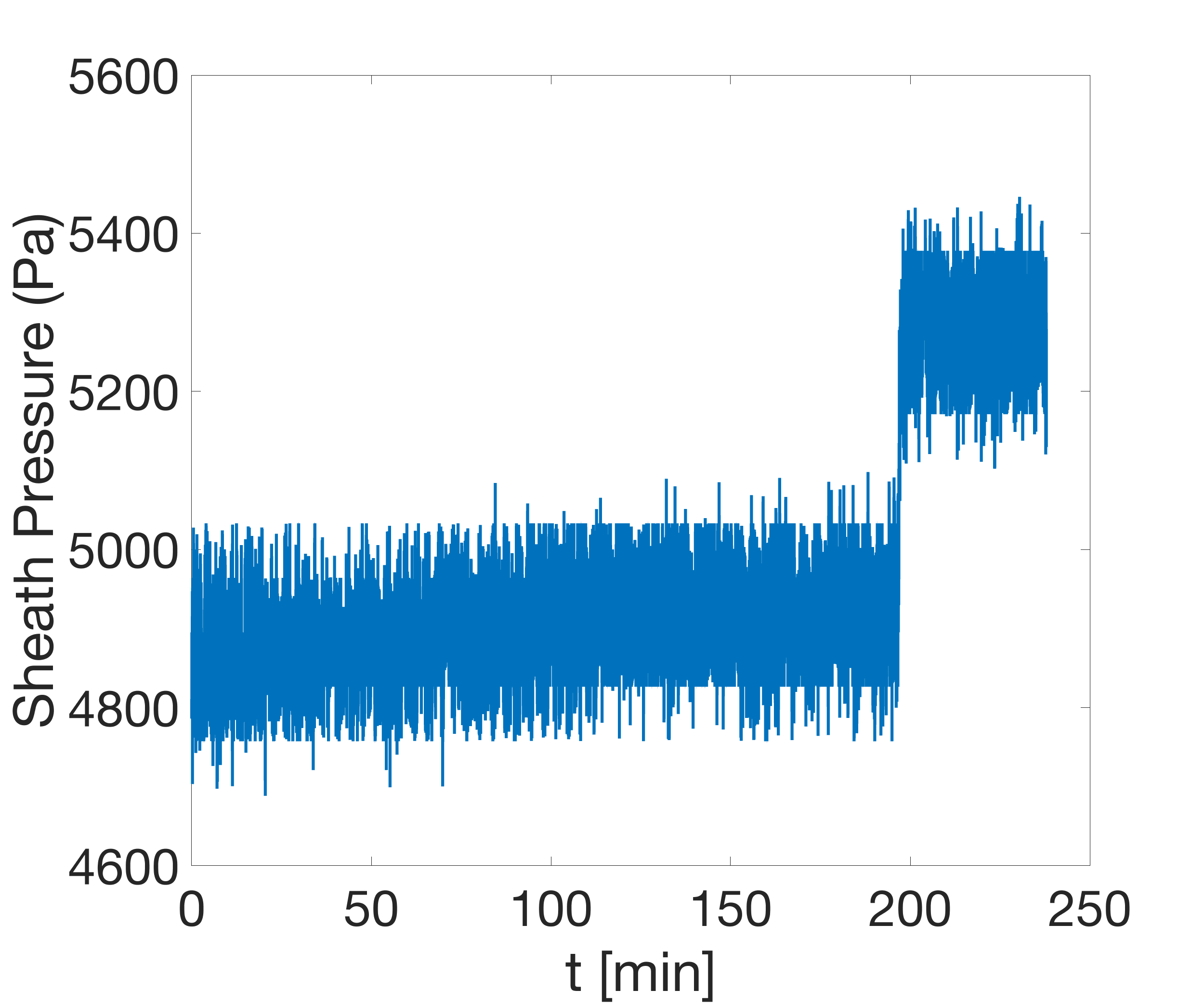}
        \caption{}
    \end{subfigure}
    \caption{Recorded outputs for the experiment in Section \ref{sec:exp2_estimation} include: (a) linewidth, $\lw$, (b) overspray, $\ov$, (c) deposited material flow rate, $\Qink$, (filtered with a 40-second moving average), (d) output carrier pressure, $\Pcarr$ and (e) output sheath pressure, $\Psh$.}
    \label{fig:experiment2_outputs}
\end{figure*}

After validating the accuracy of the estimation process, the digital twin is used to explain observed variations in AJ output. In this experiment, a 90-minute print is conducted with multiple input changes introduced at discrete time points shown in Figure \ref{fig:experiment2_inputs}, with the measured outputs shown in Figure \ref{fig:experiment2_outputs}. In this experiment, noticeable drifts in linewidth and overspray are observed during intervals where the inputs remain constant.

\begin{figure*}[!htb]
    \centering
    \begin{subfigure}[t]{0.3\textwidth}
        \centering
        \includegraphics[height=1.2in]{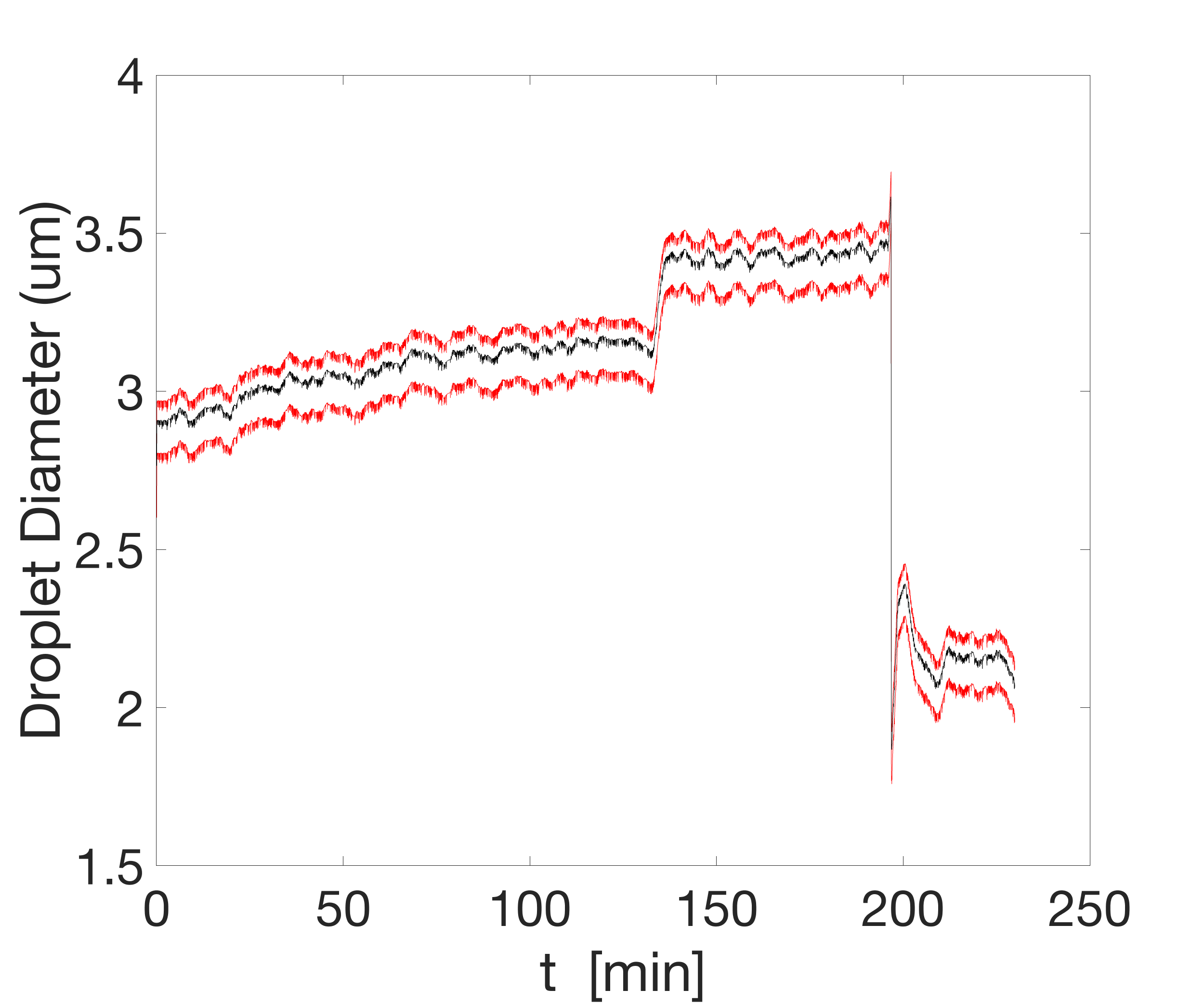}
        \caption{}
    \end{subfigure}%
    ~ 
    \begin{subfigure}[t]{0.3\textwidth}
        \centering
        \includegraphics[height=1.2in]{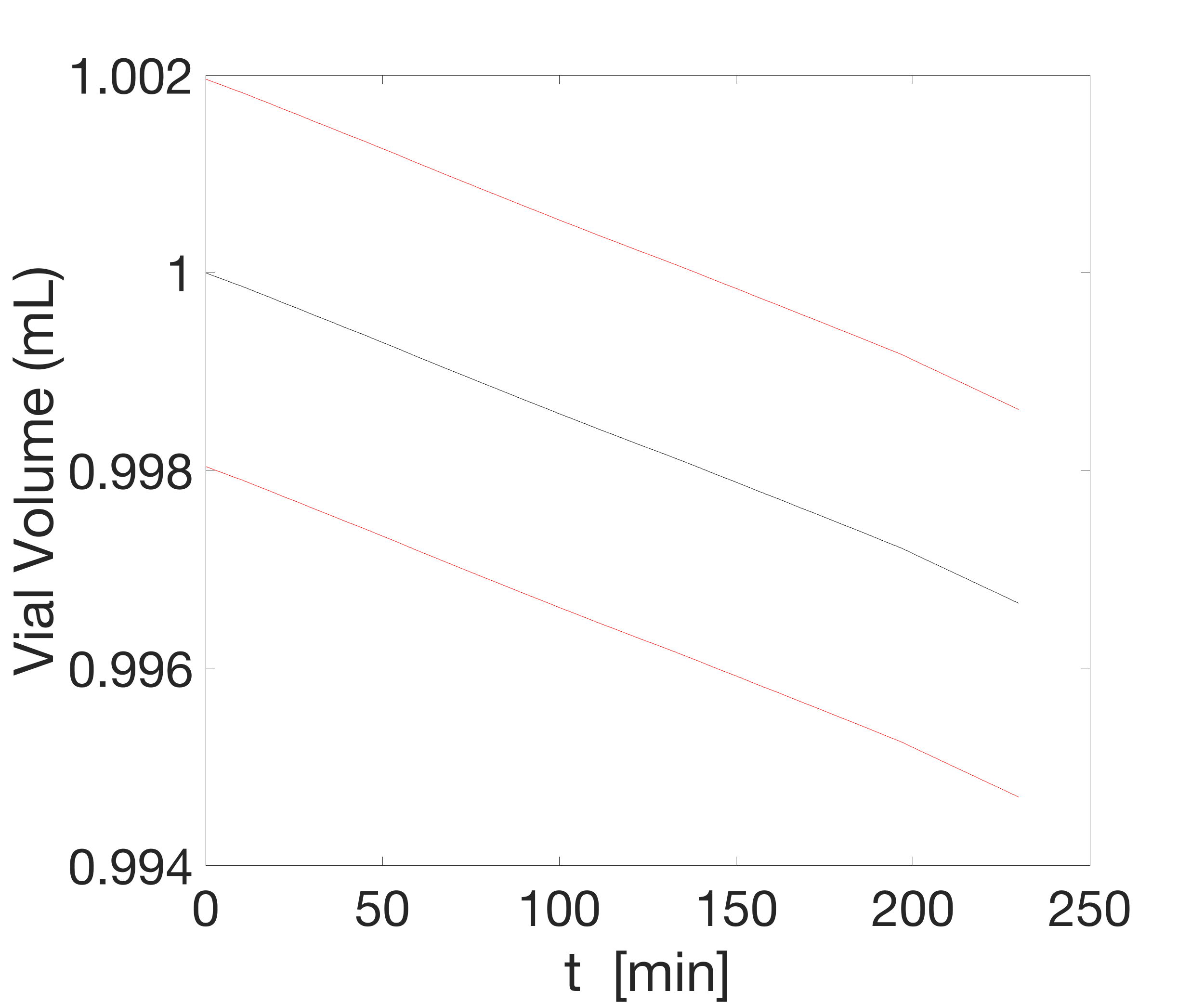}
        \caption{}
    \end{subfigure}
    ~ 
    \begin{subfigure}[t]{0.3\textwidth}
        \centering
        \includegraphics[height=1.2in]{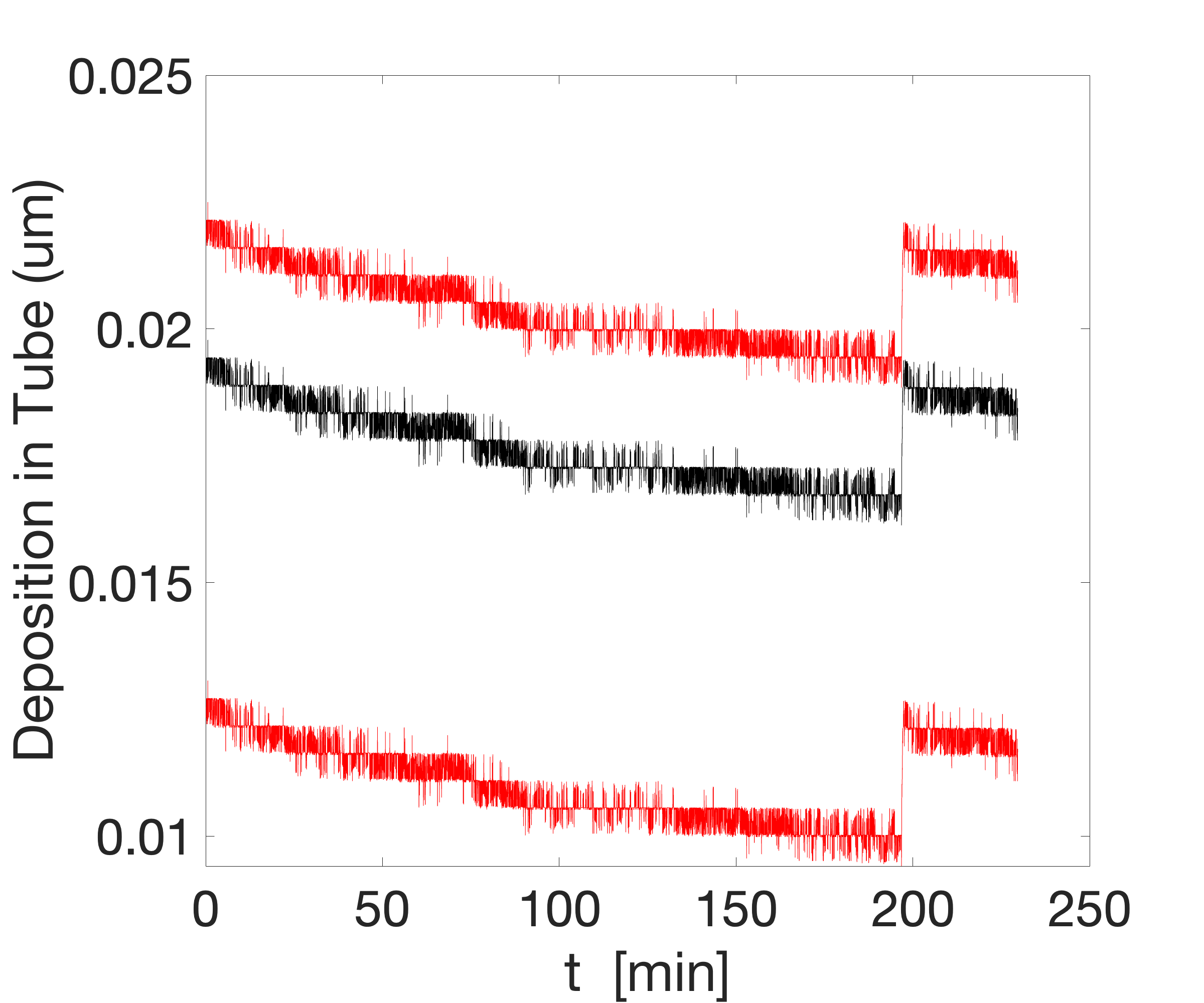}
        \caption{}
    \end{subfigure}

    \begin{subfigure}[t]{0.3\textwidth}
        \centering
        \includegraphics[height=1.2in]{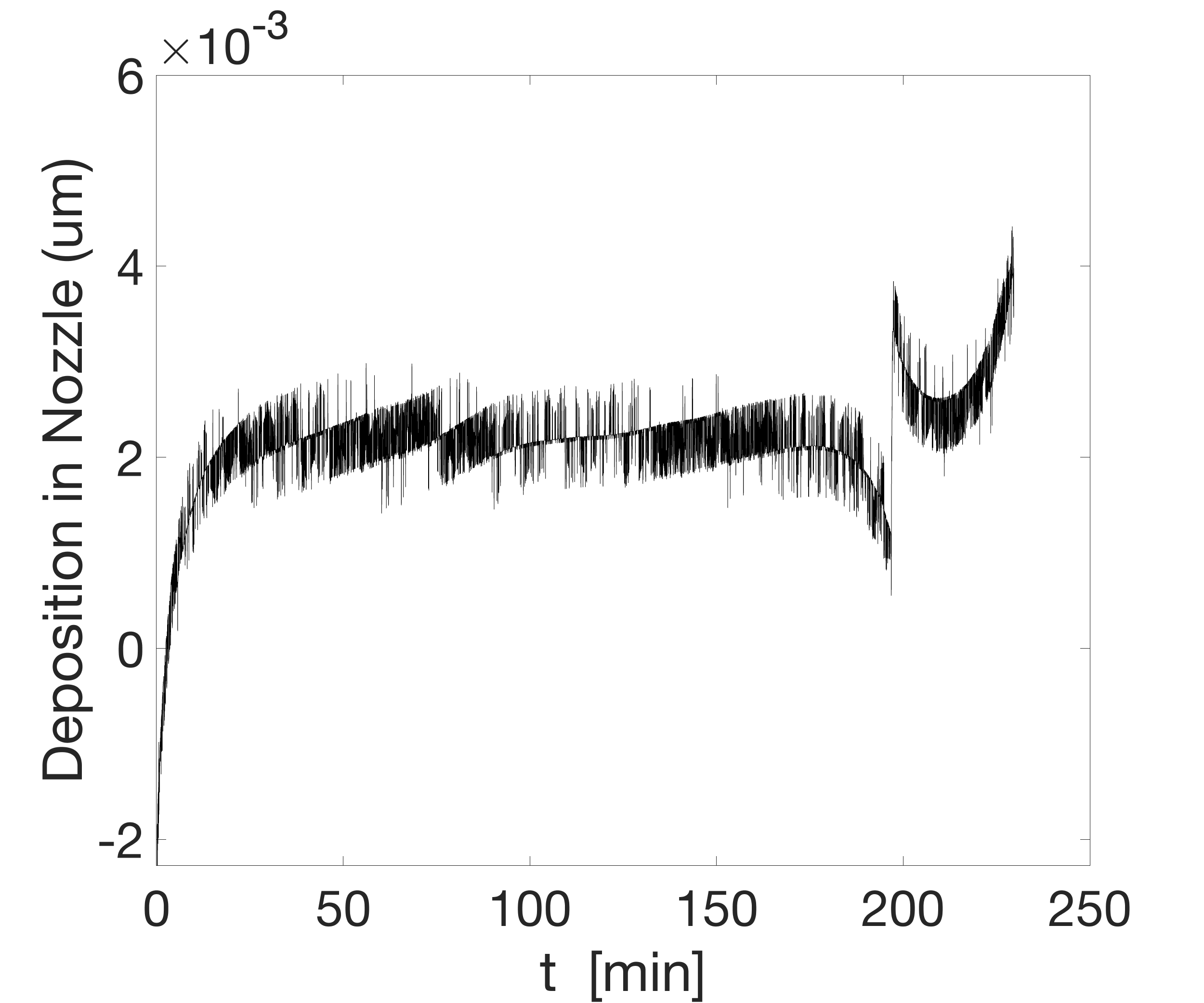}
        \caption{}
    \end{subfigure}
    ~
    \begin{subfigure}[t]{0.3\textwidth}
        \centering
        \includegraphics[height=1.2in]{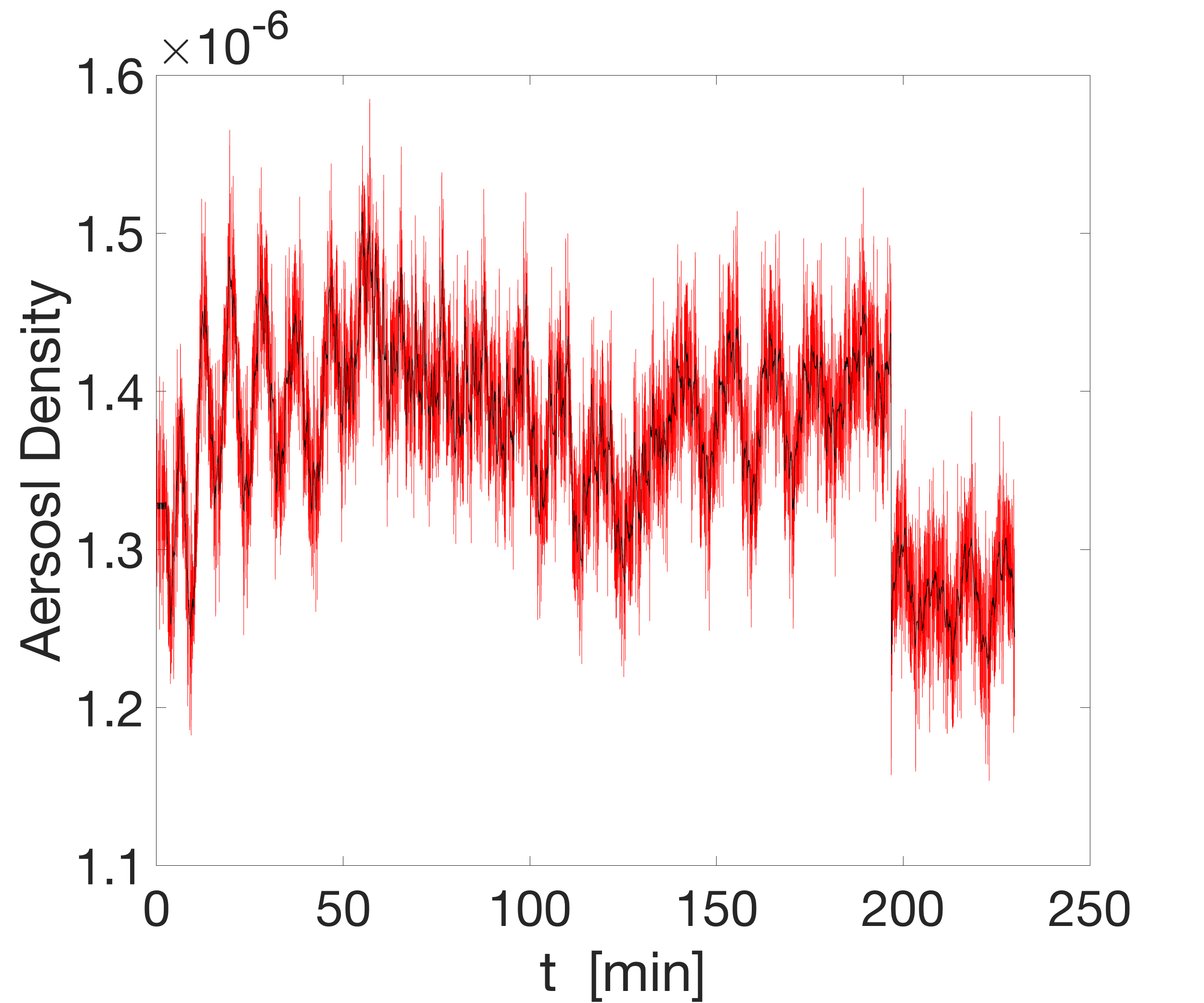}
        \caption{}
    \end{subfigure}
    \caption{Closed-loop prediction of the digital twin for the experiment in Section \ref{sec:exp2_estimation} using EM reveals trends in the latent state parameters that influence changes in the outputs including for (a) median aerosol droplet diameter out of the atomizer, $\DA$, (b) ink solution volume in vial, $\Vl$, (c) ink deposition in the tube, $\DT$, (d) ink deposition in the nozzle, $\DN$, (e) fraction of aerosol droplets in the carrier stream, $\AD$. The black curve represents the mean inferred state, while the red curves indicate the 90\% confidence interval. The confidence bounds for (d) are not shown as they are on the order of 1 \textmu m.
}
    \label{fig:digital_twin_states_exp2}
\end{figure*}

To investigate the source of these variations, the digital twin is used to infer the internal system states over time. The results, shown in Figure \ref{fig:digital_twin_states_exp2}, suggest that changes in the median droplet diameter and aerosol volume fraction are the root cause of variations in linewidth, overspray and material flow rate. The material flow rate strongly depends on the carrier gas and ultrasonic atomizer settings because these parameters directly influence droplet formation and transport. For example, higher carrier flow rates result in more material being transported, while ultrasonic power controls the atomization that affects droplet concentration \cite{rajan2001correlations}. Additionally, the median aerosol droplet diameter is observed to vary in response to changes in the input conditions, even though this dependency is not explicitly represented in the transition model. The evidence presented in Section \ref{sec:exp1_estimation} suggests that this drift is real, even though it was not explicitly measured for this experiment. This indicates that the transition model may not fully capture the underlying physics governing droplet formation and transport. Through closed-loop state estimation, the inferred droplet diameter adapts dynamically with input changes to compensate for variations in linewidth and overspray, effectively bridging the gap between the simplified model and the observed system behavior. To account for these effects within the transition model, this result suggests the need to expand the droplet diameter transition dynamics to include influences from the carrier flow rate. Future work will focus on learning this relationship by extending the parameter set $\theta$ to capture these dependencies more accurately.

The other implication of these inferred fluctuations in the aerosol diameter is that a crucial aspect of process control for AJ printing is understanding and controlling aerosol generation, which is closely linked to ink characteristics and rheology \cite{secor_vapor}.

\subsection{Experiment 3: Updating Model State Parameters for Better Prediction}
\label{sec:exp3_parameter_estimation}

This experiment demonstrates the digital twin’s ability to reflect changes in the physical AJ system. The digital twin periodically updates its transition model parameters using newly collected data, enabling it to track real-time changes and improve the accuracy of its open-loop predictions.

During the experiment, the AJ system undergoes two input changes, as detailed in Appendix~\ref{app:experiment3_details}, Figure \ref{fig:exp3_inputs}: a step change in the carrier flow rate at 37.6~min, followed by a change in the sheath flow rate at 68.5 min. It is worth noting that significant drift occurs before the input carrier flow change (see Figure \ref{fig:exp3_drift}). Based on the inferred states of the machine shown in Figure \ref{fig:exp3_inferred}, this is primarily due to shifts in the aerosol droplet diameter and aerosol density. This has been a consistent result throughout this study and motivates the focus for future work.

Prior to any input changes ($t<37.6$~min), the digital twin uses the EM algorithm within the EKF  framework (detailed in Section 7) to update the transition model parameters, $\theta$, in response to observed system behavior.

After this calibration phase, the updated model is used to predict the linewidth, overspray, and aerosolized ink flow rate following the input changes. These predictions are made in an open-loop manner, relying solely on the updated model and the new inputs, without performing any EKF updates.

\begin{figure*}[h!]
    \centering
    \begin{subfigure}[t]{0.45\textwidth}
        \centering
        \includegraphics[width=\columnwidth]{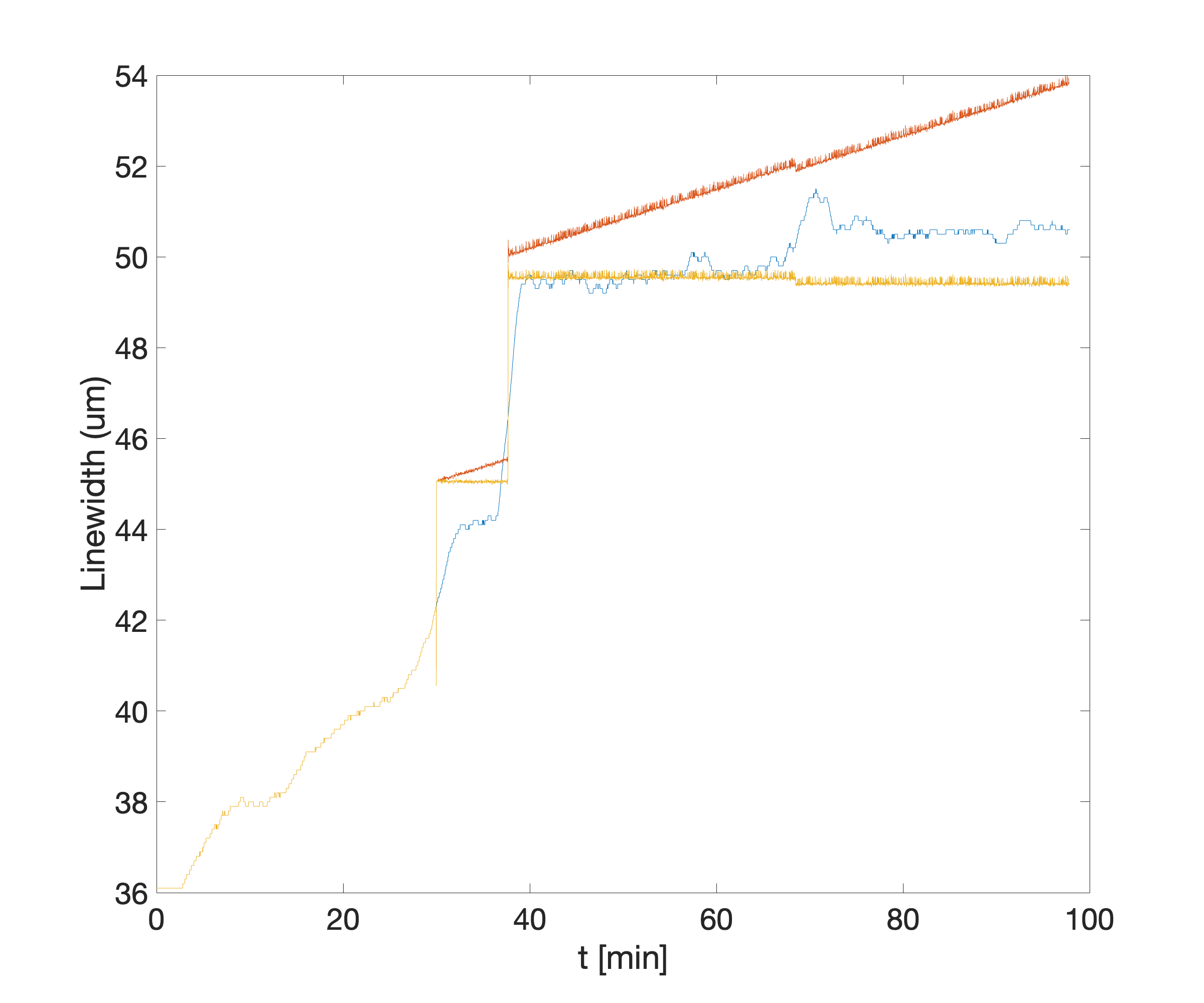}
        \caption{}
    \end{subfigure}%
    \hfill
    \begin{subfigure}[t]{0.45\textwidth}
        \centering
        \includegraphics[width=\columnwidth]{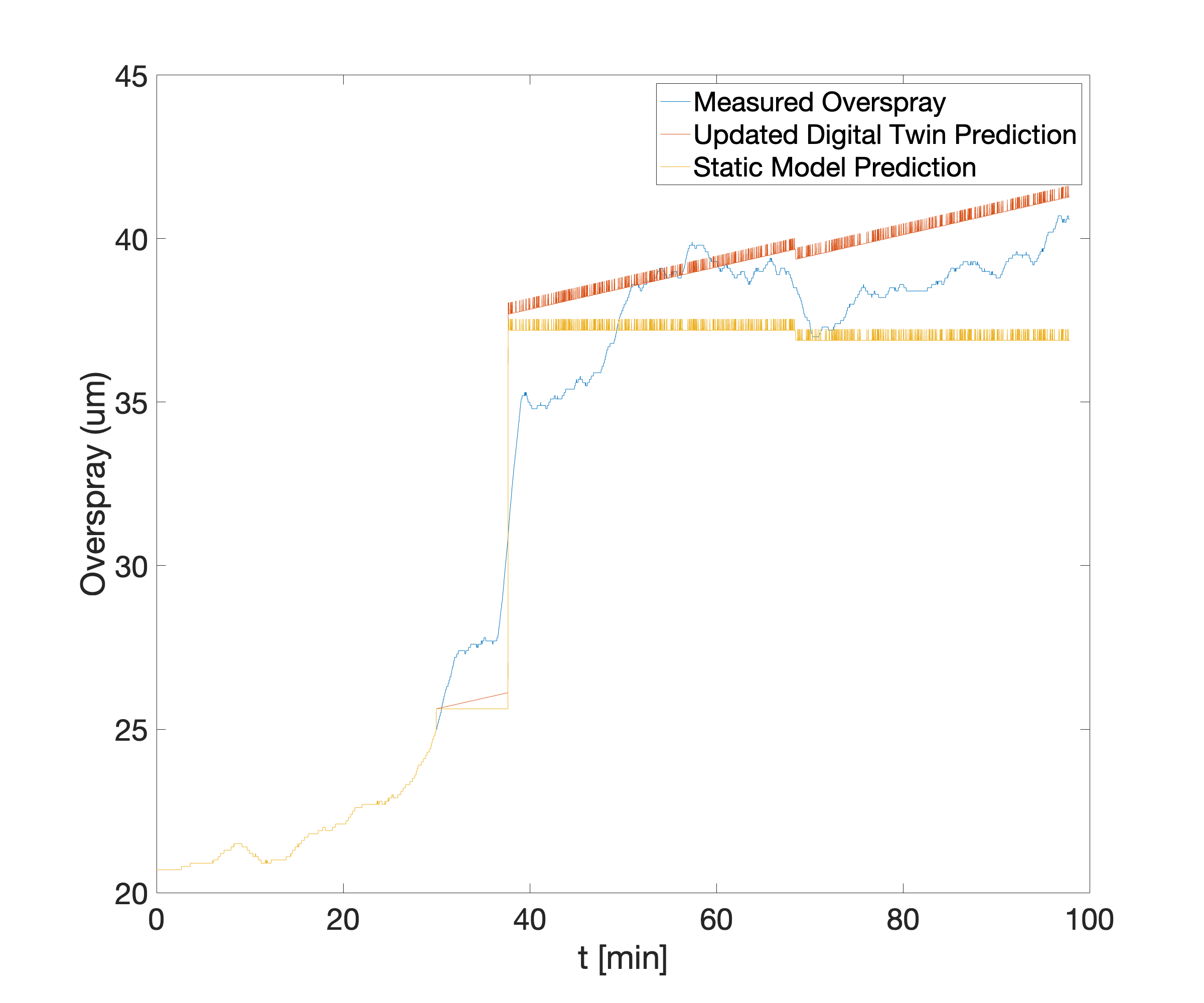}
        \caption{}
    \end{subfigure}
    \caption{The AJ system in the experiment in Section \ref{sec:exp3_parameter_estimation} experiences step changes in carrier flow at 37.6~min and sheath flow at 68.5~min as shown in Figure \ref{fig:exp3_inputs}. Open-loop prediction of (a) linewidth and (b) overspray using static physics-derived model of the latent states compared (orange line) to the open-loop prediction of the digital twin that is updated using Expectation-Maximization (red line).}
    \label{fig:open_loop_prediction_exp_3}
\end{figure*}

The results, shown in Figure \ref{fig:open_loop_prediction_exp_3}, demonstrate that the updated digital twin provides more accurate prediction compared to the baseline physics-derived model. While the static model fails to capture the system’s variations after the input change, the updated digital twin correctly predicts the resulting trends in linewidth and overspray across future time points.

These findings highlight the value of continuously learning from operational data to mirror the current state and model parameters of the system for more accurate forecasting. 

\section{Conclusion}
This work presents a digital twin model of the AJ process designed to identify the root causes of variability in print quality and predict process drifts. The digital twin framework integrates computer vision, physics-based modeling, and probabilistic estimation into a comprehensive end-to-end solution. At its core lies a physics-based macro-model of the AJ system, whose states and model parameters are continuously updated using probabilistic estimation techniques. This allows the digital twin to evolve alongside the operation of the physical machine, providing improved predictions of inconsistencies and variability in print quality. The effectiveness of the digital twin is demonstrated through accurate predictions of process drifts and shifts in line quality arising from input changes not captured by static physics-based models. The work establishes a scalable digital twin implementation applicable to other manufacturing techniques to enhance both control and process insight.

Future developments of the digital twin framework could focus on incorporating ink-specific properties to enable accurate predictions of line quality across a range of ink formulations. This includes expanding the CNN model used to estimate material flow rate to account for different substrates and inks. Future developments of the digital twin framework may focus on incorporating ink-specific properties to improve the accuracy of line quality predictions across different ink formulations and substrates. In particular, future work aims to include the material volume fraction, $\phi_m$, as part of the state-variable set to capture temporal variations in the concentration of solid material within the ink solution. In the current implementation, $\phi_m$ is held constant, as its inclusion within the state-space formulation results in a low-observable system, limiting the system’s ability to uniquely estimate all state variables from available measurements. Incorporating $\phi_m$ dynamically will therefore require additional measurements capable of providing information related to ink composition or flow characteristics. Including $\phi_m$ in the dynamical system model would allow the digital twin to represent changes in ink composition that directly influence flow behavior, deposition rate, and final line morphology.
Finally, future work might integrate the digital twin within a model-predictive control framework for automated, closed-loop control of the AJ process. This work could automatically adjust process parameters to maintain desired line quality by compensating for variations in changes in latent state. These advancements could promote the AJ digital twin as a robust tool that minimizes process variations across operating conditions and make AJ printing a more reliable manufacturing technique.

\section{Funding Statement}

A portion of this research was sponsored by the Army Research Laboratory and was accomplished under Cooperative Agreement Number W911NF-20-2-0175.  The views and conclusions contained in this document are those of the authors and should not be interpreted as representing the official policies, either expressed or implied, of the Army Research Laboratory or the U.S. Government. The U.S. Government is authorized to reproduce and distribute reprints for Government purposes notwithstanding any copyright notation herein.
\section{Author Contribution}
A.A. and  J.R. collected the data, designed the algorithms and coded the experimental results.
All authors were involved in preparing and reviewing the manuscript.

\clearpage

\newpage
\appendix
\setcounter{figure}{0}
\renewcommand\thefigure{S\arabic{figure}}    
\setcounter{table}{0}
\renewcommand\thetable{S\arabic{table}}    
\section*{Supplementary Material}
\section{Variations in Latent States}
\label{app:variations_experiment}

\begin{figure}[b]
\centering
\begin{subfigure}{0.49\textwidth}
    \centering
    \includegraphics[height=1.8in]{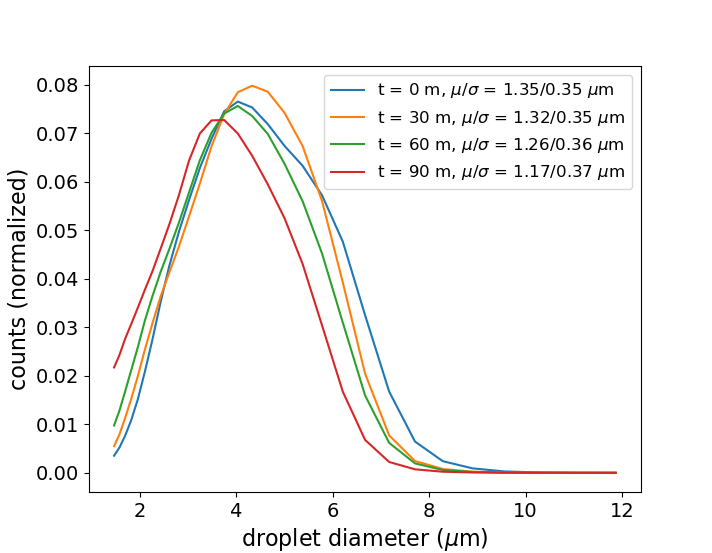}
    \caption{}
\end{subfigure}
\hfill 
\begin{subfigure}{0.49\textwidth}
    \centering
    \includegraphics[height=1.8in]{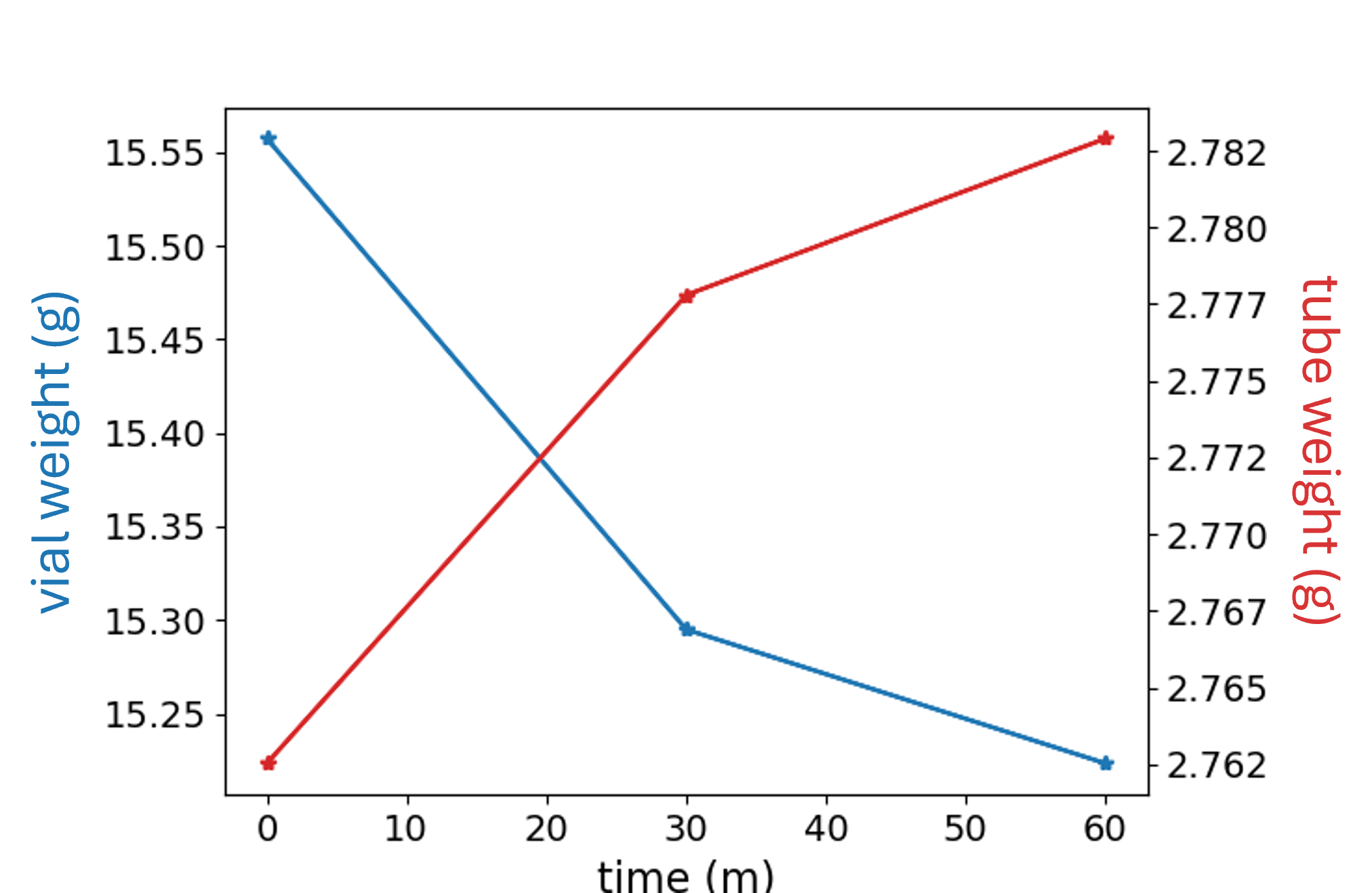}
    \caption{}
\end{subfigure}
\bigskip
\caption{Measure of hidden states for a print, a) Histogram measurement of aerosol droplet diameters taken at various points in time during a print, b) Tube and vial weights measured during a print}
\label{fig:variations_in_latent_states}
\end{figure}

To motivate the use of a digital twin, several latent state variables were directly measured during a single AJ print (specifically, the median aerosol droplet diameter, vial level, nozzle deposition, and tube deposition) using the procedures detailed in Appendix \ref{app:experimental_setup_validation}. To measure the latent states, measurements were taken during the print periodically as the AJ process was halted. These latent variables are typically assumed to be constant throughout a print and are often treated as static parameters in physics-based models. However, experimental observations in Figure \ref{fig:variations_in_latent_states} reveal clear and consistent drifts in these internal states over time. These variations correlate strongly with changes in output line quality, including linewidth, overspray, and ink flow rate, highlighting the need for a digital twin capable of tracking hidden system behavior in real time.

\begin{figure}
    \centering
    \includegraphics[width=0.9\columnwidth]{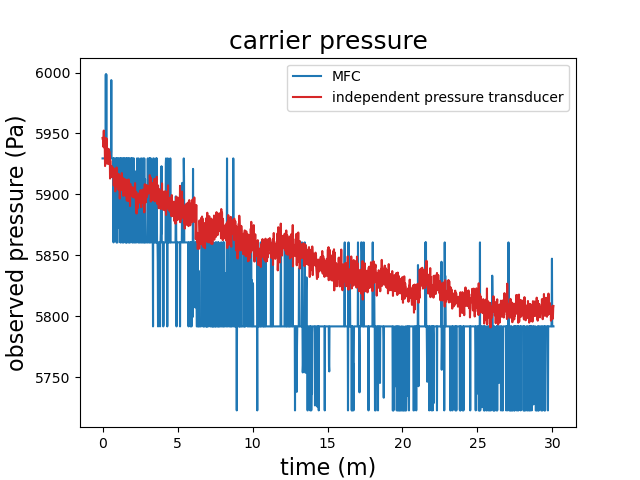}
    \caption{Drift in the observed carrier pressure which causes drift in the tube deposition inference. Data saved from MFC with validation from independent pressure transducer. Bias from pressure transducer is removed to provide a stronger visual comparison.}
    \label{fig:carrier_drift}
\end{figure}

Drift in the ink deposition in the carrier tube was consistently observed by the state estimation algorithm. This is largely due to a drift in the observed carrier pressure. In figure \ref{fig:carrier_drift}, the drift in the carrier pressure is shown a portion of the experiment detailed in section \ref{sec:exp1_estimation}. To verify the drift, an independent pressure transducer (TE Connectivity M3041-000005-5PG) was attached near the carrier MFC. The output of this measurement closely matches that recorded by the MFC.

\section{Computer Vision Algorithm to Extract Line Quality Parameters}
\label{app:cv_algorithm}

\begin{algorithm}
    \caption{Linewidth and overspray extraction}
    \begin{algorithmic}[1]
		\For{each frame in inspection video}
            \State Filter frame using Canny edge detection
            \State Use Hough transform on filtered image to detect line 
            \State Rotate image so line is perfectly horizontal
			\For {Each batch of columns in frame}
                \State Calculate average and standard deviation for each row of pixels
				\State Find two local minima
				\For{Each local min}
                    \State Find first peak moving out toward edge of frame
                    \If {Peak exists}
                        \State Linewidth edge found
                    \Else 
                        \State Find first peak from standard deviation profile
                    \EndIf
                \EndFor
                \For{Each linewidth edge found above}
                    \State Find overspray based on threshold
                \EndFor
			\EndFor
		\EndFor
	\end{algorithmic}
\end{algorithm}
\FloatBarrier
\subsection{Experimental data and validation}

\begin{figure*}
    \centering
    \begin{subfigure}[t]{0.49\columnwidth}
        \includegraphics[width=1\textwidth]{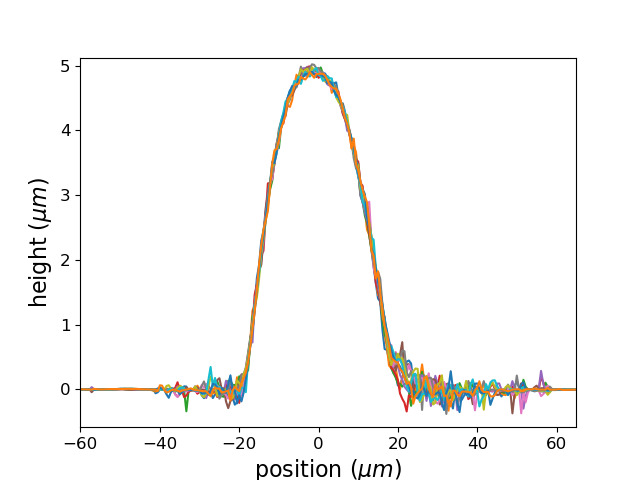}
        \caption{}
        \label{fig:lw_variance_proof}
    \end{subfigure}
    \begin{subfigure}[t]{0.49\columnwidth}
        \includegraphics[width=1\textwidth]{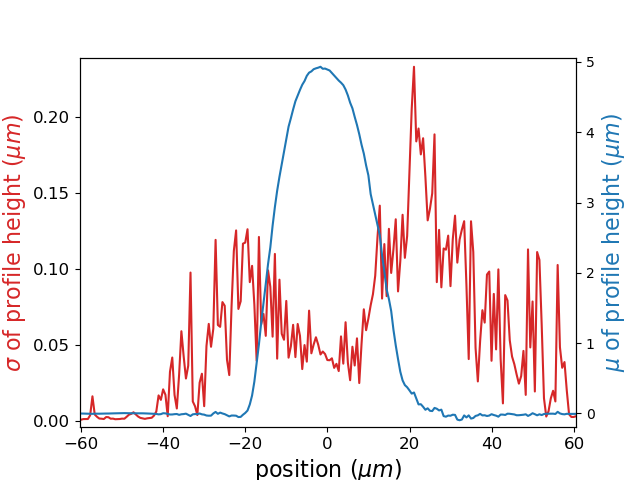}
        \caption{}
        \label{fig:lw_variance_prof_mu_std}
    \end{subfigure}
    \begin{subfigure}[b]{0.95\columnwidth}
        \includegraphics[width=1\textwidth]{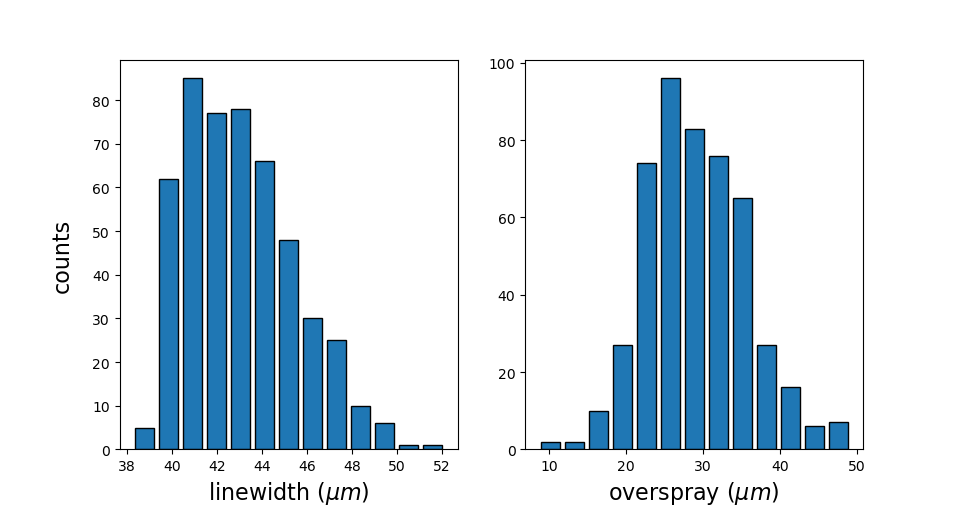}
        \caption{}
        \label{fig:lw_variance_hist}
    \end{subfigure}
    \caption{a) Adjacent profiles taken with optical profilometer show the degree of variation in linewidth and overspray, b) Mean and standard deviation across height profiles shown in part (a), c) histogram of linewidth and overspray calculated on 495 profiles measured with the optical profilometer}
\end{figure*}

The apparently random measurement variation of the linewidth and overspray plots is validated with data taken from an optical profilometer of the printed structure. Figure \ref{fig:lw_variance_proof} shows 12 adjacent profiles superimposed on one another. The xy step size is just over 500 nm, so the total window length is around 6~\textmu{m}. The linewidth algorithm is highly sensitive to the location of the shoulders in the profile. A shoulder in this sense is defined as the point were the profile is no longer monotonically increasing/decreasing. Shoulders are shown in Figure \ref{fig:lw_def} at the points labeled beginning and end of linewidth. The locations of these shoulders appears to have random variation. This is further shown in Figure \ref{fig:lw_variance_prof_mu_std}, in which the mean and standard deviation of the profiles in Figure \ref{fig:lw_variance_proof} are plotted as a function of position. The plot exhibits peaks in standard deviations at the shoulders of the profile. This increased variation in height translates to variation in linewidth. Finally, the linewidth and overspray was calculated on the profiles measured in a 250~\textmu{m} window. The histogram of these plots are shown in Figure \ref{fig:lw_variance_hist}. The random variation in linewidth and overspray appears to follow a log-normal distribution, where $\mu = 3.8, \sigma = 0.06$ for the linewidth, and $\mu = 3.3, \sigma = 0.23$ for overspray. This translates to a mean of 43.1~\textmu{m} and standard deviation of 5.8~\textmu{m} for linewidth and a mean of 29.1~\textmu{m} and standard deviation of 47.1~\textmu{m} for overspray. Note that these are the standard deviation values for a step of roughly 0.5~\textmu{m}, whereas the window for the linewidth vs. time plots has a window of 25~\textmu{m}. 

\begin{figure}
    \centering
    \includegraphics[width=1\columnwidth]{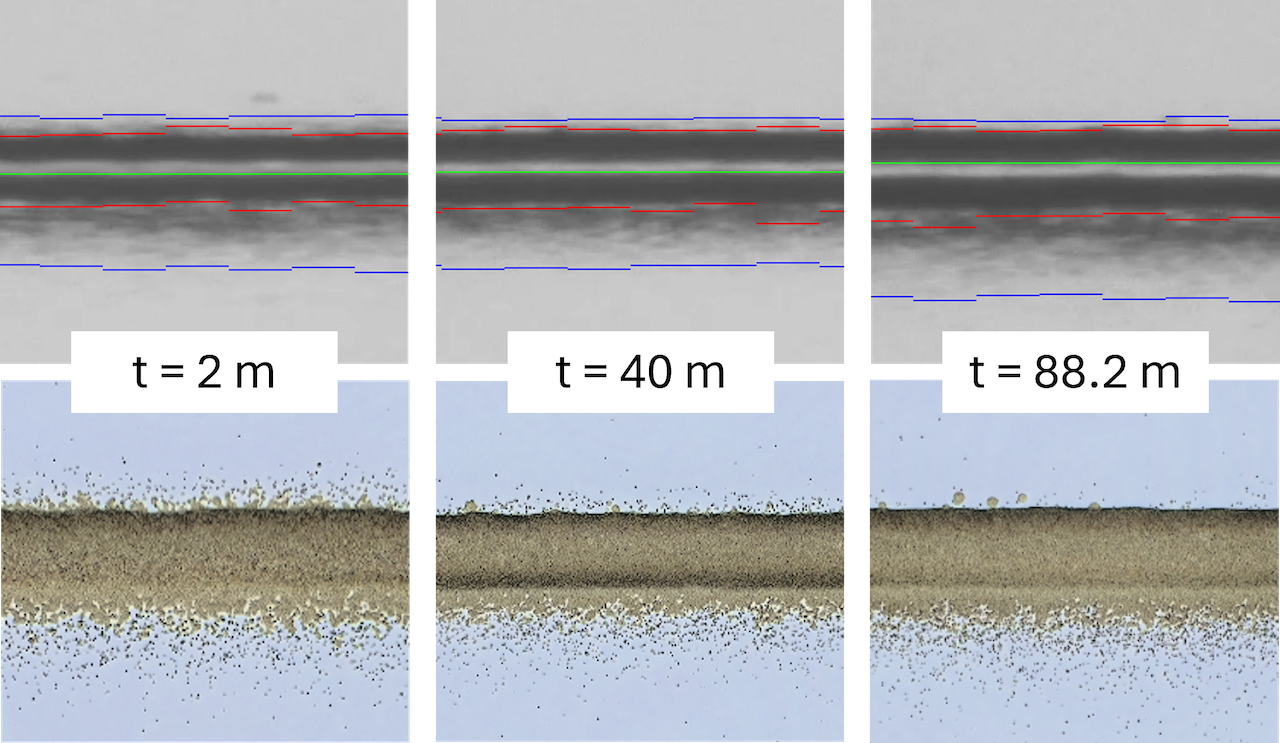}
    \caption{Annotated images of prints at different times throughout the print, with accompanying images from optical profilometry. For the experiment referred to in section 8.1}
    \label{fig:sup_dat_8_1}
\end{figure}

\begin{figure}
    \centering
    \includegraphics[width=1\columnwidth]{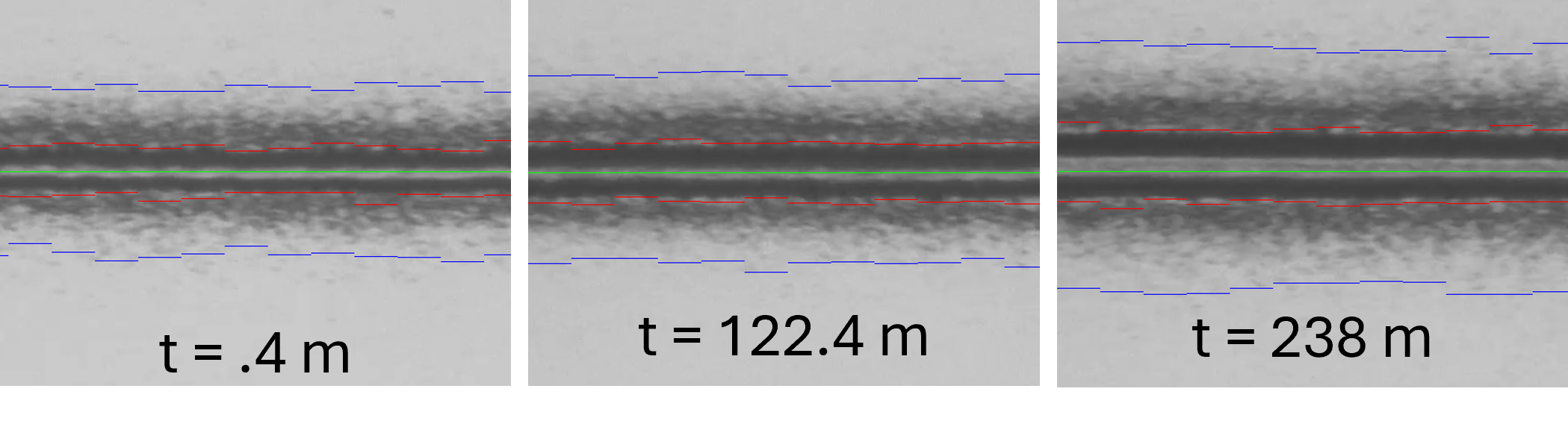}
    \caption{Annotated images of prints at different times throughout the print. For the experiment referred to in section 8.1.1}
    \label{fig:sup_dat_8_1_1}
\end{figure}

\begin{figure}
    \centering
    \includegraphics[width=0.9\columnwidth]{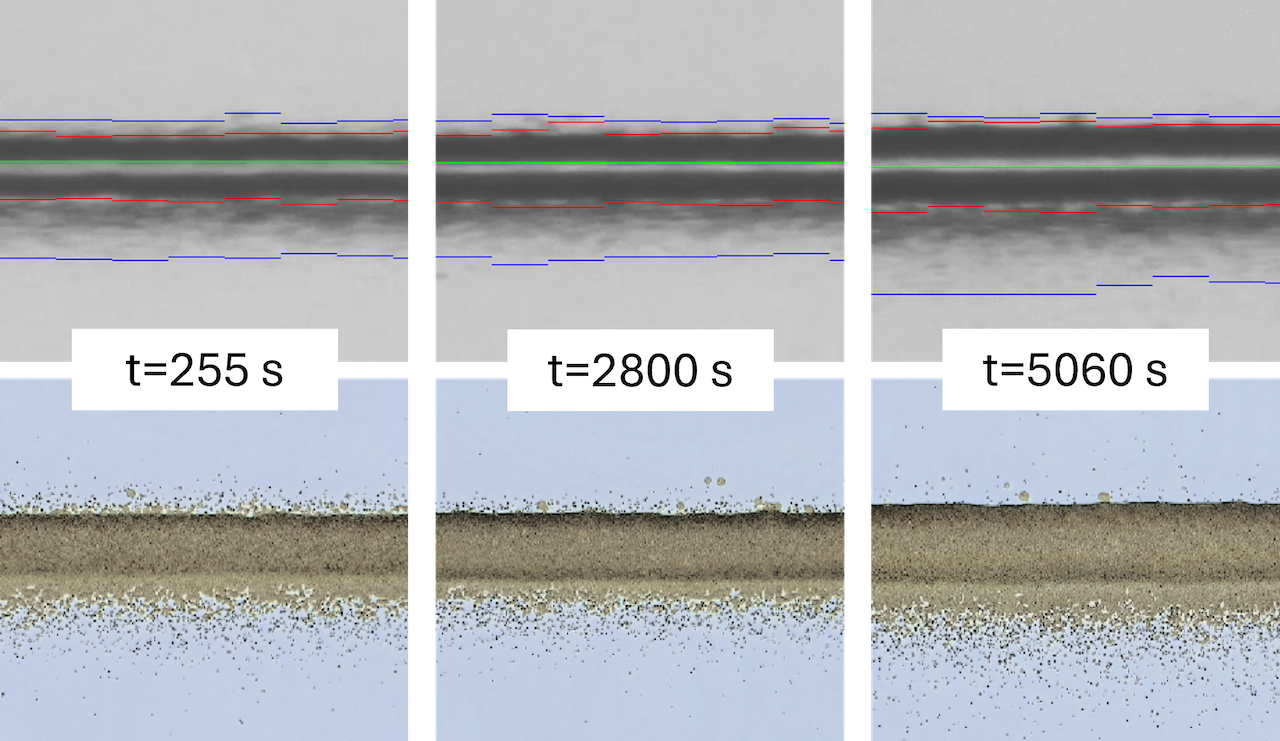}
    \caption{Image data of print. Top row depicts annotated frames from alignment camera and bottom row shows detailed images from the confocal microscope. Images in the same column roughly correspond to the same location on the print surface. Images were taken from print referred to in section 8.2}
    \label{fig:sup_dat_8_2}
\end{figure}

Further image data of the prints from sections \ref{sec:exp1_estimation}, \ref{sec:exp2_estimation}, and \ref{sec:exp3_parameter_estimation} are shown in figures \ref{fig:sup_dat_8_1}, \ref{fig:sup_dat_8_1_1}, and \ref{fig:sup_dat_8_2}. The figures show frames from the alignment camera video which have been annotated using the computer vision algorithm. For Figures \ref{fig:sup_dat_8_1} and \ref{fig:sup_dat_8_2}, the annotated images are shown alongside images from the confocal microscope of the corresponding location on the print surface. 

\FloatBarrier

\subsection{Aerosol Generation out of Ultrasonic Atomizer}
\label{app:aerosol_generation}
The probability density of an aerosol with diameter $\DA$ is given as a log-normal distribution as follows:
\begin{equation}
    p(\DA) = \frac{1}{\DA \sigma_{\DP}\sqrt{2\pi}} \exp \left[ -\frac{\ln(\DA)-\mu_{\DP}}{2\sigma_{\DP}^2} \right],
    \label{eq:aerosol_droplet_distribution}
\end{equation}
where $\mu_{\DP}$ and  $\sigma_{\DP}$ are the mean and standard deviation of $\ln(\DA)$.  
These parameters are expressed as a function of median $\DP$, fixing the standard deviation of $\DA$ to $1/4$ of $\DP$  \cite{secor2018principles}.

\subsubsection{Ink Deposition in Tube}
\label{app:tube_deposition}

Aerosolized droplets of the ink solution are transported through the carrier tube to the AJ nozzle. During this process, aerosol droplets have been observed to deposit onto the inner walls of the carrier tube. While the build up caused by this deposition is slow, it can lead to gradual changes in the print process that impact line quality and process stability.

To account for this phenomenon, the accumulation of aerosol droplets within the carrier tube is modeled as a latent variable. To simplify computations, the ink is assumed to be uniformly deposited along the inner walls of the tube. This accumulation is measured through the change in the tube’s effective radius, $\DT$, where a value of $\DT=0$ corresponds to zero deposition and $\DT=.785$~mm models a fully blocked carrier tube with a .785~mm inner radius (denoted as $\tubeR$).

The deposition rate of ink within the carrier tube (i.e., $\frac{d \DT}{dt}$) can be characterized using survival probabilities, which represent the likelihood that an aerosol droplet successfully traverses the entire length of the tube. Two primary mechanisms reduce the survival probability: diffusion, where lighter droplets disperse from the center of the carrier stream and adhere to the tube walls, and gravitational settling, where heavier droplets are pulled downward by gravity and accumulate on the lower surface.

The effect of diffusion and gravitational settling mechanisms on the distribution of aerosol droplets are modeled by a probability distribution function, $\Fdiff$ and $\Fgrav$  respectively. These mechanisms were previously derived using analytical models for general aerosol droplets in \cite{thomas1958gravity, secor2018principles}. Applying the work in \cite{thomas1958gravity,secor2018principles} to study the deposition in the carrier tube of the Optomec AJ machine, the diffusion and gravitational settling mechanisms are functions of the ink viscosity ($\eta$), carrier flow rate ($\Qcarr$) and aerosol droplet size ($\DA$). The survival probabilities affect the distribution of aerosol droplets out of the atomizer which follows a log-normal distribution, $p(\DA)$ \eqref{eq:aerosol_droplet_distribution}, with a median aerosol size of $\DP$.

The survival probability of a droplet with a diameter of $d_a$ due to gravitational settling is given by:
\begin{equation}
\Fgrav(d_a,\Qcarr,\RT)=\frac{2}{\pi}\left(\alpha\beta+\sin^{-1}{\beta}-2\alpha^3\beta\right),
\end{equation}
\begin{equation}
    \alpha=\left(\frac{3L\mu_{TS}}{8\mu_a(R_T-\RT)}\right)^{1/3}, \beta=\sqrt{1-\alpha^2}, 
\mu_a=\frac{\Qcarr}{\pi (R_T-\RT)^2},\ \mu_{TS}=\frac{\rho_pd_p^2gC_c}{18\eta_g}
\end{equation}
The parameters of the equation and their values for the Optomec AJ and silver-ink solution are provided in Table \ref{tab:carrier_tube_graviational_parameters}.

\begin{table}[h]
\centering
\begin{tabular}{||c c c||} 
 \hline
 Parameter & Description & Value \\ 
 \hline
 $L_T$ & Length of Tube & $0.4572$~m \\ 
 \hline
$R_T$ & Radius of Tube & $7.89\times 10^{-4}$~m \\
 \hline
 $d_a$ & Median diameter of ink droplets &  \\
 \hline
 $\rho_p$ & Droplet density & 5804~kg/m$^3$ \\
 \hline
 $\eta_g$ & Viscosity of carrier gas & $72\times 10^{-3}$ \\ 
 \hline
 $g$ & Gravitational acceleration & 9.8~m/s$^2$ \\ 
 \hline
 $\Qcarr$ & Carrier flow rate &  \\ 
\hline
 $C_c$ & Slip correction factor &  \\ 
 \hline
\end{tabular}
\caption{Parameters for Optomec Aerosol Jet Printer setup for printing silver ink using N$_2$ carrier gas}
\label{tab:carrier_tube_graviational_parameters}
\end{table}

\begin{figure}
    \centering
    \includegraphics[width=0.9\columnwidth]{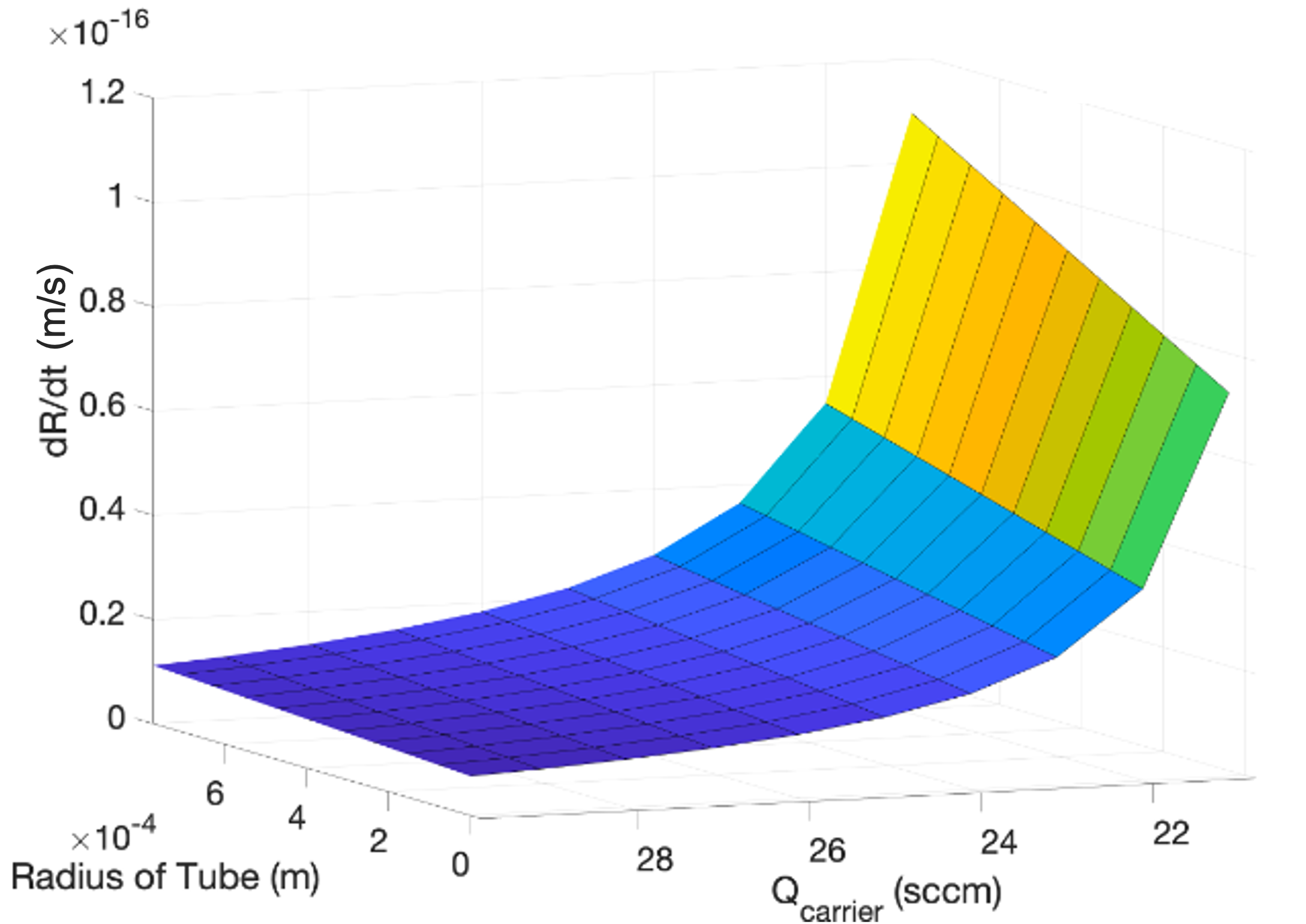}
    \caption{Rate of ink deposition in the carrier tube in \eqref{eq:dr_dt_carrier_supp} as a function of the radius of the tube (representing deposition) and carrier flow.}
    \label{fig:carrier_deposition_rate}
\end{figure}

For the silver-ink solution, the plot of the survival probability due to gravitational settling is shown in Figure  \ref{fig:carrier_deposition_rate}.

The second effect in the tube transport is due to diffusion of the ink droplets which impinge on the side of the tube. The survival probability of an ink droplet with diameter $d_a$ is modeled by the following equation:
\begin{align}
    \Fdiff\left(d_a,\Qcarr \right)=x_c^2\left(2-x_c^2\right) \\ 
    \left(1-x_c\right)^2\left(1-x_c^2\right)=\frac{\pi L_T}{\Qcarr}D(d_a),
\end{align}
where $x_c$ is the critical radius and $D(d_a)$ is the aerosol diffusion coefficient as a function of the ink droplet diameter, $d_a$, provided by the Stokes-Einstein equation:
\begin{equation}
    D\left(d_a\right)=\frac{kTC_c}{2\pi\mu d_a}.
    \label{eq:stokes_einstein}
\end{equation}

However, for the silver-ink solution, the survival probability of droplets between 1~\textmu{m} and 5~\textmu{m} is nearly 1, making the effect of drift on deposition negligible. In this case, deposition is approximated solely by $\Fgrav$. We therefore define $\Fdep^T = (1-\Fgrav)$. The rate of volume of ink deposited in the carrier tube, denoted $Q_{dep}^T$, is then defined as:

\begin{equation}
\begin{aligned}
    Q_{dep}^T(t) = \int_{\DA} \Fdep^T\left(\DA, \DT(t), \Qcarr(t) \right) \; p(\DA)\; \Qdrop d(\DA).
    \label{eq:ink_deposition_rate_carrier}
\end{aligned}
\end{equation}
%
%
where $\Qdrop$ is the ink flow rate given as $\Qdrop(t) = \AD(t) \Qcarr(t)$ with $\AD$ as the latent state variable for volume fraction of aerosol droplets in Table \ref{tab:dt_state_variables}.

Assuming the volume of ink deposition uniformly coats the tube, the change in the carrier tube radius due to ink deposition is:
\begin{equation}
\begin{aligned}
 \frac{d\DT}{dt} =&  \frac{Q_{dep}^T(t)}{2\pi (\tubeR - \DT \left(t\right))L_T}
 \label{eq:dr_dt_carrier_supp}
\end{aligned}
\end{equation}
where  $L_T$ is the length of the coated carrier tube. The rate of change in the radius of the tube is plotted in Figure \ref{fig:carrier_deposition_rate} with respect to the change in radius of the tube, $\DT (t)$, and the carrier gas flow rate, $\Qcarr(t)$.

\subsubsection{Ink Deposition in Nozzle}
\label{app:nozzle_deposition}
Aerosol droplets flowing through the AJ system can deposit onto the AJ nozzle, leading to clogging and potential disruptions in printing quality. While this can be rare it is a known occurrence to AJ users. The rate of ink deposition on the nozzle is analyzed by the survival probability of a single aerosol droplet successfully passing through it.

As discussed in the previous section, survival probabilities are generally influenced by both diffusion and gravitational settling. However, since the nozzle is oriented vertically, gravitational effects do not impact droplet survival and the probability of an aerosol droplet with size $d_a$ depositing on the nozzle is governed by the diffusion mechanism, denoted as $\Fdiff$, and is given by: 
\begin{align}
    \Fdiff\left(d_a,\Qcarr+\Qsh \right)=x_c^2\left(2-x_c^2\right) \\ 
    \left(1-x_c\right)^2\left(1-x_c^2\right)=\frac{\pi L_N}{\Qcarr+\Qsh}D(d_a),
\end{align}
where $x_c$ is the critical radius and $D(d_a)$ is the aerosol diffusion coefficient as a function of the ink droplet diameter, $d_a$, provided by the Stokes-Einstein equation in \eqref{eq:stokes_einstein}. This probability is a function of key process parameters, including the carrier and sheath gas flow rates ($\Qcarr$, $\Qsh$), the median aerosol droplet diameter ($\DP$), the length of the nozzle, $L_N$, and the nozzle radius ($\DT$).
The rate of ink deposited in the nozzle is given as
\begin{equation}
\begin{aligned}
    Q_{dep}^{N}(t) = \int_{\DA}\Fdep^N(\DA, \DN(t), \Qcarr(t)+\Qsh(t) \\ \times p_{log}(\DA)\Qdrop(t)\; d(\DA).
\end{aligned}
\end{equation}
where $p_{log}(\DA)$ is the log-normal distribution of aerosolized particles out of the ultrasonic atomizer and $\Fdep^N = 1 - \Fdiff$. Note, since the survival probability through the carrier tube is nearly 1 (i.e., values of $\Fdep^{carr}$ of 0), the distribution of aerosol droplets out of the carrier tube is assumed to follow the log-normal distribution out of the ultrasonic atomizer. 


Similarly the aerosol droplets are assumed to uniformly deposit around the inner wall of the nozzle. As a result, the change in the nozzle radius is described as 

\begin{equation}
 \frac{d\DN}{dt} =  \frac{1}{2\pi (\nozzleR - \DN(t))L_N}\int_{\DA} \Fdep^{N}(\DA)\; \Qdrop(t) \; d(\DA).
\end{equation}
where $\nozzleR$ is the nominal radius of the nozzle at 35~\textmu{m}.

The plots derived from this equation are shown in Figure \ref{fig:nozzle_dep}:

\begin{figure}[h]
    \centering
    \includegraphics[width=1.0\columnwidth]{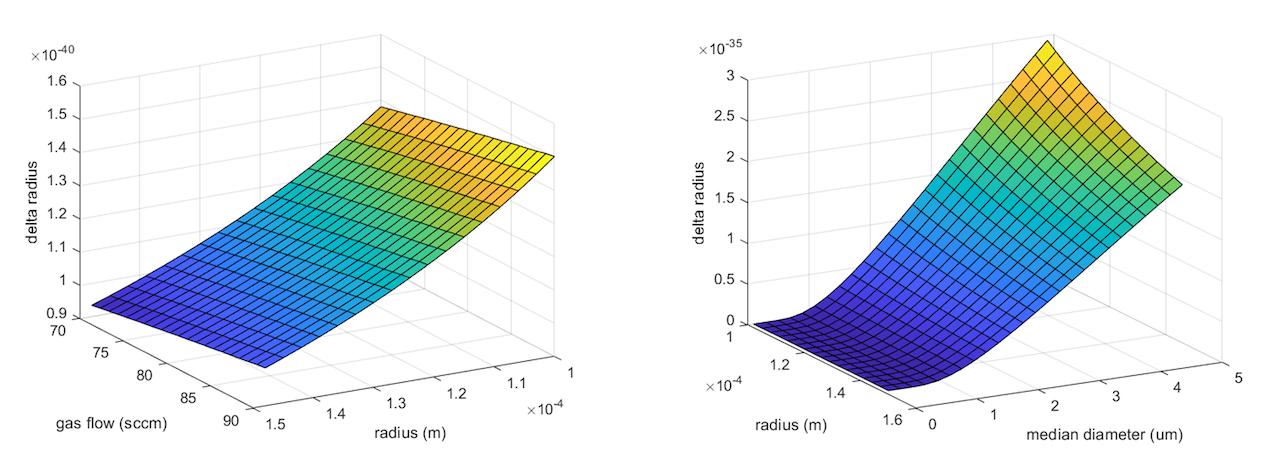}
    \caption{Rate of ink deposition in the nozzle. Note that while units displayed are not mks, the calculation for nozzle radius was made with mks units for input.}
    \label{fig:nozzle_dep}
\end{figure}


\subsection{Estimating the Droplet Generation Model}
\label{app:effective_droplet_generation}

The net droplet generation rate, $H(\cdot)$, is expressed in units of volume per unit time (\textmu{m}$^3$/s). To determine an analytical function of $H(\cdot)$ a Box-Behnken design of experiments is used to determine the response as a function of the following independent variables: carrier flow rate ($\Qcarr$), volume of ink solution in the vial ($\Vl$), and atomization power ($\PA$). Three levels for each independent variable were selected and are shown below in Table  \ref{tab:doe_levels}. 

\begin{table*}[]
    \centering
    \begin{tabular}{|c|c|}
        \hline
         Independent Variable & Levels  \\
         \hline
         Carrier flow rate ($\Qcarr$) &  15, 25, 35 [sccm] \\
         \hline
         Volume of ink solution in the vial ($\Vl$) & 0.5, 1, 1.5 [mL] \\
         \hline
         Atomization power ($\PA$) & 300, 370, 440 [mA] \\
         \hline
    \end{tabular}
    \caption{Levels used for inputs in experimental determination of $H(\cdot)$}
    \label{tab:doe_levels}
\end{table*}
In this work, the analytical form of $H(\cdot)$ is assumed to be quadratic and therefore the experiments are run with 13 unique inputs, with the center point in the process space repeated twice to quantify experimental standard deviation. The experimental inputs are chosen such that they lie at the midpoints of edges in the three-dimensional process space. The experiment consisted of printing lines 2.5 cm in length at each of the predetermined process inputs. For each print, the platen temperature was set to 70~$^{\circ}$C and sheath flow to 50 sccm. The process speed was constant for the duration of the print set at 2 mm/s, meaning the print for each operating point lasted for 12 seconds. For process configurations where the material rate coming out of the nozzle was low, the process speed was slowed down to improve the post processing measuring accuracy. After changing the operating points, the AJ was allowed to print for a set amount of time before printing the 2.5 cm line to allow the internal states of the tool to settle before beginning the experiment.  \par
After printing, the profiles from the silver lines were measured at three places using a Tencor p15 surface profilometer. The net droplet generation rate was then calculated by integrating the area under the measured profiles and multiplying by the process speed. The droplet generation rate was then averaged across the three profilometer scans, giving the value of $H(\cdot)$ at that point. The data was then fit to a quadratic model using ordinary least squares with L1 regularization and a penalty parameter $\alpha$=9. To maximize the model accuracy, the coefficients with p-values greater than .05 were pruned, leaving the following model:
\begin{align*}
H(\cdot) =  & 3.3\times 10^2 \cdot\Qcarr^2+2.7\times 10^4 \cdot\Vl \cdot\Qcarr+66\cdot\Qcarr\cdot\PA + \\ &2.1\times 10^2 \cdot\PA\cdot\Vl-5.2\times 10^5 \cdot\Vl -4.8\times 10^4 \cdot\Qcarr-1.1\times 10^3 \cdot\PA+7.7\times 10^5 \cdot
\label{eq:model}
\end{align*}

\section{Computational Fluid Dynamic Simulation of Aerosol Jet Printer}
\label{app:cfd_simulation}

\begin{figure}
    \centering
    \includegraphics[width=0.2\columnwidth]{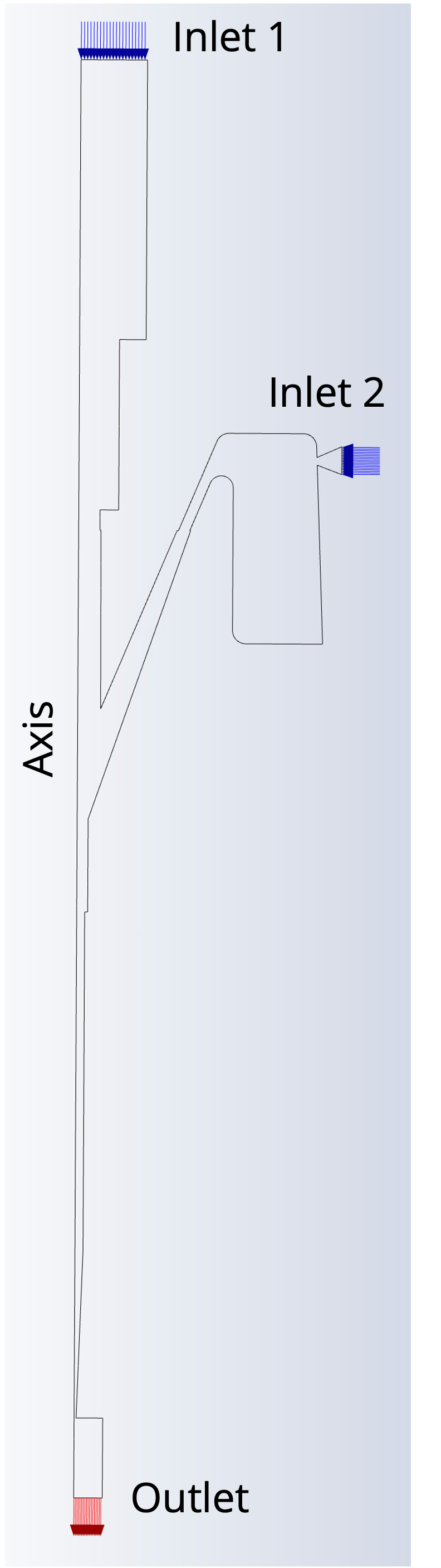}
    \caption{Axisymmetric model of Aerosol Jet Printer}
    \label{fig:axisymmetric_model}
\end{figure}

The trajectory of ink droplets within the AJ system is simulated using computational fluid dynamics (CFD) in ANSYS Fluent. To simplify the model and mitigate numerical issues that may cause asymmetries, an axisymmetric representation of the AJ machine, shown in Figure \ref{fig:axisymmetric_model}, is utilized. This model is then rotated around its central axis to approximate a full three-dimensional system. 

The model consists of three primary boundaries: Inlet1, Inlet2, and the Outlet. Inlet1 represents the carrier gas boundary, which is also responsible for injecting ink droplets into the AJ machine. Inlet2 introduces the sheath gas, which focuses the ink stream, while the Outlet corresponds to the substrate boundary where the ink droplets are deposited.

A multiphase flow simulation in ANSYS Fluent is defined to accurately capture aerosol jet dynamics. In this model, nitrogen (N$_2$) serves as the carrier gas, defining the continuous phase, while ink droplets are treated as discrete particles transported along the fluid field. The carrier gas flow is modeled as a continuum using the Navier-Stokes equations, whereas the ink droplets follow an Euler-Lagrange approach, where each droplet’s trajectory is individually computed based on velocity field gradients.

Since ink droplets occupy a very low volume fraction in the system, the interactions between discrete droplets are assumed to be negligible. This assumption is validated by micro-well experiments, which show that the ink-to-N$_2$ volume ratio is on the order of $10^{-8}$. As a result, the ink droplets interact with the fluid phase but do not significantly influence the nitrogen gas flow, thereby simplifying the computation without sacrificing accuracy.

\subsection{ANSYS Model Setup}

\begin{figure}
    \centering
    \includegraphics[width=\columnwidth]{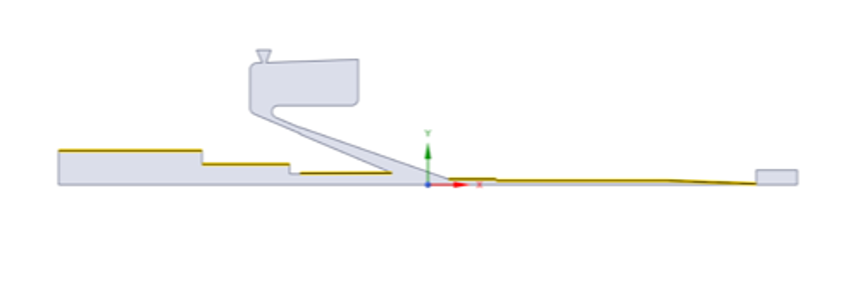}
    \caption{Nozzle deposition is modeled by modifying the geometry of the axisymmetric model. For a deposition of $\Delta R_{nozzle}$, the edges highlighted in yellow are moved closer to the axis by an identical amount.}
    \label{fig:nozzle_deposition_model}
\end{figure}

The ANSYS Fluent simulation begins with an axisymmetric representation of the AJ nozzle based on \cite{chen2018effect}. Ink deposition is modeled by reducing the nozzle radius to account for accumulation on the inner walls, with the highlighted edges in the axisymmetric model (Figure \ref{fig:nozzle_deposition_model}) decreased in steps of $\DN = 7.5$~\textmu{m}. In the absence of deposition, the nozzle tip has a minimum radius of 75~\textmu{m}. This procedure produces ten distinct geometries, each simulated under varying carrier flow, sheath flow, and median droplet diameter conditions to evaluate the effect of nozzle deposition on linewidth and overspray (Figure \ref{fig:line_quality_v_nozzle_deposition}).

With the geometry defined, the simulation of aerosol droplets is performed in two steps: first fluid flow is computed and then a discrete number of aerosol droplets are injected.

\begin{figure}
    \centering
    \includegraphics[width=0.9\columnwidth]{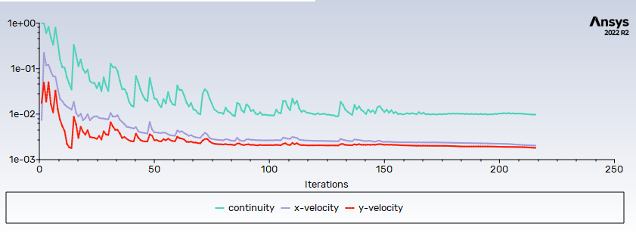}
    \caption{Plot of scaled residuals during ANSYS simulation iterations}
    \label{fig:scaled_residuals}
\end{figure}

The gas flow through the AJ, defined as N$_2$ gas, is injected at the carrier gas stream and the sheath gas stream mass flow inlets (Inlet1 and Inlet2). This defines a boundary condition for the N$_2$ gas and is simulated to model fluid flow of the N$_2$ gas through the AJ machine and onto the substrate as shown in Figure \ref{fig:ansys_velocity}. The simulations are solved until the continuity error of conservation of mass is less than a tolerance of 0.01. The continuity error over iterations is shown in Figure \ref{fig:scaled_residuals}.

\begin{figure}
    \centering
    \includegraphics[width=0.6\columnwidth]{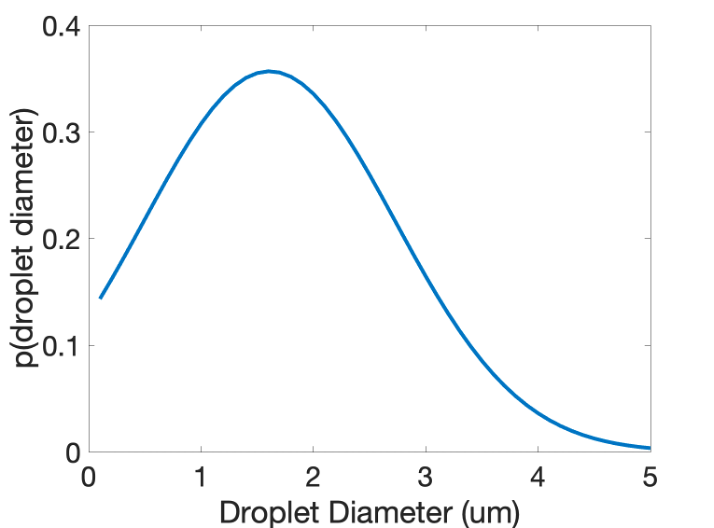}
    \caption{Distribution of ink droplet diameters out of atomizer from \cite{secor2018principles}}
    \label{fig:atomizer_diameter_distribution}
\end{figure}

\begin{figure}
    \centering
    \includegraphics[width=0.6\columnwidth]{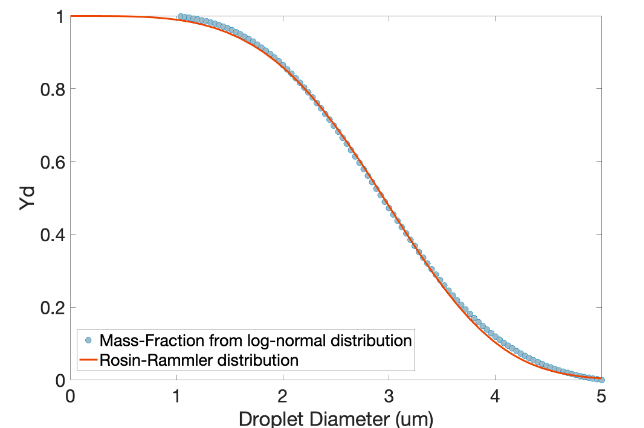}
    \caption{Mass fraction of ink droplet diameters into the Aerosol Jet printer is fit onto a Rosin-Rammler distribution.}
    \label{fig:mass_fraction_rosin_rammler}
\end{figure}

Once the fluid flow solution is obtained, ink droplets, modeled as discrete particles, are introduced from the carrier inlet, represented by Inlet1. These discrete particles represent aerosolized droplets with a specified size distribution. Aerosol droplet sizes are known to follow a log-normal distribution from the ultrasonic atomizer \cite{secor2018principles}, as illustrated in Figure \ref{fig:atomizer_diameter_distribution}. However, ANSYS Fluent does not natively support log-normal distributions for discrete particle diameters. To address this, the droplet size distribution is approximated using a Rosin-Rammler distribution, which represents the mass fraction, $Y_d$, of discrete particles as:

\begin{equation}
Y_d(d_a) = e^{-(\DA/\bar{d}_a)^n},
\end{equation}
where  $\bar{d}_a$  is the mean droplet diameter, and $n$ is a spread parameter chosen to best fit the log-normal distribution. The accuracy of this approximation is demonstrated in Figure \ref{fig:mass_fraction_rosin_rammler}, which compares the Rosin-Rammler fit to the log-normal distribution of ink droplet diameters.

After defining the initial conditions and particle size distribution, the trajectory of each aerosol droplet is simulated using an Euler-Lagrange approach. To simplify the the computational model, the method assumes negligible particle interactions and a low particle-to-carrier gas volume ratio.

\begin{figure}
    \centering
    \includegraphics[width=0.6\linewidth]{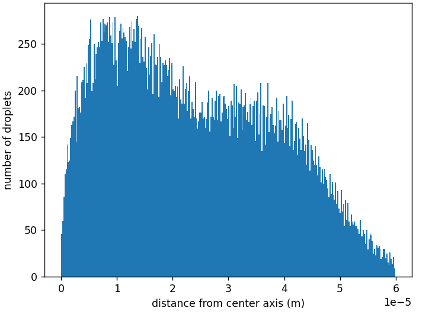}
    \caption{The output of the CFD simulation provides the radial position of the aerosol droplets from the center of the substrate for 5000 particles. The plot illustrates the number of particles as a function of radial distance.}
    \label{fig:radial_output_fluent}
\end{figure}

The simulation output provides the radial distribution of aerosol droplets on the substrate (i.e., Outlet), as shown in Figure \ref{fig:radial_output_fluent}. This data is then processed to define an expected linewidth and overspray values that closely align with experimental measurements.

\subsubsection{Estimating Line Quality Parameters}
\label{app:estimating_line_params_from_ansys}

Because the droplet behavior after impinging on the surface is not explicitly modeled by the simulation, the output of the above CFD analysis is a series of aerosolized droplets of various sizes at locations along the radial axis. This represents the radial locations of the droplets after impinging on the print surface. The goal is to define linewidth and overspray based on CFD results that capture the secondary effects of droplet behavior, closely approximating the features of the printed silver lines. This approach enables the use of CFD analysis to assess how latent states and inputs influence the final printed output.

\begin{figure}[h!]
    \centering
    \includegraphics[width=0.9\linewidth]{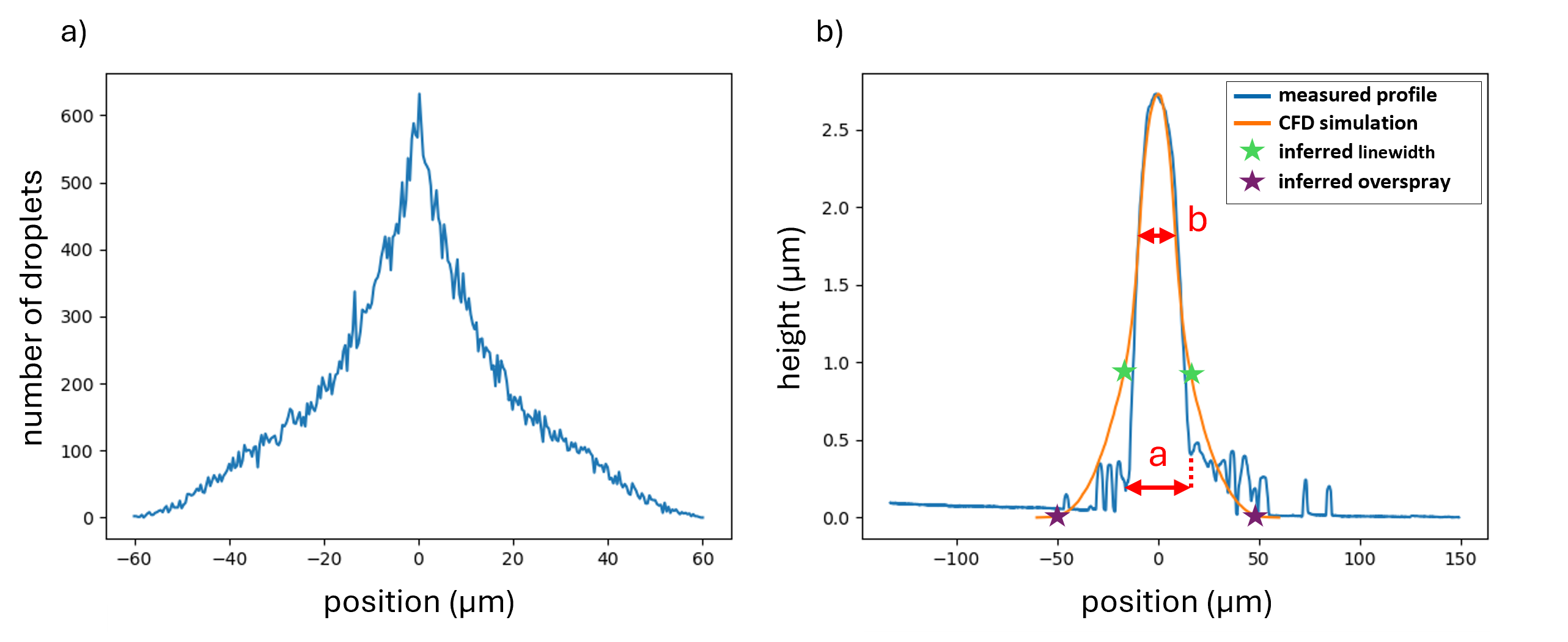}
    \caption{a) Number of droplets along cross section of line. b) Transformed CFD profile overlayed on physical profile. Points where linewidth and overspray are inferred are shown with stars.}
    \label{fig:profile_droplets}
\end{figure}

The output of the CFD simulation represents the impingement of aerosol droplets onto a static platen. However, in aerosol jet printing, the printed trace is formed by moving the platen beneath the stationary nozzle. To simulate this dynamic process using static CFD data, the droplet positions are first transformed from an axisymmetric coordinate system to a 2D Cartesian plane by randomly assigning an azimuthal angle from $[0, 2\pi]$ to each droplet. Then, a printed line is formed by summing the droplet positions along one dimension, effectively emulating the motion of the nozzle in a straight line. This results in the histogram shown in Figure \ref{fig:profile_droplets}~a) that describes the number of droplets along the axis perpendicular to motion of the platen.

Next, to accurately compare the CFD simulation output with experimental data, the histogram of droplet distributions in an incremental simulated line is converted into a height profile representing the volume of aerosol droplets along the axis. Each droplet is assumed to impinge on the print surface and spread by a constant factor \cite{droplet_spread_factor}, with its mass distributed evenly over this footprint. The total mass deposited at a given point is obtained by summing the contributions of all droplets. The droplet spreading factor is set to 3.8, consistent with previous studies \cite{droplet_spread_factor}. The resulting profile, alongside a measured profile, is shown in Figure \ref{fig:profile_droplets}~b). This volumetric representation enables the definition of quantitative line quality metrics, providing insight into how hidden parameters influence print quality.

As shown in Figure \ref{fig:profile_droplets}~b), the output of the CFD simulation closely matches experiment near the center of the lines. The divergence between the simulated and experimental results occur toward the edge of the line, where the effect of droplet behavior on the print surface becomes more prominent. This discrepancy arises because the CFD simulation does not capture post-impact liquid interactions or the cumulative forces from impinging droplets. To ensure that the CFD simulations accurately reflect real-life prints, the simulated linewidth and overspray are corrected to match experimental observations. This correction is based on identifying the point where the CFD-generated height profile and the measured profile begin to diverge, denoted as width $b$ in Figure \ref{fig:profile_droplets}~b). Based on comparisons between CFD profiles and measured data, $b$ is defined at 80\% of the peak value.

The ratio of the width $b$ to the experimental linewidth, shown in Figure \ref{fig:profile_droplets}b as width $a$, is evaluated across a wide range of experimental conditions. Analysis of experimental data across diverse operating conditions shows that $b/a = 2.1$ with a standard deviation of $\sigma = 0.32$. 

Therefore, to estimate the experimental linewidth from simulations, $b_{\rm CFD}$ is first determined as the width corresponding to 80\% of the maximum peak as:
\begin{equation}
    b_{CFD} = \max_{r_1,r_2} |r_1 - r_2|, s.t.\;\; h(r) > 0.8 \; h_{max} \;\forall~r\in[r_1,r_2]
\end{equation}
where $h_{max} = \max_r h(r)~\forall~r\in [\infty, \infty]$. $h(r)$ represents the height profile from the CFD simulation as a function of $r$, denoting the distance perpendicular from the center of the simulated line.

The width, $b_{CFD}$, is then multiplied by a constant ratio of $b/a = 2.1$ to approximate the experimental linewidth given by:
\begin{equation}
    L_{w_{CFD}} = b_{CFD}\cdot b/a
\end{equation}
This correction quantifies the gap between CFD predictions and experimental measurements, enabling more accurate estimation of printed feature dimensions.

Overspray in the CFD profile is defined as the point where the profile reaches 1\% of the height at each linewidth edge. This threshold was selected empirically by comparing CFD-generated profiles with experimentally measured cross-sectional profiles of printed traces across a wide range of process conditions.

This choice is justified by the CFD simulation’s ability to accurately capture the spatial distribution of aerosol droplets as they exit the nozzle and strike the substrate. While post-impingement interactions are not explicitly modeled, the simulation reliably predicts the initial droplet landing positions, particularly at the outer edges of the overspray region where droplet density is low and secondary effects, such as liquid interactions and surface tension-driven flow, are minimal.

\FloatBarrier
\section{Modeling Fluidic Resistance of Aerosol Jet Printer}
\label{app:fluidic_resistance}
\begin{figure}
    \centering
    \includegraphics[width=1\columnwidth]{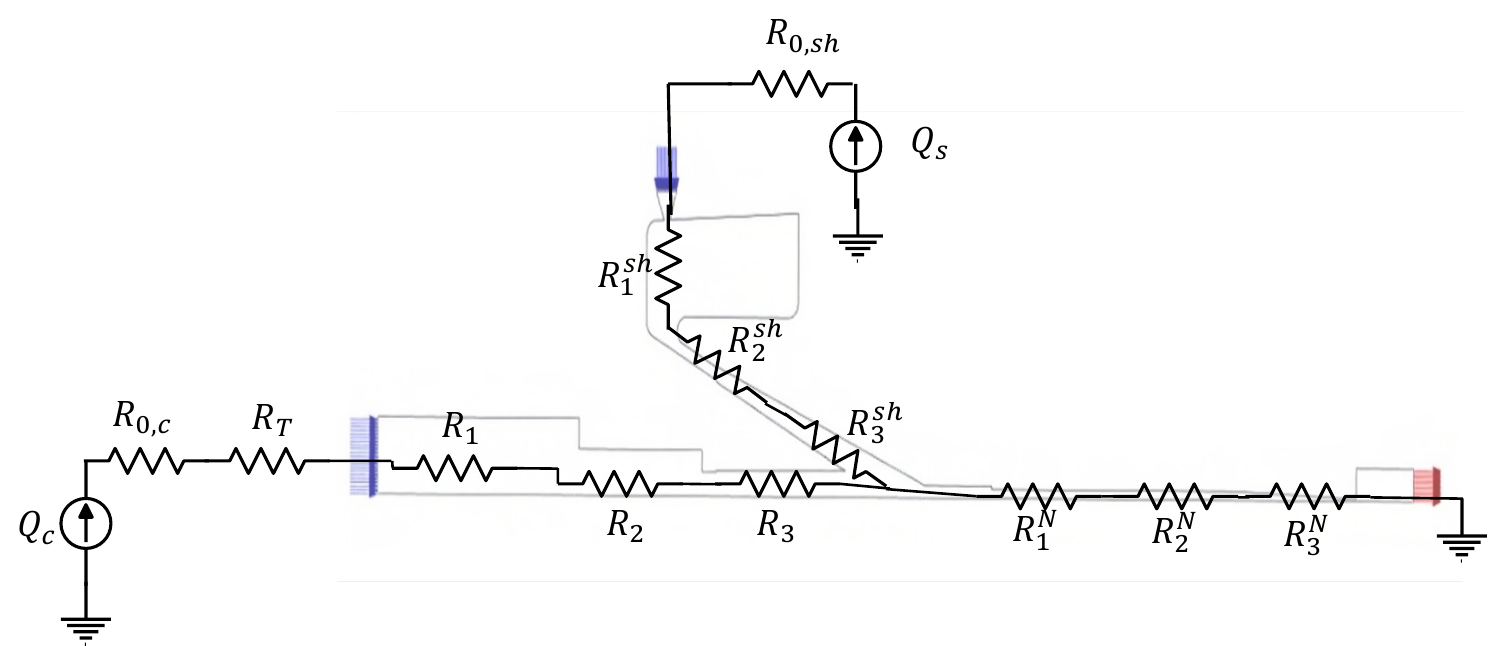}
    \caption{A resistive respresentation of the AJ models the fluidic resistance through the printer with input flows, $\Qcarr$ and $\Qsh$. Individual resistances are overlayed on top of the AJ part to which they correspond. Additionally, $R_T$ models the resistance of the carrier tube, and $R_{0,sh}, R_{0,c}$ correspond to the resistances of the tubes from the sheath and carrier MFCs, respectively.}
    \label{fig:resistive_model_AJ}
\end{figure}

A fluidic resistance model of the AJ system is developed to relate measured gas pressures to gas flow rates. This model is derived by calculating the fluidic resistance of each segment of the AJ system, including the axisymmetric geometry shown in Figure~\ref{fig:axisymmetric_model} and the tubing connecting the MFCs and ultrasonic atomizer.
The resulting resistive network is illustrated in Figure~\ref{fig:resistive_model_AJ}, with the computed resistance values for each segment provided in Table \ref{tab:resistance_values}.

\begin{table*}[]
    \centering
    \begin{tabular}{||c c c||}
    \hline
        Physical Element &  Resistance Name & Resistance [Pa-s/m$^3$]\\
        \hline
        Carrier Tube & $R_T$ &\eqref{eq:resistance_carrier_tube} \\
        \hline  
        Carrier Gas Inlet & $R_1$ &$1.1972\times10^4$ \\
        \hline
        Carrier Gas Inlet & $R_2$ &$5.6305\times10^4$ \\
        \hline
         Carrier Gas Inlet & $R_3$ &$6.9590\times10^5$ \\
        \hline 
          Sheath Gas Inlet & $R_1^{sh}$ &$2.185\times10^5$ \\
        \hline  
          Sheath Gas Inlet & $R_2^{sh}$ & $175.66$ \\
        \hline   
          Sheath Gas Inlet & $R_3^{sh}$ & $1.28 \times 10^{6}$ \\
        \hline  
           Nozzle & $R_1^{N}$  &$5.36\times10^6$ \\
        \hline  
           Nozzle & $R_2^{N}$ &$6.07\times10^{7}$ \\
        \hline  
        Nozzle & $R_3^{N}$ & \eqref{eq:resistance_nozzle_tip} \\
        \hline  
    \end{tabular}
    \caption{Fluidic resistance values of each section of the AJ in Figure \ref{fig:resistive_model_AJ}, including the carrier gas inlet, sheath gas inlet, and nozzle.}
    \label{tab:resistance_values}
\end{table*}
The resistance of the carrier tube, $\tubeResistance$, is a function of the ink deposition in the tube, $\DT$, defined as:
\begin{align}
      \tubeResistance(\DT) &= \frac{8 \eta L_T}{\pi (\tubeR - \DT)^4} \\
      \label{eq:resistance_carrier_tube}
\end{align}
where $\eta$ is the viscosity of the ink (17.5 Pa$\cdot$s), $L_T$ is the length of the tube and $\tubeR$ is the radius of the tube which reduces linearly with ink deposition, $\DT$.

\begin{figure}
    \centering
    \includegraphics[width=0.4\columnwidth]{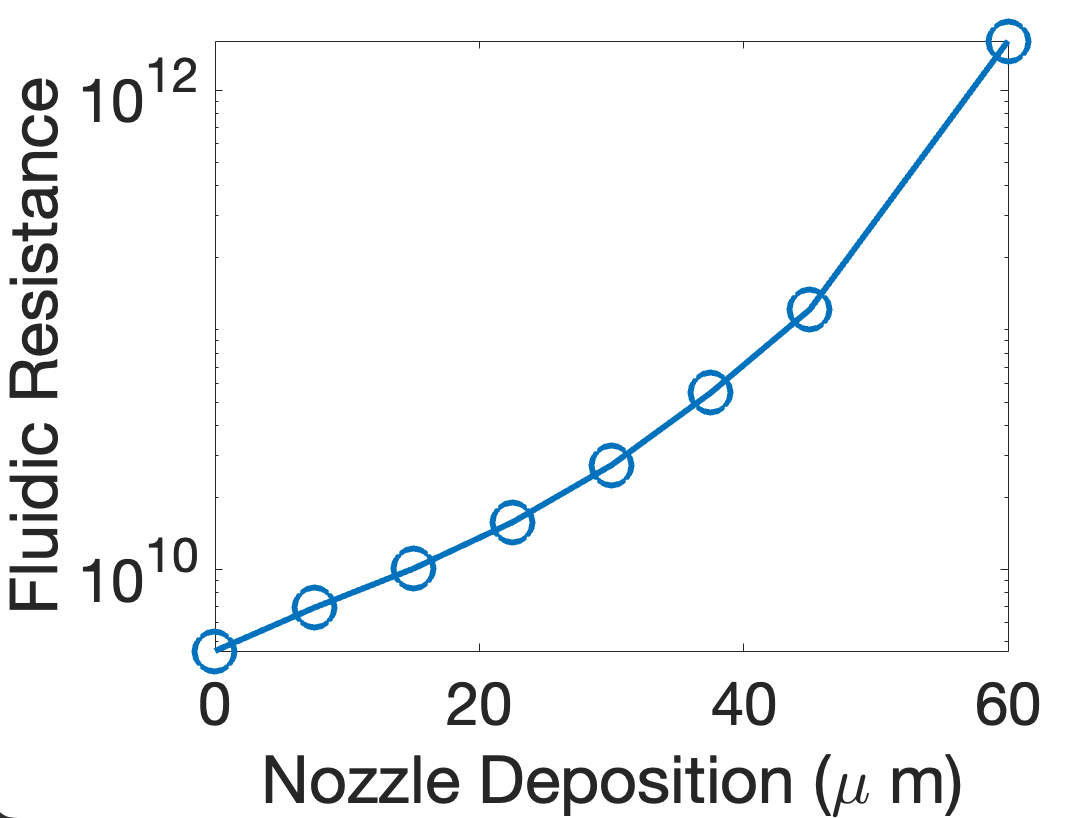}
    \caption{Results of ANSYS simulation for fluidic resistance of nozzle tip}
    \label{fig:fluidic_resistance_nozzle_tip}
\end{figure}

The fluidic resistance of the nozzle tip, $R_3^N$, depends on ink deposition in the nozzle, $\DN$, and requires a nonlinear analysis of the Navier-Stokes equations \cite{manual2009ansys}. An analytical relationship between $R_3^N$ and $\DN$ is derived by varying the ink deposition in the nozzle and measuring the resulting pressure using ANSYS Fluent \cite{manual2009ansys}. The results are shown in Figure \ref{fig:fluidic_resistance_nozzle_tip}, and the best polynomial fit for the fluidic resistance is a 7th-order polynomial:
\begin{equation}
    R_3^N(\DN) = \alpha_0 +\sum_{i=1}^7 \alpha_i \DN^i,
    \label{eq:resistance_nozzle_tip}
\end{equation}
where the values for $\alpha_i$ are provided in Table \ref{tab:polyfit_resistance_nozzle_deposition}.

\begin{table}[]
    \centering
    \begin{tabular}{||c|c||}
    \hline
        Coefficient & Value \\
        \hline
        $\alpha_0$ & $4.57\times10^9$ \\
        $\alpha_1$ & $2.75\times10^9$ \\
        $\alpha_2$ & $-7.96\times10^8$ \\
        $\alpha_3$ & $9.73\times 10^7$ \\
        $\alpha_4$ & $-5.81\times10^6$ \\
        $\alpha_5$ & $1.83\times 10^5$ \\
        $\alpha_6$ & $-2.92\times 10^3$ \\
        $\alpha_7$ & $18.67$\\
        \hline
    \end{tabular}
    \caption{Coefficients of 7th order polynomial fit for modeling fluidic resistance of the nozzle tip, $R_3^N$. Units of the coefficients are consistent with the m-k-s system.}
    \label{tab:polyfit_resistance_nozzle_deposition}
\end{table}

The total resistance from the carrier port is defined as:
\begin{equation}
    R_{c} = R_T + R_1 + R_2 + R_3 + R_1^N + R_2^N + R_3^N,
    \label{eq:carrier_resistance}
\end{equation}
and the resistance from the sheath port is defined as:
\begin{equation}
    R_{sh} = \frac{R_1^{sh} + R_2^{sh}+R_3^{sh}}{2} + R_1^N + R_2^N + R_3^N.
\end{equation}

The relation between carrier pressure readings, flow and nozzle ($\DN$) and tube ($\DT$) deposition is given by:
\begin{equation}
    \Pcarr = \Qcarr \times R_{c}(\DN, \DT) + \Qsh(R_1^N + R_2^N + R_3^N(\DN)),
    \label{eq:carrier_pressure_eq}
\end{equation}
for the carrier pressure. The sheath pressure is given by
\begin{equation}
    \Psh = \Qsh\times R_{sh}(\DN) + \Qcarr(R_1^N + R_2^N + R_3^N(\DN)).
    \label{eq:sheath_pressure_eq}
\end{equation}

\section{RTS Smoother Algorithm}
\label{app:rts_smoother}

\begin{algorithm}[H]
\caption{Rauch--Tung--Striebel (RTS) Smoother}
\label{alg:rts_smoother}
\begin{algorithmic}[1]
\State{Input: Initial state estimate $\hat{x}_0$, error covariance $P_0$, transition function $f_d(x, u)$, output function $g_d(x, u)$, transition noise covariance $\Sigma_\xi$, output noise covariance $\Sigma_w$, inputs $\{u_k\}$, measurements $\{y_k\}$, time window $T$}

\State \textbf{Forward Kalman filter pass} \State{\quad$\hat x_{k|k}$, $P_{k|k} \gets$ Algorithm \ref{alg:ekf_alg} }

\Statex
\State \textbf{Backward RTS smoothing pass}
\For{$k = n-1:0$}
  \State $J_k \gets P_{k|k} F_k^\top \big(P_{k+1|k}\big)^{-1}$ 
  \State $\hat x_{k|n} \gets \hat x_{k|k} + J_k \big(\hat x_{k+1|N} - \hat x_{k+1|k}\big)$
  \State $P_{k|n} \gets P_{k|k} + J_k \big(P_{k+1|n} - P_{k+1|k}\big) J_k^\top$
\EndFor
\end{algorithmic}
\end{algorithm}

\subsection{Covariance Matrix Values of Digital Model}
\label{app:covariance}
The diagonal entries of the covariance matrices, $\Sigma_\xi=\text{diag}([\sigma^2_{\DA}, \sigma^2_{\Vl}, \sigma^2_{\DT}, \sigma^2_{\DN}, \sigma^2_{\AD}])$ and $\Sigma_w=\text{diag}([\sigma^2_{\lw}, \sigma^2_{\ov}, \sigma^2_{\Pcarr}, \sigma^2_{\Psh}, \sigma^2_{\Qink}])$, represent the process noise variances associated with the transition model in \eqref{eq:transition model} and the output model in \eqref{eq:output_model}, respectively. The standard deviations are estimated by analyzing deviations between measured and predicted state and output trajectories under steady operating conditions. The values of the covariance matrix for the transition and output models are provided in Table \ref{tab:transition_covariance_values} and Table \ref{tab:output_covariance_values}, respectively.

In future work, the covariance matrices should be dynamically adjusted using the expectation–maximization, allowing the measurement and state uncertainties to be iteratively estimated from data to better reflect time-varying system behavior.
\begin{table}[h!]
\parbox{.45\linewidth}{
    \centering
    \begin{tabular}{|p{1cm}|p{2cm}|}
        \hline
        $\sigma_{\DA}$ & 0.1 \textmu m/s \\
         \hline
        $\sigma_{\Vl}$ & $1e-3$ mL/s\\
         \hline
        $\sigma_{\DT}$ & $1e-3$ \textmu m/s\\
         \hline
        $\sigma_{\DN}$ & $3.35e-3$ \textmu m/s\\
         \hline
         $\sigma_{\AD}$ & $1e-8$ 1/s\\
         \hline
    \end{tabular}
    \caption{Standard deviation values of the transition function (used to form the covariance matrix, $\Sigma_\xi$)}
    \label{tab:transition_covariance_values}
}
\hfill
\parbox{.45\linewidth}{
    \centering
    \begin{tabular}{|p{1cm}|p{2cm}|}
        \hline
        $\sigma_{\lw}$ & 3 \textmu m\\
         \hline
        $\sigma_{\ov}$ & 5 \textmu m\\
         \hline
        $\sigma_{\Pcarr}$ & $10$  sccm\\
         \hline
        $\sigma_{\Psh}$ &10 sccm\\
         \hline
         $\sigma_{\Qink}$ &$1e-5$ sccm\\
         \hline
    \end{tabular}
    \caption{Standard deviation values of the output function (used to form the covariance matrix, $\Sigma_w$)}
    \label{tab:output_covariance_values}
    }
\end{table}

\section{Experimental Setup}
\label{app:experimental_setup_validation}
The accuracy of the digital twin state estimation is validated through manual measurements of the latent states at discrete time point. The deposition in tube and nozzle were measured by separately weighing each component in the tube and nozzle assembly using a high-precision scale (Radwag AS 60/220 R2, .01 mg resolution). This measurement was typically taken before and after printing to measure the mass of the ink accumulated during the print. Similarly, the change in ink vial level is measured before and after printing. 
The volume fraction of aerosol droplets in carrier stream, $\AD$, is measured by post-processing the wafer using a 3D optical profilometer (Keyence VKX3000). The total material printed during a given time-window is calculated by integrating the height of the printed surface across the field of view as:
\begin{equation}
    V_{ink} = \inkdensity^{-1} \cdot\iint \limits_{xy} h_s(x,y)dydx
\end{equation} 
where $h_s(x,y)$ is the height of the printed material at a point $x,y$ on the substrate surface and $\inkdensity = .087$ and is defined as the volume percentage of silver in the ink solution.

The aerosol volume fraction, $\AD$, is then determined by:
\begin{equation}
    \AD(t) = \frac{V_{ink}}{V_c}\quad 
\end{equation}
where $V_c$ is the volume of carrier gas given by
\begin{equation}
    V_{c} = \Qcarr \cdot t_i
\end{equation} 
and $t_i=|\Delta x|/\nu$ is the print time during the selected window that is calculated by dividing the distance traveled along the print path, $|\Delta x|$, by the print speed, $\nu$. This assumes that the print path is parallel to the $x$ axis.

The aerosol size distribution was periodically measured during a print using an aerodynamic particle sizer (APS 3321, TSI Incorporated) with a measurement range from 0.5 to 20~\textmu{m}. This measurement was performed by first turning off sheath flow, carrier flow, and ultrasonic atomization and disconnecting the transport tube from the deposition head. The tube is then inserted into the inlet of the APS 3321, at which point the flows and atomization are turned back on to the same setpoints. The machine is allowed to settle for two minutes before data is collected using the APS 3321, at which point everything is turned off and the tube is reconnected to the deposition head. The setpoints are again turned on and the machine settles for two minutes before resuming the print. The resulting data from the APS 3321 is a histogram of the aerosol droplet diameters. However, this histogram does not accurately reflect the true distribution of the aerosol droplet sizes because the water in the droplets is rapidly evaporated by the ambient air drawn in by the APS 3321. The measured histogram will therefore filter larger particles. This effect is accounted for by calculating the amount of droplet evaporation using the Stefan-Fuchs model \cite{superhodrophobic_printing,fuchs}. 

\FloatBarrier

\section{Inferring AJP States Experiment}
\label{app:experiment_1_details}

\begin{table}[h!]
    \centering
    \begin{tabular}{|c|c|c|c|}
    \hline
        Measurement Time & Open-Loop Prediction & Closed-Loop Prediction & Measurement Value\\ \hline
        30 min & 2.8235 & 3.5328 & 3.6904 \\ \hline
        60 min & 2.6216 & 3.5532 & 3.5057 \\ \hline
        80 min & 2.4197 & 3.3265 & 3.4391 \\ \hline
    \end{tabular}
    \caption{Measurement of median aerosol particle diameter, $\DA$, at discrete time-points.}
    \label{tab:median_diameter_ex1}
\end{table}
\begin{table}[h!]
    \centering
    \begin{tabular}{|c|c|c|c|}
    \hline
        Measurement Time & Open-Loop Prediction & Closed-Loop Prediction & Measurement Value\\ \hline
        30 min & 0.9999 & 0.9995 &   0.99985 \\ \hline
        60 min & 0.9998 & 0.999 &  0.9996 \\ \hline
        90 min & 0.9997 & 0.9985 & 0.9994 \\ \hline
    \end{tabular}
    \caption{Measurement of change in ink solution volume in vial, $\Vl$, at discrete time-points.}
    \label{tab:vial_level_ex1}
\end{table}
\begin{table}[h!]
    \centering
    \begin{tabular}{|c|c|c|c|}
    \hline
        Measurement Time & Open-Loop Prediction & Closed-Loop Prediction & Measurement Value\\ \hline
        30 min & $2.35 \times 10^{-5}$ & $8.49 \times 10^{-5}$ &  0\\ \hline
        60 min & $4.765 \times 10^{-5}$ & $4.982 \times 10^{-5}$ &  0\\ \hline
        90 min & $7.11 \times 10^{-5}$ & $4.736 \times 10^{-5}$ &  0\\ \hline
    \end{tabular}

    \caption{Measurement of ink deposition in tube, $\DT$, at discrete time-points.}
    \label{tab:tube_deposition_ex1}
\end{table}
\begin{table}[h!]
    \centering
    \begin{tabular}{|c|c|c|c|}
    \hline
        Measurement Time & Open-Loop Prediction & Closed-Loop Prediction  & Measurement Value\\ \hline
        30 min & 0 & $3.161\times10^{-8}$ &  $3.05\times10^{-8}$ \\ \hline
        60 min & 0 & $3.21\times10^{-8}$ & $3.10\times10^{-8}$ \\ \hline
        90 min & 0 & $3.257\times10^{-8}$ & $3.15\times10^{-8}$ \\ \hline
    \end{tabular}
    \caption{Measurement of ink deposition in nozzle, $\DN$, at discrete time-points.}
    \label{tab:nozzle_deposition_ex1}
\end{table}
\begin{table}[h!]
    \centering
    \begin{tabular}{|c|c|c|c|}
    \hline
        Measurement Time & Open-Loop Prediction & Closed-Loop Prediction & Measurement Value\\ \hline
        30 min & $2.0127\times10^{-7}$  & $3.022\times10^{-6}$  & $3.09\times10^{-7}$ \\ \hline
        60 min & $2.169\times10^{-7}$ & $1.238\times10^{-6}$  &$3.106\times10^{-7}$ \\ \hline
        90 min & $2.108\times10^{-7}$ & $3.939\times10^{-7}$ &$3.65\times10^{-7}$ \\ \hline
    \end{tabular}
    \caption{Measurement of aerosol volume fraction in carrier stream, $\AD$, at discrete time-points.}
    \label{tab:aerosol_density_ex1}
\end{table}
\FloatBarrier

\section{Ablation Experiment}
\label{app:exp1_ablation_experiment}

\begin{figure}
    \centering
    \begin{subfigure}[t]{0.3\textwidth}
        \centering
        \includegraphics[height=1.2in]{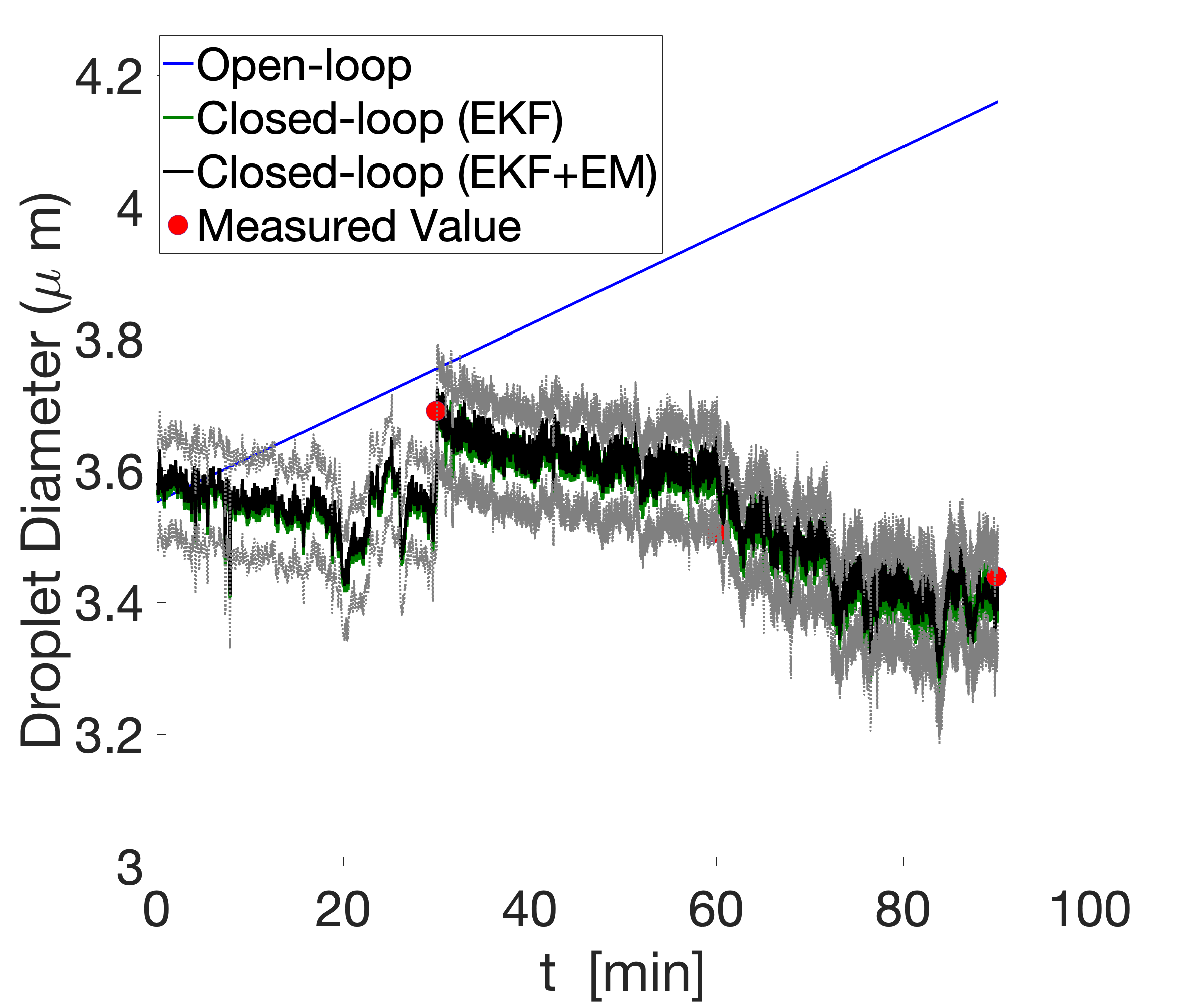}
       \caption{}
    \end{subfigure}%
    ~ 
    \begin{subfigure}[t]{0.3\textwidth}
        \centering
        \includegraphics[height=1.2in]{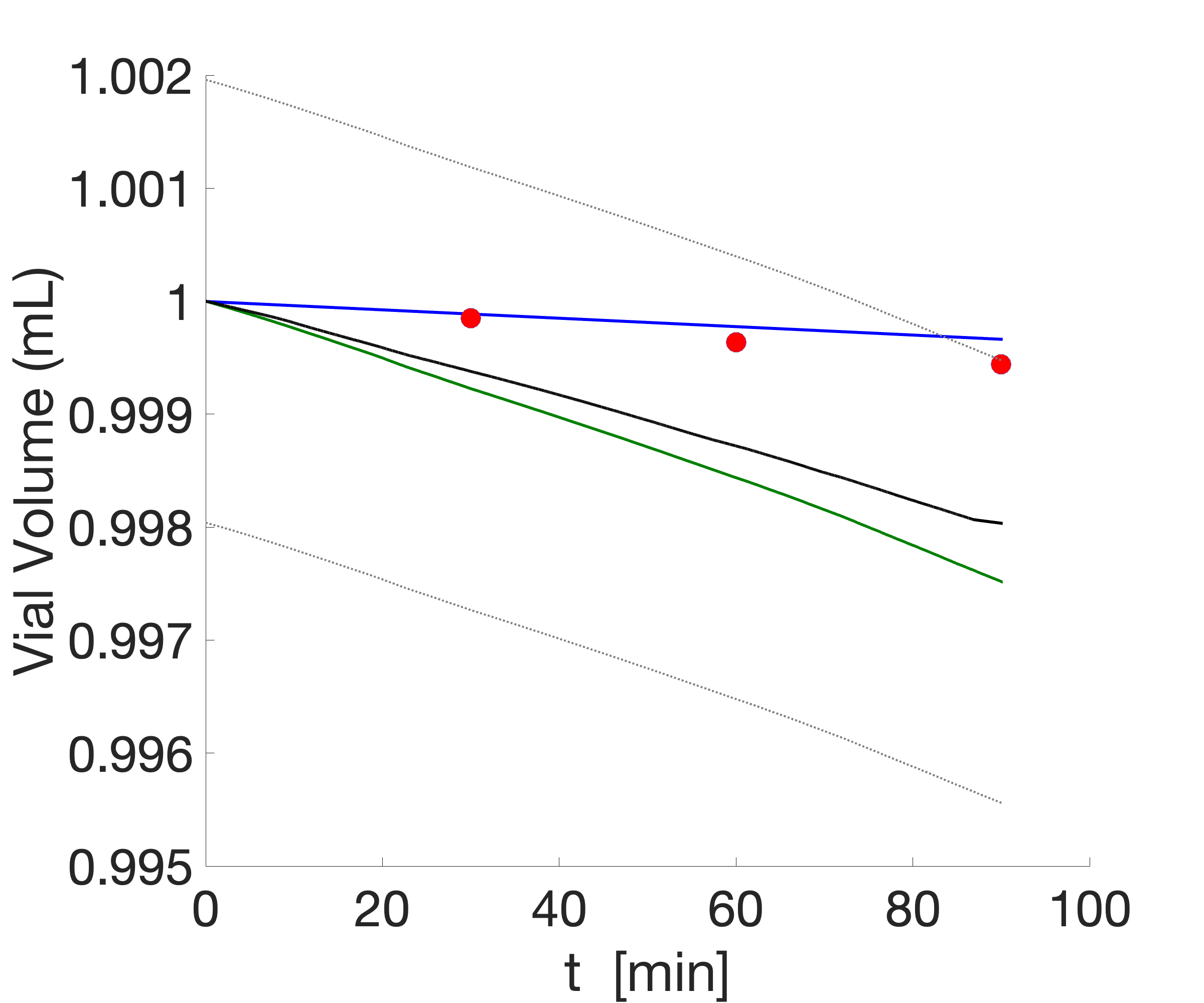}
       \caption{}
    \end{subfigure}
        ~
    \begin{subfigure}[t]{0.3\textwidth}
        \centering
        \includegraphics[height=1.2in]{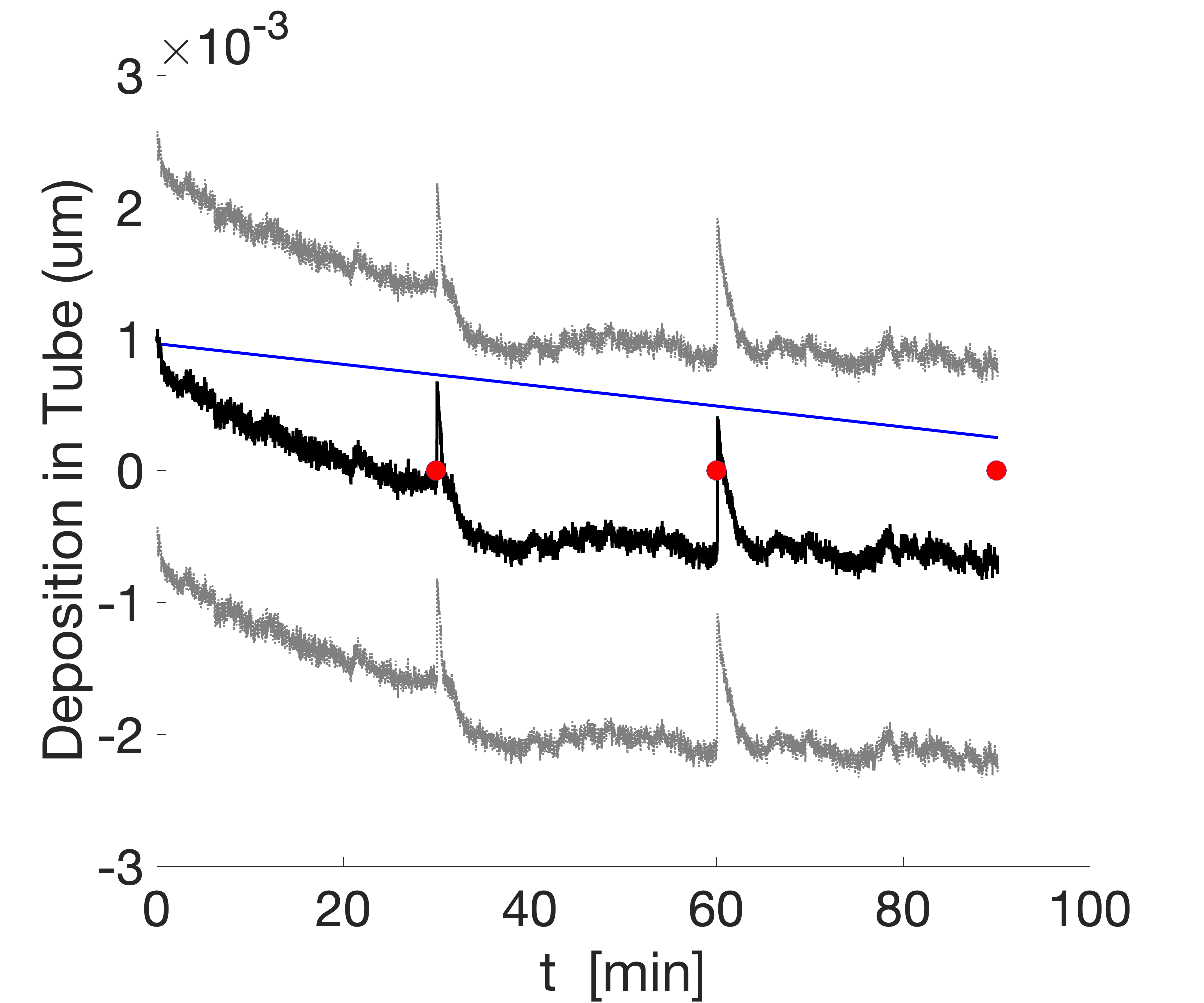}
       \caption{}
    \end{subfigure}
    
        \begin{subfigure}[t]{0.3\textwidth}
        \centering
        \includegraphics[height=1.2in]{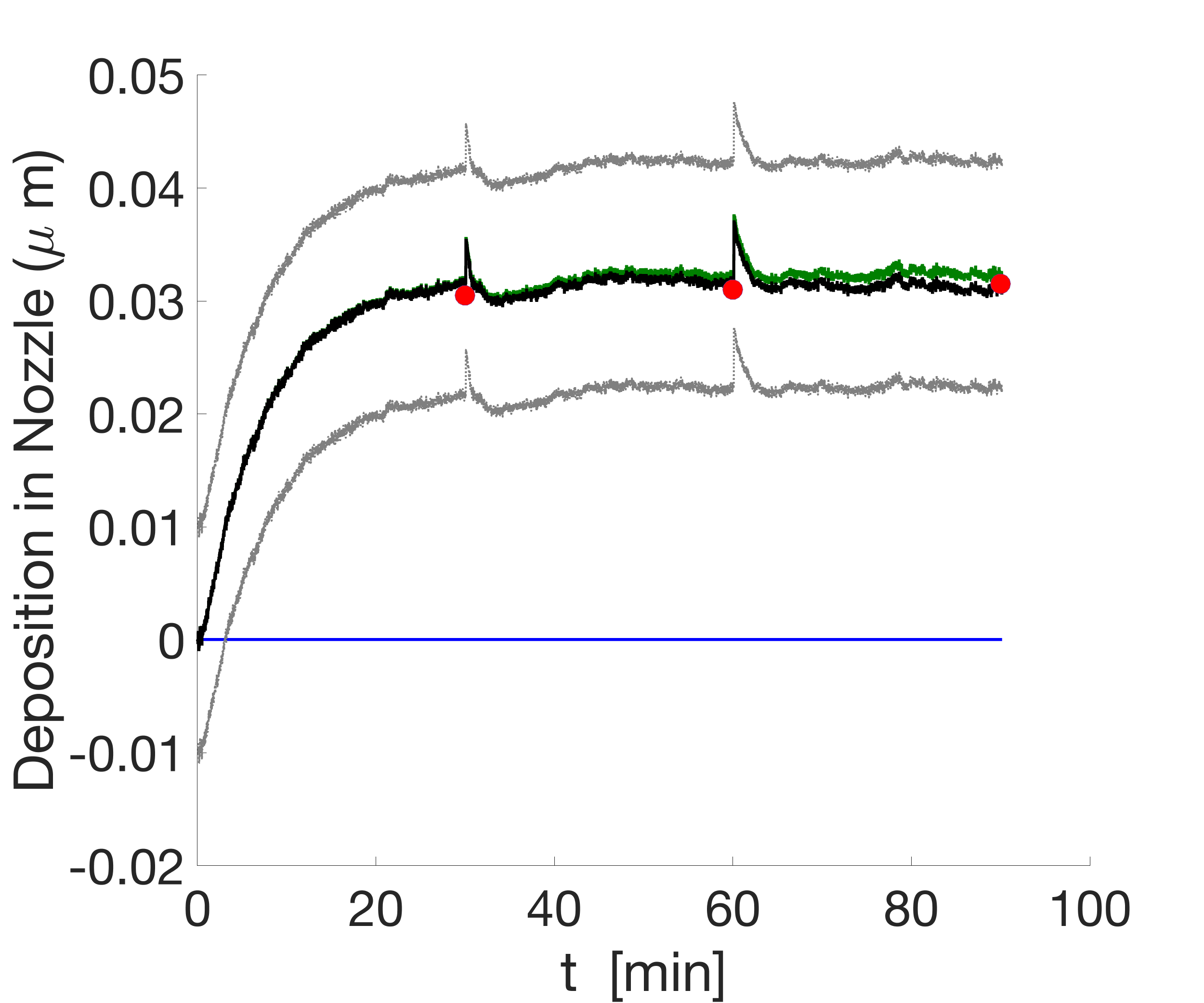}
        \caption{}
    \end{subfigure}
        ~ 
    \begin{subfigure}[t]{0.3\textwidth}
        \centering
        \includegraphics[height=1.2in]{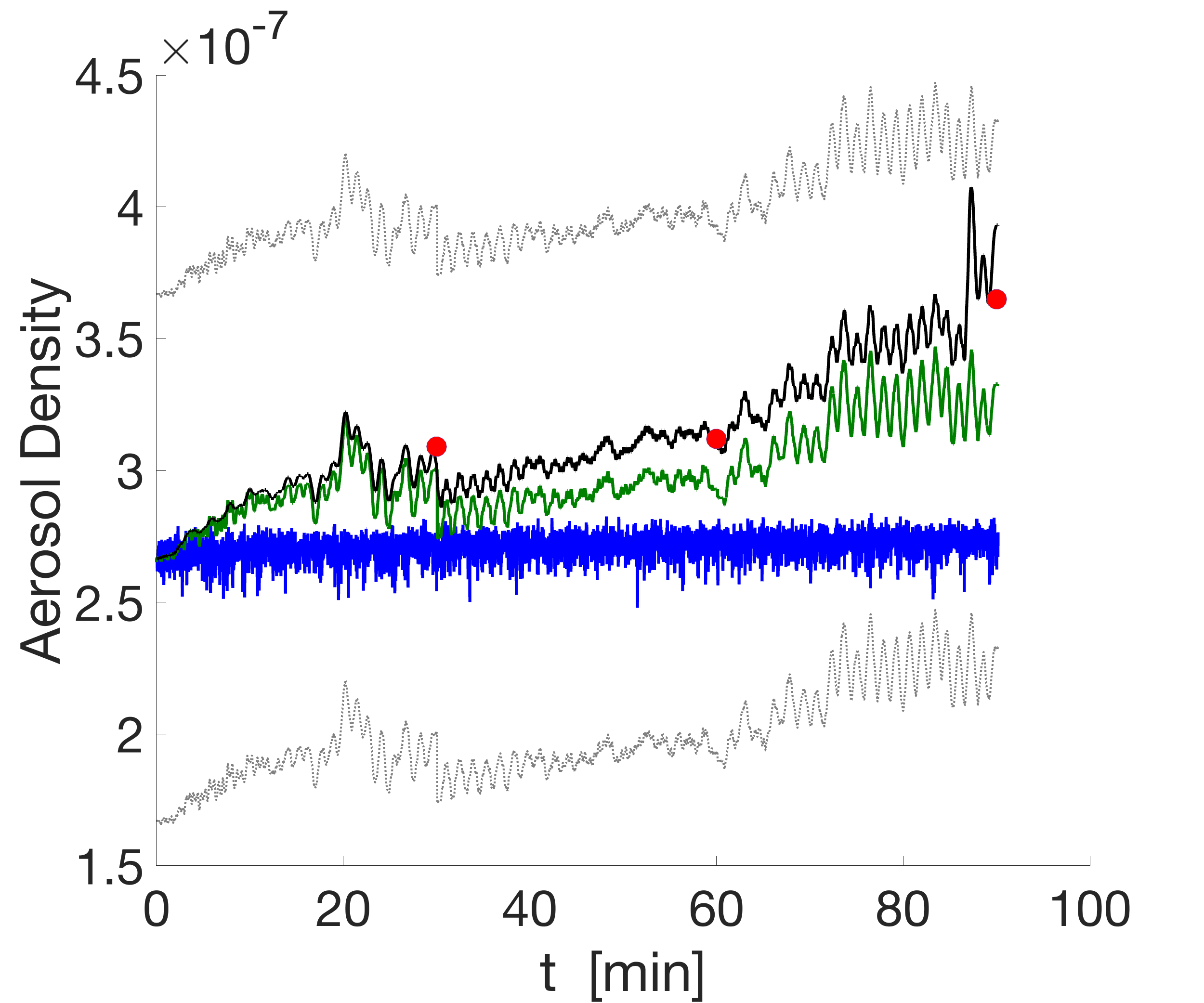}
        \caption{}
    \end{subfigure}
    \caption{Open-loop prediction for experiment in Section \ref{sec:exp1_estimation} using static physics-derived model of the latent states compared to the closed-loop prediction of the digital twin updated using EM, closed loop prediction using EKF and the measured values of latent states at discrete time-points. The estimated latent states include (a) aerosol droplet diameter out of atomizer, $\DA$, (b) ink solution volume in the vial, $\Vl$, (c) ink deposition in the tube, $\DT$, (d) ink deposition in the nozzle, $\DN$, (e) volume fraction of aerosol droplets in carrier stream, $\AD$. The measurement of the latent states are within the confidence bounds ($\pm 2\sigma$) of the closed-loop estimation, shown by the dotted (gray) lines.}
    \label{fig:open_loop_prediction_exp_1_ablation}
\end{figure}

To illustrate the benefit of updating digital model parameters using EM over the standard EKF, the predicted states are compared in Figure \ref{fig:open_loop_prediction_exp_1_ablation}. The model of $\AD$, initially obtained through experimental fitting, is refined over time using incoming data. This data-driven updating allows the model to adapt to changing device behavior and operating conditions.

\begin{figure}
    \centering
    \begin{subfigure}[t]{0.3\textwidth}
        \centering
        \includegraphics[height=1.2in]{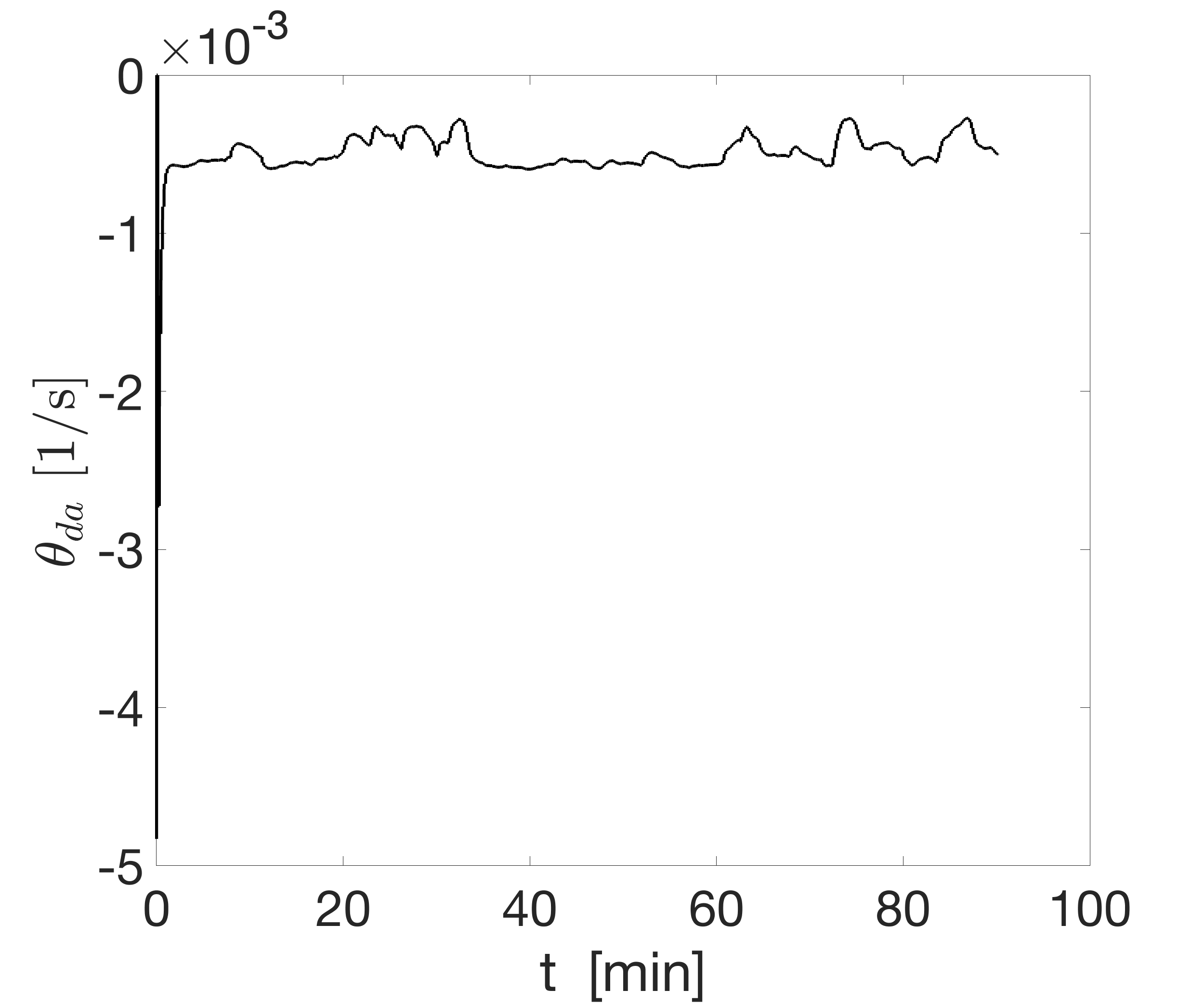}
           \caption{}
    \end{subfigure}%
    ~ 
    \begin{subfigure}[t]{0.3\textwidth}
        \centering
        \includegraphics[height=1.2in]{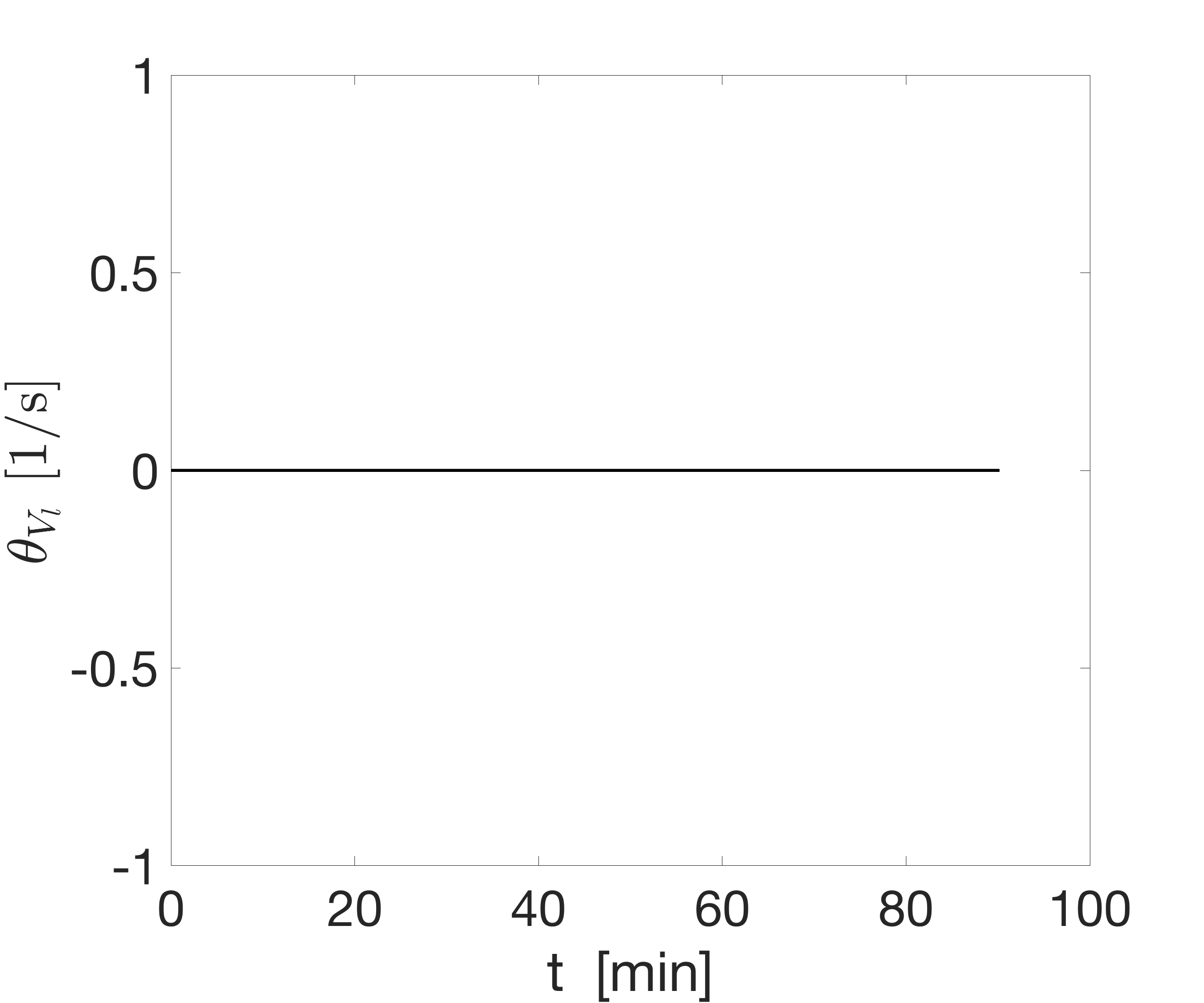}
            \caption{}
    \end{subfigure}
        ~
    \begin{subfigure}[t]{0.3\textwidth}
        \centering
        \includegraphics[height=1.2in]{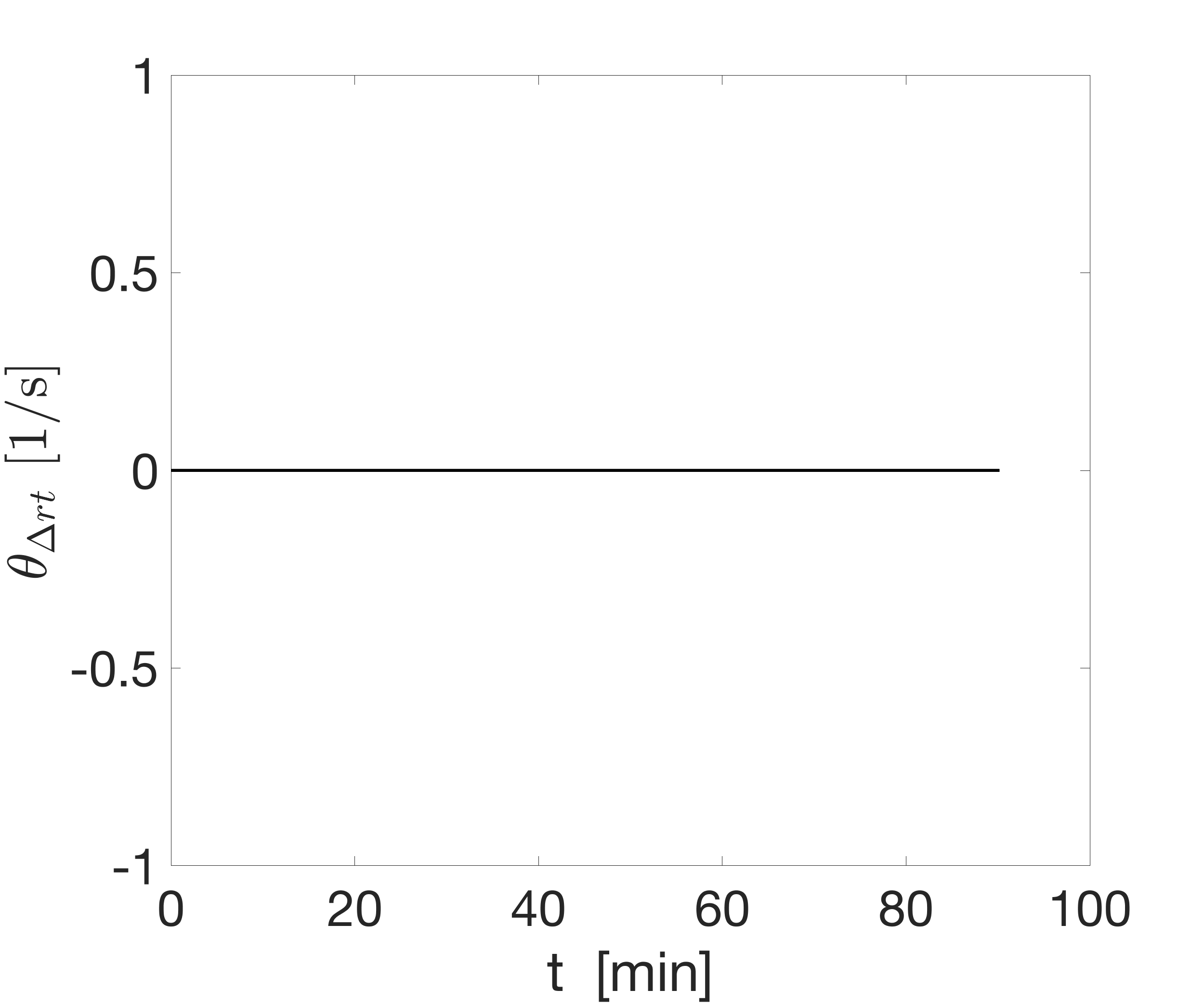}
           \caption{}
    \end{subfigure}
    
        \begin{subfigure}[t]{0.3\textwidth}
        \centering
        \includegraphics[height=1.2in]{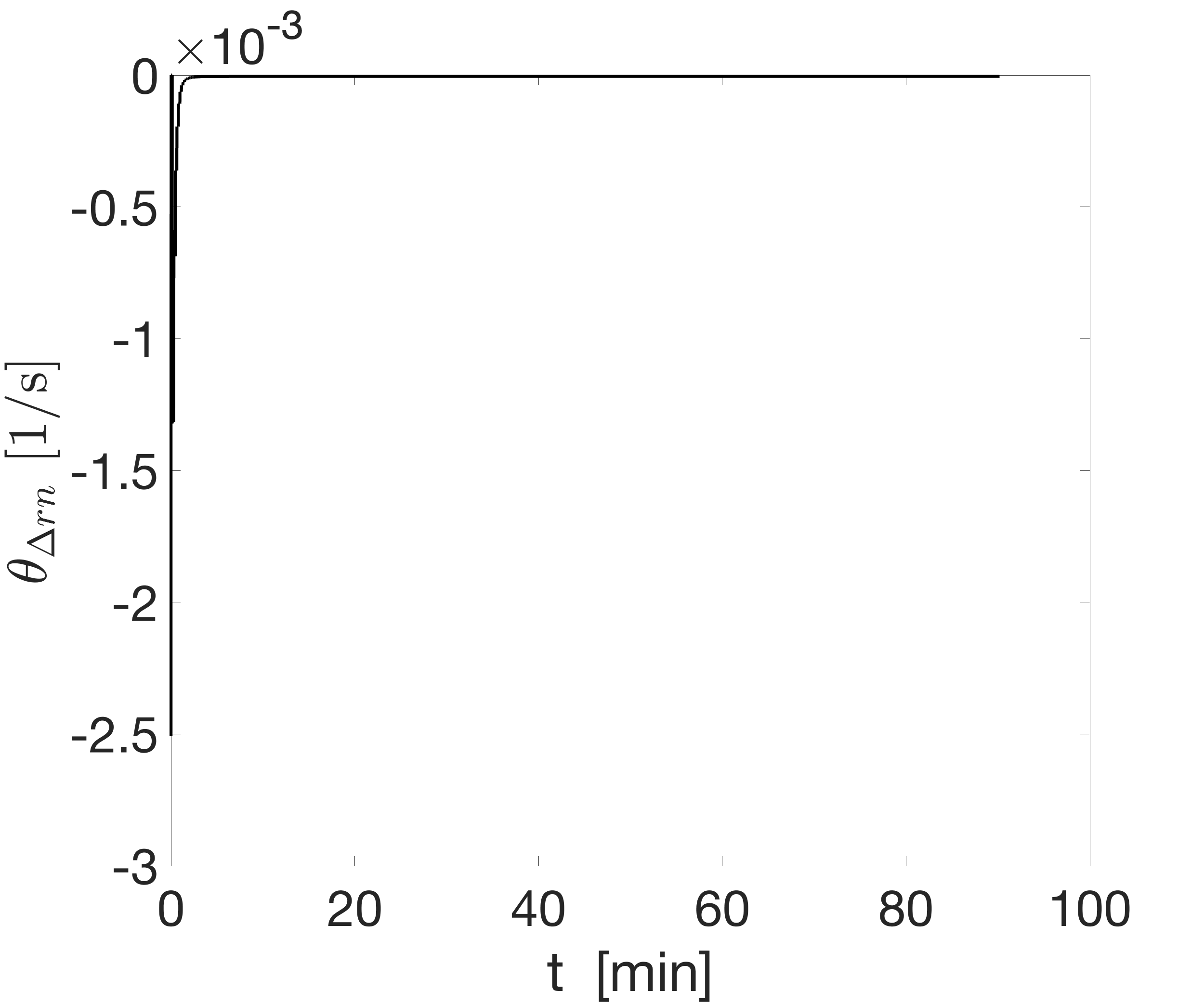}
            \caption{}

    \end{subfigure}
        ~ 
    \begin{subfigure}[t]{0.3\textwidth}
        \centering
        \includegraphics[height=1.2in]{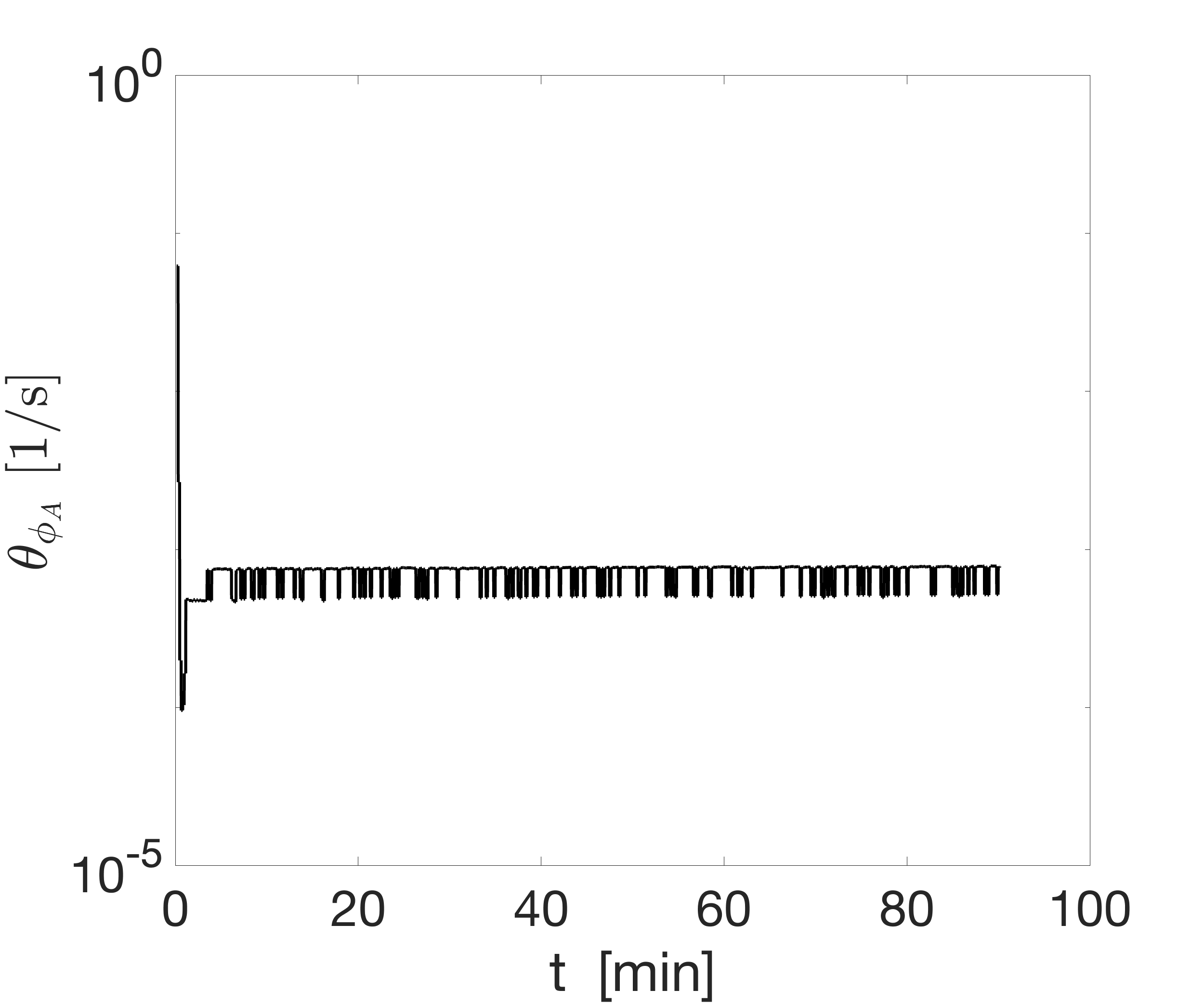}
        \caption{}
    \end{subfigure}
    \caption{The EM algorithm iteratively updates the model parameters, $\theta$, to capture evolving device behavior for the experiment in Section \ref{sec:exp1_estimation} These include parameters associated with the transition models for (a) median particle diameter, $\theta_{\DA}$, (b) vial volume, $\theta_{\Vl}$, (c) ink deposition in the tube, $\theta_{\DT}$, (d) ink deposition in the nozzle, $\theta_{\DN}$, and (e) aerosol density, $\theta_{\AD}$. }
    \label{fig:open_loop_prediction_exp_1_theta}
\end{figure}

The EM algorithm iteratively updates the model parameters, $\theta$, to adapt its dynamics to better reflect the behavior of the measured system. As shown in Figure~\ref{fig:open_loop_prediction_exp_1_theta}, this process leads to gradual adjustments in parameter estimates, $\theta$, over time. The most significant updates occur in the parameters associated with the transition models of the median particle diameter ($\theta_{\DA}$), ink deposition in the nozzle ($\theta_{\DN}$), and aerosol density ($\theta_{\AD}$).

Although literature generally reports that the median aerosol droplet diameter remains stable under constant operating conditions, the invasive measurements collected in this study indicate fluctuations in diameter. The EM algorithm captures this behavior through modifying the value of $\theta_{\DA}$, which effectively modifies the transition model to reflect the behavior of $\theta_{\DA}$ given a relatively small set of observations. This provides an empirical correction that accounts for potential epistemic uncertainty in the model, arising from incomplete understanding of droplet formation dynamics and their sensitivity to perturbations in operating conditions.

For the aerosol density, $\AD$, the transition model was originally derived from controlled experiments based on several simplifying assumptions, including uniform atomization and steady-state aerosol generation. These assumptions, combined with variations in the characterization experiments, contribute to uncertainty in the model structure. Through EM updates, the algorithm learns a value for $\theta_{\AD}$, which compensates for unmodeled phenomena.

Together, these results highlight how the EM algorithm can be used not only for parameter estimation but also as a diagnostic tool to reveal areas of uncertainty within the physics-based model. While the learned parameters should not be interpreted as definitive physical mechanisms, they provide a data-informed adjustment that reduces model bias and aligns predictions more closely with observed trends. They also offer valuable insights into where and how the model could be refined to better represent the observed system behavior.


\FloatBarrier

\section{Updating the Digital Twin Parameters Experiment}

\begin{figure}
    \centering
    \begin{subfigure}[t]{0.3\textwidth}
        \centering
        \includegraphics[height=1.2in]{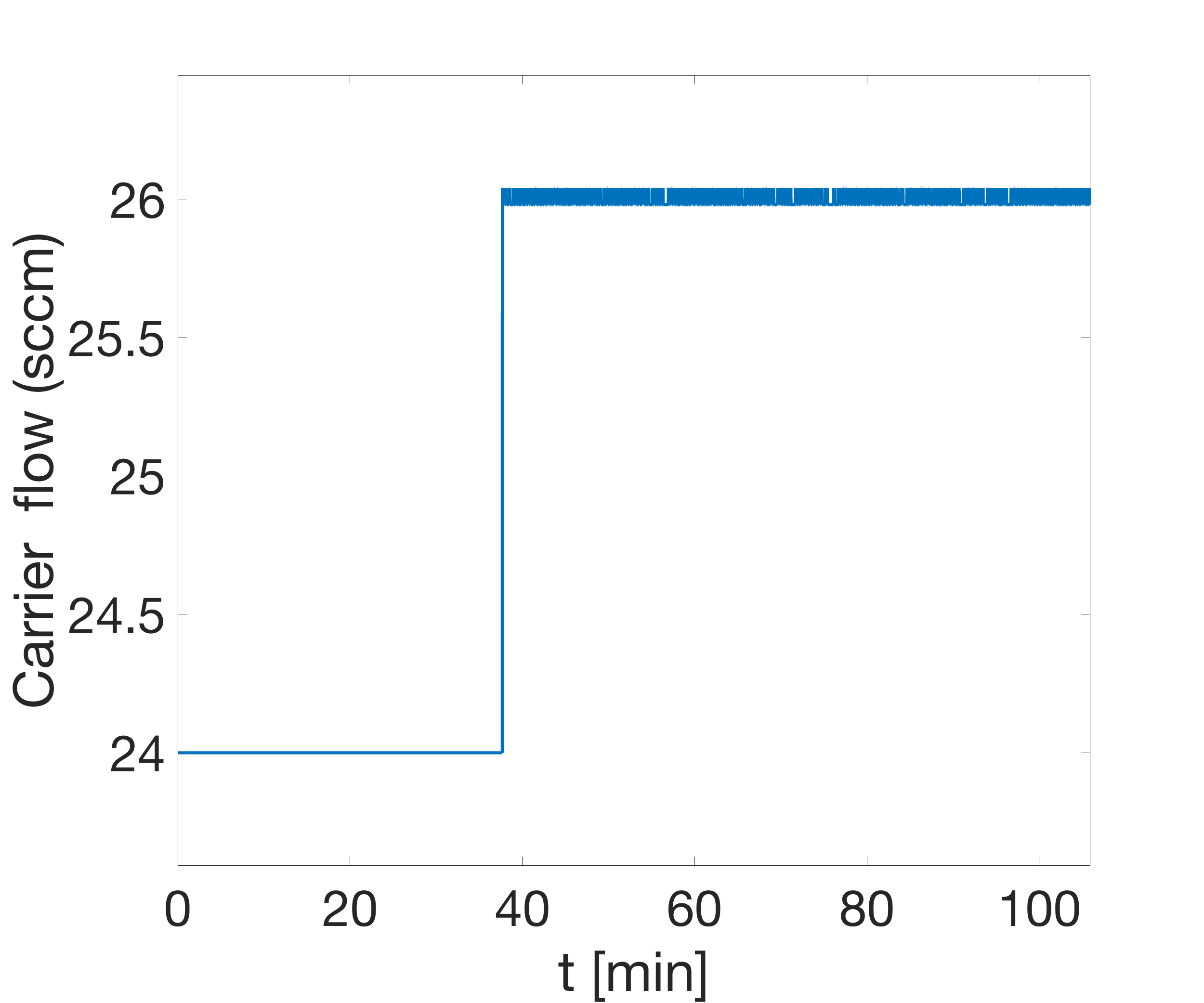}
        \caption{}
    \end{subfigure}%
    ~ 
    \begin{subfigure}[t]{0.3\textwidth}
        \centering
        \includegraphics[height=1.2in]{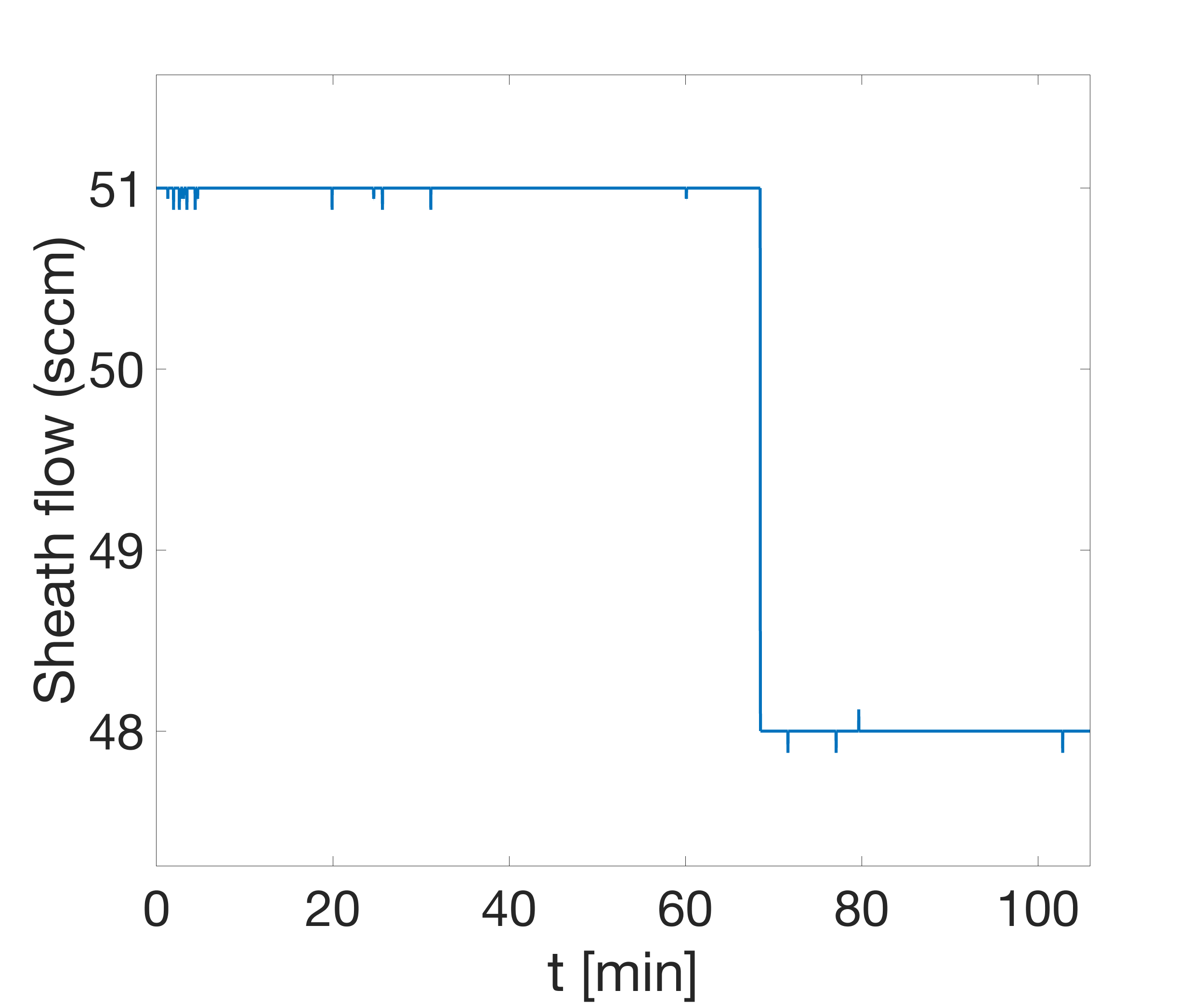}
        \caption{}
    \end{subfigure}
        ~ 
    \begin{subfigure}[t]{0.3\textwidth}
        \centering
        \includegraphics[height=1.2in]{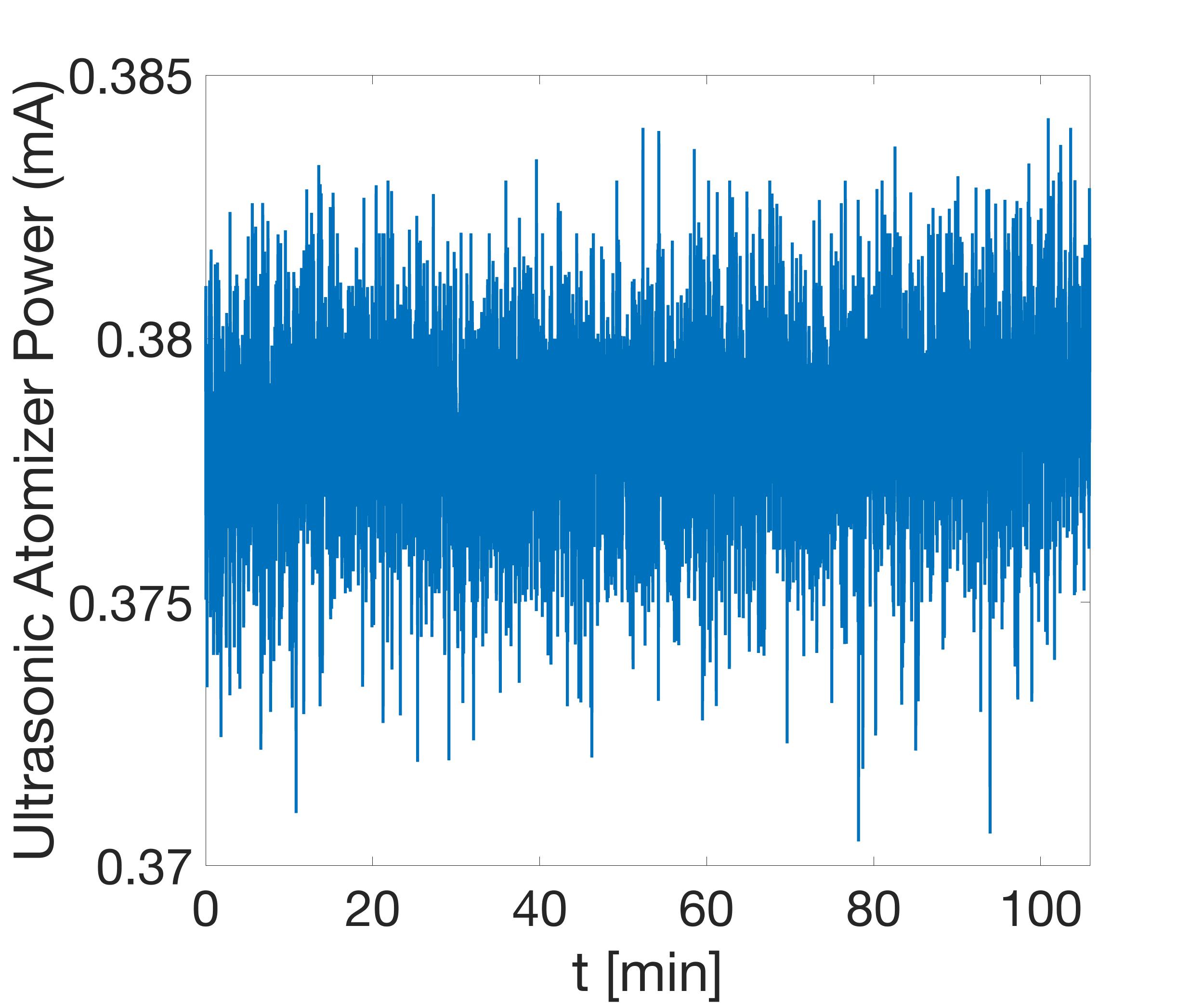}
        \caption{}
    \end{subfigure}
    \caption{Inputs of the experiment in Section \ref{sec:exp3_parameter_estimation} include: (a) carrier flow rate, $\Qcarr$, is initially set at 24 sccm and undergoes a step change at 37.6 min to 26~sccm, (b) sheath flow rate, $\Qsh$, is initially set at 51~sccm and undergoes a step change at 68.5 min to 48 sccm, (c) ultrasonic atomizer current, $\PA$, is set at a constant 0.375~mA; with variations are due to limitations in the atomizer's ability to maintain constant current. }
    \label{fig:exp3_inputs}
\end{figure}

\begin{figure}
    \centering
    \begin{subfigure}[t]{0.3\textwidth}
        \centering
        \includegraphics[height=1.2in]{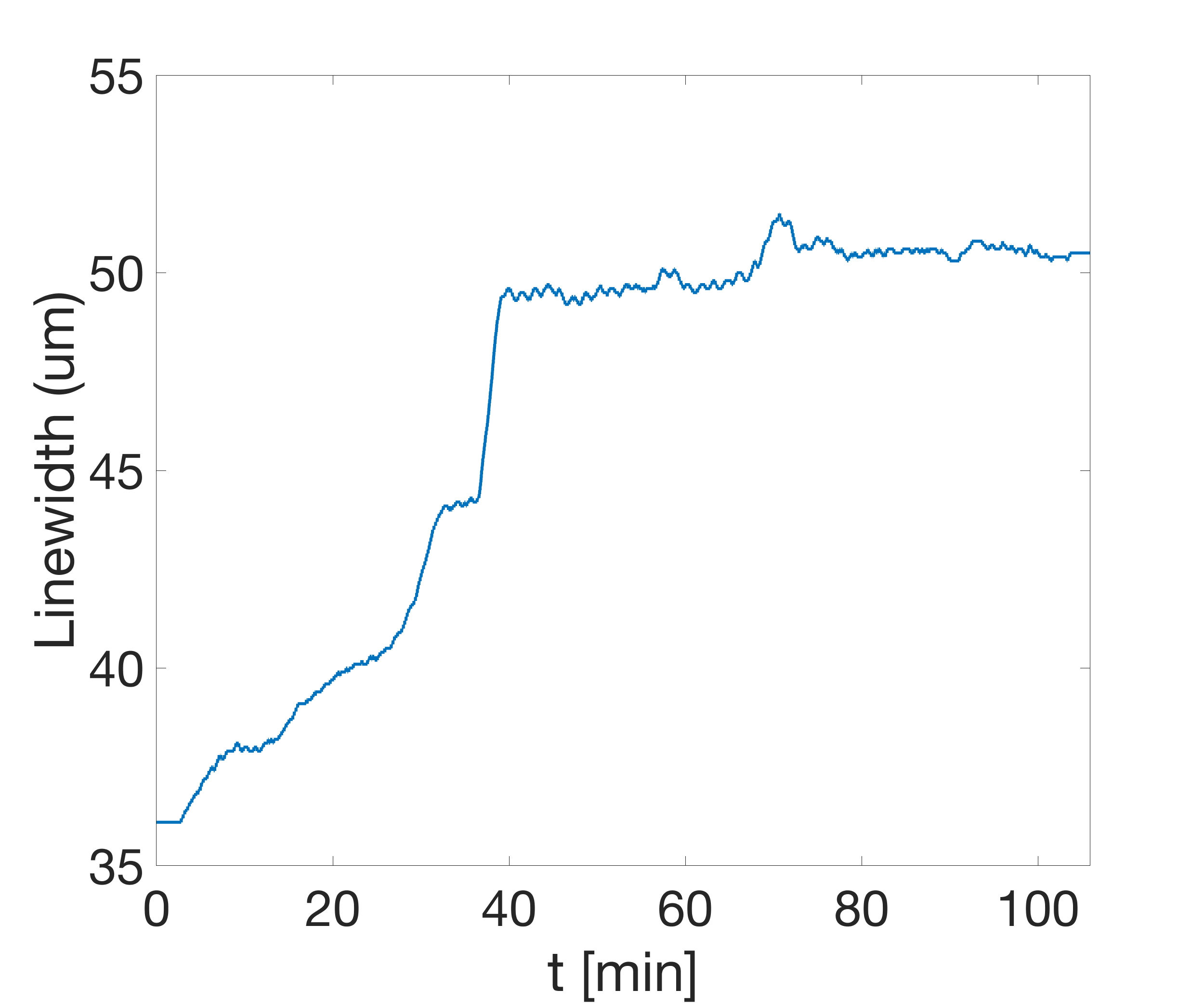}
        \caption{}
    \end{subfigure}%
    ~ 
    \begin{subfigure}[t]{0.3\textwidth}
        \centering
        \includegraphics[height=1.2in]{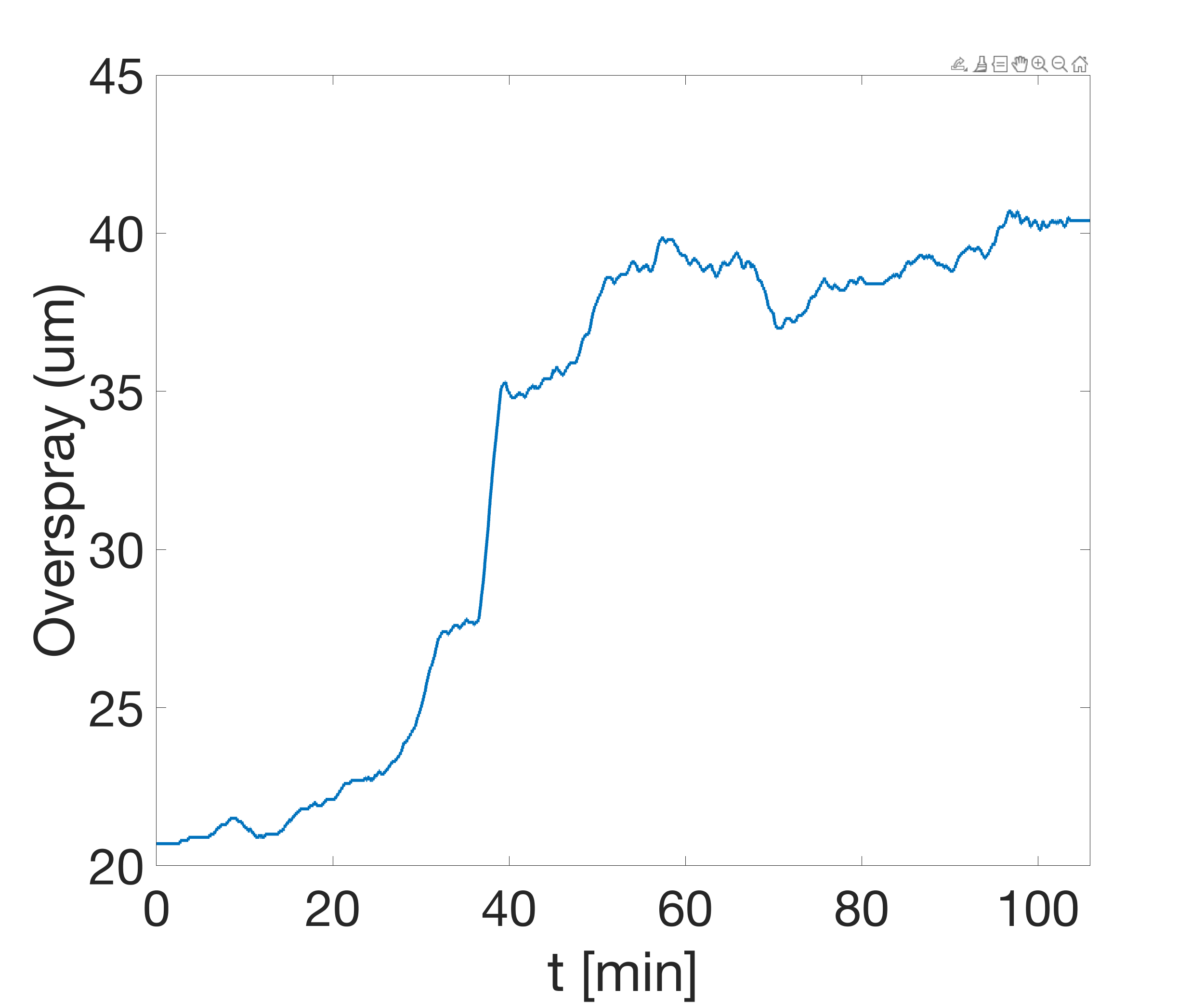}
        \caption{}
    \end{subfigure}
        ~ 
    \begin{subfigure}[t]{0.3\textwidth}
        \centering
        \includegraphics[height=1.2in]{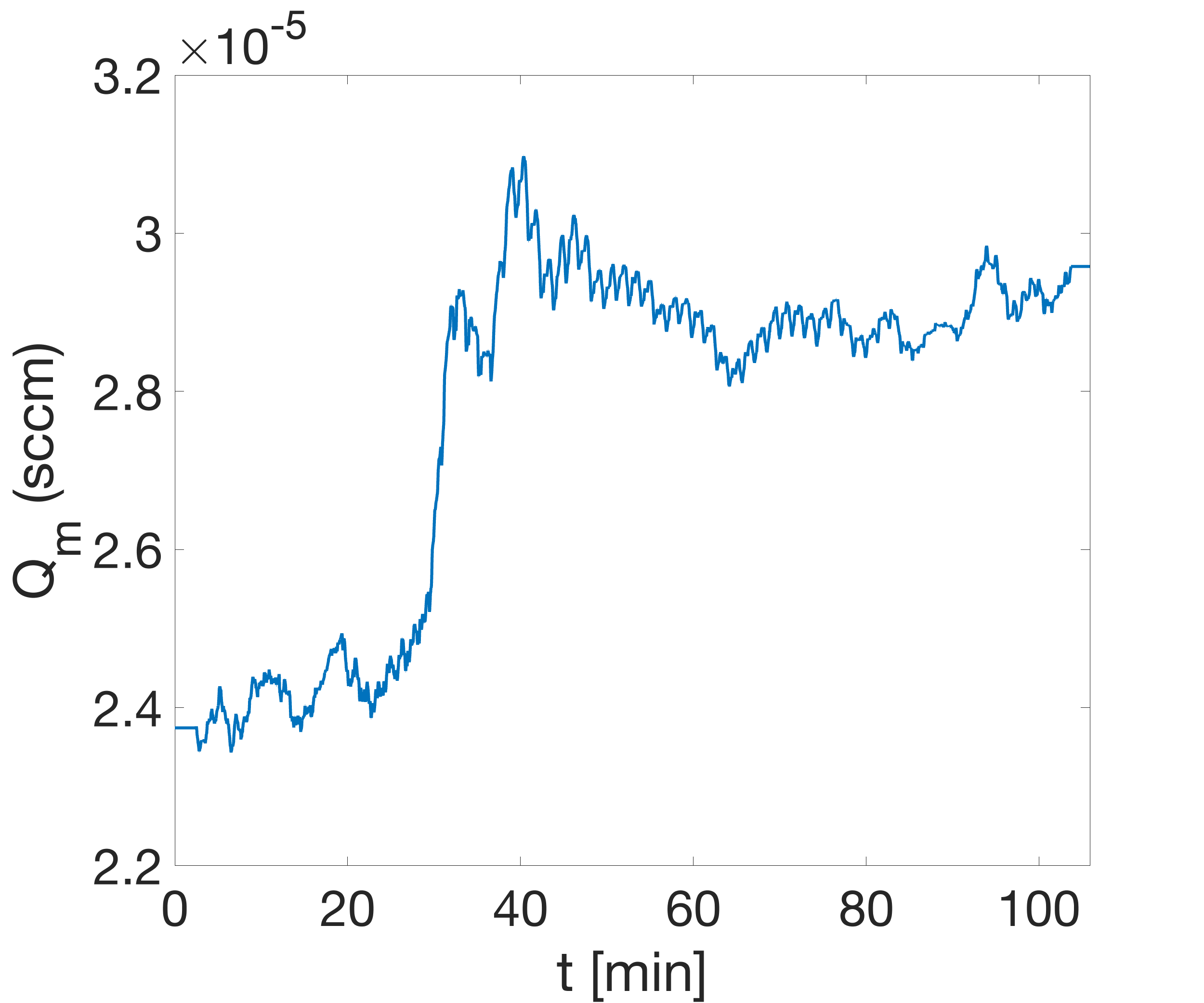}
        \caption{}
    \end{subfigure}
    \newline
    \begin{subfigure}[t]{0.3\textwidth}
        \centering
        \includegraphics[height=1.2in]{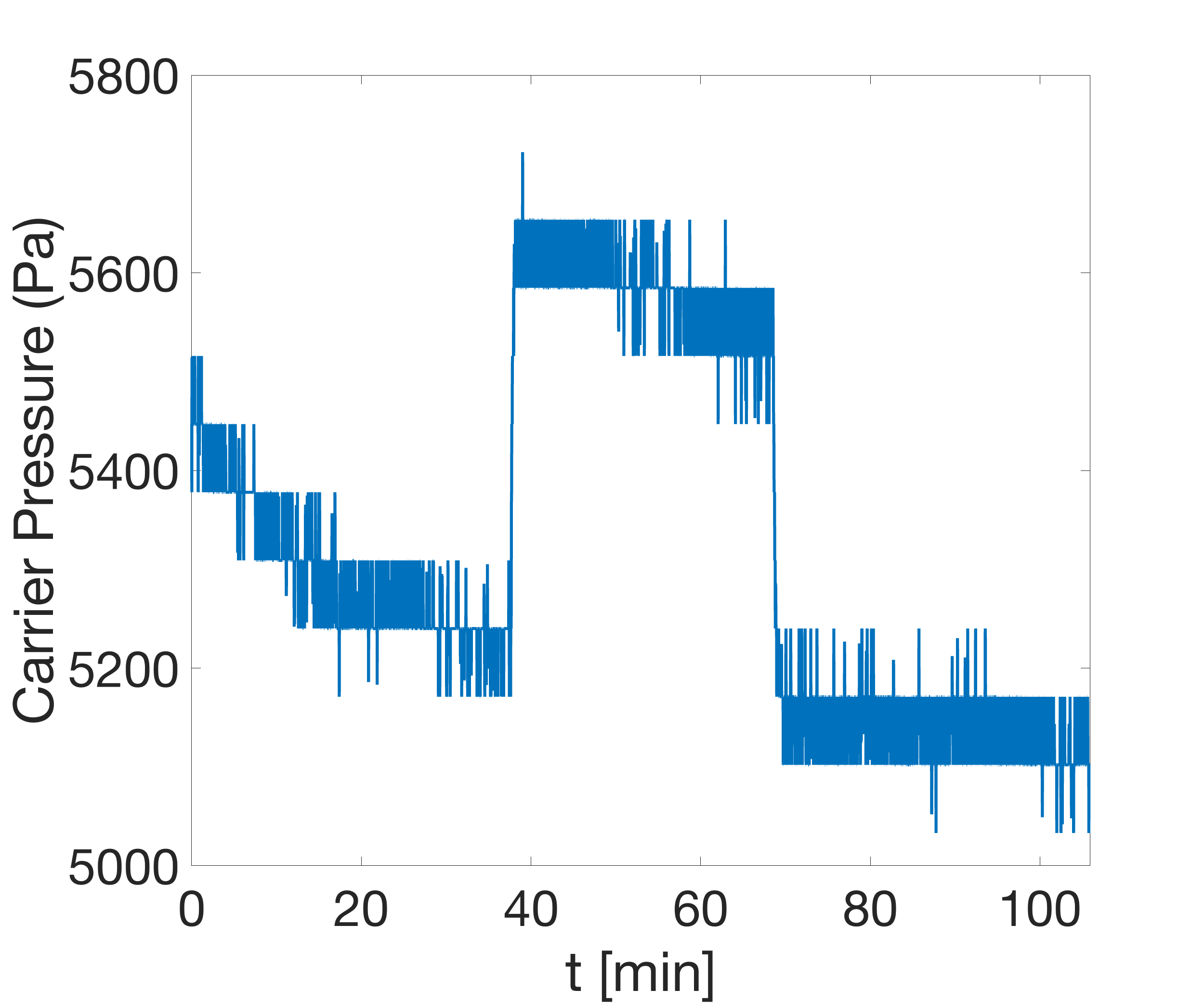}
        \caption{}
    \end{subfigure}
        \begin{subfigure}[t]{0.3\textwidth}
        \centering
        \includegraphics[height=1.2in]{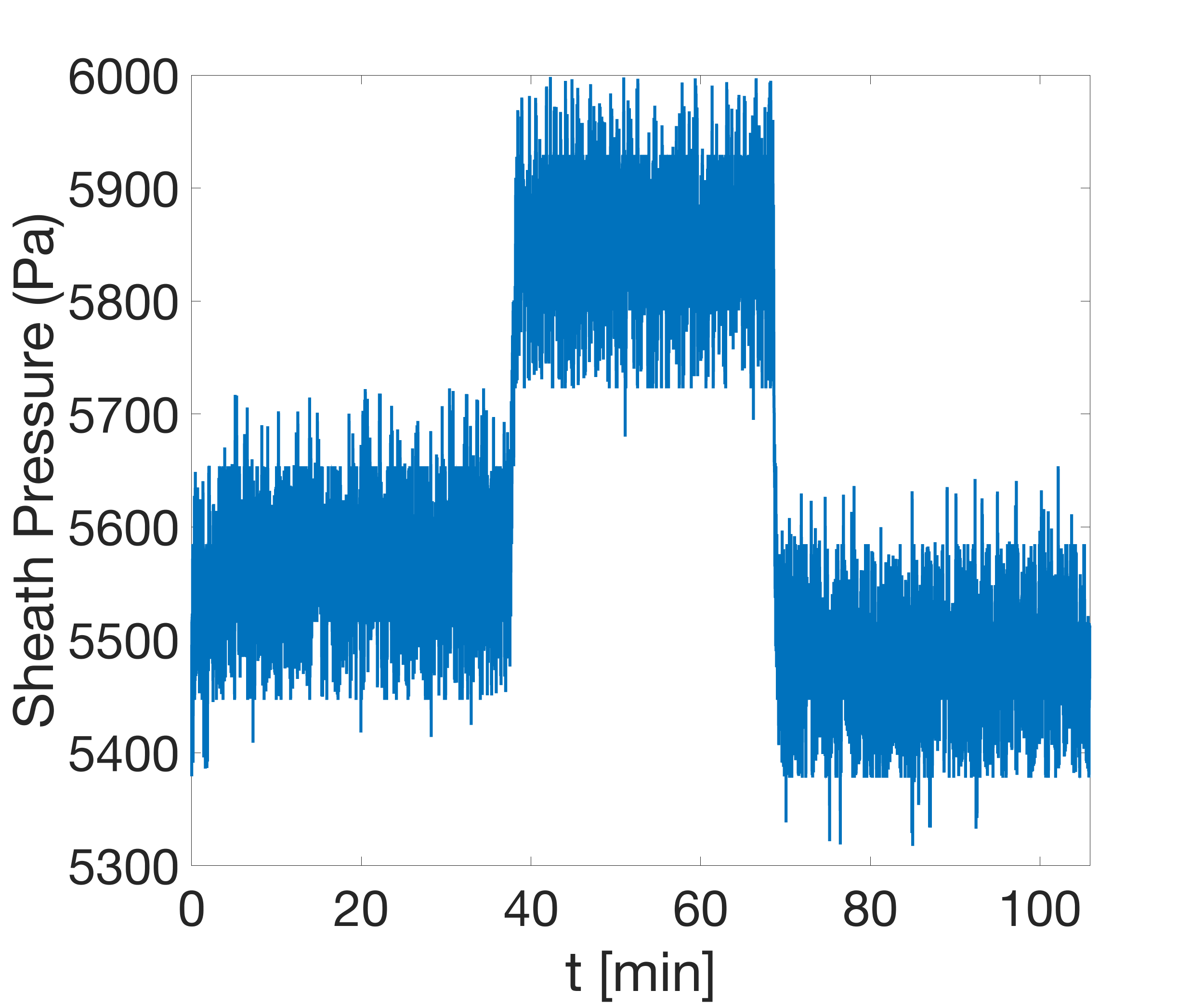}
        \caption{}
    \end{subfigure}
    \caption{Output of the parameter estimation method for the experiment in Section \ref{sec:exp3_parameter_estimation} include: (a) linewidth, $\lw$, (b) overspray, $\ov$, (c) deposited material flow rate, $\Qink$, (d) output carrier pressure, $\Pcarr$, (e) output sheath pressure, $\Psh$.}
    \label{fig:exp3_outputs}
\end{figure}

We present more data relating to the experiment discussed in section \ref{sec:exp3_parameter_estimation}. Figure \ref{fig:exp3_inputs} shows the inputs used in the experiment. As described in the text, the atomization current remains constant, and the sheath and carrier flow inputs are both stepped between two operating points. The recorded outputs are shown in Figure \ref{fig:exp3_outputs}.
\label{app:experiment3_details}
\FloatBarrier

\begin{figure}
    \centering
    \includegraphics[height=1.2in]{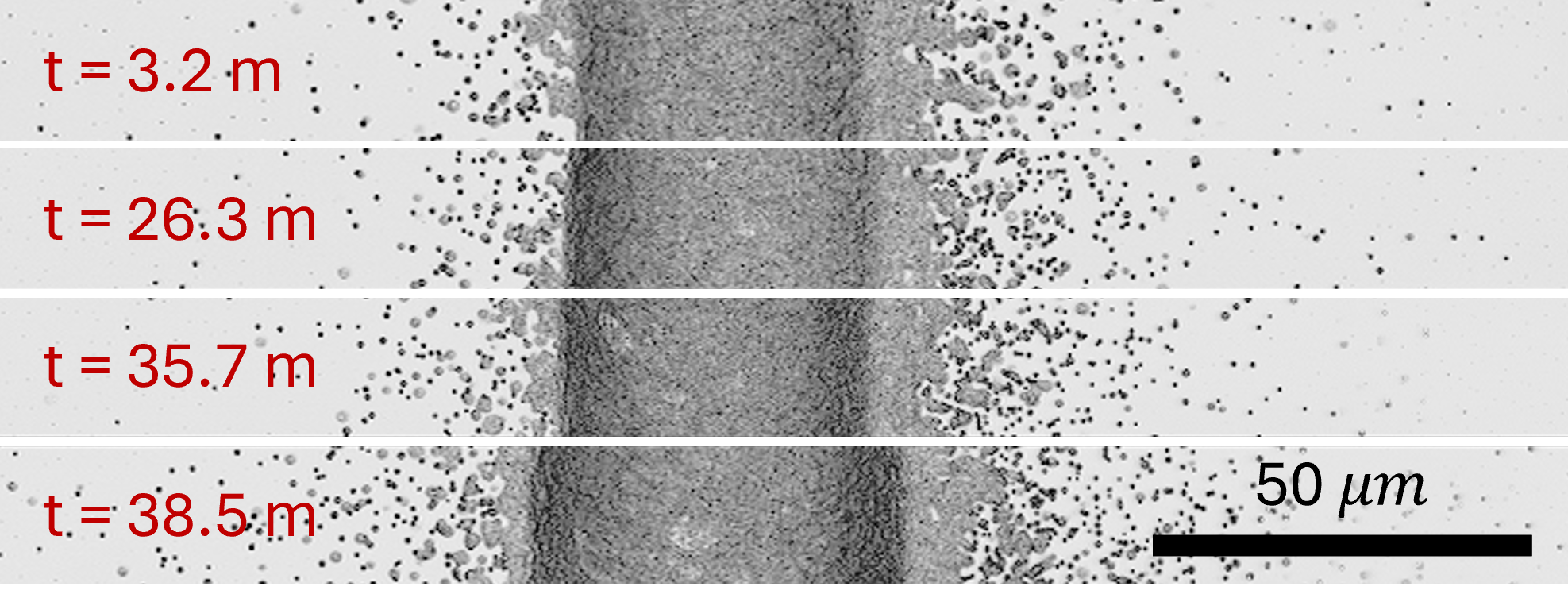}
    \caption{Optical verification of drift in linewidth from experiment detailed in Section \ref{sec:exp3_parameter_estimation}. Images were taken with optical profilometer and include various times up to just after the point where the input shift in carrier flow occurred}.
    \label{fig:exp3_drift}
\end{figure}

\begin{figure}
    \centering
    \begin{subfigure}[t]{0.3\textwidth}
        \centering
        \includegraphics[height=1.2in]{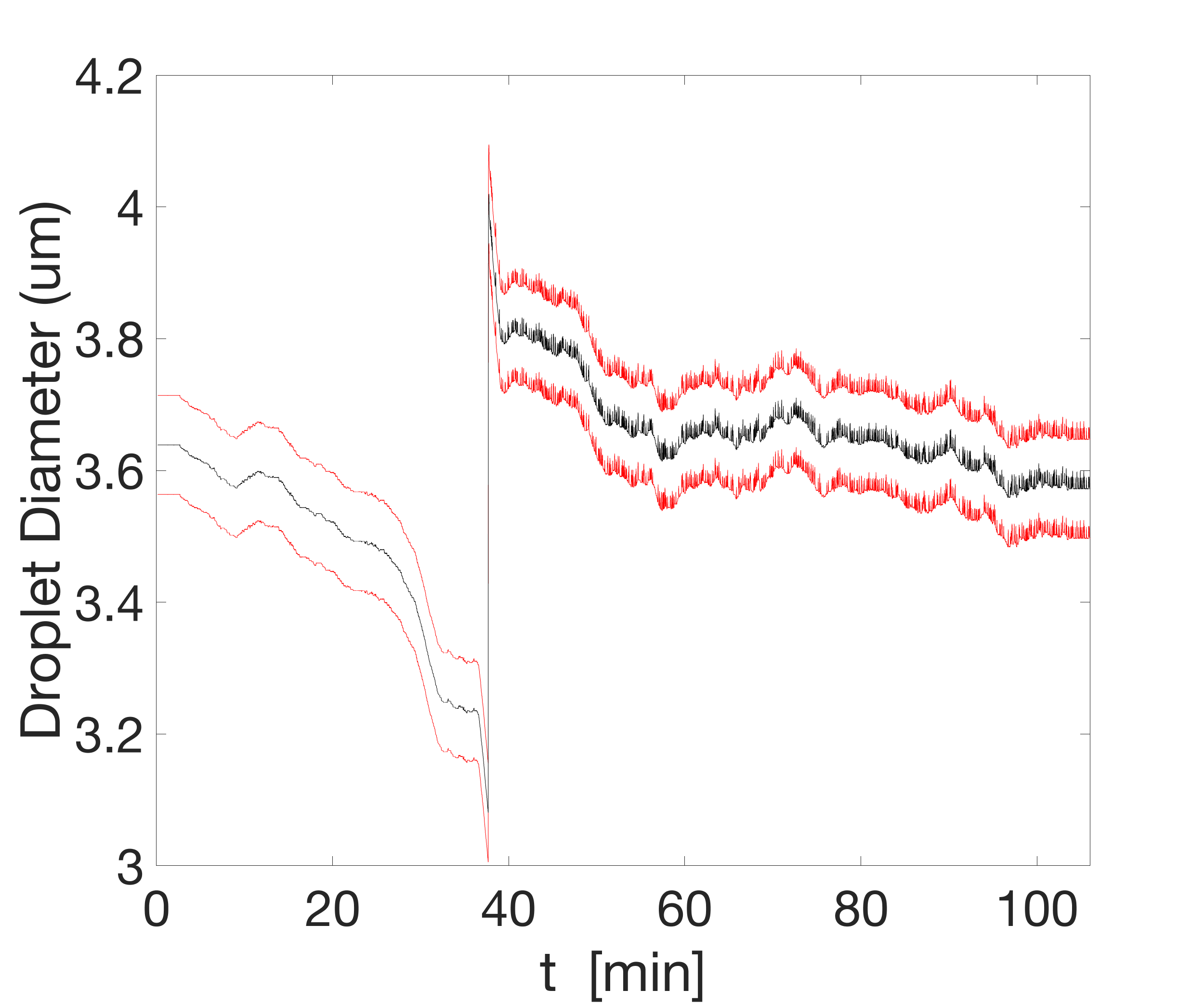}
       \caption{}
    \end{subfigure}%
    ~ 
    \begin{subfigure}[t]{0.3\textwidth}
        \centering
        \includegraphics[height=1.2in]{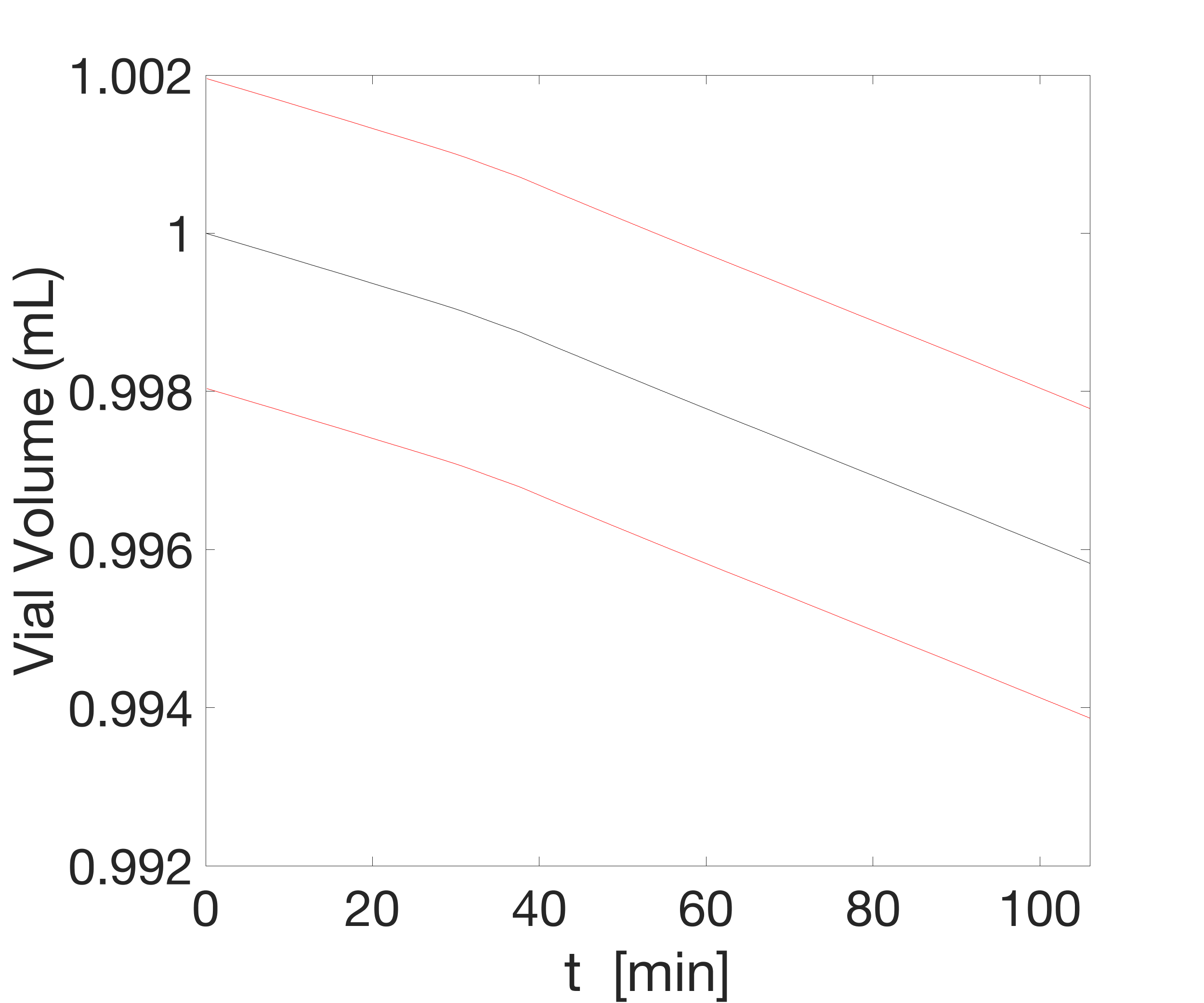}
       \caption{}
    \end{subfigure}
    ~ 
    \begin{subfigure}[t]{0.3\textwidth}
        \centering
        \includegraphics[height=1.2in]{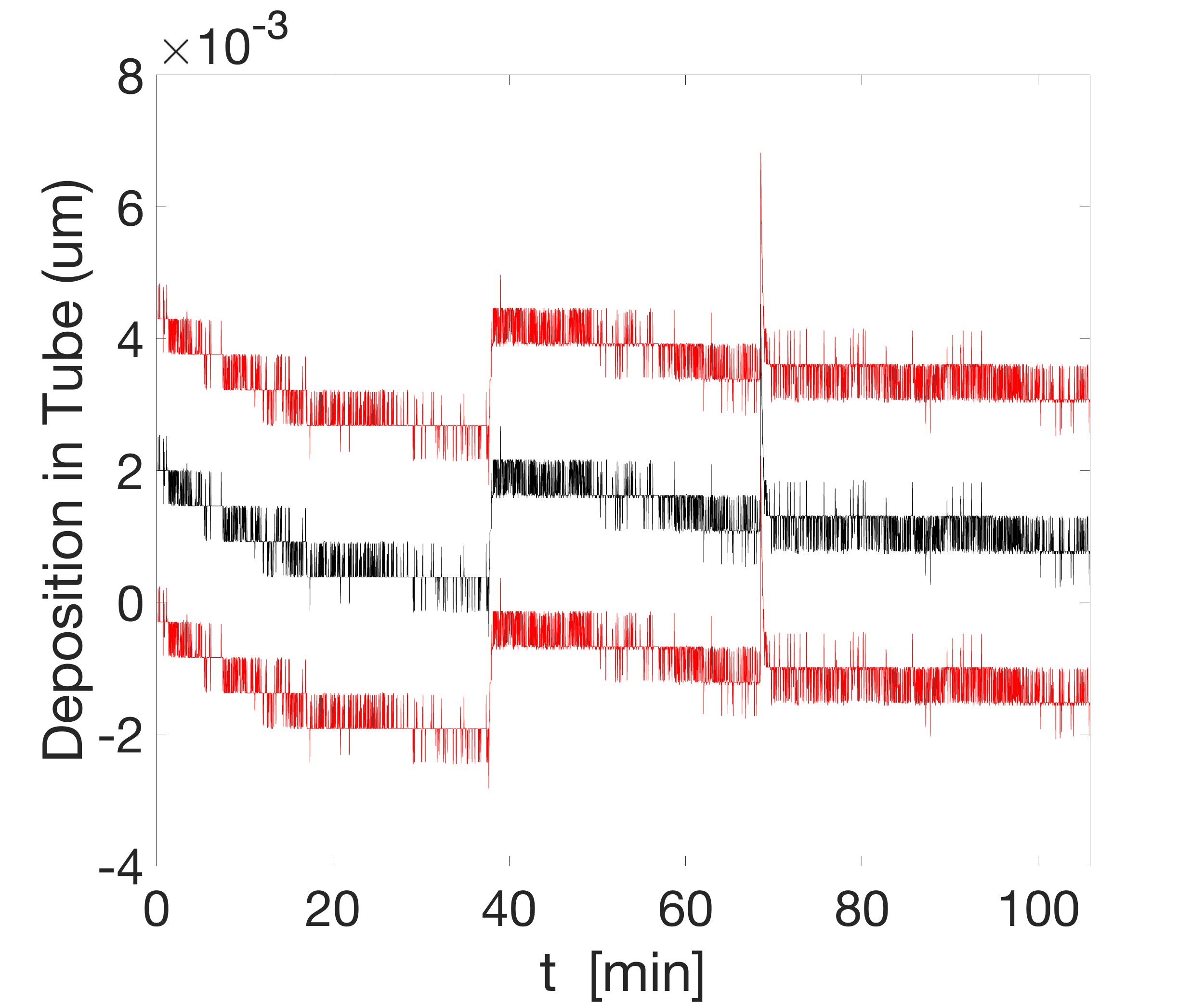}
       \caption{}
    \end{subfigure}

    \begin{subfigure}[t]{0.3\textwidth}
        \centering
        \includegraphics[height=1.2in]{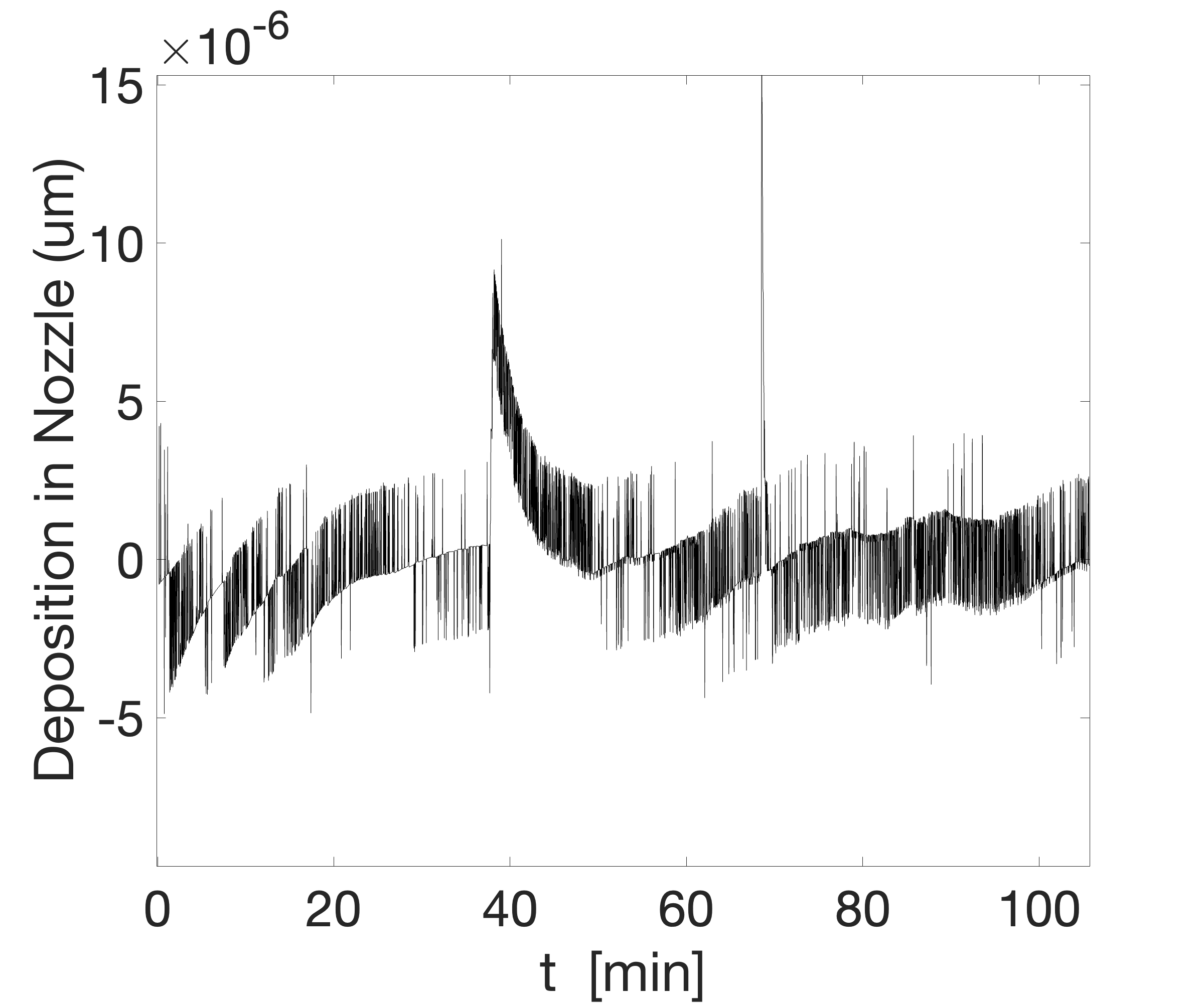}
        \caption{}
    \end{subfigure}
    ~
    \begin{subfigure}[t]{0.3\textwidth}
        \centering
        \includegraphics[height=1.2in]{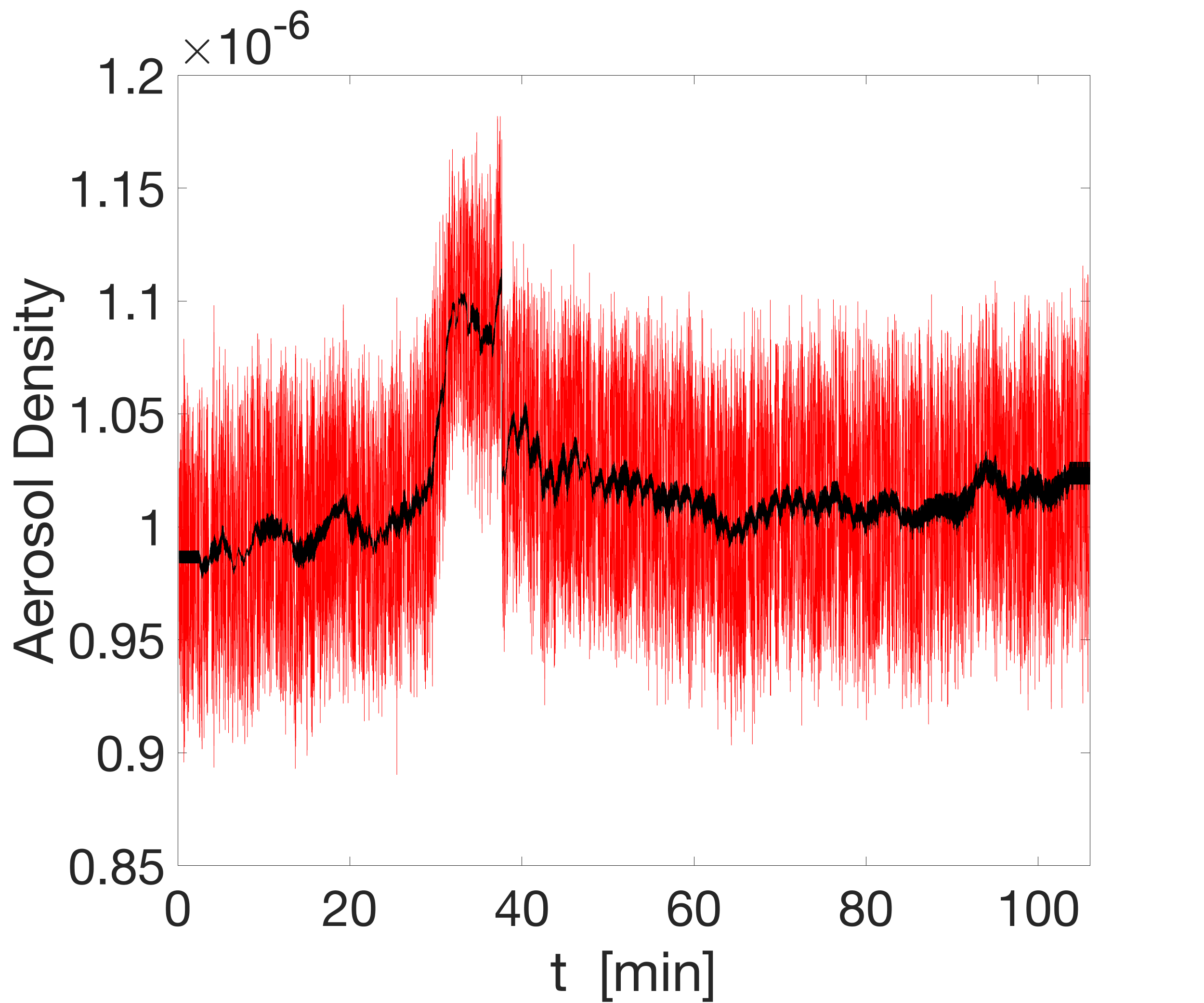}
       \caption{} 
    \end{subfigure}
    \caption{Closed-loop prediction of the digital twin using EM reveals trends in the latent state parameters that influence changes in the outputs including for the experiment in section \ref{sec:exp3_parameter_estimation}. The inferred latent states include: (a) median aerosol droplet diameter out of the atomizer, $\DA$, (b) ink solution volume in vial, $\Vl$, (c) ink deposition in the tube, $\DT$, (d) ink deposition in the nozzle, $\DN$, (e) fraction of aerosol droplets in the carrier stream, $\AD$. The black curve represents the mean inferred state, while the red curves indicate the 90\% confidence interval. The confidence bounds for (d) are not shown as they are on the order of 10~nm.
    }
    \label{fig:exp3_inferred}
    \end{figure}

As evidenced in Figure \ref{fig:exp3_outputs}, significant drift occurs before the first input step. To investigate this further, and bolster the evidence for the accuracy of the linewidth algorithm, we scanned the wafer using an optical profilometer. Images from a select locations, including after the input step, are shown in Figure \ref{fig:exp3_drift}. The images indicate significant drift before the input step, with the most dramatic drift occurring after the step. To explain this drift, we refer to the inferred states, depicted in figure \ref{fig:exp3_inferred}. 

\clearpage
\newpage
\bibliographystyle{unsrt}
\bibliography{main} 

\begin{thebibliography}{10}

\bibitem{uya2023design}
Afahaene~Edet Uya, Sanjee Lamsal, Srikanth Itapu, Frank~X Li, Pedro Cortes, Gianfranco Trovato, Lili Dong, and Vamsi Borra.
\newblock Design and fabrication of terahertz (thz) antenna using aerosol jet printing.
\newblock In {\em 2023 IEEE 23rd International Conference on Nanotechnology (NANO)}, pages 157--161. IEEE, 2023.

\bibitem{o2015aerosol}
Michael O'Reilly, Michael~J Renn, and David Sessoms.
\newblock Aerosol jet{\textregistered} enabled 3d antenna and sensors for iot applications.
\newblock In {\em International Symposium on Microelectronics}, volume 2015, pages 000203--000209. International Microelectronics Assembly and Packaging Society, 2015.

\bibitem{jahan2024aerosol}
Sanjida Jahan, Chunshan Hu, Bin Yuan, Sandra~M Ritchie, and Rahul Panat.
\newblock Aerosol jet 3d printing of gold micropillars and their behavior under compressive loads.
\newblock {\em Additive Manufacturing}, 92:104385, 2024.

\bibitem{feng2019aerosol}
James~Q Feng and Michael~J Renn.
\newblock Aerosol jet{\textregistered} direct-write for microscale additive manufacturing.
\newblock {\em Journal of Micro-and Nano-Manufacturing}, 7(1):011004, 2019.

\bibitem{saleh2022cmu}
Mohammad~Sadeq Saleh, Sandra~M Ritchie, Mark~A Nicholas, Hailey~L Gordon, Chunshan Hu, Sanjida Jahan, Bin Yuan, Rriddhiman Bezbaruah, Jay~W Reddy, Zabir Ahmed, et~al.
\newblock Cmu array: A 3d nanoprinted, fully customizable high-density microelectrode array platform.
\newblock {\em Science Advances}, 8(40):eabj4853, 2022.

\bibitem{ali2021sensing}
Md~Azahar Ali, Chunshan Hu, Sanjida Jahan, Bin Yuan, Mohammad~Sadeq Saleh, Enguo Ju, Shou-Jiang Gao, and Rahul Panat.
\newblock Sensing of covid-19 antibodies in seconds via aerosol jet nanoprinted reduced-graphene-oxide-coated 3d electrodes.
\newblock {\em Advanced Materials}, 33(7):2006647, 2021.

\bibitem{deiner2017inkjet}
L~Jay Deiner and Thomas~L Reitz.
\newblock Inkjet and aerosol jet printing of electrochemical devices for energy conversion and storage.
\newblock {\em Advanced Engineering Materials}, 19(7):1600878, 2017.

\bibitem{paulsen2012printing}
Jason~A Paulsen, Michael Renn, Kurt Christenson, and Richard Plourde.
\newblock Printing conformal electronics on 3d structures with aerosol jet technology.
\newblock In {\em 2012 Future of Instrumentation International Workshop (FIIW) Proceedings}, pages 1--4. IEEE, 2012.

\bibitem{deiner2019aerosol}
L~Jay Deiner, Thomas Jenkins, Thomas Howell, and Michael Rottmayer.
\newblock Aerosol jet printed polymer composite electrolytes for solid-state li-ion batteries.
\newblock {\em Advanced Engineering Materials}, 21(12):1900952, 2019.

\bibitem{saleh20183d}
Mohammad~Sadeq Saleh, Jie Li, Jonghyun Park, and Rahul Panat.
\newblock 3d printed hierarchically-porous microlattice electrode materials for exceptionally high specific capacity and areal capacity lithium ion batteries.
\newblock {\em Additive Manufacturing}, 23:70--78, 2018.

\bibitem{smithdrift}
Chess~Boughey Michael~Smith, Yeon Sik~Choi and Sohini Kar-Narayan.
\newblock Controlling and assessing the quality of aerosol jet printed features for large area and flexible electronics.
\newblock {\em Flexible and Printed Electronics}, 2:015004, 2017.

\bibitem{salary2021computational}
Roozbeh Salary, Jack~P Lombardi, Darshana~L Weerawarne, Prahalada Rao, and Mark~D Poliks.
\newblock A computational fluid dynamics investigation of pneumatic atomization, aerosol transport, and deposition in aerosol jet printing process.
\newblock {\em Journal of Micro-and Nano-Manufacturing}, 9(1):010903, 2021.

\bibitem{tafoya2020understanding}
Rebecca~R Tafoya and Ethan~B Secor.
\newblock Understanding and mitigating process drift in aerosol jet printing.
\newblock {\em Flexible and Printed Electronics}, 5(1):015009, 2020.

\bibitem{chen2018effect}
Guang Chen, Yuan Gu, Harvey Tsang, Daniel~R Hines, and Siddhartha Das.
\newblock The effect of droplet sizes on overspray in aerosol-jet printing.
\newblock {\em Advanced Engineering Materials}, 20(8):1701084, 2018.

\bibitem{RAMESH2023312}
Srikanthan Ramesh, Zhiheng Xu, Iris~V. Rivero, and Denis~R. Cormier.
\newblock Computational fluid dynamics and experimental validation of aerosol jet printing with multi-stage flow focusing lenses.
\newblock {\em Journal of Manufacturing Processes}, 95:312--329, 2023.

\bibitem{full_process_ajp_modeling}
Yufeng Jin, Hao Yi, Huajun Cao, and Xianshan Dong.
\newblock Full-process aerosol jet printing modelling: achieving high-fidelity simulation via coupling jetting and deposition.
\newblock {\em Virtual and Physical Prototyping}, 20(1):e2516665, 2025.

\bibitem{ma17133179}
Haining Zhang, Haifeng Xu, Lin Cui, Zhenggao Pan, Pil-Ho Lee, Min-Kyo Jung, and Joon-Phil Choi.
\newblock An extensive study of the influence of key flow variables on printed line quality outcomes during aerosol jet printing using coupled three-dimensional numerical models.
\newblock {\em Materials}, 17(13), 2024.

\bibitem{secor_vapor}
Rebecca~R Tafoya and Ethan~B Secor.
\newblock Understanding and mitigating process drift in aerosol jet printing.
\newblock {\em Flexible and Printed Electronics}, 5(1):015009, feb 2020.

\bibitem{secor_heating}
Bella~I. Guyll, Logan~D. Petersen, Cary~L. Pint, and Ethan~B. Secor.
\newblock Enhanced resolution, throughput, and stability of aerosol jet printing via in line heating.
\newblock {\em Advanced Functional Materials}, 34(28):2316426, 2024.

\bibitem{zhang_aidriven}
Haining Zhang, Yongrae Kim, Lin Cui, Seung~Ki Moon, and Joon~Phil Choi.
\newblock Ai-driven process optimization framework for enhancing print quality in aerosol jet printing.
\newblock {\em International Journal of Precision Engineering and Manufacturing-Green Technology}, 12:853--867, 2025.

\bibitem{secor_light_scattering}
Jeremy~D. Rurup and Ethan~B. Secor.
\newblock Predicting deposition rate and closing the loop on aerosol jet printing with in-line light scattering measurements.
\newblock {\em Advanced Engineering Materials}, 25(12):2201919, 2023.

\bibitem{grieves2017digital}
Michael Grieves and John Vickers.
\newblock Digital twin: manufacturing excellence through virtual factory replication.
\newblock {\em White paper, NASA}, 2017.

\bibitem{tao2018digital}
Fei Tao, Qinglin Qi, Ang Liu, and Andrew Kusiak.
\newblock Digital twin and big data towards smart manufacturing and industry 4.0: 360 degree comparison.
\newblock {\em IEEE Access}, 6:3585--3593, 2018.

\bibitem{national2023foundational}
Engineering National Academies~of Sciences and Medicine.
\newblock {\em Foundational research gaps and future directions for digital twins}.
\newblock The National Academies Press (US), 2023.

\bibitem{grieves2016digital}
Michael Grieves and John Vickers.
\newblock Digital twin: Mitigating unpredictable, undesirable emergent behavior in complex systems.
\newblock In {\em Transdisciplinary perspectives on complex systems: New findings and approaches}, pages 85--113. Springer, 2016.

\bibitem{li2021digital}
Luning Li, Sohaib Aslam, Andrew Wileman, and Suresh Perinpanayagam.
\newblock Digital twin in aerospace industry: A gentle introduction.
\newblock {\em IEEE Access}, 10:9543--9562, 2021.

\bibitem{kritzinger2018digital}
Werner Kritzinger, Matthias Karner, Georg Traar, Jan Henjes, and Wilfried Sihn.
\newblock Digital twin in manufacturing: A categorical literature review and classification.
\newblock {\em Ifac-PapersOnline}, 51(11):1016--1022, 2018.

\bibitem{zolin2020digital}
DS~Zolin and EN~Ryzhkova.
\newblock Digital twins for electric grids.
\newblock In {\em 2020 International Russian Automation Conference (RusAutoCon)}, pages 175--180. IEEE, 2020.

\bibitem{iso2020automation}
ISO/DIS 23247-1.
\newblock Automation systems and integration—digital twin framework for manufacturing—part 1: overview and general principles, 2020.

\bibitem{song2023digital}
Zhao Song, Christoph~M Hackl, Abhinav Anand, Andre Thommessen, Jonas Petzschmann, Omar Kamel, Robert Braunbehrens, Anton Kaifel, Christian Roos, and Stefan Hauptmann.
\newblock Digital twins for the future power system: An overview and a future perspective.
\newblock {\em Sustainability}, 15(6):5259, 2023.

\bibitem{opencv_library}
G.~Bradski.
\newblock {The OpenCV Library}.
\newblock {\em Dr. Dobb's Journal of Software Tools}, 2000.

\bibitem{paszke2019pytorch}
A~Paszke.
\newblock Pytorch: An imperative style, high-performance deep learning library.
\newblock {\em arXiv preprint arXiv:1912.01703}, 2019.

\bibitem{manual2009ansys}
User Manual.
\newblock Ansys fluent 12.0.
\newblock {\em Theory Guide}, 67, 2009.

\bibitem{resnet}
Kaiming He, Xiangyu Zhang, Shaoqing Ren, and Jian Sun.
\newblock Deep residual learning for image recognition, 2015.

\bibitem{lathuiliere2019comprehensive}
St{\'e}phane Lathuili{\`e}re, Pablo Mesejo, Xavier Alameda-Pineda, and Radu Horaud.
\newblock A comprehensive analysis of deep regression.
\newblock {\em IEEE transactions on pattern analysis and machine intelligence}, 42(9):2065--2081, 2019.

\bibitem{secor2018principles}
Ethan~B Secor.
\newblock Principles of aerosol jet printing.
\newblock {\em Flexible and Printed Electronics}, 3(3):035002, 2018.

\bibitem{ferreira2007box}
SL~Costa Ferreira, RE~Bruns, Hadla~Sousa Ferreira, Geraldo~Domingues Matos, JM~David, GC~Brand{\~a}o, EG~Paranhos da~Silva, LA~Portugal, PS~Dos~Reis, AS~Souza, et~al.
\newblock Box-behnken design: An alternative for the optimization of analytical methods.
\newblock {\em Analytica chimica acta}, 597(2):179--186, 2007.

\bibitem{ribeiro2004kalman}
Maria~Isabel Ribeiro.
\newblock Kalman and extended kalman filters: Concept, derivation and properties.
\newblock {\em Institute for Systems and Robotics}, 43(46):3736--3741, 2004.

\bibitem{rauch1965maximum}
Herbert~E Rauch, F~Tung, and Charlotte~T Striebel.
\newblock Maximum likelihood estimates of linear dynamic systems.
\newblock {\em AIAA journal}, 3(8):1445--1450, 1965.

\bibitem{rajan2001correlations}
Raghavachari Rajan and Aniruddha~B Pandit.
\newblock Correlations to predict droplet size in ultrasonic atomisation.
\newblock {\em Ultrasonics}, 39(4):235--255, 2001.

\bibitem{thomas1958gravity}
Jess~W Thomas.
\newblock Gravity settling of particles in a horizontal tube.
\newblock {\em Journal of the Air Pollution Control Association}, 8(1):32--34, 1958.

\bibitem{droplet_spread_factor}
J.~B. Lee, N.~Laan, K.~G. de~Bruin, G.~Skantzaris, N.~Shahidzadeh, D.~Derome, J.~Carmeliet, and D.~Bonn.
\newblock Universal rescaling of drop impact on smooth and rough surfaces.
\newblock {\em Journal of Fluid Mechanics}, 786:R4, 2016.

\bibitem{superhodrophobic_printing}
Ke~Zhong, Jace Rozsa, Dinesh~K. Patel, Lining Yao, Gary~K. Fedder, and Mohammad~F. Islam.
\newblock Aerosol jet printing of superhydrophobic surfaces.
\newblock {\em Advanced Materials Technologies}, 10(10):2401878, 2025.

\bibitem{fuchs}
S.~Nešić and J.~Vodnik.
\newblock Kinetics of droplet evaporation.
\newblock {\em Chemical Engineering Science}, 46(2):527--537, 1991.

\end{thebibliography}

\end{document}